%% file: thesis.tex
\documentclass[12pt,a4paper]{book}
\input{./text/style}

\author{By\\Najmul Haque}

\begin{document}

\input{./text/front}

\cleardoublepage
\phantomsection
\tableofcontents

{\doublespacing

\chapter*{Synopsis}
\addcontentsline{toc}{chapter}{Synopsis}
\input{./text/synopsis}

}

{\onehalfspacing


}

{\doublespacing

\cleardoublepage
\phantomsection
\addcontentsline{toc}{chapter}{List of Figures}
\listoffigures

\mainmatter
\pagestyle{fancy}
\fancyhead{}
\fancyfoot{}
\lhead{\leftmark}
\setlength{\headheight}{14.5pt}
\cfoot{\thepage}
\input{./text/intro}

\input{./text/htl}

\input{./text/1loop}
\input{./text/2loop}
\input{./text/3loop}

\input{./text/dilepton}

\input{./text/sum}
\appendix
\input{./text/aleph}
\onehalfspacing

\input{./text/bib}
}


\end{document}

%% file: text/style.tex
\usepackage{amsmath,amssymb,amsfonts,bm,textcomp,titlesec}
\usepackage{graphicx,subfigure,color,cite,etoolbox,epsfig}
\usepackage{setspace,changepage}
\usepackage[toc,page,titletoc]{appendix}
\usepackage{enumitem,environ,tikz}
\usepackage{tabularx,colortbl,fancyhdr}
\usepackage[margin=10pt,font=small,labelfont=bf]{caption}
\apptocmd{\thebibliography}{\csname phantomsection\endcsname\addcontentsline{toc}{chapter}{\bibname}}{}{}

\titleformat{\chapter}[display]
{\normalfont\bfseries\filcenter}
{\LARGE\MakeUppercase{\chaptertitlename} \thechapter}
{1ex}
{\titlerule[2pt]
\vspace{2ex}%
\LARGE}
[\vspace{1ex}%
{\titlerule[2pt]}]

\usepackage[colorlinks=true,linktocpage=true,linkcolor=blue,citecolor=blue]{hyperref}
\usepackage[margin=3.0cm]{geometry}
\def\be{\begin{eqnarray}}
\def\ee{\end{eqnarray}}
\def\lb{\left(}
\def\rb{\right)}
\def\hmu{\hat\mu}
\def\nn{\nonumber\\}
\def\sumintb{\sum\!\!\!\!\!\!\!\!\!\int\limits}
\def\sumintf{\sum\!\!\!\!\!\!\!\!\!\!\int\limits}
\def\sumintbb{\sum\!\!\!\!\!\!\!\!\!\int\limits}
\def\sumintbf{\sum\!\!\!\!\!\!\!\!\!\!\!\int\limits}
\def\sumintff{\sum\!\!\!\!\!\!\!\!\!\!\!\int\limits}
\def\sumintbbb{\sum\!\!\!\!\!\!\!\!\!\!\!\int\limits}
\def\sumintbbf{\sum\!\!\!\!\!\!\!\!\!\!\!\!\!\int\limits}
\def\sumintbff{\sum\!\!\!\!\!\!\!\!\!\!\!\!\!\int\limits}
\def\sumintfff{\sum\!\!\!\!\!\!\!\!\!\!\!\!\!\int\limits}
\def\Za{\frac{\zeta'(-1)}{\zeta(-1)}}
\def\Zc{\frac{\zeta'(-3)}{\zeta(-3)}}
\def \Slash{\slash \!\!\!\!}
\def \del{\partial}

\def\nn{\nonumber\\}

\def\L{\ln\frac{\hat\Lambda}{2}}
\def\Lg{\ln\frac{\hat\Lambda_g}{2}}
\def\[{\left[}
\def\]{\right]}

%% file: text/front.tex
\begin{center}
 
{\Large {\bf {\LARGE 
{ \doublespacing
Some Applications of Hard Thermal Loop Perturbation Theory in Quark Gluon Plasma}}}

}

\vskip 0.70cm
{\bf {\em By}} 
\vskip -0.2cm
{\bf {\large NAJMUL HAQUE}}

{\bf {\large PHYS05200804003}}
\vskip 0.5cm
{\bf {\large Saha Institute of Nuclear Physics}}
\vfill
\vfill
\vskip 2.8cm
{\bf {\em {\large A thesis submitted to the
\vskip 0.05cm
Board of Studies in Physical Sciences
\vskip 0.05cm
In partial fulfillment of requirements
\vskip 0.05cm
For the Degree of 
}}}
\vskip 0.05cm
{\large{\bf{DOCTOR OF PHILOSOPHY}}}
\vskip 0.1cm
{\bf {\em of}}
\vskip 0.1cm
{\bf {\large HOMI BHABHA NATIONAL INSTITUTE}}
\vfill
\includegraphics[height=3.5cm, width=3.5cm]{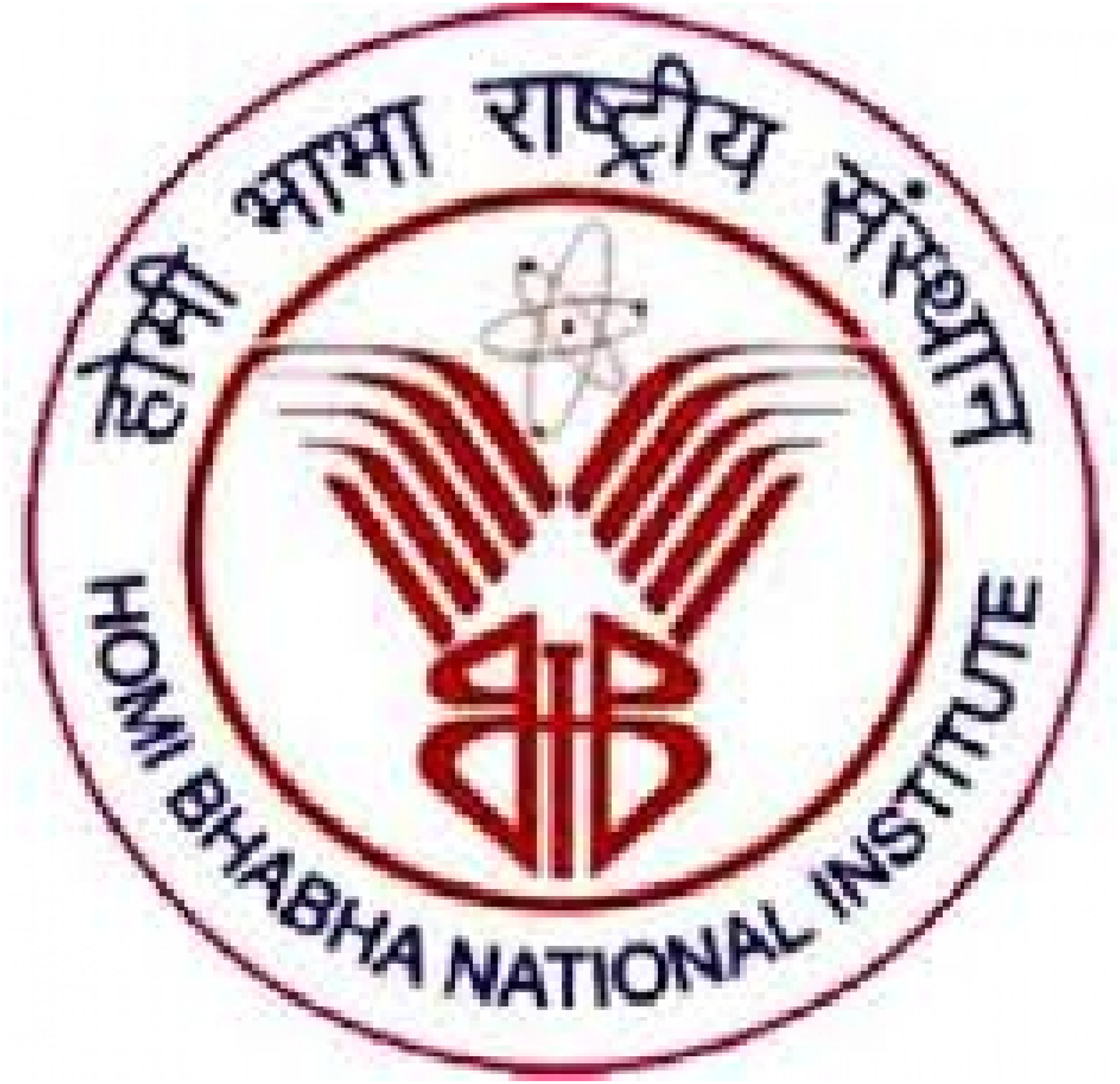}
\vfill
{\bf {\large July, 2014}}
\vfill
\end{center}

\frontmatter
\pagestyle{plain} 
\centerline{{\bf{\LARGE Homi Bhabha National Institute}}}
\vskip 0.3cm
\centerline{{\bf {\large Recommendations of the Viva Voce Board}}}
\vskip 0.3cm

{\onehalfspacing 
As members of the Viva Voce Board, we certify that we have read the
dissertation prepared by {\bf NAJMUL HAQUE} entitled {\bf Some Applications
of Hard Thermal Loop Perturbation Theory in Quark Gluon Plasma} and
recommend that it maybe accepted as fulfilling the dissertation
requirement for the Degree of Doctor of Philosophy.

}
\vskip .8cm
\underline{\hspace{12.0cm}} Date:
\vskip -0.2cm 
Chair - Prof. Asit Kumar De
\vskip .8cm
\underline{\hspace{12.0cm}} Date:
\vskip -0.2cm 
Guide/Convener - Prof. Munshi Golam Mustafa
\vskip .8cm
\underline{\hspace{12.0cm}} Date:
\vskip -0.2cm 
Member 1 - Prof. Avaroth Harindranath
\vskip .8cm
\underline{\hspace{12.0cm}} Date:
\vskip -0.2cm 
Member 2 - Prof. Pradip Kumar Roy
\vskip .8cm
\underline{\hspace{12.0cm}} Date:
\vskip -0.2cm 
External Examiner - Prof. Saumen Datta
\vskip .80cm
%
{\onehalfspacing 
\hspace{0.7cm} Final approval and acceptance of this dissertation is
contingent upon the candidate's submission of the final copies of the
dissertation to HBNI.
\vskip -0.2cm
\hspace{0.7cm} I hereby certify that I have read this dissertation
prepared under my direction and recommend that it may be accepted as
fulfilling the dissertation requirement.

}

{\doublespacing 
\vskip .6cm 

{\bf Date:} 
\vskip -0.0cm 
{\bf Place:} \hspace{7cm} Guide:{\underline{ \hspace{5.0cm}}}
\vskip -0.4cm 
\hspace{9.65cm}Prof. Munshi Golam Mustafa
\newpage
\cleardoublepage
\centerline{{\bf {\large STATEMENT BY AUTHOR}}}
\vskip 1.00cm
%
%
This dissertation has been submitted in partial fulfillment of
requirements for an advanced degree at Homi Bhabha National Institute
(HBNI) and is deposited in the Library to be made available to borrowers
under rules of the HBNI.
\vskip 0.6cm
Brief quotations from this dissertation are allowable without special
permission, provided that accurate acknowledgement of source is made.
Requests for permission for extended quotation from or reproduction of
this manuscript in whole or in part may be granted by the Competent
Authority of HBNI when in his or her judgment the proposed use of the
material is in the interests of scholarship. In all other instances,
however, permission must be obtained from the author.
\vskip 1.5cm

$~$\hspace{11cm}Najmul Haque
\newpage
\cleardoublepage
~
\vskip 1.2cm
\centerline{{\bf{\large{DECLARATION}}}}
\vskip 1.2cm
I, hereby declare that the investigation presented in the thesis has been
carried out by me. The work is original and has not been submitted
earlier as a whole or in part for a degree / diploma at this or any
other Institution / University.
\vskip 2.5cm
\rightline{Najmul Haque \hspace{0.9cm}}
\newpage
\cleardoublepage
\centerline{{\bf{\large{List of Publications arising from the thesis}}}}

{\bf{\Large Journal}}
\begin{enumerate}
\item {Three-loop HTLpt thermodynamics at finite
temperature and chemical potential}, {\it Najmul Haque, Aritra Bandyopadhyay, Jens O. Andersen, Munshi G. Mustafa,
Michael Strickland and Nan Su,} {\bf{JHEP {\bf 1405}, 027 (2014)}}.

\item {Three-loop HTLpt Pressure and Susceptibilities at Finite Temperature and Density,
{\it {Najmul Haque, Jens O. Andersen, Munshi G. Mustafa, Michael Strickland, Nan Su}},
{\bf{Phys.\ Rev. D89, 061701(R) (2014) }}}.

\item{Quark Number Susceptibilities from Two-Loop Hard Thermal Loop Perturbation Theory,
{\it {Najmul Haque, Munshi G. Mustafa, Michael Strickland}},
{\bf{JHEP {\bf 1307}, 184 (2013)}}}.

\item {Two-loop HTL pressure at finite temperature and chemical potential,
{\it {Najmul Haque, Munshi G. Mustafa, Michael Strickland,}}
{\bf{Phys.\ Rev.\ D {\bf 87}, 105007 (2013)}}}.
\item {Conserved Density Fluctuation and Temporal Correlation Function in HTL Perturbation Theory,
{\it {Najmul Haque, Munshi G. Mustafa,  Markus H. Thoma,}}
{\bf{Phys.\ Rev.\ D {\bf 84}, 054009 (2011)}}}.
\item {Low Mass Dilepton Rate from the Deconfined Phase,
{\it {Carsten Greiner , Najmul Haque, Munshi G. Mustafa, Markus H. Thoma,}}
{\bf{Phys.\ Rev.\ C {\bf 83b}, 014908 (2011)}}}.
\end{enumerate}

 {\bf{\Large Chapters in books and lectures notes}}
 \hspace{1cm} N. A. \\
\newpage
 {\bf{\Large Conferences}}
\begin{itemize}
\item {Quark Number Susceptibility and Thermodynamics in HTL approximation\\
{\it {Najmul Haque, Munshi G. Mustafa,}}\\
{\bf{Nucl.\ Phys.\ A {\bf 862-863}, 271 (2011)}}}.
\end{itemize}

 {\bf{\Large Others}}
\begin{itemize}
 \item {A Modified Hard Thermal Loop Perturbation Theory\\
{\it {Najmul Haque, Munshi G. Mustafa}} \\
{\bf{arXiv: 1007.2076 [hep-ph]}}}.
\end{itemize}
\vskip 3cm
\rightline{Najmul Haque \hspace{0.9cm}}
\newpage
\centerline{{\bf {\large DEDICATIONS}}}
\vskip 5cm
\centerline{{\it {\large Dedicated to my daughter.}}}

\newpage
\cleardoublepage
~
\vskip -.50cm
\centerline{{\bf{\large ACKNOWLEDGEMENTS}}}
\vskip 0.30cm

First of all, I would like to express my deep gratitude to my supervisor Munshi Golam Mustafa.
He has introduced me to the Thermal Field Theory and guided me throughout the whole PhD tenure in
academic as well as non-academic purpose.  
I am greatly grateful to my collaborator Michael Strickland who always has tried to give me a hand whenever needed.
Gratitude also goes to my collaborators from my institute and outside the institute. Special acknowledgement
goes to my colleague/collaborator Somdeb Chakraborty who has introduced me to the applications of AdS/CFT 
duality to Quark Gluon Plasma. I am also grateful to my other collaborators: Jens Andersen, Nan Su, Markus Thoma,
Carsten Greiner, Shibaji Roy, Aritra Bandyopadhyay, Shankhadeep Chakrabortty, Binoy Krishna Patra, Lata Thakur and Uttam Kakade.

I appreciate the fruitful and enlightening interactions with Palash B. Pal, Asit K. De, Purnendu Chakraborty,
 Swapan Majhi. Also special acknowledgement goes to our divisional head Asit K. De for
non-academic supports during my stay in SINP. 

The time at SINP could not be more lively and lovely without my dear colleagues and 
thanks to them : Raktim Abir, Ramanuj Banerjee, Pratyay Banerjee, Srijit Bhattacharjee, Anirban Biswas, 
Soumyajyoti Biswas, Pritibhajan Byakti, Baishali Chakraborty, Mainak Chakraborty, 
Sovan Chakraborty, Dipankar Das, Jayanta Das, Chowdhury Aminul Islam, Santanu Maiti, Arindam Mazumdar,
Amaresh Metya, Manas Mondal, Santanu Mondal, Shyamal Mondal, Rana Nandi, Tapan Naskar, Lab Saha, 
Niladri Sarkar, Sreemoyee Sarkar, Satyajit Seth.

The financial supports from DAE are gratefully acknowledged.

Last but not least, I am indebted to the countless supports from my family over years specially
from my wife and my daughter.

}

%% file: text/synopsis.tex
In recent years substantial experimental and theoretical efforts have been undertaken to investigate 
the versatile physics issues involved in ultra-relativistic heavy-ion collisions, i.e., collisions of
atomic nuclei in which center-of-mass energy per nucleon is much larger than the nucleon rest mass. 
The principal goal of this initiative is to explore the phase structure of the underlying theory of
strong interactions - Quantum Chromodynamics (QCD) - by creating in the laboratory the new state of matter
``Quark-Gluon Plasma (QGP)''. This new state of matter is predicted to exist under extreme conditions like at
high temperatures and/or densities, when a phase transition takes place from a hadronic to a deconfined state 
of quarks and gluons. In nature such new states are believed to have existed and still may be encountered on large
scales in at least two astrophysical aspects: i) in the evolution of early universe where a few tens of microseconds
after the \textquoteleft big bang\textquoteright\  a  transient stage of strongly interacting matter prevailed at
temperatures $10^{12}{\rm K} (\sim 200 {\rm MeV})$ with a very small net baryon numbers; ii) in the interior of
neutron stars where mass densities are likely to exceed $10^{15}{\rm gm}/{\rm cm}^3$ about four times the central 
density of nuclei while the surface temperatures are as low as $10^5$K or less. However, these astrophysical objects
are by far remote in space and time so that their use for the study of QGP are quite difficult.
This makes us to turn attention for a consistent study of QGP in the laboratory through high energy heavy-ion collisions. 

The collider experiments currently dedicated to this search are the Relativistic Heavy Ion Collider
(RHIC) at Brookhaven National Laboratory (BNL) and the Large Hadron Collider (LHC) at the
European Organization for Nuclear Research (CERN). Future experiments are planned at
the Facility for Antiproton and Ion Research (FAIR) at the Gesellschaft f$\ddot{\rm u}$r Schwerionenforschung (GSI) facility.
These experiments at RHIC BNL and LHC CERN have provided us wealth of information~\cite{Arsene:2004fa,Adcox:2004mh,Back:2004je,
Adams:2005dq} in understanding the properties of hot and dense matter and the theoretical predictions. On the other
hand recent numerical lattice QCD (LQCD) has given us information on various thermodynamic properties of the matter {\it viz.},
critical temperature~\cite{Fodor:2004nz,Gupta:2011wh}, nature of phase transition, 
the equation of state~\cite{Boyd:1996bx,Borsanyi:2010cj,Borsanyi:2012uq,Borsanyi:2012cr}, various susceptibilities~\cite{Borsanyi:2011sw},
transport coefficients, spectral properties of pseudo-scalar and vector meson resonances etc. at finite temperature and density. 
Ultimately, one would again expect to validate this by characterizing the QGP in terms of its experimentally observed properties. 
The commissioning of RHIC and LHC, and various experiments performed therein have ushered in a new era. The analysis of 
the data has yielded a rich abundances of results that possibly indicate a glimpse of the predicted plasma phase of QCD.
This calls for a better theoretical understanding of the particle properties in a hot and dense medium.

Quantum chromodynamics (QCD) exhibits a rich phase structure and the equation of state (EoS) which describes the matter can
be characterized by different degrees of freedom depending upon the temperature and the chemical potential. The determination
of the equation of state (EoS) of QCD matter is extremely important to QGP phenomenology. There are various effective models
to describe the EoS of strongly interacting matter; however, one would prefer to utilize systematic first-principles QCD methods.
The currently most reliable method for determining the EoS is lattice QCD. Finite temperature lattice QCD calculations are now
quite sound; however, due to the sign problem, it is not straightforward to extend such calculations to finite
baryon chemical potential. In practice, it is possible to obtain information about the
behavior of the thermodynamic functions at small baryon chemical potential by making
a Taylor expansion of the partition function around zero chemical potential and extrapolating the result.
This requires the calculation of various quark-number susceptibilities evaluated at zero
chemical potential. Since extrapolations based on a finite number of Taylor coefficients can only be trusted
within the radius of convergence of the expansion, it would be nice to have an alternative
framework for calculating the finite temperature and chemical potential QCD thermodynamic potential and associated quantities.
Perturbative QCD (pQCD) is an alternative framework which 
can be applied at high temperature and/or chemical potentials where the strong coupling $(g^2\!=\!4\pi \alpha_s )$ is small in
magnitude and non-perturbative effects are expected to be small. However, due to infrared singularities in the gauge sector,
the perturbative expansion of the finite-temperature and density QCD partition function breaks down at order $g^6$ requiring
non-perturbative input albeit through a single numerically computable number~\cite{Kajantie:2002wa}.
Up to order $g^6 \ln(1/g)$ it is possible to calculate the necessary coefficients using analytic (resummed)
perturbation theory~\cite{Vuorinen:2003fs} at finite temperature and chemical potential.

The hard thermal loop perturbation theory (HTLpt) is a state-of-the-art resummed perturbation theory with a given mass prescription that
reorganizes the usual perturbation theory at finite temperature/density quantum chromodynamics. In hard thermal loop (HTL) approximation
the loop expansion and coupling expansion are not symmetrical as higher order diagrams contribute to the lower order one. As a result
some of the quantities, such as equation of states, various susceptibilities, trace anomaly, speed of sound etc., calculated within a
given loop order using the hard thermal loop approximation suffer incompleteness in the corresponding perturbative order. At each order
in HTLpt the result is an infinite series in $g$, the strong coupling. Using  the mass expansion one keeps terms through order $g^5$
(if one uses only LO mass prescription) all loop-orders of HTLpt in order to make the calculation tractable.  At leading order (LO)
one obtains only the correct perturbative coefficients for $g^0$ and $g^3$ terms when one expands in a strict power series in $g$.
At next-to-leading order (NLO)  one obtains the correct $g^0,\ g^2$ and $g^3$ coefficients whereas at next-to-next leading order
(NNLO) one obtains the correct $g^0,\ g^2,\ g^3,\ g^4$ and $g^5$ coefficients.

In the thesis I shall discuss a systematic computation of QCD pressure at finite temperature and finite chemical
potential in one-loop~\cite{Haque:2010rb}, two-loop~\cite{Haque:2012my} and three-loop~\cite{Haque:2013sja,Haque:2014rua} Hard Thermal Loop
perturbation theory. The corresponding results will be compared with recent lattice QCD data.

Fluctuations of conserved quantities have been considered as important and relevant probes of Quark Gluon Plasma
formation in relativistic heavy-ion collision. These fluctuations measure the response to system with an external field.
In particular Quark Number Susceptibility(QNS) defines the response of conserved quark number density when
we change the chemical potential of the system. Quark Number Susceptibility can be related to the charge fluctuations of the system
and is therefore of direct experimental relevance. This thesis will also discuss a very systemic computation of Quark
Number Susceptibilities in one-loop~\cite{Haque:2010rb,Haque:2011iz,Haque:2011vt}, two-loop~\cite{Haque:2013qta} and 
three-loop~\cite{Haque:2013sja,Haque:2014rua}
Hard Thermal Loop perturbation theory. Also in three-loop HTLpt case,
in general, off diagonal susceptibilities are non-zero, so I shall also extend NNLO HTLpt to calculate
various order diagonal and off-diagonal susceptibilities. For NNLO calculation we will use the two-loop perturbative mass
prescription for debye mass, the effective order of $g$ becomes higher than $g^5$. The results of various order HTLpt are gauge
invariant and in particular, NNLO results are complete in $g^5$. This result is completely analytic that does not require any free fit
parameter beside renormalization scale. I shall also extend three-loop HTLpt to calculate other thermodynamical quantities,
which are relevant to deconfined state of matter, {\it viz.} energy density,
trace anomaly, speed of sound,  entropy density etc. The various quantities agree very well
with lattice QCD results within error bars  down to 200 MeV. 

In addition to calculations of the thermodynamic quantities, hard-thermal-loop perturbation 
theory can be use to calculate various physical quantities which are relevant
to the deconfined state of matter. The dilepton rate is a good probe to study the deconfined state of matter as the mean
free path for real of virtual photons are large and accordingly thermal dileptons  
have been theoretically proposed as a signal of QGP a long time ago~\cite{McLerran:1984ay}.
So, this thesis will also analyze the dilepton rates from HTLpt  with various non-perturbative results obtained using
Gluon condensate in the
quark propagator, $\rho$ meson-quark interaction
in an effective model and lattice QCD in  \cite{Greiner:2010zg}. The results will also be contrasted with  in-medium Hadron gas rate.
Based on this, a more realistic way of the quark-hadron duality hypothesis can be advocated than it was done in the literature.

%% file: text/intro.tex
\chapter{Introduction}
\label{chapter:introduction}
It is now well-known that colored quarks and gluons are confined within color
singlet hadronic bound states by strong interactions~\cite{Gross:1973id}. The
theory which describes the behavior of quarks and gluons is known as 
Quantum Chromodynamics (QCD). Much of the support for QCD derives from its ability
to produce the almost noninteracting behavior of quarks at
short distances~\cite{Gross:1973id}. This feature of the
theory, known as asymptotic freedom, explains the approximate
scaling observed in the deep inelastic scattering of leptons off hadrons and leads to
many quantitative predictions of scaling deviations at high energy. The success of these predictions, as well as
many other confirmations of the predictions of perturbative QCD at short distances
has greatly increased the confidence in the theory.
\begin{figure}[tbh]
\begin{center}
\includegraphics[width=11cm]{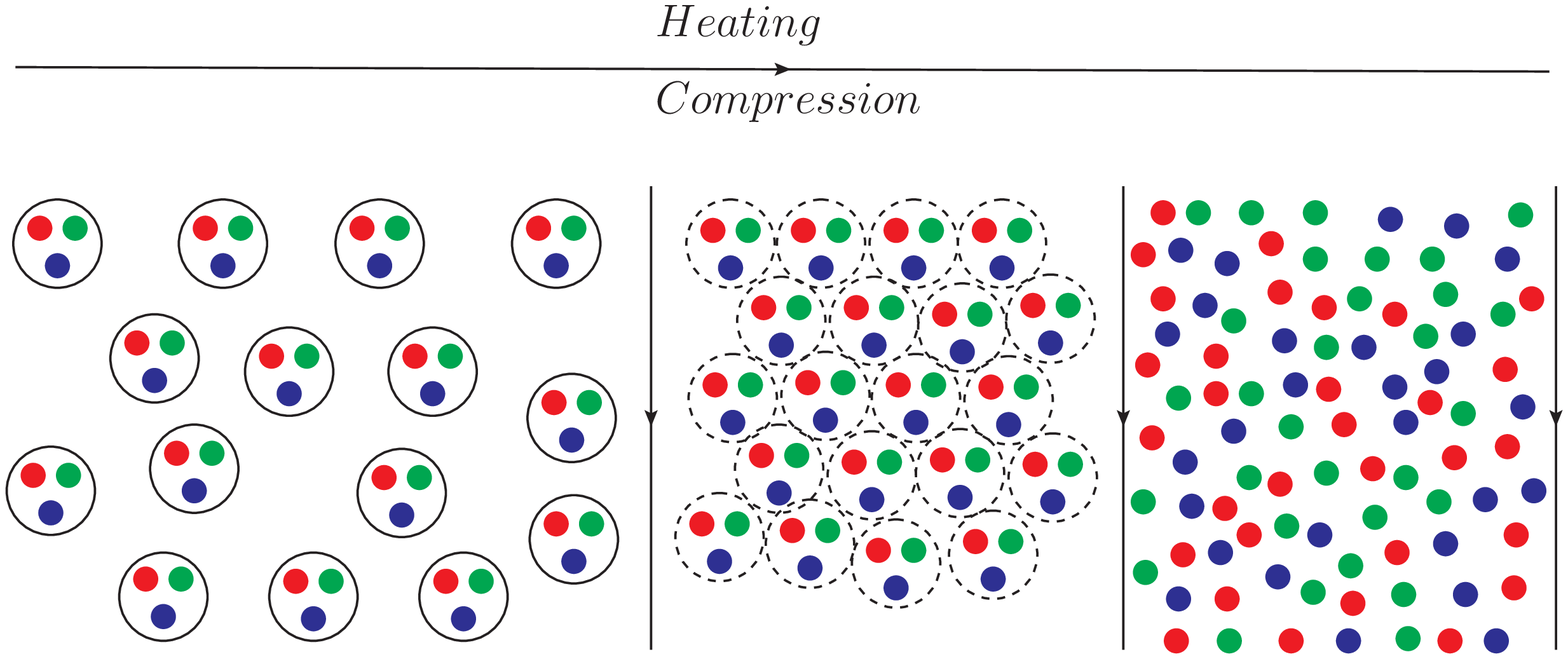}
\end{center}
\caption{Nuclear matter at extreme conditions.}
\label{hot_dense}
\end{figure}

So it is worth to apply the knowledge of QCD to explore the properties of hadronic
matter in extreme environments, such as at high temperature and/or at high baryonic density. As nuclear matter is 
heated and compressed hadrons occupy more and more of the available space within nucleus as schematically depicted in 
Fig.~(\ref{hot_dense}). Eventually they start to overlap
and quarks and gluons confined initially begin to percolate between hadrons thus being liberated. Under this conditions
quarks and gluons are no longer remain confined within hadrons and a new state of matter known as Quark Gluon Plasma (QGP)
is produced. There are three places where one might look for this deconfined state of matter, {\em viz.}
 
i) in the evolution of early universe where a few tens of microseconds after
the \textquoteleft big bang\textquoteright\ a transient stage of strongly interacting matter prevailed
at temperatures $10^{12}{\rm K} (\sim 200 {\rm MeV})$ with a very small net baryon numbers; 

ii) in the interior of neutron stars where mass densities are likely to exceed $10^{15}{\rm gm}/{\rm cm}^3$
which is about four times the central density of nuclei while the surface temperatures are as low as $10^5$K or less;

iii) in the collision of heavy ions at very high energy per nucleon, in which states of high
density and temperature might be produced. 

\begin{figure}[tbh]
\begin{center}
\includegraphics[width=9cm]{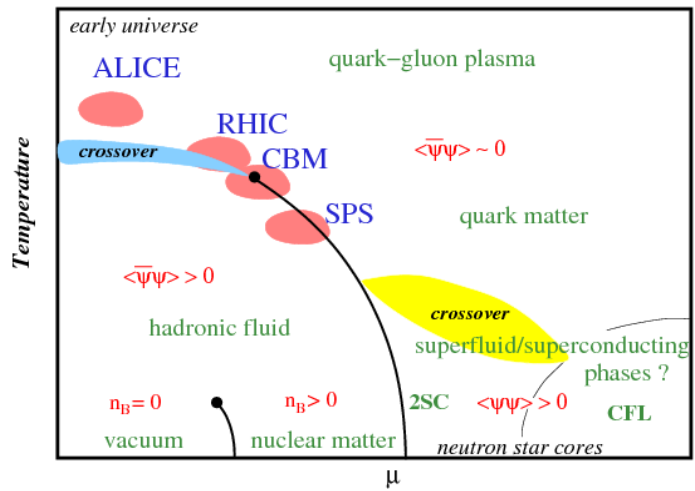}
\end{center}
\caption{Schematic view of QCD phase diagram.}
\label{qcd_phase_diag}
\end{figure}

The above mentioned three situations have been schematically depicted in QCD
phase diagram in Fig.~(\ref{qcd_phase_diag}). Nevertheless, the \textquoteleft big bang\textquoteright\ is far remote
in time and the astrophysical objects are far remote in space and their use for the study of QGP are quite difficult.
This makes us to turn attention for a consistent study of QGP in the laboratory through high energy heavy-ion collisions. 

In heavy-ion-collisions experiments, two heavy nuclei are accelerated to ultra-relativistic speeds
and directed towards each other. During the collision, kinetic energy of the Lorentz contracted nuclei
are deposited at the collision region that  produce a very hot and dense 
\textquotedblleft fireball\textquotedblright. This fireball expands hydrodynamically under its own pressure, and cools
while expanding. This expansion continues till all of the hot and dense fireball is converted into
the hadronic phase. Upon completion of phase conversion, the hadronic matter continues to expand until
the mean free path of the hadrons becomes larger than the dimensions of the system and they loose thermal contact.
This is called freeze-out. At the freeze-out temperature, the hadrons cease to interact with each other and they stream
freely away to be detected in the experiments.

In recent years substantial experimental efforts have been
undertaken to investigate the versatile physics issues involved in
ultra-relativistic heavy-ion collisions, i.e., collisions of atomic nuclei in
which center-of-mass energy per nucleon is much larger than the nucleon rest
mass. The principal goal of this initiative is to explore the phase structure of QCD
by creating QGP in the laboratory; in other words, how QCD works in extreme conditions. 
Heavy-ion experiments have been performed at 
Alternating Gradient Syncroton in Brookhaven National Laboratory (BNL), USA with
$E_{\rm lab}\sim 14$ GeV/nucleon continued at Super Proton Syncroton (SPS) in the European
Organization for Nuclear Research (CERN) with $E_{\rm lab}\sim 200$ GeV/nucleon.
At present the relativistic heavy-ion collision experiments in Relativistic
Heavy Ion Collider (RHIC) at BNL with $E_{\rm cm}\sim200$ GeV/nucleon (and with beam energy scan
down to $7.7$ GeV/nucleon) and Large Hadron Collider (LHC) at CERN~\cite{Carminati:2004fp,Alessandro:2006yt}
with $E_{\rm cm}\sim2.76$ and $5.5$
TeV/nucleon  are operational. Further, at the forthcoming fixed target experiment at the Facility for
Antiproton and Ion Research (FAIR)~\cite{Friman:2011zz} in Gesellschaft f\"ur
Schwerionenforschung (GSI), Germany as well as heavy-ion-collisions experiment at Nuclotron-based Ion Collider fAcility (NICA) in 
Joint Institute for Nuclear Research(JINR), Russia also plan to scan energy ranges
from 10-45 GeV/nucleon and  4-11 GeV/nucleon respectively.
These past experiments in  CERN-SPS~\cite{Heinz:2000bk}, ongoing experiments in 
BNL-RHIC~\cite{Arsene:2004fa,Adcox:2004mh,Back:2004je,Adams:2005dq,
Adare:2009qk,Adler:2006yt,Adare:2006ti,Adcox:2002au,Chujo:2002bi} and CERN-LHC~\cite{Aamodt:2010pa,Aamodt:2010jd,
Chatrchyan:2011sx,Aamodt:2010cz,Aamodt:2010pb,ALICE:2011ab}
have provided us wealth of information in understanding the properties of hot
and dense matter and the theoretical predictions. 

%

Theoretical study of hot and dense nuclear matter which produced in early universe
just after big bang or expected to exist in astrophysical objects like neutron star 
or produced in heavy-ion experiments are very important.  
QCD exhibits a rich phase structure and the equation of
state (EoS) which describes that the matter can be characterized by different degrees
of freedom depending upon the temperature and the chemical potential. The determination
of the EoS of QCD matter is extremely important to QGP
phenomenology. There are various effective models 
to describe the EoS and various order conserved density fluctuations of strongly interacting matter, {\em e.g.},
the Nambu-Jona-Lasinio (NJL)~\cite{Hatsuda:1994pi,Kunihiro:1991qu,Hatta:2002sj,Sasaki:2006ws}, 
Polyakov-loop extended Nambu-Jona-Lasinio (PNJL)~\cite{Fukushima:2003fm,Fukushima:2003fw,Ratti:2005jh,Ghosh:2006qh,
Ghosh:2007wy,Mukherjee:2006hq,Roessner:2006xn,Sasaki:2006ww,Bhattacharyya:2010wp,Bhattacharyya:2010jd,
Bhattacharyya:2010ef,Bhattacharyya:2011na,Bhattacharyya:2012up,Steinheimer:2014kka,Ratti:2007jf},
Renormalization Group approach~\cite{Schaefer:2006ds,Schaefer:2009ui}, quasi-particle 
models~\cite{Peshier:1994zf,Peshier:1995ty,Peshier:1999ww,Peshier:2002ww,Bluhm:2007nu,Bluhm:2007cp,
Bannur:2006ww,Bannur:2007tk,Gardim:2009mt},
Polyakov-loop extended quark meson (PQM) model~\cite{Skokov:2010uh,
Herbst:2013ail,Herbst:2013ufa}, AdS/CFT correspondence and holographic QCD~\cite{Kim:2006ut,Kim:2010ag,Kim:2010zg,Cui:2011wb}
have been used to calculate various thermodynamic functions and various order quark number susceptibilities.

However, one would prefer to
utilize systematic first-principles QCD methods.
The currently most reliable method for determining the EoS, various order quark number susceptibilities and
other relevant quantities for deconfined matter is lattice
QCD~\cite{lqcd1,lqcd2,Borsanyi:2010cj,Borsanyi:2011sw,Berges:2004hn,
Borsanyi:2012ve,Borsanyi:2012cr,Borsanyi:2012uq,Borsanyi:2013bia,Borsanyi:2012xf
,Borsanyi:2012rr,Ratti:2010kj,Borsanyi:2007bf,
Borsanyi:2011bm,Borsanyi:2011kg,Borsanyi:2010zh,Aoki:2006we,Borsanyi:2011zzc,Borsanyi:2010bp,Aoki:2006br,Aoki:2009sc,
Borsanyi:2013cga,Datta:2010sq,Gupta:2014qka,Gavai:2008zr,Datta:2009tj,Gavai:2005da,Gavai:2004se,
Gavai:2003nn,Gavai:2003mf,Gavai:2002kq,Gavai:2001ie,Gavai:2001fr,Gavai:2014lia,Gavai:2011uk,Banerjee:2008ii,Gavai:2008ea,
Allton:2005gk,Bazavov:2009zn,Bernard:2004je,Petreczky:2009at,Gottlieb:1987ac,Gottlieb:1988cq,Gavai:1989ce,Petreczky:2009cr,
Cheng:2009zi,Petreczky:2012rq,Borsanyi:2013hza,Sharma:2013hsa,Karsch:2000ps,Bazavov:2013dta,Bazavov:2013uja,Bazavov:2012vg,
Bazavov:2012jq,Cheng:2008zh}.
At this point in time lattice calculations can be
performed at arbitrary temperature, however, due to the sign problem, it is not straightforward to 
extend such calculations to finite baryon chemical potential. In practice, it is possible to obtain
information about the behavior of the thermodynamic functions at small baryon chemical potential~\cite{Brambilla:2014aaa}
by making a Taylor expansion of the partition function around $\mu_B=0$ and extrapolating the result.
This requires the calculation of various quark-number susceptibilities
evaluated at zero chemical potential.
Since extrapolations based on a finite number of Taylor coefficients can only be trusted at small chemical potential,
it would be nice to have an alternative framework for calculating the finite temperature and chemical potential
QCD thermodynamic potential and associated quantities.  This is important in light of the ongoing beam energy scan at the 
Relativistic Heavy Ion Collider (RHIC) and the forthcoming experiments at the Facility for Antiproton and Ion Research (FAIR).
As an alternative to lattice QCD calculations, one natural option is to compute the thermodynamic potential using perturbation theory.

Perturbative QCD (pQCD) can be applied at high temperature and/or chemical potentials where
the strong coupling $(g^2\!=\!4\pi \alpha_s )$ is small
in magnitude and non-perturbative effects are expected to be small; however,
one does not know a priori how large the temperature should be for this method to result in a good approximation to reality. The
calculation of thermodynamic functions at finite temperature
and/or finite chemical potential using perturbative approach has a long history.
In 1977 free energy for electron~\cite{Freedman:1976dm} and also for
quark~\cite{Freedman:1976ub} have been computed at zero temperature
and finite chemical potential up to order $g^4$ where  the
coefficient of $g^4$ term was obtained numerically. 
Later this calculation was reproduced in~\cite{Vuorinen:2003fs} analytically. 
In 1978, QCD free energy  at finite temperature
and zero chemical potential was calculated in~\cite{Shuryak:1977ut,Chin:1978gj} up to order $g^2$.
In 1979, QCD free energy at finite temperature and finite chemical potential was
extended in~\cite{Kapusta:1979fh} up to order $g^3$.
In 1983, QCD free energy up to order $g^4\log g$ at finite temperature
and finite chemical potential was
calculated in~\cite{Toimela:1982hv}.
In early 1990s the free energy
at finite temperature but at zero chemical potential was calculated to order
$g^4$ for massless scalar $\phi^4$ theory~\cite{Frenkel:1992az,Arnold:1994ps,
Arnold:1994eb}, quantum electrodynamics (QED)~\cite{Parwani:1994xi,Arnold:1994ps,Arnold:1994eb}
and QCD~\cite{Arnold:1994ps,Arnold:1994eb}, respectively. The corresponding 
calculations to order $g^5$ were obtained soon afterwards
\cite{Parwani:1994zz,Braaten:1995cm,Parwani:1994je,Andersen:1995ej,
Zhai:1995ac,Braaten:1995ju,Braaten:1995jr}. Later this calculation has been extended 
to calculate the QCD free energy at finite temperature but at zero chemical
potential up to order $g^6$ in Ref.\cite{Kajantie:2002wa} where the
coefficient of $g^6\log g$ term was computed analytically but the coefficient
of the $g^6$ term was fitted from lattice QCD data. In 2003, the free energy
at finite chemical potential~\cite{Vuorinen:2003fs,Ipp:2006ij} and hence quark number susceptibility
~\cite{Vuorinen:2002ue} has been extended in up to order $g^6\log g$. The results of QCD pressure
at finite temperature and chemical potential up to order $g^6\log g$ from Ref.~\cite{Vuorinen:2003fs}
will be discussed in Sec.~(\ref{press_vuorinen}).
For massless scalar theories
the perturbative free energy is now known to order $g^6$ \cite{Gynther:2007bw}
and order $g^8 \log g$ \cite{Andersen:2009ct}.

Unfortunately, for all the above-mentioned theories the resulting weak-coupling
approximations, truncated order-by-order in the
coupling constant, are poorly convergent unless the coupling constant is tiny.
Therefore, a straightforward perturbative expansion
in powers of $\alpha_s$ for QCD does not seem to be of any quantitative use even
at temperatures many orders of magnitude higher
than those achievable in heavy-ion collisions. Also, due to infrared
singularities in the gauge sector, the perturbative expansion
of the finite-temperature and density QCD partition function breaks down at 
order $g^6$ requiring non-perturbative input albeit through a single numerically
computable number~\cite{Linde:1980ts}. 

The poor convergence of finite-temperature perturbative expansions of
thermodynamic functions stems from the fact that at high
temperature the classical solution is not described by massless gluonic states.
Instead one must include plasma effects such as
the screening of electric fields and Landau damping via a self-consistent resummation~\cite{Braaten:1989mz}.
The inclusion of plasma effects can be achieved by reorganizing perturbation
theory. 
 
There are several ways of systematically reorganizing the finite-temperature
perturbative expansion~\cite{Blaizot:2003tw,Kraemmer:2003gd,Andersen:2004fp,
Chiku:1998kd,Braaten:1989kk,Braaten:1991gm,Blaizot:1999ap,Blaizot:2000fc,
Blaizot:2001vr,Peshier:1994zf,Peshier:1995ty,Peshier:1999ww,
Peshier:2002ww,Andersen:2002jz,Jiang:2010jz,Blaizot:1999ip}. In this
thesis I will focus on the hard-thermal-loop (HTL) perturbation theory
method~\cite{Andersen:1999fw,Andersen:1999sf,Andersen:1999va,Chakraborty:2001kx,Chakraborty:2003uw,
Haque:2010rb,Haque:2011iz,Haque:2011vt,Andersen:2012wr,Andersen:2002ey,Andersen:2003zk,Haque:2012my,
Haque:2013qta,Andersen:2009tc,Andersen:2010ct,Andersen:2009tw,Andersen:2010wu,Andersen:2011sf,
Andersen:2011ug,Haque:2013sja,Haque:2014rua,Blaizot:2002xz,Chakraborty:2002yt,Su:2012iy}. The HTL perturbation theory is inspired by variational 
perturbation theory~\cite{Yukalov:1976pm,Stevenson:1981vj,
Duncan:1988hw,Duncan:1992ba,Sisakian:1994nn,Janke:1995zz}. HTL perturbation theory is a gauge-invariant
extension of screened perturbation theory
(SPT)~\cite{Karsch:1997gj,Andersen:2000yj,Andersen:2001ez,Andersen:2008bz},
 which is a perturbative reorganization for finite-temperature massless scalar
field theory. In the SPT approach, one introduces a 
single variational parameter which has a simple interpretation as a thermal
mass. In SPT a mass term is added to and subtracted from
the scalar Lagrangian, with the added piece kept as part of the free Lagrangian
and the subtracted piece associated with the 
interactions.
 The mass parameter is then required to satisfy a variational equation which is
obtained by a principle of minimal sensitivity. 
This naturally led to the idea that one could apply a similar technique to gauge
theories by adding and subtracting a mass in 
the Lagrangian. However, in gauge theories, one cannot simply add and subtract a
local mass term since this would violate gauge 
invariance. As a result, a gauge-invariant generalization of SPT called
hard-thermal-loop (HTL) perturbation theory
was developed~\cite{Braaten:1989mz} by Braaten and Pisarski in 1990 by
distinguishing soft $(p\sim gT)$ and hard $(p\sim T)$ momenta
scale. HTL perturbation theory is a reorganization of usual perturbation theory
at finite temperature where higher order
diagrams contribute to the lower order one. In HTL pertrbation theory one needs
to add and subtract HTL improvement term
which modifies the propagators and vertices self-consistently so that the
reorganization is manifestly gauge 
invariant~\cite{Braaten:1991gm}.

In the thesis I shall discuss a systematic computation of QCD thermodynamics 
in leading order as well as beyond leading order at finite temperature and finite chemical potential
using HTL perturbation theory. Conserved density fluctuations and quark number susceptibility 
will be discussed in leading order withing HTL perturbation theory. In next-to-leading order
QCD pressure and various order quark number susceptibilities will be discussed. In next-to-next-leading order,
the computation of all the thermodynamical quantities {\em viz.} pressure,
various order quark number susceptibilities, energy density, entropy density, speed of sound, trace anomaly etc.
will be discussed. The corresponding HTL perturbation theory results in all order 
will be compared with recent lattice QCD data.

Besides the thermodynamic calculations, this thesis will also be focused to analyzed the dilepton 
rate from hot and dense nuclear matter. As the electromagnetic probes, such as real photon and dileptons, are a 
particular example of {\textquoteleft  circumstantial evidence\textquoteright}, and accordingly thermal dileptons have been 
theoretically proposed long time ago~\cite{McLerran:1984ay}. At SPS
energies~\cite{Agakishiev:1995xb,Agakishiev:1997au,Masera:1995ck,Drees:1998rn}
there was an indication for an enhancement of the dilepton 
production at low invariant mass ($0.2\le M(\mbox{GeV}) \le 0.8$ ) compared 
to all known sources of electromagnetic decay of the hadronic particles 
and the contribution of a radiating simple hadronic fireball (for
comprehensive reviews see Refs.~\cite{Rapp:1999ej,Rapp:2009yu,Cassing:1999es}) . 
One of the possible explanations of this is the modification of 
the in-medium properties of the vector meson ({\em viz.,} $\rho$-meson) by
rescattering in a hadronic phase along with only the lowest order perturbative
rate, {\em i.e.}, $q\bar q$ annihilation from a
QGP~\cite{Rapp:1999ej,Rapp:2009yu,Cassing:1999es,Brown:1991kk,Friman:1997tc,
Rapp:1997fs,Rapp:1995zy,Gale:1993zj,Rapp:1999qu,Klingl:1997kf,Peters:1997va,
Cassing:1997jz,Post:2000qi,Srivastava:1996vi,Pal:1996xn,Srivastava:1996wr,Pal:1998jr,
Srivastava:1996qd,Alam:1999sc,Alam:2001ar}. Also at RHIC energies~\cite{Adare:2009qk}
a substantial amount of excess of electron pairs was reported in the
low invariant mass region. Models taking into account
in-medium properties of hadrons with various ingredients (see for
details~\cite{Dusling:2007su,Bratkovskaya:2008bf}) can not explain the data from RHIC in the
range $0.15 \le M (\mbox{GeV})\le 0.5$, whereas they fit the SPS data
more satisfactorily, indicating that a possible non-hadronic source
becomes important at RHIC.
 
On the other hand, the higher order perturbative calculations~\cite{Aurenche:1998nw} 
are also not very reliable at temperatures 
within the reach of the heavy-ion collisions. Moreover, perturbative 
calculations of the dilepton rate seem not to converge even in 
small coupling ($g$) limit.  Nevertheless, the lowest order perturbative 
$q\bar q$ annihilation is the only dilepton rate from the QGP phase that
is extensively used in the literatures. However, at large invariant
mass this contribution should be dominant but not at low invariant 
mass, where nonperturbative effects should play an important role. 
Unfortunately, the lattice data~\cite{Karsch:2001uw} due to its limitations 
also could not shed any light on the low mass dileptons.  However, the 
lattice calculations~\cite{Boyd:1996bx,Allton:2003vx,Allton:2005gk,
Bazavov:2009zn,Bernard:2004je,Petreczky:2009at,Petreczky:2010xg} provide evidence for 
the existence of nonperturbative effects associated with the bulk 
properties of the deconfined phase, in and around the deconfinement
temperature, $T_c$. Also, indications have been found that the 
QGP at RHIC energies behaves more as a strongly coupled liquid 
than a weakly coupled gas~\cite{Thoma:2004sp}. Thus,  a nonperturbative 
analysis of the dilepton rate from the deconfined phase is essential. 

The dilepton emission at low invariant mass from the deconfined phase is 
still an unsettled issue in heavy-ion collisions at SPS and RHIC energies 
and, in particular, would be an important question for LHC energies and 
for compact baryonic matter formation in future FAIR energies, 
and also for the quark-hadron duality~\cite{Rapp:1999ej,Rapp:2009yu,Rapp:1999us} that 
entails a reminiscence to a simple perturbative lowest order quark-antiquark 
annihilation rate~\cite{Cleymans:1986na}. In this thesis we reconsider the dilepton 
production rates within the perturbative QCD, and non-perturbative models 
based on lattice inputs and phenomenological
$\rho - q$ interaction in the deconfined phase. The analysis suggests that 
the nonperturbative dilepton rates are indeed important at the low 
invariant mass regime.  

Below we give a brief introduction to statistical physics, QCD at finite temperature
and density, pQCD pressure up to $g^6\log g$ and asymptotic nature of QCD which will be necessary to study various
quantities in the following chapters.
\section{Statistical physics and quantum partition function}
For a relativistic system which can freely exchange energy and particles with
its surroundings, the most important function in
thermodynamics is the grand canonical partition function
\be
Z \;=\; \sum_{\rm states} e^{-\beta\lb{\cal E}_i-\mu N\rb}
    \;=\; \sum_{\rm states} \langle {\cal E}_i | e^{-{\cal\beta (H-\mu N)}} | {\cal E}_i
\rangle
    \;=\; {\rm Tr}\,e^{-{\cal \beta (H-\mu N)}}\;.
\label{Z}
\ee
Here ${\cal E}_i$ is the energy of the state $| {\cal E}_i \rangle$, $N$ is the number of particles
 and ${\cal H}$ is the Hamiltonian of the system. The inverse temperature of 
the system is denoted by $\beta=1/T$, and $\mu$ is the chemical potential of the
particles in the system. All of the thermodynamic properties 
can be determined from~(\ref{Z}). For example, the pressure, entropy density,
particle density and  energy density are given by
\be
{\cal P} \!\!&=&\!\! \frac{\partial (T \log Z)}{\partial V} \;, \\
{\cal S} \!\!&=&\!\! \frac{1}{V}\frac{\partial (T \log Z)}{\partial T}=  \frac{\partial {\cal P}}{\partial T}\;, \\
n_i\!\!&=&\!\! \frac{1}{V}\frac{\partial (T \log Z)}{\partial \mu_i}=  \frac{\partial {\cal P}}{\partial\mu_i}\;, \\
{\cal E} \!\!&=&\!\! - {\cal P} + T{\cal S}+\sum\limits_{i=1}^{N_f}\mu_i n_i\;,
\ee 
where $V$ is the volume of the system and $N_f$ is the number of flavors of the system. Typically, the width $L$ of a system is
much larger than the inverse temperature, (i.e. $L\gg2\pi/T$), such that one can
use the infinite volume limit to describe the thermodynamics of a finite volume to
good approximation. In all calculations performed in this thesis, this infinite
volume limit is taken. Then it turns out that $\log Z$ becomes 
proportional to $V$, such that the pressure becomes
\be
{\cal P} \;=\; \frac{T \log Z}{V} \;.
\ee
The extension to field theory is straightforward. If ${\cal H}$ is the
Hamiltonian of a quantum field theory in $d$-dimensional space and hence
$(d+1)$-dimensional space-time, then the partition function~(\ref{Z}) is
\be
Z \;=\; {\rm Tr}\,e^{-{\cal H}/T} \;=\; \int {\cal
D}\varphi\,e^{-\int_0^{1/T}\!\!d\tau\int d^dx\,{\cal L}(\varphi)}\;,
\label{Z-QFT}
\ee
with ${\cal L}$ the Lagrangian density of the theory and \emph{periodic}
boundary conditions 
\be
\varphi(0,{\bf x}) \;=\; \varphi(1/T,{\bf x})\;.
\label{periodic_phi}
\ee
for bosonic fields $\varphi$. For fermionic fields, it turns out that to
implement Pauli statistics one must impose \emph{anti-periodic} boundary
conditions
\be
\varphi(0,{\bf x}) \;=\; - \varphi(1/T,{\bf x})\;.
\label{antiperiodic_phi}
\ee
\section{QCD at finite temperature}
Quantum Chromodynamics is a gauge theory for the strong interaction describing
the interactions between quarks and gluons.
The QCD Lagrangian density in Minkowski space can be written as
\be
{\cal L}_{\rm QCD}&=&
-\frac{1}{2}{\rm Tr}\left[F_{\mu\nu}F^{\mu\nu}\right]
+\sum_i\bar{\psi}_i \left[i\gamma^\mu D_\mu -\gamma_0\mu_i- m_i\right] \psi_i
\nn
&&+\ {\cal L}_{\rm gf}
+{\cal L}_{\rm ghost} + \Delta{\cal L}_{\rm QCD}\;.
\label{qcd_lag}
\ee
where $\Delta{\cal L}_{\rm QCD}$ contains counterterms necessary to cancel the ultraviolet
divergences in perturbative calculations. The gluon field strength is $F_{\mu\nu}=
\partial_{\mu}A_{\nu}-\partial_{\nu}A_{\mu} -ig[A_{\mu},A_{\nu}]$. The gluon
field is
$A_\mu = A_\mu^a t^a$, with generators $t^a$ of the fundamental representation
of SU(3) normalized so that ${\rm Tr}\,t^a t^b=
\delta^{ab}/2$. In the quark sector there is an explicit sum over the $N_f$
quark flavors with masses $m_i$ and $D_\mu = \partial_\mu - i g A_\mu$ is the
covariant derivative in the fundamental representation. The Lagrangian~(\ref{qcd_lag})
is mathematically simple and beautiful, however in order to carry out a physical
calculation with it perturbatively, a gauge fixing is needed to remove unphysical degrees of
freedom. The ghost term ${\cal L}_{\rm ghost}$ depends on the choice of the
gauge-fixing term ${\cal L}_{\rm gf}$. One popular choice for the gauge-fixing
term that depends on an arbitrary gauge parameter $\xi$ is the general covariant
gauge:
\be
{\cal L}_{\rm gf}\!\!&=&\!\!
-\frac{1}{\xi}{\rm Tr}\left[\left(\partial^{\mu}A_{\mu}\right)^2\right] \;.
\ee
The corresponding ghost term in the general covariant gauge reads
\be
{\cal L}_{\rm ghost} \;=\; - \bar\eta^a \partial^2 \eta^a + g f^{abc} \bar\eta^a
\partial^\mu(A_\mu^b \eta^c)\;,	
\ee
where $\eta$ and $\bar\eta$ are anti-commuting ghosts and anti-ghosts
respectively and $f^{abc}$ is structure constant of SU(3).

The finite temperature QCD partition function is obtained by a Wick rotation of
the theory from Minkowski space to Euclidean space.
It is achieved by the substitution $t=i\tau$ with $t$ being the Minkowski time
and $\tau$ being the Euclidean one. The resulting
Euclidean partition function is
\be
Z \;=\; \int {\cal D}A_\mu {\cal D}\bar{\psi} {\cal D}\psi {\cal D}\bar{\eta}
{\cal D}\eta \exp\left[-\int_0^{1/T}\!\!d\tau\int d^3x
\,{\cal L}_{\rm QCD}^E\right]\;,
\ee
with ${\cal L}_{\rm QCD}^E$ the Wick-rotated Lagrangian density. Feynman rules
are exactly the same as in zero-temperature field
theory except that the imaginary time $\tau$ is now compact with extent $1/T$.
To go from $\tau$ to frequency space, one should 
perform a Fourier series decomposition rather than a Fourier transform. The only
difference with zero-temperature Feynman rules
will then be that loop frequency integrals are replaced by loop frequency sums:
\begin {equation}
  \int \frac{d^4P}{(2\pi)^4} \; \rightarrow \; T \sum_{\omega_n} \int
\frac{d^3p}{(2\pi)^3}
\label{intro_fs}
\end {equation}
with the sum over discrete imaginary-time frequencies known as Matsubara
frequencies
\be
  \omega_n \!\!& = &\!\! 2 n \pi T \hspace{3cm}\mbox{bosons} \;, \\
  \omega_n \!\!& = &\!\! (2n+1) \pi T-i\mu \hspace{1cm}\mbox{fermions} \;.
\ee
to implement the periodic or anti-periodic boundary conditions in Eq.~(\ref{periodic_phi})
and Eq.~(\ref{antiperiodic_phi}) respectively.

We define the dimensionally regularized bosonic and fermionic sum-integrals as
\be
  \sumintb_{P}& \;\equiv\; &
  \left(\frac{e^{\gamma_E}\Lambda^2}{4\pi}\right)^\epsilon\;
  T\sum_{P_0=2n\pi T}\:\int \frac{d^{3-2\epsilon}p}{(2 \pi)^{3-2\epsilon}}\;,\\ 
  \sumintf_{\{P\}}& \;\equiv\; &
  \left(\frac{e^{\gamma_E}\Lambda^2}{4\pi}\right)^\epsilon\;
  T\sum_{P_0=(2n+1)\pi T-i\mu}\:\int \frac{d^{3-2\epsilon}p}{(2 \pi)^{3-2\epsilon}}\;,
\label{sumint-def}
\ee
where $3-2\epsilon$ is the dimension of space, $\gamma_E\approx0.577216$ the Euler-Mascheroni
constant commonly known as Euler gamma, $\Lambda$ is an arbitrary momentum scale, $P=(P_0,p)$
is the bosonic loop momentum, and $\{P\}$ is the fermionic loop momentum. The factor
$(e^{\gamma_E}/4\pi)^\epsilon$ is introduced so that, after minimal subtraction 
of the poles in $\epsilon$ due to ultraviolet divergences, $\Lambda$ coincides 
with the renormalization scale of the $\overline{\rm MS}$ renormalization scheme.
Note that we are denoting four momentum in Euclidean space
as $P=(P_0,p)$ and in Minkowski space as $P=(p_0,p)$.

Sometimes we use shorthand notation for $d=3-2\epsilon$ dimensional integration as
\be
\int\limits_{\bf p}\equiv \int \frac{d^{3-2\epsilon}p}{(2 \pi)^{3-2\epsilon}}.
\ee  

So one needs to evaluate the frequency sum(\ref{intro_fs}) for bosonic case as 
\be
T\sum\limits_{p_0}f(p_0=i\omega_n=2n\pi i T)=\frac{T}{2\pi i}\oint\limits_C
dp_0\frac{\beta}{2}f(p_0)\coth\frac{\beta p_0}{2}\ ,
\label{freq_boson}
\ee
where the contour $C$ is as shown Fig.~(\ref{contour1}a):
\begin{figure}[tbh]
\begin{center}
\includegraphics[height=8cm]{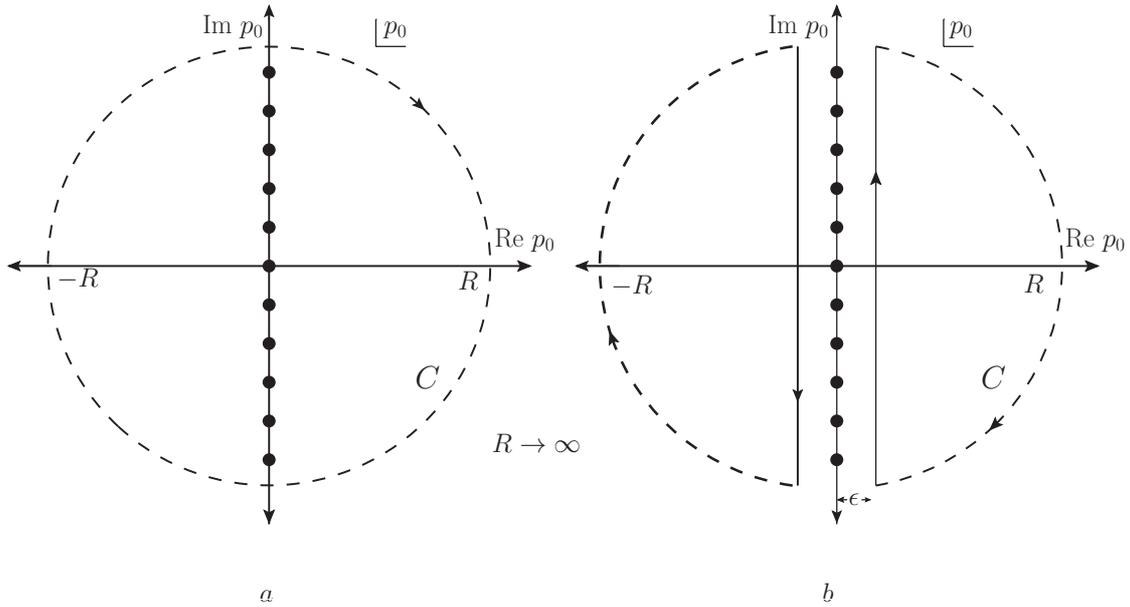}
\end{center}
\caption{Contour for bosonic frequency sum.}
\label{contour1}
\end{figure}
The function $\frac{\beta}{2} \coth\frac{\beta p_0}{2}$ has poles at $p_0 = 2\pi n i T $ and is everywhere
else bounded and analytic. The contour can be deformed as Fig.~(\ref{contour1}b).

So the frequency sum in Eq.~(\ref{freq_boson}) can be rewritten as
\be
T\sum\limits_{p_0}f(p_0=i\omega_n=2n\pi i T)&=&\frac{1}{2\pi i}\int\limits_{-i\infty+\epsilon}^{i\infty+\epsilon}
dp_0f(p_0)\frac{1}{2}\coth\frac{\beta p_0}{2}
\nn
&+&\frac{1}{2\pi i}\int\limits_{i\infty-\epsilon}^{-i\infty-\epsilon}
dp_0f(p_0)\frac{1}{2}\coth\frac{\beta p_0}{2}
\nn
&=&\frac{1}{2\pi i}\int\limits_{-i\infty+\epsilon}^{i\infty+\epsilon}
dp_0f(p_0)\left[\frac{1}{2}+\frac{1}{e^{\beta p_0}+1}\right]
\nn
&-&\frac{1}{2\pi i}\int\limits_{-i\infty-\epsilon}^{i\infty-\epsilon}
dp_0f(p_0)\left[\frac{1}{2}+\frac{1}{e^{\beta p_0}+1}\right]
\ee
Below we demonstrate one examples of frequency sum for bosonic momentum:
\be
\sum\limits_{p_0=2n\pi i T}\frac{1}{P^2}&=&\sum\limits_{p_0=2n\pi i T}\frac{1}{p_0^2-p^2}
=\sum\limits_{p_0=2n\pi i T}\frac{1}{2p}\left[\frac{1}{p_0-p}-\frac{1}{p_0+p}\right]
\nn
&=&\frac{1}{2p}\left[-\left(\frac{1}{2}+\frac{1}{e^{\beta p}-1}\right)+\left(\frac{1}{2}+\frac{1}{e^{-\beta p}-1}\right)\right]
\nn
&=&-\frac{1}{p}\left[\frac{1}{2}+n_B(p)\right],
\label{results_boson_freq}
\ee
where $n_B(p)=1/\left(\exp(\beta p)-1\right)$ is Bose-Einstein distribution function. When one calculates $(d+1)$ dimensional
sum-integrals, the first term of the frequency sum~(\ref{results_boson_freq}) vanishes due to dimensional regularization and 
gets the finite results for this particular sum-integration.

\begin{figure}[tbh]
\begin{center}
\includegraphics[height=7cm]{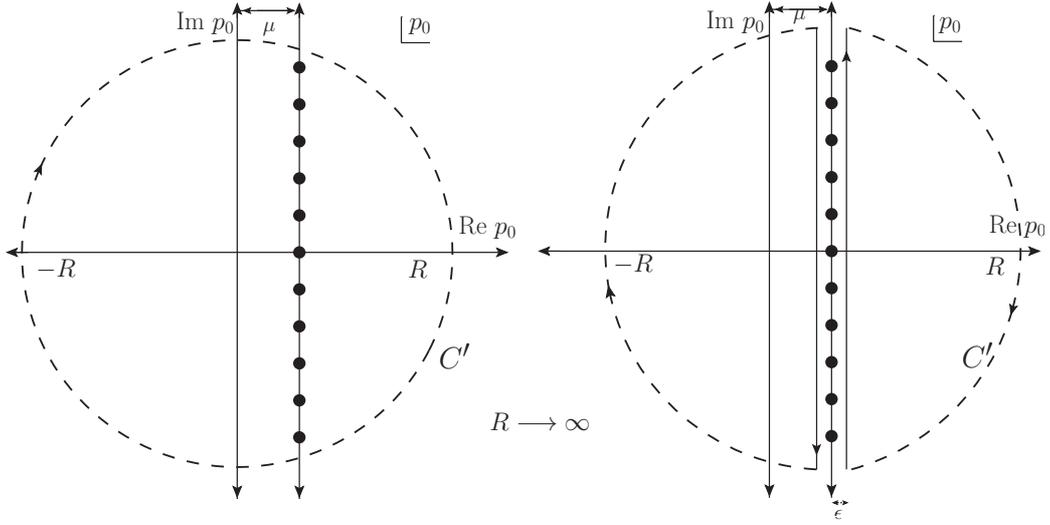}
\end{center}
\caption{Contour for fermionic frequency sum.}
\label{contour2}
\end{figure}

One can also evaluate the frequency sum(\ref{intro_fs}) for fermionic case in the similar manner as
\be
T\sum\limits_{p_0}f(p_0=i\omega_n=(2n+1)\pi i T + \mu)=\frac{T}{2\pi i}\oint\limits_{C'}
dp_0\frac{\beta}{2}f(p_0)\tanh\frac{\beta(p_0-\mu)}{2}
\label{freq_fermion}
\ee
where the contour $C'$ is as shown in the Fig.~(\ref{contour2}).
The function $\frac{\beta}{2} \tanh\frac{\beta (p_0-\mu)}{2}$ has poles at $p_0 = (2n+1)\pi i T +\mu$ and is everywhere
else bounded and analytic. The contour $C'$ can be deformed in the similar manner as in the bosonic case to avoid the pole
on the contour. Then the frequency sum in Eq.~(\ref{freq_fermion}) can be rewritten as
\be
T\sum\limits_{p_0}f(p_0=i\omega_n=(2n+1)\pi i T+\mu)&=&\frac{1}{2\pi i}\int\limits_{-i\infty+\mu+\epsilon}^{i\infty+\mu+\epsilon}
dp_0f(p_0)\frac{1}{2}\tanh\frac{\beta p_0}{2}
\nn
&&\hspace{-2cm}+\frac{1}{2\pi i}\int\limits_{i\infty+\mu-\epsilon}^{-i\infty+\mu-\epsilon}
dp_0f(p_0)\frac{1}{2}\tanh\frac{\beta(p_0-\mu)}{2}
\nn
&&\hspace{-2cm}=\frac{1}{2\pi i}\int\limits_{-i\infty+\mu+\epsilon}^{i\infty+\mu+\epsilon}
dp_0f(p_0)\left[\frac{1}{2}+\frac{1}{e^{\beta(p_0-\mu)}+1}\right]
\nn
&&\hspace{-2cm}-\frac{1}{2\pi i}\int\limits_{-i\infty+\mu-\epsilon}^{i\infty+\mu-\epsilon}
dp_0f(p_0)\left[\frac{1}{2}+\frac{1}{e^{\beta(p_0-\mu)}+1}\right].\hspace{1cm}
\ee
Below we demonstrate one examples of frequency sum for  fermionic momentum:
\be
\sum\limits_{p_0=2n\pi i T}\frac{1}{P^2}&=&\sum\limits_{p_0=2n\pi i T}\frac{1}{p_0^2-p^2}
=\sum\limits_{p_0=2n\pi i T}\frac{1}{2p}\left[\frac{1}{p_0-p}-\frac{1}{p_0+p}\right]
\nn
&=&\frac{1}{2p}\left[-\left(\frac{1}{2}-\frac{1}{e^{\beta( p-\mu)}+1}\right)+\left(\frac{1}{2}
-\frac{1}{e^{-\beta (p+\mu)}+1}\right)\right]
\nn
&=&-\frac{1}{2p}\left[1-n_F(p)\right],
\ee
where
$n_F(p)=[e^{\beta(p-\mu)}+1]^{-1}+[e^{\beta(p+\mu)}+1]^{-1}=[n_F^-(p)+n_F^+(p)]$.
%
%

After performing the frequency sum, one is left with dimensionally regularized spatial momentum integration,
which will be discussed in the subsequently chapter. However, all other frequency sums can be evaluated in similar way as
discussed above.

\section[Perturbative pressure in QCD]{Perturbative pressure in QCD up to order $g^6\log g$}
\label{press_vuorinen}
\begin{figure}[t]
\subfigure{
\includegraphics[width=7.5cm,height=8cm]{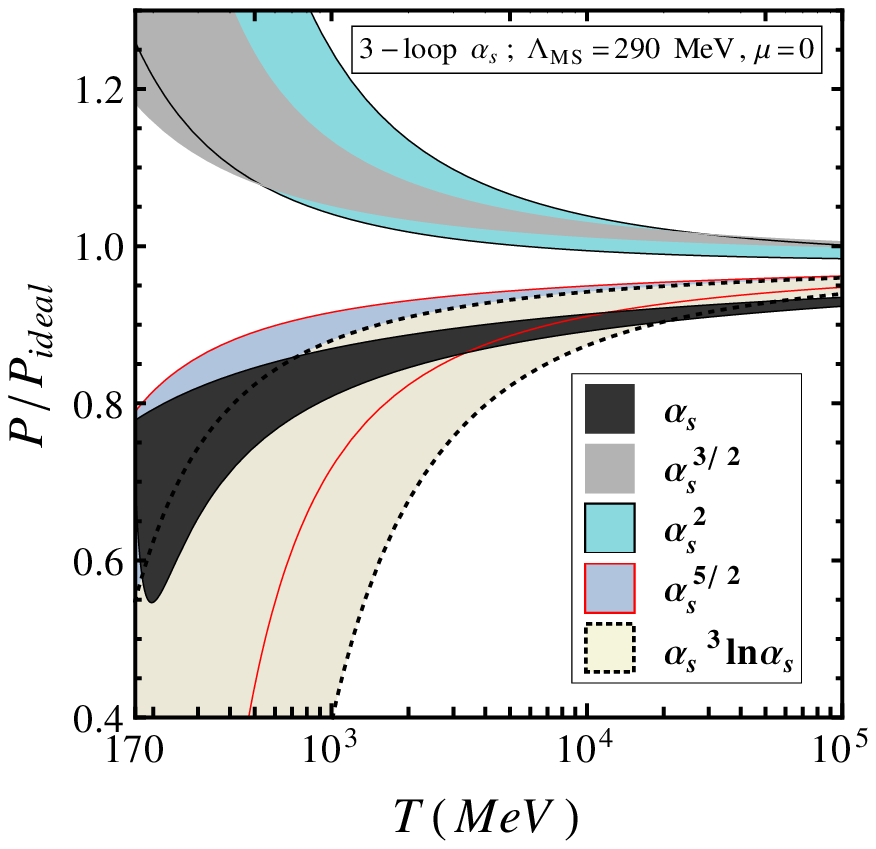}}
\hspace{-.5cm}
\subfigure{
\includegraphics[width=7.5cm,height=8cm]{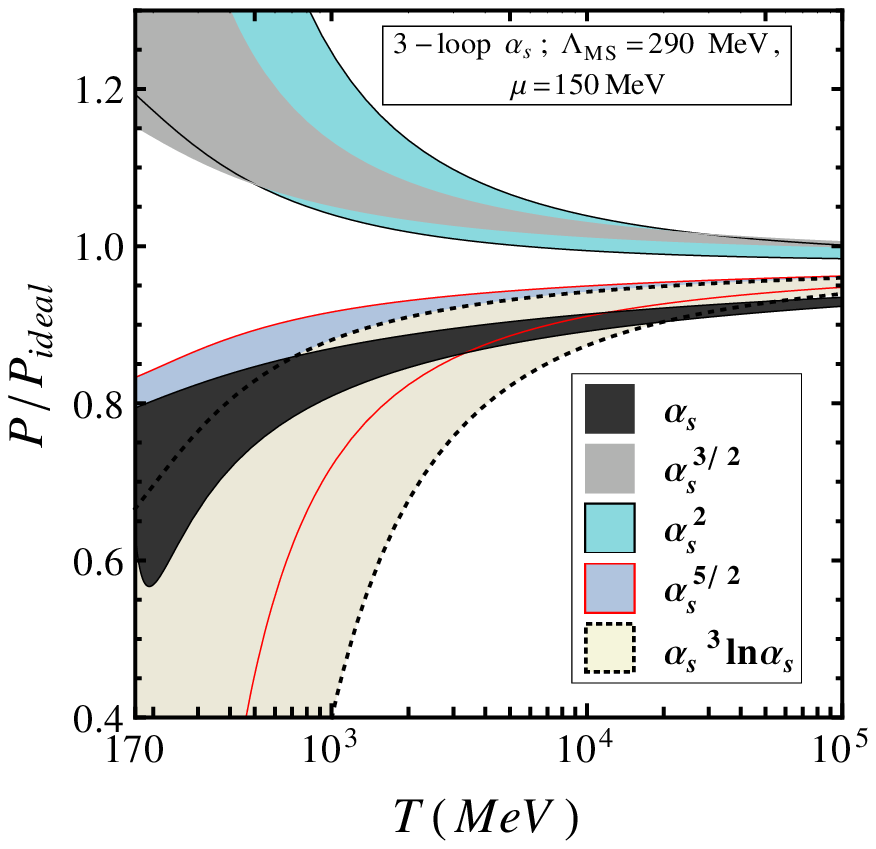}}
\caption[pQCD pressure as a function of the temperature.
Successive perturbative approximations are shown through order $\alpha_s^3\ln\alpha_s$ for vanishing $\mu$ (left) 
and for non-vanishing $\mu$ (right).]{The $N_f=3$ pQCD pressure specified in Eq.~(\ref{freeg}) as a function of the temperature.
Successive perturbative approximations are shown through order $\alpha_s^3\ln\alpha_s$ for vanishing $\mu$ (left) 
and for non-vanishing $\mu$ (right).  The shaded bands indicate the variation of the pressure as the $\overline{\rm MS}$
renormalization scale is varied around a central value of $\Lambda=2\pi \sqrt{T^2+\mu^2/\pi^2}$
\cite{Vuorinen:2002ue,Vuorinen:2003fs,Rebhan:2003wn,Cassing:2007nb,Gardim:2009mt,Kurkela:2009gj} 
by a factor of two.
We use $\Lambda_{\overline{\rm MS}}=290$ MeV based on recent lattice calculations~\cite{Bazavov:2012ka} 
of the three-loop running of $\alpha_s$.
}
\label{pertfig}
\end{figure}

In perturbation theory one can explicitly separate the contributions coming from the soft sector (momenta 
on the order of $g T$ where $g^2 = 4 \pi \alpha_s$) and the hard sector (momenta 
on the order of $T$) using effective field theory/dimensional reduction methods 
\cite{Braaten:1995cm,Braaten:1995jr,Kajantie:2002wa,Blaizot:2003iq,Andersen:2004fp,Vuorinen:2004rd,Laine:2006cp}.  
After doing this one finds that the hard-sector contributions, which form a power series in even powers of $g$, 
converge reasonably well; however, the soft sector perturbative series, which contains odd powers
of $g$, is poorly convergent. Below we present the result of QCD free energy ${\cal F} (T,\mu)$ for $N_f$ 
flavor in perturbative series
up to order up to order $g^6\log g$ {\em i.e.}, $\alpha_s^{3}\ln (\alpha_s)$.
\begin{eqnarray}
{\cal F} &=& - \frac{8 \pi^2}{45} T^4 \,
\biggl[ {\cal F}_0
+ {\cal F}_2  \frac{\alpha_s}{\pi}
+ {\cal F}_3  \left( \frac{\alpha_s}{\pi} \right)^{3/2}
\!\!  + {\cal F}_4  \left( \frac{\alpha_s}{\pi} \right)^2
\nn
&&
\hspace{2.5cm}+\ {\cal F}_5  \left( \frac{\alpha_s}{\pi} \right)^{5/2}
\!\! + {\cal F}_6\left( \frac{\alpha_s}{\pi} \right)^3 +\cdots \biggr] , 
\label{freeg}
\end{eqnarray}
where we have specialized to the case $N_c=3$ and
\be
{\cal F}_0 &=& 1 + \frac{21}{32}N_f \left(1+\frac{120}{7}\hat\mu^2 
+\frac{240}{7}\hat\mu^4 \right)  \, ,
\ee
\be
{\cal F}_2 &=& - \frac{15}{4} \left[ 1 + \frac{5 N_f}{12}\left(1+\frac{72}{5}\hat\mu^2 
+\frac{144}{5}\hat\mu^4 \right)  \right]\;,
\ee
\be
{\cal F}_3 &=& 30 \left[ 1 + \textstyle\frac{1}{6}\left(1 + 12\hat\mu^2\right)
N_f \right]^{3/2}
\ee
\be
 {\cal F}_4 &=&  237.223 + \left(15.963 + 124.773\ \hat\mu^2 -319.849\hat\mu^4 \right) N_f
 \nn
&-& 
  \left( 0.415 + 15.926\ \hat\mu^2 + 106.719\ \hat\mu^4\right)N_f^2 \nonumber
 \nn &+& 
 \frac{135}{2} \left[ 1 + \textstyle\frac{1}{6}\left(1+12\hat\mu^2\right)N_f  \right]
         \log \left[ \frac{\alpha_s}{\pi}
         \left(1 + \textstyle\frac{1}{6}\left(1+12\hat\mu^2\right)N_f \right) \right] 
 \nn &-& 
 \frac{165}{8} \left[1+\frac{5}{12}\left(1+\frac{72}{5}\hat\mu^2 
 +\frac{144}{5}\hat\mu^4 \right) N_f\right]\left(1 -\frac{2}{33} N_f\right)\log{\hat \Lambda}\; ,
\ee

\be
 {\cal F}_5 &=& -\left( 1 + \frac{1+12\hat\mu^2}{6} N_f\right)^{1/2}
 	\Bigg[ 799.149 + \left(21.963 - 136.33\ \hat \mu^2 + 482.171\ \hat\mu^4 \right)N_f 
 \nonumber\\&& \hspace{2mm}
        + \left(1.926 + 2.0749\ \hat\mu^2 - 172.07\ \hat\mu^4\right) N_f^2\Bigg] 
 \nonumber\\ && \hspace{4mm}
        +\ \frac{495}{2} \left(1+\frac{1+12\hat \mu^2}{6} N_f\right)\left(1 -\frac{2}{33} N_f\right)\log{\hat \Lambda}
 \; , 
\ee
\be
  {\cal F}_6 &=& -\Bigg[659.175 + \left(65.888 -341.489\ \hat\mu^2 + 1446.514\ \hat\mu^4\right)N_f 
\nn
&+&
\left(7.653 + 16.225\ \hat \mu^2 - 516.210\ \hat \mu^4\right) N_f^2
-\frac{1485}{2}\left(1+\frac{1+12\hat\mu^2}{6}N_f\right)
\nn
&\times&\left(1-\frac{2}{33}N_f\right)\log{\hat\Lambda} \Bigg]
\log\left[\frac{\alpha_s}{\pi}\left(1+\frac{1+12 \hat \mu^2}{6}N_f\right)4\pi^2\right]\nn
&-& 475.587\log\left[\frac{\alpha_s} {\pi}\ 4\pi^2C_A\right], 
\ee
where here and throughout all hatted quantities are scaled by $2\pi T$, e.g. $\hat\mu = \mu/(2\pi T)$, 
$\Lambda$ is the modified minimum subtraction ($\overline{\rm MS}$)  
renormalization scale,  and $\alpha_s=\alpha_s(\hat \Lambda)$ is the running coupling. 
At finite $T$ the central value of the renormalization scale is usually chosen to be $2\pi T$. However,
at finite $T$ and $\mu$  we  use the  central scale $\Lambda = 2 \pi \sqrt{ T^2 + (\mu/\pi)^2}$~\cite{Vuorinen:2002ue,
Vuorinen:2003fs,Rebhan:2003wn,Cassing:2007nb,Gardim:2009mt,Kurkela:2009gj}.
In Fig.~(\ref{pertfig}) we plot the ratio of the pressure to an ideal gas of quarks and gluons.
The figure clearly demonstrates the poor convergence of the naive perturbative series 
and the increasing sensitivity of the result to the renormalization scale 
as successive orders in the weak coupling expansion are included.  

The poor convergence in soft sector which contains the odd power of $g$ in perturbative series
 suggests that in order to improve
 the convergence of the resulting perturbative approximation one should treat the soft sector
 non-perturbatively, or at least resum soft corrections to the pressure.  There have been works
 in the framework of dimensional reduction which effectively perform such soft-sector resummations
 by not truncating the soft-scale contributions in a power series in $g$, see {\em e.g.}
 \cite{Kajantie:2002wa,Blaizot:2003iq,Laine:2006cp}.  This method seems to improve the
 convergence of the perturbation series and provides motivation to find additional analytic methods
 to accomplish soft-sector resummations.

\section{Beta function and asymptotic freedom}

The beta function $\beta(\alpha_s)$ of a quantum field theory encodes the
dependence of a coupling parameter $\alpha_s$ on the
energy scale $\Lambda$ of a given physical process. The coupling satisfy the
following relation~\cite{Beringer:1900zz,vanRitbergen:1997va}:
\be
\Lambda^2\frac{\partial \alpha_s}{\partial \Lambda^2}=
\beta(\alpha_s)=-\alpha_s^2\lb b_0+b_1\alpha_s+b_2\alpha_s^2+b_3\alpha_s^3
+\cdots\rb
\label{beta}
\ee
where $b_0,b_1,b_2$ and $b_3$ are referred to as beta-function coefficient in
one, two, three and four loop respectively with 
\be
b_0\!\!\!&=&\!\!\!\frac{11c_A-2N_f}{12\pi}\nn
b_1\!\!\!&=&\!\!\!\frac{17c_A^2-N_f\lb5c_A+3c_F\rb}{24\pi^2}\nn
b_2\!\!\!&=&\!\!\!\frac{1}{128\pi^3}\left[\frac{2857}{27}c_A^2+N_f\lb2c_F^2-\frac{205}{9}
c_Ac_F-\frac{1415}{27}c_A^2\rb\right.\nn
&&
+\left.N_f^2\lb\frac{22}{9}c_F+
\frac{79}{27}c_A\rb\right]\
\ee
where $c_A=N_c$ is number of colors, $c_F=(N_c^2-1)/2N_c$.

This dependence on the energy scale is known as the running of the coupling
parameter, and theory of this kind of scale-dependence
in quantum field theory is described by the renormalization group which refers
to a mathematical apparatus that allows one to 
investigate the changes of a physical system as one views it at different
distance scales.

To lowest order in the coupling constant a beta function is either positive
indicating the growth of charge at short distance
or negative indicating the decrease of charge at short distance. Until 1973,
only examples of the former were known~\footnote
{'t Hooft reported a similar discovery at the Marseille conference on
renormalization of Yang-Mills fields and applications to
particle physics in 1972 without publishing it.}. The discovery that only
non-Abelian gauge theories allow for
a negative beta function is usually credited to Gross and
Wilczek~\cite{Gross:1973id}, and to Politzer~\cite{Politzer:1973fx}. The
solution
to~({\ref{beta}}) for QCD in one loop level reads
\be
\alpha_s(\Lambda) \;=\; \frac{g(\Lambda)^2}{4\pi} \;=\; \frac{2\pi}{\left(11 -
\frac{2}{3}N_f\right) \log\left(\Lambda / 
\Lambda_{\rm QCD}\right)}\;,\label{1loop_running}
\ee
which clearly shows \emph{asymptotic
freedom}~\cite{Gross:1973id,Politzer:1973fx}, i.e. $\alpha_s \rightarrow 0$ as
$\Lambda \rightarrow \infty$.
The parameter $\Lambda_{\rm QCD}$ is a scale above which the theory works as
``chosen'' by the  world in which we live.

\begin{figure}[tbh]
\begin{center}
\includegraphics[width=9cm]{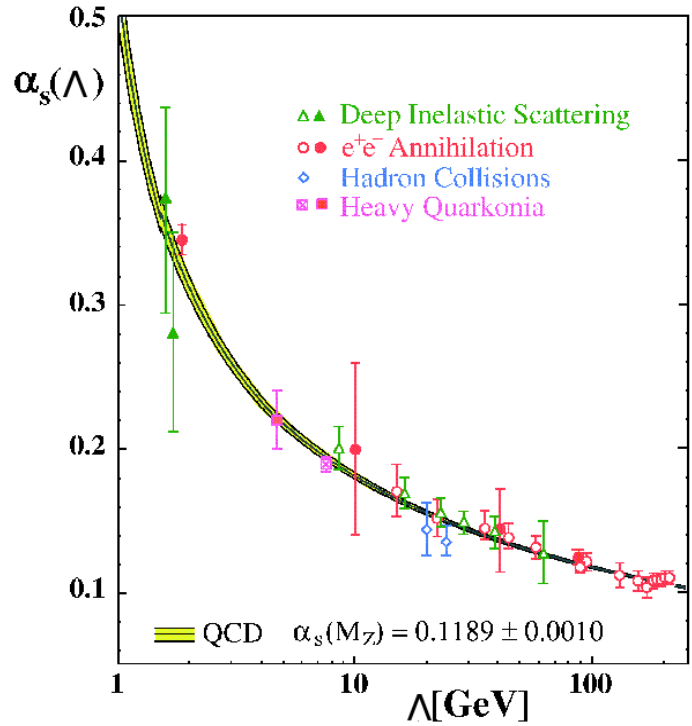}
\end{center}
\caption[QCD running coupling as a function of energy scale]{QCD running coupling as a function of energy scale.
This figure is adapted from Ref.~\cite{Bethke:2006ac}.}
\label{alphas_plot}
\end{figure}
In three loop level, the solution to~({\ref{beta}}) is
\be
\alpha_s(\Lambda)&=&\frac{1}{b_0 t}\left[1-\frac{b_1}{b_0^2}\frac{\ln
t}{t}+\frac{b_1^2(\ln^2 t-\ln t-1)+b_0b_2}{b_0^4 t}\right.
\nn
&&\left.-\frac{b_1^3\lb\ln^3 t-\frac{5}{2}\ln^2 t-2\ln
t+\frac{1}{2}\rb+3b_0b_1b_2\ln t-\frac{1}{2}b_0^2b_3}{b_0^6t^3}+\cdots\right] \label{3loop_running}
\ee
with $t=\ln \frac{\Lambda^2}{\Lambda_{\overline {MS}}}$.
It is well known that QCD exhibits confinement at large distances or low
energies which terminates the validity of perturbation
theory due to the infrared growth of the coupling. However it is precisely the
asymptotic freedom that ensures the possibility
of a perturbative treatment for the ultraviolet sector of the theory which sets
the stage to study the high temperature phase
of non-Abelian theory in this thesis.

\section{Scope of the thesis}

The thesis is organized as follows:
In Chapter~\ref{chapter:htl}, we discuss the limitations of bare perturbation theory and a brief
introduction to HTL perturbation theory.
In Chapter~\ref{chapter:1loop} we discuss conserved density fluctuation and temporal correlation function
in hard thermal loop perturbation theory.
In Chapter~\ref{chapter:2loop} we discuss thermodynamic 
functions {\em viz} pressure and quark number susceptibilities using HTL perturbation
theory in two loop level. The Chapter~\ref{chapter:3loop}
is devoted to the study of all possible three-loop thermodynamic functions at finite temperature 
and finite chemical potential using HTL perturbation
theory for QCD. Low mass dilepton rate from deconfined state of matter will be discussed
in Chapter~\ref{chapter:dilepton}. We summarize in Chapter~\ref{chapter:sum} together with
a brief outlook of HTL perturbation theory.

%% file: text/htl.tex
\chapter{Hard Thermal Loop Perturbation Theory}
\label{chapter:htl}
In this chapter, and in the rest of the thesis, we consider thermal field theories at high temperature, which means temperature
much higher than all zero-temperature masses or any mass scales generated at zero temperature. 

We can apply finite temperature field theory to calculate various thermodynamic quantities perturbatively. But naive perturbation
theory have some serious problems. One of the problem is in the computation of gluon damping rate. If one uses naive perturbation
theory to calculate gluon damping rate, one gets gauge dependent results and it is also negative in some gauges. This was sometimes
interpreted as signal of plasma instability. Yet one knows from general field theoretical arguments that the positions of the pole
of the gluon propagator, whose imaginary part gives the gluon damping rate, is gauge-independent.  

Another problem in naive perturbation theory is infrared divergence. It has been known for many years that for naive perturbative
expansion of finite temperature, partition function  breaks down due to infrared divergences. To cure those problem of naive perturbation
theory, a consistent perturbative expansion requires the resummation of an infinite subset of diagrams from all orders of
perturbation theory. Hard thermal loop (HTL) perturbation theory is one of such resummation. In HTL perturbation theory, we define
two scales of momenta {\em viz.} hard and soft. When the momentum $p\sim T$, it is called hard scale and when momenta $p\sim gT$,
this is called soft scale with $g$ coupling constant of the theory. We discuss the HTL perturbation theory in case of scalar 
as well as gauge theory in the next section in details.
\section{Scalar field theory}
In this section we discuss about the simplest interacting scalar field theory, namely
a single massless scalar field with a $\phi^4$ interaction. The Lagrangian for such scalar
field in  Euclidean space can be written as
\be
{\mathcal L}\;=\;\frac{1}{2}(\partial_{\mu}\phi)^2+\frac{1}{24}{g^2}\phi^4\;.
\label{sl}
\ee
The Lagrangian density in Eq.~(\ref{sl}) can be divided into a free part and an interacting
part as
\be
{\mathcal L}_{\rm free}\!\!&=&\!\!\frac{1}{2}(\partial_{\mu}\phi)^2\;,\\
{\mathcal L}_{\rm int}\!\!&=&\!\!\frac{1}{24}g^2\phi^4\;.
\ee
Radiative corrections are then calculated in a loop expansion which is equivalent to a power series in $g^2$. We shall see that the 
perturbative expansion breaks down at finite temperature and the weak-coupling expansion becomes an expansion in $g$ rather than $g^2$.

\begin{figure}[htb]
\begin{center}
\includegraphics[height=2cm]{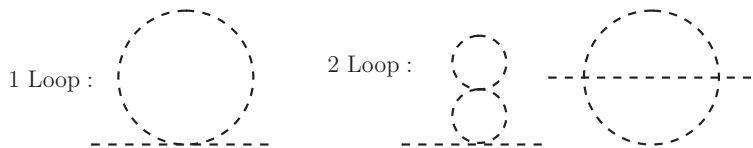}
\end{center}
\caption[One and two-loop scalar self-energy]{One- and two-loop scalar self-energy graphs.}
\label{thmass}
\end{figure}

We will first calculate the self-energy by evaluating the relevant diagrams. The Feynman diagrams that contribute to the
self-energy up to two loops are shown in Fig.~(\ref{thmass}).

The one-loop diagram is independent of the external momentum and the resulting integral expression is
\be\nonumber
\Pi^{(1)}\!\!&=&\!\!\frac{1}{2}g^2\sumintb_P\frac{1}{P^2} \; , \\ \nonumber
\!\!&=&\!\!\frac{1}{24}g^2T^2 \; , \\
\!\!&\equiv&\!\!m^2 \;,
\label{fm}
\ee
where the superscript indicates the number of loops. The notation $P=(P_0,{\bf p})$ represents the Euclidean four-momentum.
The Euclidean energy $P_0$ has discrete values: $P_0=2n\pi T$ for bosons and
$P_0=(2n+1)\pi T$ for fermions, where $n$ is an integer. Eq.~(\ref{fm}) represents the leading order thermal mass of scalar field with
$\phi^4$ interaction. The sum-integral over $P$ is defined in Eq.~(\ref{sumint-def}), represents a summation 
over Matsubara frequencies and integration of spatial momenta in $d=3-2\epsilon$ dimensions~\footnote{For an introduction to 
thermal field theory and the imaginary time formalism see Refs.~\cite{kapusta} and \cite{lebellac}.}. The above sum-integral has ultraviolet
power divergences that are set to zero in dimensional regularization. We are then left with the finite result~(\ref{fm}),
which shows that thermal fluctuations generate a mass for the scalar field of order $gT$.
This thermal mass is analogous to the Debye mass which is well-known from the non-relativistic QED plasma.

\begin{figure}[tbh]
\begin{center}
\includegraphics[height=2.8cm,width=12cm]{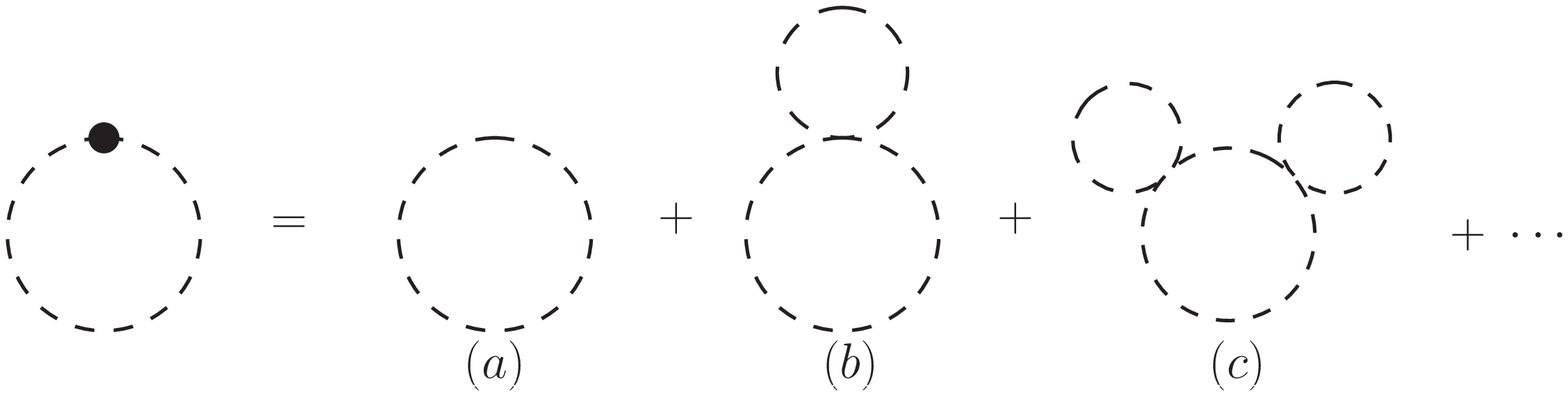}
\caption[Bubble diagrams contributing to the scalar self-energy]{Bubble diagrams contributing to the scalar self-energy.}
\label{2lself}
\end{center}
\end{figure}

We next focus on the two-loop diagrams and first consider the double-bubble in Fig.~(\ref{2lself}b). 

\noindent
This diagram is also independent of the external momentum and gives the following sum-integral
\be
\Pi^{(2b)}\;=\;-\frac{1}{4}g^4\sumintbb_{PQ}\frac{1}{P^2}\frac{1}{Q^4} \;.
\ee
This integral is infrared divergent. The problem stems from the middle loop with two propagators. In order
to isolate the source of the divergence, we look at the contribution from the zeroth Matsubara mode to the $Q$ integration
\be
-\frac{1}{4}g^4\sumintb_P\frac{1}{P^2}T\int\limits_{\bf q}\frac{1}{q^4} \;.
\label{g3}
\ee 
The integral over $q$ in Eq.~(\ref{g3}) behaves like $1/q$ and
linearly infrared divergent as $q \rightarrow 0$. This infrared divergence indicates that naive perturbation theory breaks
down at finite temperature. However, in practice this infrared divergence is screened by a thermally generated mass and we must
somehow take this into account. The thermal mass can be incorporated by using an effective propagator:
\be
\Delta(P)\;=\;\frac{1}{P^2+m^2}\;,
\label{impprop}
\ee
where $m$ can be obtained from Eq.~(\ref{fm}) as $m=gT/\sqrt{24} \ll T$.

\begin{figure}[htb]
\begin{center}
\includegraphics{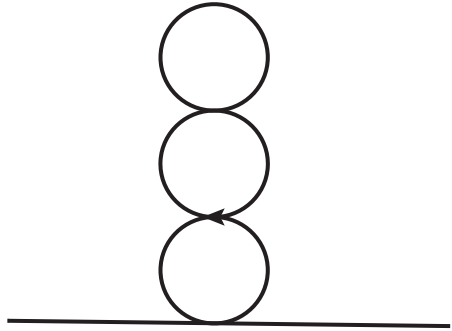}
\end{center}
\caption[Three-loop scaler self-energy diagram]{Subleading three-loop self-energy diagram.}
\label{fig:3l}
\end{figure}

If the momenta of the propagator is of order $T$ or {\it hard}, clearly the thermal mass is a perturbation and can be omitted.
However, if the momenta of the propagator is of order $gT$ or {\it soft}, the thermal mass is as large as the bare inverse 
propagator and cannot be omitted. The mass term in the propagator~(\ref{impprop}) provides an infrared cutoff of order $gT$.
The contribution from~(\ref{g3}) would then be
\be
-\frac{1}{4}g^4\sumintb_P\frac{1}{P^2}T\int_{\bf q}\frac{1}{(q^2+m^2)^2}
\;=\;-\frac{1}{4}g^4\left(\frac{T^2}{12}\right)\left(\frac{T}{\pi m}\right)
+{\mathcal O}\left(g^4mT\right)\;.
\ee
Since $m\sim gT$, this shows that the double-bubble contributes at order $g^3T^2$ to the self-energy and not at order $g^4T^2$
as one might have expected. Similarly, one can show that the diagrams with any number of bubbles like Fig.~(\ref{2lself}c) are all
of order $g^3$. Clearly, naive perturbation theory breaks down since the order-$g^3$ correction to the thermal mass receives 
contributions from all loop orders. On the other hand, the three-loop diagram shown in Fig.~(\ref{fig:3l}), is of order $g^4T^2$ 
and thus sub-leading. Therefore, we only need to resum a subset of all possible Feynman graphs in order to obtain a consistent 
expansion in $g$.

\vspace{6mm}

If we use the effective propagator to recalculate the one-loop self-energy, we obtain
\be\nonumber
\Pi^{(1)}(P)\!\!&=&\!\!\frac{1}{2}g^2\sumintb_P\frac{1}{P^2+m^2} \\ \nonumber
\!\!&=&\!\!\frac{1}{2}g^2\left[T\int\limits_{\bf p}\frac{1}{p^2+m^2}+\sumintb_P^{\prime}\frac{1}{P^2}
+{\mathcal O}\left(m^2\right)\right] \\
\!\!&=&\!\!\frac{g^2}{24}T^2\left[1-\frac{g\sqrt{6}}{4\pi}+{\mathcal O}\left(g^2\right)\right]\;.
\ee
where here, and in the following, the prime on the sum-integral indicates that we have excluded the $n=0$ mode from the sum over
the Matsubara frequencies. The order $g^3$ corresponds to the summation of the bubble diagrams in Fig.~(\ref{2lself}), which can be
verified by expanding the effective propagator in Eqn.~(\ref{impprop}) around $m=0$. Thus by taking the thermal mass into account,
one is resumming an infinite set of diagrams from all orders of perturbation theory.

TThe one loop scalar self-energy in Eq~(\ref{fm}) is the first example of a {\it hard thermal loop} (HTL).
In HTL perturbation theory, loop corrections are $g^2T^2/P^2$ times the corresponding tree-level amplitude, where $P$ 
is a momentum that characterizes the external lines. From this definition, it is clear that,
whenever the external momentum $P$ is hard {\em i.e.} $P\sim T$, the loop correction is suppressed by $g^2$ and
is thus a perturbative correction. However, when the external momentum $P$ is soft {\em i.e.} $P(\sim gT)$,
the HTL correction is same order of tree-level amplitude and is therefore as important as the tree-level contribution to 
the amplitude. These loop corrections are called ``hard'' because the relevant integrals are dominated by momenta of order $T$.
Also note that the hard thermal loop in the two-point function is finite since it is exclusively due to thermal fluctuations. 
Quantum fluctuations do not enter. Both properties are shared by all hard thermal loops.
\subsection*{n- point functions in scalar theory}
In the previous section we have discussed about two-point function in scalar field theory. If one calculate 
higher order $n-$ point functions in scalar theory, one can show that the one-loop correction to the four-point function
at high temperature behaves as~\cite{Pisarski:1990ds}
\be
\Gamma^{(4)}\;\propto\;g^4\log\left(T/p\right)\;,
\ee
where $p$ is the external spatial momentum. Thus the loop correction to the four-point function increases
logarithmically with temperature $(T)$.
It is therefore always down by $g^2\log(1/g)$ as the external momentum $p$ is soft in HTL approximation, 
and one can conclude that it is sufficient to use a bare vertex. More generally, one can show that the only
hard thermal loop in scalar field theory is the tadpole diagram in Fig.~(\ref{thmass}) and resummation is 
taken care of by including the thermal mass in the propagator. In gauge theories, the situation is much 
more complicated than the scalar theory as we shall discuss in the next section.
\section{Gauge theories}
\label{gres}
In the previous section, we discussed about HTL resummation for scalar theory with $\phi^4$ interaction. For scalar theories,
the resummation simply amounts to include the thermal mass in the propagator. The higher order functions such as four point
vertex depends logarithmically on the temperature, corrections to the bare vertex are always down by powers of $g^2\log{g}$.
In gauge theories, the situation is much more complicated. The equivalent HTL self-energies are no longer local,
but depend in a nontrivial way on the external momentum.
In addition, it is also necessary to use effective vertices that also depend on the external momentum. It turns out that all hard
thermal loops are gauge-fixing independent~\cite{Braaten:1989mz,Frenkel:1989br,Braaten:1989kk,Braaten:1990az,Taylor:1990ia,Kobes:
1990dc}. This was shown explicitly in covariant gauges, Coulomb gauges, and axial gauges. All the $n$- point functions 
also satisfy tree-level like Ward identities.
Furthermore, there exists a gauge invariant effective Lagrangian in gauge theory, found independently by Braaten and Pisarski 
\cite{Braaten:1991gm} and by Taylor and Wong \cite{Taylor:1990ia}, that generates all of the hard thermal loop $n$-point functions.
From a renormalization group point of view this is an effective Lagrangian for the soft scale $gT$ that is obtained by integrating
out the hard scale $T$.
\subsection{Polarization tensor}
In this section we discuss in some detail the hard thermal loop for the vacuum polarization tensor $\Pi^{\mu\nu}$ for photon. 
Later we extend it for gluon also.
The hard thermal loop in the photon propagator was first calculated by Silin more than forty years ago \cite{silin}.
 \begin{figure}[tbh]
 \begin{center}
 \includegraphics[width=6cm]{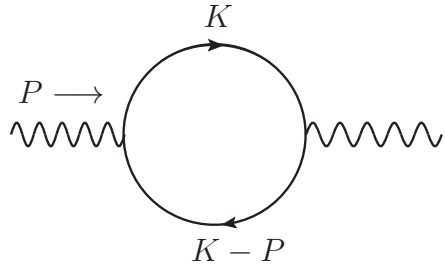}
 \end{center}
 \caption[One-loop photon self-energy]{One-loop photon self-energy diagram.}
 \label{qed}
 \end{figure}
The Feynman diagram for the one-loop self-energy photon is shown in Fig.~(\ref{qed}) and results can be expressed in Euclidean space
as 
\be
\Pi^{\mu\nu}(P)\;=\;e^2\,\sumintf_{\{K\}}{\rm Tr}\left[\frac{K\!\!\!\!/\gamma^{\mu}(K\!\!\!\!/-P\!\!\!\!/)\gamma^{\nu}}{K^2(K-P)^2}
\right]\;,
\ee
where $P=(P_0,p)$ is momentum of the external photon and ${\rm Tr}$ denotes the trace over Dirac matrices.
After taking the trace over Dirac matrices, the self-energy in $d=3$ spatial dimension can be written as
\be\nonumber
\Pi^{\mu\nu}(P)\!\!&=&\!\!8e^2\sumintf_{\{K\}}\frac{K^{\mu}K^{\nu}}{K^2(K-P)^2}-4\delta^{\mu\nu}e^2\sumintf_{\{K\}}\frac{1}{K^2}
\\ &&
+\;2\delta^{\mu\nu}P^2e^2\sumintf_{\{K\}}\frac{1}{K^2(K-P)^2}-4e^2\sumintf_{\{K\}}\frac{K^{\mu}P^{\nu}+K^{\nu}P^{\mu}}{K^2(K-P)^2} \;.
\label{pimunu_trace}
\ee

As we are interested in the high-temperature limit, we may assume that $K \gg P$
because the leading contribution in $T$ to the loop integral is given by the region $K \sim T$. With this assumption, the 
self-energy in Eq.~(\ref{pimunu_trace}) simplifies to
\be
\Pi^{\mu\nu}(P)\;=\;8e^2\sumintf_{\{K\}}\frac{K^{\mu}K^{\nu}}{K^2(K-P)^2}-4\delta^{\mu\nu}e^2\sumintf_{\{K\}}\frac{1}{K^2}\;.
\label{htl-self}
\ee
We first consider the spatial components of $\Pi^{\mu\nu}(P)$. The sum over Matsubara frequencies can be evaluated using
\be
&&\hspace{-1cm} T\sum_{\{K_0\}}\frac{1}{K^2(P-K)^2}= \frac{1}{4k|{\bf p}-{\bf k}|}\Bigg\{\nonumber \\ 
\!\!&&\!\! \Big(1-n_F(k)-n_F(|{\bf p}-{\bf k}|)\Big)\left[\frac{-1}{iP_0-k-
|{\bf p}-{\bf k}|}+\frac{1}{iP_0+k+|{\bf p}-{\bf k}|}\right] \nonumber \\ 
&+&\!\!\! \Big(n_F(k)-n_F(|{\bf p}-{\bf k}|)\Big)\left[\frac{-1}{iP_0-k+|{\bf p}-{\bf k}|}+\frac{1}{iP_0+k-|{\bf p}-{\bf k}|}
\right]\!\Bigg\} \! ,\hspace{1.6cm}
\ee 
which is derived from a contour integral in the complex energy plane. The second term in Eq.~(\ref{htl-self}) 
is rather simple. We obtain 
\be
\Pi^{ij}(P)\!\!\!&=&\!\!\!-2e^2\delta^{ij}\int_{\bf k}\frac{1}{k}\left(1-2n_F(k)\right)+2e^2\int_{\bf k}\frac{k^ik^j}{
k|{\bf k}-{\bf p}|} \nn 
\!\!\!\!\!\!&\times&\!\!\!\!\!\!
\;\Bigg\{\!\Big(1-n_F(k)-n_F(|{\bf k}-{\bf p}|)\Big)\!\!\left[\frac{-1}{iP_0-k-|{\bf k}-{\bf p}|}+
\frac{1}{iP_0+k+|{\bf k}-{\bf p}|}\right] \nn
\!\!\!\!\!&+&\!\!\!\!\! \;\Big(n_F(k)-n_F(|{\bf k}-{\bf p}|)\Big)\!\!\left[\frac{-1}{iP_0-k+|{\bf k}-{\bf p}|}
+\frac{1}{iP_0+k-|{\bf k}-{\bf p}|}
\right]\Bigg\}\! , \hspace{1.5cm}
\label{omself}
\ee 
where $n_F(x)=1/(\exp(\beta x)+1)$ represents the Fermi-Dirac distribution function. The zero-temperature part of Eq.~(\ref{omself})
is logarithmically divergent in the ultraviolet region. This term depends on the external momentum and is canceled by standard 
zero-temperature wave-function renormalization. We next consider the terms that depend on temperature. In the case that the
loop momentum is soft, the Fermi-Dirac distribution functions can be approximated by a constant. The contribution from the
integral over the magnitude of $k$ is then of order $g^3$ and subleading. When the loop momentum is hard, one can expand the
terms in the integrand in powers of the external momentum. We can then make the following approximations 
\be
n_F(|{\bf k}-{\bf p}|)\!\!&\approx&\!\!n_F(k)-\frac{dn_F(k)}{dk}{\bf p}\!\cdot\!\hat{\bf k}\;,\\
|{\bf k}-{\bf p}|\!\!&\approx&\!\!k-{\bf p}\!\cdot\!\hat{\bf k}
\;,
\ee
where $\hat{\bf k}={\bf k}/k$ is a unit vector. Thus the angular integration decouples from the 
integral over the magnitude $k$. This implies
\be
\Pi^{ij}(P)\!\!&=&\!\!-\frac{2e^2}{\pi^2}\int_0^{\infty}dk\;k^2\,\frac{dn_F(k)}{dk}\int\frac{d\Omega}{4\pi}
\frac{-iP_0}{{-iP_0+\bf p}\!\cdot\!\hat{\bf k}} \hat{k}^i\hat{k}^j \;,
\nonumber \\
\!\!&=&\!\! \frac{e^2 T^2}{3}\int\frac{d\Omega}{4\pi}\frac{-iP_0}{{-iP_0+\bf p}\!\cdot\!\hat{\bf k}} \hat{k}^i\hat{k}^j \;.
\label{pi_ij_euclidean}
\ee
The results in Eq.~(\ref{pi_ij_euclidean}) can be analytically continued to the Minkowski space by replacing $iP_0\rightarrow p_0$ as
\be
\Pi^{ij}(P)
\!\!&=&\!\! \frac{e^2 T^2}{3}\int\frac{d\Omega}{4\pi}\frac{p_0}{{p_0-\bf p}\!\cdot\!\hat{\bf k}} \hat{k}^i\hat{k}^j \;.
\label{pi_ij}
\ee
Note that the momentum in Minkowski space is denoted with $P=(p_0,p=|\bf p|)$ whereas momentum in Euclidean space is denoted 
with  $P=(P_0,p=|\bf p|)$.
The other components of the self-energy tensor $\Pi^{\mu\nu}(P)$ are derived in the same manner or obtained
using the transversality of polarization tensor:
\be
P_{\mu}\Pi^{\mu\nu}(P)\;=\;0\;.
\ee
One find the other components of the self energy tensor from ~\cite{lebellac,Su:2011zv} as 
\be
\Pi^{00}(P)\!\!&=&\!\! \frac{e^2 T^2}{3} \left(\int\frac{d\Omega}{4\pi}\frac{p_0}{p_0-{\bf p}\!\cdot\!\hat{\bf k}} +1 \right)\;, 
\\ 
\Pi^{0j}(P)\!\!&=&\!\! \frac{e^2 T^2}{3} \int\frac{d\Omega}{4\pi}\frac{-p_0}{p_0-{\bf p}\!\cdot\!\hat{\bf k}}\hat{k}^j\;.
\ee
In $d$ dimensions, we can compactly write the self-energy tensor as
\be
\label{a1}
\Pi^{\mu\nu}(P)\;=\;m_D^2\left[{\mathcal T}^{\mu\nu}(P,-P)-n^{\mu}n^{\nu}\right]\;,
\label{scomp}
\ee
where $n$ specifies the thermal rest frame is canonically given by $n = (1,{\bf 0})$. We have defined 
\be
m_D^2\;=\;-4(d-1)e^2\sumintf_{\{K\}}\frac{1}{K^2}\;=\;\frac{e^2 T^2}{3}\;,
\ee
and the tensor ${\mathcal T}^{\mu\nu}(P,Q)$, which is defined only for momenta that satisfy $P+Q=0$, is
\be
{\mathcal T}^{\mu\nu}(P,-P)\;=\;\left \langle Y^{\mu}Y^{\nu}\frac{P\!\cdot\!n}{P\!\cdot\!Y}\right\rangle_{\bf\hat{y}} \;.
\label{T2-def}
\ee
The angular brackets indicate averaging over the spatial directions of the light-like vector $Y=(1,\hat{\bf y})$. 
The tensor ${\cal T}^{\mu\nu}$ is symmetric 
in $\mu$ and $\nu$ and satisfies the ``Ward identity''
\be
P_{\mu}{\cal T}^{\mu\nu}(P,-P)\;=\;P\!\cdot\!n\;n^{\nu}\;.
\label{ward-t2}
\ee
The self-energy tensor $\Pi^{\mu\nu}$ is therefore also symmetric in $\mu$ and $\nu$ and satisfies
\be
P_{\mu}\Pi^{\mu\nu}(P)\!\!&=&\!\!0\;,\\
\label{contr}
g_{\mu\nu}\Pi^{\mu\nu}(P)\!\!&=&\!\!-m_D^2\;.
\ee
The gluon self-energy tensor can be expressed in terms of two scalar functions, the transverse and longitudinal 
self-energies $\Pi_T$ and $\Pi_L$ as
\be
\Pi^{\mu\nu}(P)=-\Pi_T(P)T_P^{\mu\nu}+\frac{P^2}{p^2}\Pi_L(P)L_P^{\mu\nu},
\ee
where the tensor $T_P^{\mu\nu}$ and $L_P^{\mu\nu}$ are
\be
T_P^{\mu\nu}&=&g^{\mu\nu}-\frac{P^\mu P^\nu}{P^2}-\frac{n_P^\mu n_P^\nu}{n_P^2},\\
L_P^{\mu\nu}&=&\frac{n_P^\mu n_P^\nu}{n_P^2}
\label{tensor}
\ee
The four-vector $n_P^{\mu}$ is
\be
n_P^{\mu} \;=\; n^{\mu} - \frac{n\!\cdot\!P}{P^2} P^{\mu}
\ee
and satisfies $P\!\cdot\!n_P=0$ and $n^2_P = 1 - (n\!\cdot\!P)^2/P^2$. (\ref{contr}) reduces to the identity
\be
(d-1)\Pi_T(P)+\frac{1}{n^2_P}\Pi_L(P) \;=\; m_D^2 \;.
\label{PiTL-id}
\ee
We can express both self-energy functions in terms of the function ${\cal T}^{00}$, which is defined in Eq.~(\ref{T2-def}), as:
\be
\Pi_T(P)\!\!&=&\!\!\frac{m_D^2}{(d-1) n_P^2}\left[ {\cal T}^{00}(P,-P) - 1 + n_P^2  \right] \;,
\label{PiT-T}
\\
\Pi_L(P)\!\!&=&\!\!m_D^2\left[ 1- {\cal T}^{00}(P,-P) \right]\;,
\label{PiT-L}
\ee
In the tensor ${\cal T}^{\mu \nu}(P,-P)$ defined in~(\ref{T2-def}), the angular brackets indicate 
the angular average over the unit vector $\hat{\bf y}$. In almost all previous work, the angular 
average in~(\ref{T2-def}) has been taken in $d=3$ dimensions. For consistency of higher order 
corrections, it is essential to take the angular average in $d=3-2\epsilon$ dimensions and 
analytically continue to $d=3$ only after all poles in $\epsilon$ have been canceled.
Expressing the angular average as an integral over the cosine of an angle, the expression for 
the $00$ component of the tensor is
\be
{\cal T}^{00}(p,-p) \!\!&=&\!\! \frac{w(\epsilon)}{2}\int_{-1}^1dc\;(1-c^2)^{-\epsilon}\frac{p_0}{p_0-|{\bf p}|c} \;,
\label{T00-int}
\ee
where the weight function $w(\epsilon)$ is
\be
w(\epsilon)\;=\;\frac{\Gamma(2-2\epsilon)}{\Gamma^2(1-\epsilon)}\;2^{2\epsilon}
                    \;=\; \frac{\Gamma(\frac{3}{2}-\epsilon)}{\Gamma(\frac{3}{2}) \Gamma(1-\epsilon)} \;.
\label{weight}
\ee
The integral in (\ref{T00-int}) must be defined so that it is analytic at $p_0=\infty$. 
It then has a branch cut running from $p_0=-|{\bf p}|$ to $p_0=+|{\bf p}|$. If we take the 
limit $\epsilon\rightarrow 0$, it reduces to
\begin{eqnarray}
{\cal T}^{00}(P,-P)\;=\;\frac{p_0}{2|{\bf p}|}\log \frac{p_0 +|{\bf p}|}{p_0-|{\bf p}|}\;,
\end{eqnarray}
which is the expression that appears in the usual HTL self-energy functions.

In three dimensions, the self-energies $\Pi_T(P)$ and $\Pi_L(P)$
reduce to
\be
\Pi_T(P)\!\!&=&\!\!\frac{m_D^2}{2}\frac{p_0^2}{p^2}\left[1-\frac{P^2}{2 p_0 p}\log\frac{p_0+p}{p_0-p}\right]\;,
\label{redt} \\
\Pi_L(P)\!\!&=&\!\!m_D^2\left[1- \frac{p_0}{2p}\log\frac{p_0+p}{p_0-p}\right]\;.
\label{redl}
\ee

So far the HTL approximation for photon self energy tensor has been discussed. 
Now we will discuss below the HTL approximation for gluon self energy tensor.

\begin{figure}[tbh]
\begin{center}
\includegraphics[height=6cm,width=12cm]{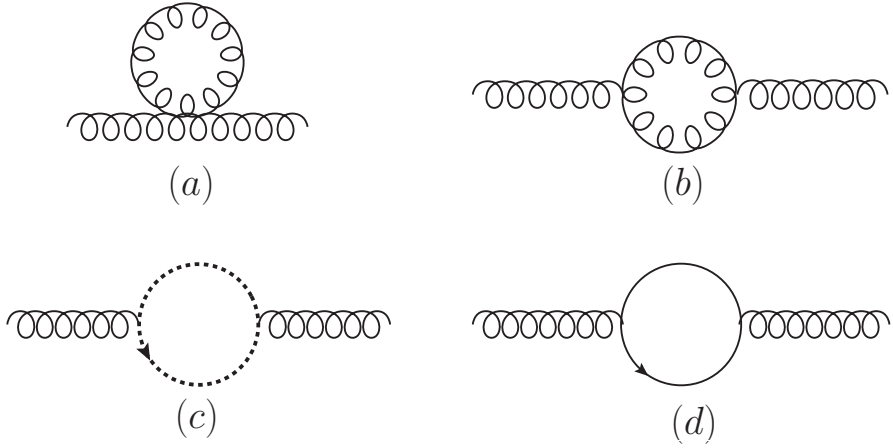}
\caption{Feynman diagrams for gluon self energy.}
\label{gluon_self}
\end{center}
\end{figure}

The hard thermal loop in the gluon self-energy was first calculated by Klimov and Weldon 
\cite{Klimov:1981ka,Klimov:1982bv,Weldon:1982aq}. In QCD, if one calculate gluon self energy, the Feynman diagrams 
that will contribute to one loop gluon self energy are shown in fig.~(\ref{gluon_self}).

If one calculate all the four diagrams, one get gluon self energy tensor in the same form as photon self energy~(\ref{scomp}),
but the Debye mass $m_D$ should be replaced by
\be
m_D^2\;=\;g^2 \left[(d-1)^2 c_A\sumintb_{K}\frac{1}{K^2}-4(d-1) s_F \sumintf_{\{K\}}\frac{1}{K^2}\right]\;,
\label{qcdmd}
\ee
where, with the standard normalization, the QCD Casimir numbers are $c_A=N_c$ and $s_F=N_f/2$. $N_c$ and $N_f$
denotes number of colors and number of flavors.  In case of three spatial dimension {\em i.e.} for $d=3$, the Debye mass
for gluon becomes
\be
m_D^2\;=\;\frac{ g^2 T^2}{3} \left[ c_A + s_F\lb1+12\hmu^2\rb \right] \; .
\label{qcdmd3}
\ee
\subsection{Gluon propagator}
\label{app:prop}

The Feynman rule for the gluon propagator in Minkowski space is
\be
i \delta^{a b} \Delta_{\mu\nu}(P) \;,
\ee
where the gluon propagator tensor $\Delta_{\mu\nu}$ depends on the choice of gauge fixing. We consider two
possibilities that introduce an arbitrary gauge parameter $\xi$:  general covariant gauge and general 
Coulomb gauge. In both cases, the inverse propagator reduces in the limit $\xi\rightarrow\infty$ to
\be
\Delta^{-1}_{\infty}(P)^{\mu\nu}\;=\;-P^2 g^{\mu \nu} + P^\mu P^\nu - \Pi^{\mu\nu}(P)\;.
\label{delta-inv:inf0}
\ee
This can also be written
\be
\Delta^{-1}_{\infty}(P)^{\mu\nu} \;=\; - \frac{1}{\Delta_T(P)} T_P^{\mu\nu} + \frac{1}{n_P^2 \Delta_L(P)} L_P^{\mu\nu}\;,
\label{delta-inv:inf}
\ee
where the transverse tensor $T_P^{\mu\nu}$ and longitudinal tensor $L_P^{\mu\nu}$ are defined in Eq.~(\ref{tensor}).
$\Delta_T(P)$ and $\Delta_L(P)$ are the transverse and longitudinal propagators related to $\Pi_L(P)$ $\Pi_T(P)$ as
\be
\Delta_T(P)\!\!&=&\!\!\frac{-1}{P^2+\Pi_T(P)}\;,
\label{Delta-T:M} \\
\Delta_L(P)\!\!&=&\!\!\frac{1}{p^2+\Pi_L(P)}\;.
\label{Delta-L:M}
\ee
The inverse propagator for general $\xi$ is
\be
\Delta^{-1}(P)^{\mu\nu}\!\!&=&\!\!\Delta^{-1}_{\infty}(P)^{\mu\nu}-\frac{1}{\xi}P^{\mu}P^{\nu}\hspace{4.4cm}\mbox{covariant}\;,
\label{Delinv:cov} \\
\!\!&=&\!\!\Delta^{-1}_{\infty}(P)^{\mu\nu}-\frac{1}{\xi}\left(P^{\mu}-P\!\cdot\!n\;n^{\mu}\right)\left(P^{\nu}-P\!\cdot\!n
\;n^{\nu}\right)\hspace{0.4cm}\mbox{Coulomb} \;.
\label{Delinv:C}
\ee
The propagators obtained by inverting the tensors in~(\ref{Delinv:C}) and~(\ref{Delinv:cov}) are
\be
\Delta^{\mu\nu}(P)\!\!&=&\!\!-\Delta_T(P)T_P^{\mu\nu} +\Delta_L(P)n_P^{\mu}n_P^{\nu} - \xi \frac{P^{\mu}P^{\nu}}
{P^4}\hspace{2cm}\mbox{covariant}\;,
\label{D-cov}
\\
\!\!&=&\!\!-\Delta_T(P)T_P^{\mu\nu}+\Delta_L(P)n^{\mu}n^{\nu}-\xi\frac{P^{\mu}P^{\nu}}{(n_P^2P^2)^2}\hspace{1.6cm}\mbox{Coulomb} \;.
\label{D-C}
\ee

It is convenient to define the following combination of propagators:
\be
\Delta_X(P) \;=\; \Delta_L(P)+\frac{1}{n_P^2}\Delta_T(P) \;.
\label{Delta-X}
\ee
Using (\ref{PiTL-id}), (\ref{Delta-T:M}), and (\ref{Delta-L:M}), it can be expressed in the alternative form
\be
\Delta_X(P) \;=\;\left[ m_D^2 - d \, \Pi_T(P) \right] \Delta_L(P) \Delta_T(P) \;,
\label{Delta-X:2}
\ee
which shows that it vanishes in the limit $m_D \to 0$. In the covariant gauge, the propagator tensor can be written
\be\nonumber
\Delta^{\mu\nu}(P)\!\!&=&\!\!\left[ - \Delta_T(P) g^{\mu \nu} + \Delta_X(P) n^\mu n^\nu \right] - \frac{n \!\cdot\! P}{p^2} 
\Delta_X(P) \left( P^\mu n^\nu  + n^\mu P^\nu \right)
\\
&&
+ \left[ \Delta_T(P) + \frac{(n \!\cdot\! P)^2}{P^2} \Delta_X(P)
        - \frac{\xi}{P^2} \right] {P^\mu P^\nu}{P^2} \;.
\label{gprop-TC}
\ee
This decomposition of the propagator into three terms has proved to be particularly convenient for explicit calculations.
For example, the first term satisfies the identity
\be
\left[- \Delta_T(P) g_{\mu \nu} + \Delta_X(P) n_\mu n_\nu \right] \Delta^{-1}_{\infty}(P)^{\nu\lambda}=
g_\mu^\lambda - \frac{P_\mu P^\lambda}{P^2}
+ \frac{n \!\cdot\! P}{n_P^2 P^2} \frac{\Delta_X(P)}{\Delta_L(P)}
        P_\mu n_P^\lambda.
\label{propid:2}
\ee
The zeros of the denominators of the Eq.~(\ref{D-cov}) or (\ref{D-C}) gives the dispersion laws for transverse and longitudinal
gluon, {\em i.e.}
\be
\Delta_T(p_0=\omega_T,p)=0,\hspace{1cm}
\Delta_L(p_0=\omega_L,p)=0.
\ee

\begin{figure}[tbh]
\begin{center}
\includegraphics[width=12cm]{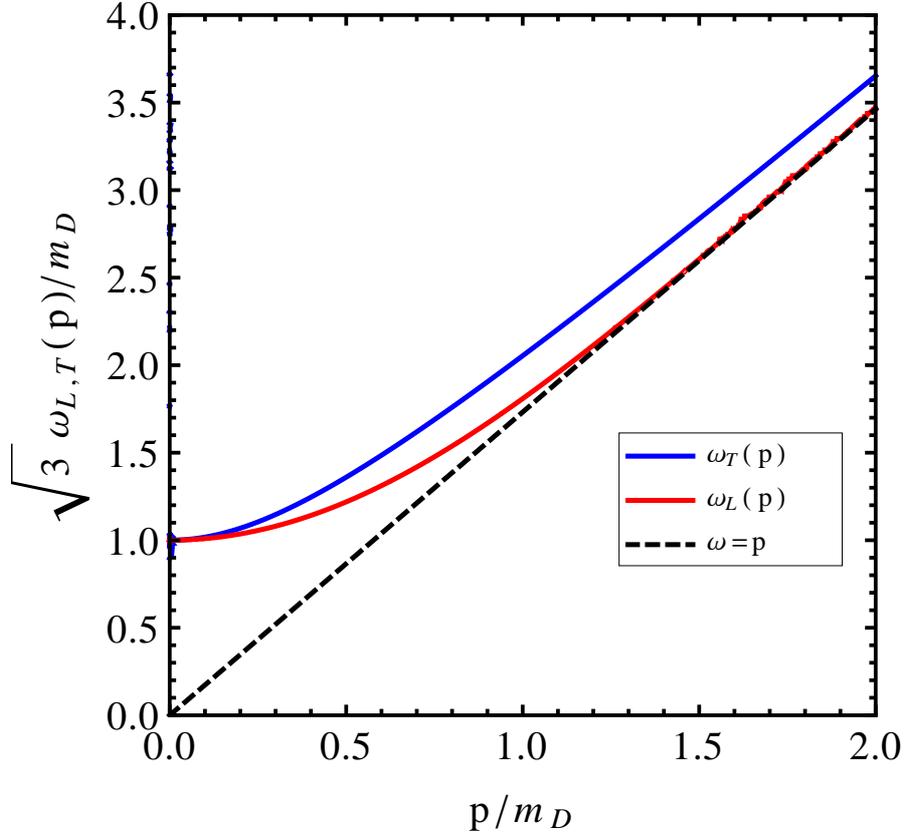}
\end{center}
\caption{Dispersion laws for photon/gluon.}
\label{disp_photon}
\end{figure}

The dispersion laws are illustrated in Fig.~({\ref{disp_photon}}).

It is possible to find approximate analytic solution of $\omega_{L,T}$ for small and large values of momentum.

For small value of momentum $(p \ll m_D)$,
\be
\omega_T&\approx&\frac{m_D}{\sqrt{3}}\lb1+\frac{9}{5}\frac{p^2}{m_D^2}-\frac{81}{35}\frac{p^4}{m_D^4}
               +\frac{792}{125}\frac{p^6}{m_D^6}\rb,\\
\omega_L&\approx&\frac{m_D}{\sqrt{3}}\lb1+\frac{9}{10}\frac{p^2}{m_D^2}-\frac{27}{280}\frac{p^4}{m_D^4}
               +\frac{9}{2000}\frac{p^6}{m_D^6}\rb
\ee
For large value of momentum $(p \gg m_D)$,
\be
\omega_T\!\!\!&\approx&\!\!\!p+\frac{m_D^2}{4p}+\frac{m_D^4}{32p^3}\[3-2\ln\frac{8p^2}{m_D^2}\]+
                  \frac{m_D^6}{128p^5}\[2\ln^2\frac{8p^2}{m_D^2}-10\ln\frac{8p^2}{m_D^2}+7\],\hspace{2cm} \\
\omega_L&\approx& p+ 2p\exp{\lb-\frac{2\lb p^2+m_D^2\rb }{m_D^2}\rb}.
\ee

\subsection{Fermionic propagator}
\begin{figure}[tbh]
\begin{center}
 \includegraphics[width=8cm]{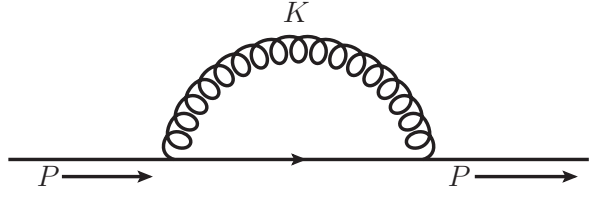}
\end{center}
\caption{One loop quark self energy.}
\label{fermion_self}
\end{figure}

We want to examine the electron (quark) propagator within HTL approximation. We must thus evaluate the Feynman diagram
in Fig.(~\ref{fermion_self}).

The full electron propagator in Euclidean space can be written as
\be
S(P)=\frac{1}{\slash\!\!\!\! P+m+\Sigma}
\ee
\subsubsection*{Evaluation of $\Sigma$}
Electron self energy in Feynman gauge is
\be
\Sigma(P)=-e^2\sumintb_K\ \frac{\gamma_\mu(\Slash K -\Slash P)\gamma^\mu}{K^2(P-K)^2}
\ee
In HTL approximation, we can neglect $P$ w.r.t. $K$. After contracting the Dirac matrices, the electron self-energy in $d=3$
dimension reduces to
\be
\Sigma(P)\approx -2e^2\sumintb_K \frac{\Slash K}{K^2(P-K)^2}
\ee
The necessary frequency sum can be performed using contour integration techniques and the results are
\be
&&\hspace{-1cm} T\sum_{K_0}\frac{1}{K^2(P-K)^2}= -\frac{1}{4k|{\bf p}-{\bf k}|}\Bigg\{\nonumber \\ 
\!\!&&\!\! \Big(1+n_B(k)-n_F(|{\bf p}-{\bf k}|)\Big)\left[\frac{1}{iP_0-k-
|{\bf p}-{\bf k}|}-\frac{1}{iP_0+k+|{\bf p}-{\bf k}|}\right] \nonumber \\ 
&+&\!\!\! \Big(n_B(k)+n_F(|{\bf p}-{\bf k}|)\Big)\left[\frac{1}{iP_0+k-|{\bf p}-{\bf k}|}+\frac{1}{iP_0-k+|{\bf p}-{\bf k}|}
\right]\!\Bigg\} \! ,\hspace{1.6cm}
\label{fermion_sumint1}
\ee 
and
\be
&&\hspace{-1cm} T\sum_{K_0}\omega_n\frac{1}{K^2(P-K)^2}= \frac{i}{4|{\bf p}-{\bf k}|}\Bigg\{\nonumber \\ 
\!\!&&\!\! \Big(1+n_B(k)-n_F(|{\bf p}-{\bf k}|)\Big)\left[\frac{1}{iP_0-k-
|{\bf p}-{\bf k}|}+\frac{1}{iP_0+k+|{\bf p}-{\bf k}|}\right] \nonumber \\ 
&-&\!\!\! \Big(n_B(k)+n_F(|{\bf p}-{\bf k}|)\Big)\left[\frac{1}{iP_0+k-|{\bf p}-{\bf k}|}+\frac{1}{iP_0-k+|{\bf p}-{\bf k}|}
\right]\!\Bigg\} \! ,\hspace{1.6cm}
\label{fermion_sumint2}
\ee 
where $n_B(x)=1/(\exp(\beta x)-1)$ and $n_F(x)=1/(\exp(\beta x)+1)$ as before.
The first term in the square bracket of Eq.~(\ref{fermion_sumint1}) leads to a behavior that is linear in $T$, and 
the corresponding term in Eq.~(\ref{fermion_sumint2}) is even more convergent; thus these terms are non-leading in $T$.
Only the second terms of both the Eqs.~(\ref{fermion_sumint1}) and ~(\ref{fermion_sumint2}) contributes to order $T^2$.
Using this knowledge, we can easily obtain the following results:
\be
\sumintb_K\ \frac{k_i}{K^2(P-K)^2}=-\frac{T^2}{16}\int\frac{d\Omega}{4\pi}\frac{\hat k_i}{P\cdot\hat K}
\ee
and
\be
\sumintb_K\ \frac{\omega_n}{K^2(P-K)^2}=-\frac{iT^2}{16}\int\frac{d\Omega}{4\pi}\frac{1}{P\cdot\hat K}
\ee
Which leads to
\be
\Sigma(P)&=&\frac{e^2T^2}{8}\int\frac{\Slash \hat K}{P\cdot\hat K}\hspace{2cm} \hat K=(-i,\hat{\bf k})
\nn
&=&\frac{e^2T^2}{8}\int\frac{\Slash Y}{P\cdot Y}\hspace{2cm}  Y=(-i,\hat{\bf y})
\nn
&=&m_f^2{\mathcal T}\!\!\!\!/(P)\;,
\label{selfq}
\ee
where
\be
\label{deftf}
{\mathcal T}^{\mu}(P) \;=\; -\left\langle\frac{Y^{\mu}}{P\cdot Y}\right\rangle_{\hat{\bf Y}}\hspace{2cm} Y\equiv(-i,{\hat{\bf y}}),
\ee
and $m_f$ is the thermal electron mass
\be
m_f^2\;=\;-3e^2\sumintf_{\{K\}}\frac{1}{K^2}=\frac{e^2T^2}{8}\;.
\ee
In Minkowski space $\Sigma(P)$ has same form as in Euclidean space as
\be
\Sigma(P)=\frac{e^2T^2}{8}\int\frac{\slash\!\!\!\! Y}{P\cdot Y}\hspace{2cm}  Y=(1,\hat{\bf y})
\label{sigma_minkowski}
\ee
The corresponding retarded dressed electron propagator for massless case reads
\be
S(P)=\frac{i}{\Slash P-\Sigma}
\label{prop_minkowski}
\ee

Now, one can evaluate electron self energy in Eq.~(\ref{sigma_minkowski}) as
\be
\Sigma(p_0,p)=\frac{m_f^2}{p_0}\gamma_0{\cal T}_P+\frac{m_f^2}{p}\gamma\cdot\hat{\bf p}
\lb1-{\cal T}_P\rb
\ee
where ${\cal T}_P$ is $00$ component of the tensor ${\cal T}^{\mu\nu}(p,-p)$
and given in Eq.~(\ref{T00-int}) as
\be
{\cal T}_P(p,-p) \!\!&=&\!\! \frac{w(\epsilon)}{2}\int_{-1}^1dc\;(1-c^2)^{-\epsilon}\frac{p_0}{p_0-|{\bf p}|c} 
\nn
&=&\left\langle\frac{p_0^2}{p_0^2-p^2c^2}\right\rangle_c
\label{tauP}
\ee

We can rewrite the inverse of electron propagator from Eq.~(\ref{prop_minkowski}) as
\be
iS^{-1}(P)=\slash\!\!\!\!P-\Sigma(P)=A_0\gamma_0-A_S\gamma\cdot\hat{\bf p}
\ee
with
\be
A_0=p_0-\frac{m_f^2}{p_0}{\cal T}_P \hspace{1cm} 
A_S=p+\frac{m_f^2}{p}\lb1-{\cal T}_P\rb
\label{A0AS_tauP}
\ee
The above expressions can be written three spatial dimension as
\be
A_0=p_0-\frac{m_f^2}{p}\ln\frac{p_0+p}{p_0-p} \hspace{1cm} 
A_S=p+\frac{m_f^2}{p}\lb1-\frac{p_0}{p}\ln\frac{p_0+p}{p_0-p}\rb
\label{A0AS}
\ee

The physical interpretation of any calculation is particularly transparent if we rewrite
\be
-iS(P)=\frac{\gamma_0-\gamma\cdot\hat{\bf p}}{2D_+(p_0,p)}+\frac{\gamma_0+\gamma\cdot\hat{\bf p}}{2D_-(p_0,p)},
\label{electron_prop}
\ee
where
\be
D_\pm(p_0,p)&=&A_0\mp A_S \nn
&=&p_0\mp p-\frac{m_f^2}{2p}\left[\left(1\mp\frac{p_0}{p}\right)\ln\frac{p_0+p}{p_0-p} \pm 2 \right]
\label{Dpm}
\ee

\begin{figure}[tbh]
\begin{center}
\includegraphics[width=12cm]{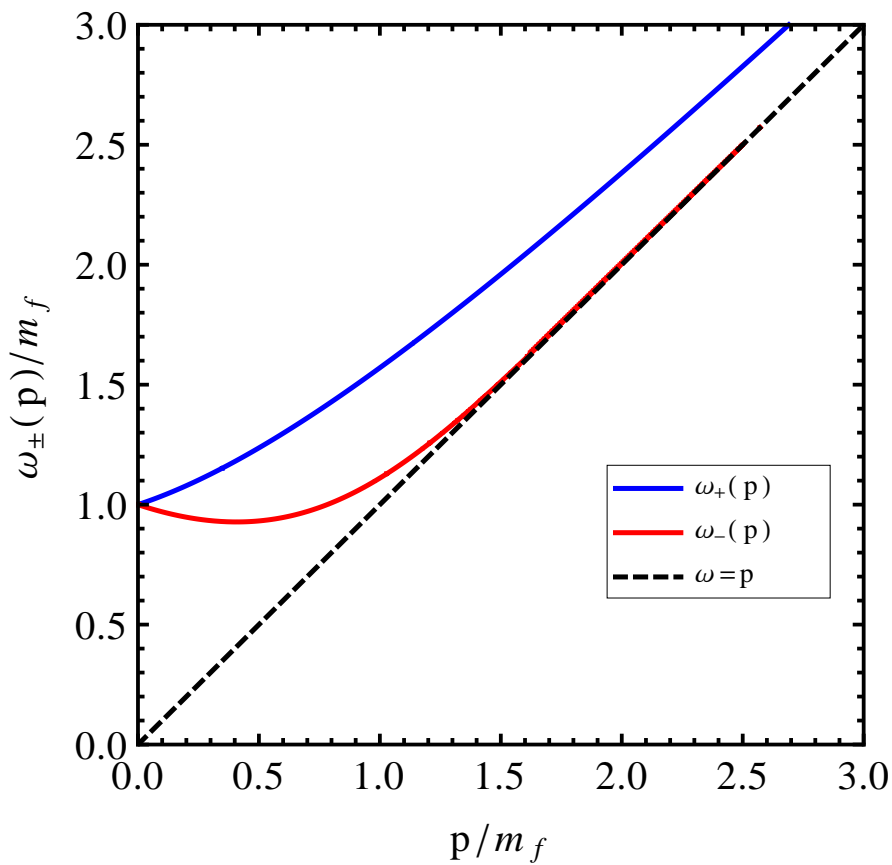}
\end{center}
\caption{Dispersion laws for fermionic excitations.}
\label{disp}
\end{figure}

As $D_\pm(p_0,p)$ has an imaginary part for space-like values of $P(p_0^2<p^2)$, it is useful to note the parity 
properties for both real and imaginary values of $D_\pm(p_0,p)$ as 
\be
\mbox{Re}\ D_+(p_0,p)&=&-\mbox{Re}\ D_-(-p_0,p)
\nn
\mbox{Im}\ D_+(p_0,p)&=&\mbox{Im}\ D_-(-p_0,p)
\label{parity}
\ee
The zeros of the denominators in Eq.~(\ref{electron_prop}) {\em i.e.} the solution of $D_\pm(p_0,p)=0$ give the position 
of the quasi-particle poles. One finds that $D_+(p_0,p)=A_0-A_S=0$ has two solutions:
\be
p_0=\omega_+(p)\hspace{2cm}
p_0=-\omega_-(p),
\ee
and, from the parity properties of Eq.~(\ref{parity}), the equation $D_-(p_0,p)=A_0+A_S=0$ has also two solutions:
\be
p_0=\omega_-(p)\hspace{2cm}
p_0=-\omega_+(p),
\ee
where $\omega_\pm(p)$ is chosen to be positive. The dispersion laws of $\omega_\pm(p)$ are illustrated in Fig.~(\ref{disp})
along-with the dispersion law of free massless fermion(dashed line).

It is possible to find approximate analytic solution of $\omega_\pm(p)$ as in the case of $\omega_{L,T}(p)$
for small and large values of momentum.

For small values of momentum $(p\ll m_f)$ the dispersion laws are
\be
\omega_+(p)&\approx& m_f + \frac{1}{3}p + \frac{1}{3}\frac{p^2}{m_f^2}-\frac{16}{135}\frac{p^3}{m_f^3},\nn
\omega_-(p)&\approx& m_f - \frac{1}{3}p + \frac{1}{3}\frac{p^2}{m_f^2}+\frac{16}{135}\frac{p^3}{m_f^3}
\ee
and for large value of momentum $(m_f\ll p \ll T)$,
\be
\omega_+(p)&\approx& p+\frac{m_f^2}{p}-\frac{m_f^4}{2p^3}\ \ln\frac{m_f^2}{2p^2}+\frac{m_f^6}{4p^5}\left[\ln^2\frac{m_f^2}{2p^2}
+\ln\frac{m_f^2}{2p^2}-1\right],\\
\omega_-(p)&\approx& p+2p \exp\lb{-\frac{2p^2+m_f^2}{m_f^2}}\rb.
\ee
\subsubsection*{Quark Self-energy(QCD)}
In the previous section, we discussed about the self energy propagator for electron. In case of QCD,
quark self energy in Feynman gauge can be written as
\be
\Sigma(P)\delta_{ij}=-\sumintb_K\ (g\gamma_\mu (t^a)_{ik})(\Slash K -\Slash P)(g\gamma^\mu(t^a)_{kj}) \frac{1}{K^2(P-K)^2}
\ee
Using the identity $(t^a t^a)_{ij}=\frac{N_c^2-1}{2N_c}\delta_{ij}=C_F\delta_{ij}$ and after doing the frequency sum, the above quark
self energy expression reduces to 
\be
\Sigma(P)\;=\;m_q^2{\mathcal T}\!\!\!\!/(P)\;,
\ee
which is the same form of electron self energy as in Eq.~(\ref{selfq}) but thermal electron mass $m_f$ should be replaced with
quark thermal mass and which can be derive as
\be
m_q^2 &=& -3C_Fg^2\sumintf_{\{K\}}\frac{1}{K^2}\;
\nn
&=&\frac{g^2T^2}{8}C_F\lb1+4\hmu^2\rb.
\label{mq_1loop}
\ee


\subsection{Three point quark gluon vertex}
In gauge theories, there are also hard thermal loops involving vertices. For instance, the one-loop correction to the
three-point function in QCD, can compactly be written as
\be
\Gamma^{\mu}(P,Q,R)\;=\;\gamma^{\mu}-m_q^2\tilde{{\mathcal T}}^{\mu}(P,Q,R)\;,
\label{qqg}
\ee
where the tensor in the HTL correction term is only defined for $P-Q+R=0$:
\be
\tilde{{\mathcal T}}^{\mu}(P,Q,R)\;=\;\left\langle Y^{\mu}\left(\frac{Y\!\!\!\!/}{(Q\!\cdot\!Y)(R\!\cdot\!Y)}\right)\right
\rangle_{\hat{\bf Y}}\;.
\ee
\begin{figure}[tbh]
 \begin{center}
  \includegraphics[width=12cm]{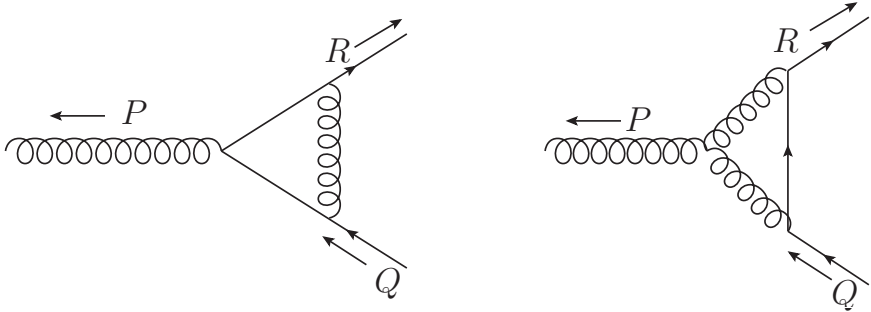}
 \end{center}
\caption{One loop correction to the three-point quark gluon vertex.}
\end{figure}

The quark-gluon vertex satisfies the Ward identity
\be
P^{\mu}\Gamma^{\mu}(P,Q,R)\;=\;S^{-1}(Q)-S^{-1}(R)\;,
\label{qward1}
\ee
where $S(Q)$ is the resummed effective fermion propagator. 

\subsection{Quark-gluon four-vertex}
We define the quark-gluon four-point vertex with outgoing gluon 
        momenta $P$ and $Q$, incoming quark
momentum $R$, and outgoing
        quark  momentum $S$. It reads
\be
\delta^{ab} \Gamma^{\mu\nu}_{abij}(P,Q,R,S) \!\!&=&\!\! 
    - g^2 m_q^2 c_F \delta_{ij} \tilde{\cal T}^{\mu\nu}(P,Q,R,S)  \\
\!\!&\equiv&\!\!g^2 c_F \delta_{ij} \Gamma^{\mu\nu}  \, ,
\label{4qgv}
\ee
There is no tree-level term. The tensor in the 
HTL correction term is only defined for $P+Q-R+S=0$,
\be\nonumber
\tilde{{\cal T}}^{\mu\nu}(P,Q,R,S)
&=&\left\langle
Y^{\mu}Y^{\nu}\left(\frac{1}{R\!\cdot\!Y}+\frac{1}{S\!\cdot\!Y}\right)
\right.\\&&\times \left.
\frac{\Slash Y}{[(R-P)\!\cdot\!Y]\;[(S+P)\!\cdot\!Y]}
\right\rangle .
\ee
This tensor is symmetric in $\mu$ and $\nu$ and is traceless.
It satisfies the Ward identity:
\be
\!\!P_{\mu}\Gamma^{\mu\nu}(P,Q,R,S)\!=\!\Gamma^{\nu}(Q,R-P,S)-
\Gamma^{\nu}(Q,R,S+P)\,.
\label{qward2}
\ee
\subsection{Three gluon vertex}
\begin{figure}[tbh!]
 \begin{center}
  \includegraphics[height=8cm,width=10cm]{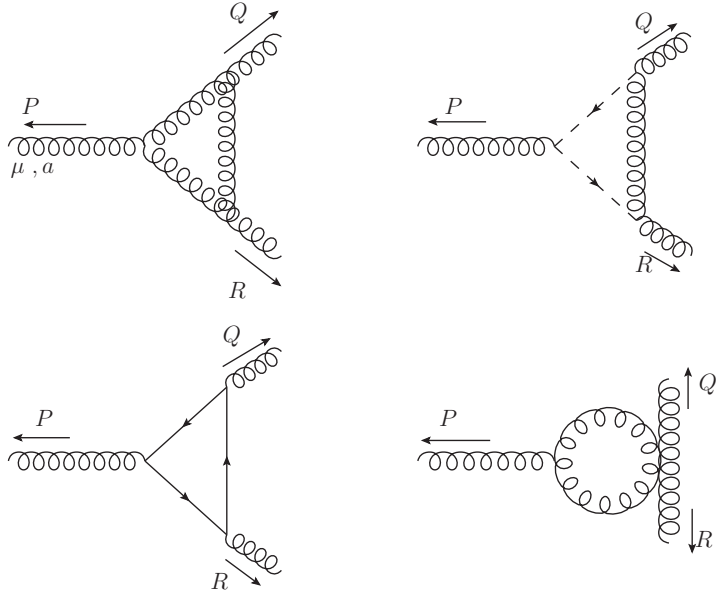}
 \end{center}
\caption{One-loop correction to three gluon vertex.}
\end{figure}
The three-gluon vertex
for gluons with outgoing momenta $P$, $Q$, and $R$,
Lorentz indices $\mu$, $\nu$, and $\lambda$,
and color indices $a$, $b$, and $c$ is
\be
i\Gamma_{abc}^{\mu\nu\lambda}(P,Q,R)=-gf_{abc}
\Gamma^{\mu\nu\lambda}(P,Q,R)\;,
\ee
%
where $f^{abc}$ are the structure constants and
the three-gluon vertex tensor is
\be
\Gamma^{\mu\nu\lambda}(P,Q,R)&=&
g^{\mu\nu}(P-Q)^{\lambda}+
g^{\nu\lambda}(Q-R)^{\mu}\nn
 &&
+\ 
g^{\lambda\mu}(R-P)^{\nu}
-m_D^2{\cal T}^{\mu\nu\lambda}(P,Q,R)\;.
\label{Gam3}
\ee
%
The tensor ${\cal T}^{\mu\nu\lambda}$ in the HTL correction term
is defined only for $P+Q+R=0$:
\be
{\cal T}^{\mu\nu\lambda}(P,Q,R) \;=\;
 - \Bigg\langle Y^{\mu} Y^{\nu} Y^{\lambda}
\left( \frac{P\!\cdot\!n}{P\!\cdot\!Y\;Q\!\cdot\!Y}
	- \frac{R\!\cdot\!n}{\!R\cdot\!Y\;Q\!\cdot\!Y} \right)
	\Bigg\rangle_{\hat\bf y}\;.
\label{T3-def}
\ee
%
This tensor is totally symmetric in its three indices and traceless in any
pair of indices: $g_{\mu\nu}{\cal T}^{\mu\nu\lambda}=0$.
It is odd (even) under odd (even) permutations of the momenta $P$, $Q$, and
$R$. It satisfies the ``Ward identity''
\be
Q_{\mu}{\cal T}^{\mu\nu\lambda}(P,Q,R) \;=\;
{\cal T}^{\nu\lambda}(P+Q,R)-
{\cal T}^{\nu\lambda}(P,R+Q)\;.
\label{ward-t3}
\ee
%
The three-gluon vertex tensor therefore satisfies the Ward identity
\be
P_{\mu}\Gamma^{\mu\nu\lambda}(P,Q,R) \;=\;
\Delta_{\infty}^{-1}(Q)^{\nu\lambda}-\Delta_{\infty}^{-1}(R)^{\nu\lambda}\;.
\label{ward-3}
\ee
\subsection{Four-gluon Vertex}
The four-gluon vertex
for gluons with outgoing momenta $P$, $Q$, $R$, and $S$,
Lorentz indices $\mu$, $\nu$, $\lambda$, and $\sigma$,
and color indices $a$, $b$, $c$, and $d$ is
\be
i\Gamma^{\mu\nu\lambda\sigma}_{abcd}(P,Q,R,S) &=&
- ig^2\big\{ f_{abx}f_{xcd} \left(g^{\mu\lambda}g^{\nu\sigma}
				-g^{\mu\sigma}g^{\nu\lambda}\right)
\nonumber\\&&\hspace{-2.667cm}
+2m_D^2\mbox{tr}\left[T^a\left(T^bT^cT^d+T^dT^cT^b
\right)\right]{\cal T}^{\mu\nu\lambda\sigma}(P,Q,R,S)
\big\}
\nonumber
\\
&&
\hspace{-2.667cm}
+ \; 2 \; \mbox{cyclic permutations}\;,
\ee
%
where the cyclic permutations are of
$(Q,\nu,b)$, $(R,\lambda,c)$, and $(S,\sigma,d)$.
The matrices $T^a$ are the fundamental representation
of the $SU(3)$ algebra with the standard normalization
${\rm tr}(T^a T^b) = \frac{1}{2} \delta^{ab}$.
The tensor ${\cal T}^{\mu\nu\lambda\sigma}$
in the HTL correction term is defined only for $P+Q+R+S=0$:
\be
{\cal T}^{\mu\nu\lambda\sigma}(P,Q,R,S) \!\!\!&=&\!\!\!
\Bigg\langle Y^{\mu} Y^{\nu} Y^{\lambda} Y^{\sigma}
\left( \frac{P\!\cdot\!n}{P\!\cdot\!Y \; Q\!\cdot\!Y \; (Q+R)\!\cdot\!Y}
\right.
\nonumber\\
\!\!\!&+&\!\!\!\!
\left.\frac{(P+Q)\!\cdot\!n}{Q\!\cdot\!Y\;R\!\cdot\!Y\;(R+S)\!\cdot\!Y}
+\frac{(P+Q+R)\!\cdot\!n}{R\!\cdot\!Y\;S\!\cdot\!Y\;(S+P)\!\cdot\!Y}\right)
\!\!\!\Bigg\rangle.\ 
\label{T4-def}
\ee
%
This tensor is totally symmetric in its four indices and traceless in any
pair of indices: $g_{\mu\nu}{\cal T}^{\mu\nu\lambda\sigma}=0$.
It is even under cyclic or anti-cyclic
permutations of the momenta $P$, $Q$, $R$, and $S$.
It satisfies the ``Ward identity''
\be
q_{\mu}{\cal T}^{\mu\nu\lambda\sigma}(P,Q,R,S)&=&
{\cal T}^{\nu\lambda\sigma}(P+Q,R,S)
-{\cal T}^{\nu\lambda\sigma}(P,R+Q,S)
\label{ward-t4}
\ee
%
and the ``Bianchi identity''
\be
{\cal T}^{\mu\nu\lambda\sigma}(P,Q,R,S)
+ {\cal T}^{\mu\nu\lambda\sigma}(P,R,S,Q)+
{\cal T}^{\mu\nu\lambda\sigma}(P,S,Q,R)=0\;.
\label{Bianchi}
\ee
%
When its color indices are traced in pairs, the four-gluon vertex becomes
particularly simple:

\be
\delta^{ab} \delta^{cd} i \Gamma_{abcd}^{\mu\nu\lambda\sigma}(P,Q,R,S)
&=& -i g^2 N_c (N_c^2-1) \Gamma^{\mu\nu,\lambda\sigma}(P,Q,R,S) \;,
\ee
%
where the color-traced four-gluon vertex tensor is
\be
\Gamma^{\mu\nu,\lambda\sigma}(P,Q,R,S)=
2g^{\mu\nu}g^{\lambda\sigma}
-g^{\mu\lambda}g^{\nu\sigma}
-g^{\mu\sigma}g^{\nu\lambda}
-m_D^2{\cal T}^{\mu\nu\lambda\sigma}(P,S,Q,R)\;.
\label{Gam4}
\ee
Note the ordering of the momenta in the arguments of the tensor
${\cal T}^{\mu\nu\lambda\sigma}$, which comes from the use of the
Bianchi identity (\ref{Bianchi}).
The tensor (\ref{Gam4}) is symmetric
under the interchange of $\mu$ and $\nu$,
under the interchange of $\lambda$ and $\sigma$,
and under the interchange of $(\mu,\nu)$ and $(\lambda,\sigma)$.
It is also symmetric under the interchange of $P$ and $Q$,
under the interchange of $R$ and $S$,
and under the interchange of $(P,Q)$ and $(R,S)$.
It satisfies the Ward identity
\be
p_{\mu}\Gamma^{\mu\nu,\lambda\sigma}(P,Q,R,S)
&=&\Gamma^{\nu\lambda\sigma}(Q,R+P,S)
-\Gamma^{\nu\lambda\sigma}(Q,R,S+P)\;.
\label{ward-4}
\ee
In QED there are, in fact, infinitely many amplitudes with hard thermal loops. To be precise, there are hard thermal loops in
all $n$-point functions with two fermion lines and $n-2$ photon lines. In non-Abelian gauge theories such as QCD, there are in
addition hard thermal loops in amplitudes with $n$ gluon lines \cite{Braaten:1989mz}.
\subsection{The HTL effective lagrangian in QCD}\label{htl_lag_sec}
The preceding results can be summarized in a compact way by writing an effective Lagrangian:
Hard-thermal-loop perturbation theory is a reorganization of in-medium perturbation theory for QCD.
The HTL perturbation theory Lagrangian density can be written as
\be
 {\cal L}=\left.({\cal L}_{\rm QCD}+{\cal L}_{\rm HTL})\right|_{g\rightarrow\sqrt{\delta}g}+\Delta{\cal L}_{\rm HTL} \, , 
\label{total_lag} 
\ee
where the QCD Lagrangian density ${\cal L}_{\rm QCD}$ is defined in Eq.~(\ref{qcd_lag}) of chapter~\ref{chapter:introduction}.
The HTL improvement term ${\cal L}_{\rm HTL}$ can be written as~\cite{Braaten:1991gm} 
\be
 {\cal L}_{\rm HTL}&=&(1-\delta)i m_q^2\bar\psi\gamma^\mu\left\langle\frac{Y_\mu}{Y\cdot\! D}\right\rangle_{\!\hat{\bf y}}\psi
\nn
&& -\frac{1}{2}(1-\delta)m_D^2 {\rm Tr}\lb F_{\mu\alpha}\left\langle\frac{Y^\alpha Y_\beta}{(Y\cdot\! D)^2}
    \right\rangle_{\!\hat{\bf y}} F^{\mu\beta}\rb \, ,
\label{htl_lag}
\ee
where $Y^\mu = (1, {\bf\hat{y}})$ is a light-like four-vector with ${\bf\hat{y}}$ being a three-dimensional unit vector
and the angular bracket indicates an average over the direction of ${\bf\hat{y}}$ and $\Delta{\cal L}_{\rm HTL}$
contains the counterterms necessary to cancel additional ultraviolet divergences generated by HTL perturbation theory.
The two parameters $m_D$ and $m_q$ 
can be identified with the Debye screening mass and the thermal quark mass, respectively, and account for screening effects.
There are various prescription to choose the parameters $m_D$ and $m_q$. In one loop QCD thermodynamic calculations 
in Chapter~\ref{chapter:1loop}, one loop perturbative mass prescription as given in Eq~(\ref{qcdmd3}) for $m_D$ and in 
Eq~(\ref{mq_1loop}) for $m_q$ , will be used. In two loop QCD thermodynamic calculations 
in Chapter~\ref{chapter:2loop}, we will see that thermodynamical quantities are correct up to $g^3$ order if one expand
for small running coupling. So one can use variational or two loop perturbative mass prescription instead of one loop
perturbative mass. The variational mass prescription, which will discuused in Chapter~\ref{chapter:2loop} in details,
will be used for NLO calculations.
But for the computation of various thermodynamic quantities in three loop HTL perturbation theory, variational mass equations
give the imaginary result for $m_D$, so we have only left with two loop perturbative mass prescription  for $m_D$
and that would be used in NNLO computation. We don't have any existing two loop perturbative $m_q$ in literature,
so we will use $m_q=0$ (which is a variational solution in NNLO) and we will test our result with one loop perturbative $m_q$ also.

HTL perturbation theory is defined by treating $\delta$ as a formal expansion parameter. By coupling the HTL improvement term
(\ref{htl_lag}) to the QCD Lagrangian (\ref{qcd_lag}), HTL perturbation theory systematically shifts the 
perturbative expansion from being around an 
ideal gas of massless particles to being around a gas of massive quasi-particles which are the appropriate physical
degrees of freedom at high temperature and/or chemical potential. The HTL perturbation theory Lagrangian (\ref{total_lag})
reduces to the QCD Lagrangian (\ref{qcd_lag}) if we set $\delta=1$. Physical observables are calculated in HTL perturbation
theory by expanding in powers of $\delta$, truncating at some specified order,
and then setting $\delta = 1$.  This defines a reorganization of the perturbative series in which the effects of 
$m_D^2$ and $m_q^2$ terms in (\ref{htl_lag}) are included to leading order but then systematically subtracted out
at higher orders in perturbation theory by the $\delta m_D^2$ and $\delta m_q^2$ terms in (\ref{htl_lag}).  
To obtain leading order (LO), next-to-leading order (NLO), and next-to-next-leading order (NNLO) results,
one expands to orders $\delta^0$, $\delta^1$, $\delta^2$, respectively. This Lagrangian is manifestly gauge invariant. 
An explicit, but tedious, computation shows that it generates the two- and three-point
functions correctly, which is sufficient to ensure that it is also correct for all $N-$ point functions.

Hard thermal loop perturbation has been extensively used to calculate various thermodynamical quantities over almost two decades. It
has been used to study one loop thermodynamic potential (or pressure) both at zero and finite chemical potential but at finite
temperature in Refs.~\cite{Andersen:1999fw,Andersen:1999sf,Andersen:1999va,Mogliacci:2013mca,Mogliacci:2014upa} 
and to study one loop conserved charge fluctuations in Refs.~\cite{Blaizot:2001vr,
Chakraborty:2001kx,Blaizot:2002xz,Chakraborty:2003uw,Jiang:2010jz,Haque:2010rb,Haque:2011iz,Haque:2011vt,Andersen:2012wr,
Su:2012bba,Mogliacci:2013mca,Mogliacci:2013iwa,Mogliacci:2014upa}. HTL perturbation theory has also been used to study 
next-to-leading order thermodynamic potential (or pressure) in Refs.~\cite{Andersen:2002ey,Andersen:2003zk,Haque:2012my} and hence
second and fourth order conserved change fluctuations in Ref.~\cite{Haque:2013qta}. Recently, HTL perturbation has been used to 
calculate three thermodynamic potential for pure gluonic medium in Refs.~\cite{Andersen:2009tc,Andersen:2010ct}, for QED in 
Ref.~\cite{Andersen:2009tw} and also for QCD medium in Ref.~\cite{Andersen:2010wu,Andersen:2011sf}. Very recently this calculations 
have been extended in Refs.~\cite{Haque:2013sja,Haque:2014rua} at finite chemical potential to calculate pressure, energy density,
entropy density, speed of sound, trace anomaly at finite temperature and finite chemical potential and various order diagonal and
off-diagonal quark number susceptibilities.

In addition to calculations of the thermodynamic potential, hard-thermal-loop perturbation theory has been
used to calculate various physical quantities which are relevant to the deconfined state of matter.  
Quantities such as the dilepton production rate~\cite{Braaten:1990wp,Greiner:2010zg}, photon production 
rate~\cite{Baier:1993zb}, single quark and quark anti-quark potentials~\cite{Mustafa:2004hf,Mustafa:2005je,
Chakraborty:2006md,Chakraborty:2007ug,Laine:2006ns,Dumitru:2007hy,Dumitru:2009ni,Dumitru:2009fy,
Thakur:2012eb,Thakur:2013nia}, fermion damping rate \cite{Pisarski:1993rf,Peigne:1993ky,Sarkar:2012fk}, 
photon damping rate~\cite{Abada:2011cc}, gluon damping rate~\cite{Braaten:1989kk,Braaten:1990it}, 
jet energy loss~\cite{Braaten:1991jj,Braaten:1991we,Thoma:1990fm,Romatschke:2003vc,Romatschke:2004au,
Mustafa:2004dr,Mustafa:2003vh,Djordjevic:2006tw,Chakraborty:2006db,Qin:2007rn,Qin:2009uh,Qin:2009gw},
plasma instabilities \cite{Mrowczynski:2000ed,Romatschke:2003ms,Romatschke:2004jh,Rebhan:2004ur,
Schenke:2006fz,Rebhan:2008uj,Attems:2012js}, and lepton asymmetry during leptogenesis 
\cite{Kiessig:2011ga,Kiessig:2011fw} have also been calculated using HTL perturbation theory.  
In the next chapter we discuss about leading order thermodynamics and conserved density 
fluctuations (quark number susceptibility) and temporal component of the Euclidean
correlation function in the vector channel within HTL perturbation theory.

%% file: text/1loop.tex
\chapter{One loop HTL thermodynamics}
\label{chapter:1loop}

In this chapter we discuss about leading order thermodynamic functions {\em viz.}
number density, entropy density, pressure and hence leading order conserved density fluctuations
(quark number susceptibility) and temporal 
component of the Euclidean correlation function in the vector channel within HTL perturbation theory. This chapter is based on:
1. {\em Quark Number Susceptibility and Thermodynamics in HTL approximation}, Najmul Haque,
Munshi G. Mustafa, {\bf Nucl. Phys. A 862-863, 271 (2011)}; 2. {\em Conserved Density Fluctuation and
Temporal Correlation Function in HTL Perturbation Theory}, Najmul Haque, Munshi G. Mustafa, Markus H. Thoma,
{\bf Phys. Rev. D 84, 054009 (2011).}
\section{Introduction}
Dynamical properties of many particle system can generally be studied by
employing an external perturbation, which disturbs the system only slightly
from its equilibrium state, and thus measuring the spontaneous 
response/fluctuations of the system to this external perturbation.
In general, the fluctuations are related to the correlation function through
the symmetry of the system, which provides important inputs for quantitative 
calculations of complicated many-body system. Also, many of the properties ({\em e.g.}
real and virtual photon production, various transport coefficients etc.) 
of the deconfined strongly interacting matter are reflected in the structure 
of the correlation and the spectral functions~\cite{Hashimoto:1992np,Boyd:1994np} of the 
vector current.

The static thermal dilepton rate describing the production of lepton pairs is
related to the spectral function in the vector current~\cite{Karsch:2000gi,Karsch:2001uw}. 
Within the Hard thermal loop perturbation theory (HTLpt) the vector spectral 
function has been obtained~\cite{Karsch:2000gi,Alberico:2006wc,Czerski:2008zz,Arnold:2000dr,
Arnold:2001ms,Arnold:2003zc,Aurenche:2002wq}, which is found 
to diverge due to its spatial part at the low 
energy regime. This is due to the fact that the HTL quark-photon vertex 
is inversely proportional to the photon energy and it sharply rises at 
zero photon energy. On the other hand, the fluctuations of conserved 
quantities, such as baryon number and electric charge, are considered 
to be a signal~\cite{Appelshauser:1998vn,Appelshauser:1999ft,Reid:1999it,
Stephanov:1998dy,Stephanov:1999zu} for quark-gluon plasma (QGP) 
formation in heavy-ion experiments. These conserved density fluctuations 
are closely related to the 
temporal correlation function in the vector channel through derivatives 
of a thermodynamic quantity associated with the symmetry, known as 
the thermodynamic sum rule~\cite{forster,Callen:1951vq,Kubo:1957mj}. It is expected that the 
temporal part of the spectral function associated with the symmetry 
should be a finite quantity and would not encounter any such infrared 
divergence unlike the spatial part at low energy.  A very recent lattice 
calculation~\cite{Ding:2010ga} has obtained the temporal part of the Euclidean
correlation function associated with the response of the conserved 
density fluctuations, which is found to be a finite quantity. In view 
of this we compute the
temporal correlation function in the vector channel from the quark number 
susceptibility associated with the quark number density fluctuations 
within the HTLpt~\cite{Haque:2010rb} and to compare it with recent lattice 
data~\cite{Ding:2010ga}. 

This chapter is organized as follows. In Sec.~(\ref{Generalities}) we briefly discuss some
generalities on correlation functions,  fluctuation and its response
(susceptibility) associated with conserved charges. In this section 
we also obtain the relation between the response of the density
fluctuation of the conserved charge and the corresponding temporal part of 
the Euclidean correlation function in the vector current. In Sec.~(\ref{loqns}) we present  
leading order quark number susceptibility (QNS) in HTL perturbation theory. Next using QNS in 
leading order (LO) within HTL perturbation theory we compute temporal part of 
the Euclidean correlation function in the vector current in 
HTL perturbation theory and compare with lattice data.  In Sec.~(\ref{result_1loop}) we present the results
of LO QNS and  and temporal correlation fnction and compare them with lattice QCD data. Finally, we conclude 
in Sec.~(\ref{conclusion_1loop}).
\section{Generalities}
\label{Generalities}
In this section we summarize some of the basic relations and also 
describe in details their important features relevant as well as 
required for our purpose.
\subsection{Correlation Functions}
The two-point correlation function~\cite{Hashimoto:1992np,Boyd:1994np,Karsch:2000gi,
Karsch:2001uw} of the vector current, $J_\mu=\\{\bar {\psi}}(\tau,{\vec {\mathbf x}})\Gamma_\mu \psi(\tau,{\vec
{\mathbf x}})$ with three point function $\Gamma_\mu$, 
is defined at fixed momentum ${\vec {\mathbf p}}$ as 
\begin{eqnarray}
G_{\mu\nu}(\tau,{\vec{\mathbf p}})=\int d^3x \ \langle J_\mu(\tau, 
{\vec {\mathbf x}}) J_\nu^\dagger(0, {\vec {\mathbf 0}})\rangle \ 
e^{i{\vec{\mathbf p}}\cdot {\vec{\mathbf x}}} \ , \label{cor_p}
\end{eqnarray}
where the Euclidean time $\tau$ is restricted to the interval
$[0,\beta=1/T]$ and three point function $\Gamma_\mu=\gamma_\mu$ for vector channel.
The thermal two-point vector correlation function 
in coordinate space, $ G_{\mu\nu}(\tau, {\vec {\mathbf x}})$, can be written as
\begin{equation}
G_{\mu\nu}(\tau, {\vec {\mathbf x}})=\langle J_\mu(\tau, {\vec {\mathbf x}})
J_\nu^\dagger(0, {\vec {\mathbf 0}})\rangle =T\sum_{n=-\infty}^{\infty} \int
\frac{d^3p}{(2\pi)^3} \ e^{-i(w_n\tau+{\vec{\mathbf p}}
\cdot {\vec{\mathbf x}})}\ G_{\mu\nu}(w_n,{\vec{\mathbf p}}) \ , 
\label{vec_cor_1l}
\end{equation}
where the Fourier transformed correlation function 
$G_{\mu\nu}(w_n,{\vec{\mathbf p}})$ is given at the discrete
Matsubara modes, $w_n=2\pi n T$.

 The imaginary part of the momentum space
correlator gives the spectral function $\sigma(\omega,{\vec{\mathbf p}})$ as
\begin{eqnarray}
G_H(w_n,{\vec{\mathbf p}}) &=& - \int_{-\infty}^{\infty} d\omega
\frac{\sigma_H(\omega,{\vec{\mathbf p}})}{iw_n-\omega} \nonumber \\ 
\ \Rightarrow \
\sigma_H(\omega,{\vec{\mathbf p}})&=& \frac{1}{\pi} {\mbox{Im}}
\ G_H(\omega,{\vec{\mathbf p}}) \ , \label{spec_mom}
\end{eqnarray}
where $H=(00,ii,V)$ denotes (temporal, spatial, vector). We have 
also introduced the vector spectral function 
as $\sigma_V=\sigma_{00}+ \sigma_{ii}$, where $\sigma_{ii}$ is the sum 
over the three space-space components and $\sigma_{00}$ is the time-time 
component of $\sigma_{\mu\nu}$.

Using (\ref{vec_cor_1l}) and (\ref{spec_mom}) in (\ref{cor_p}) 
the spectral representation of
the thermal correlation functions at fixed momentum 
can be obtained~\cite{Karsch:2000gi} as
\begin{equation}
G_H(\tau, {\vec {\mathbf p}})=\int_0^\infty\ d\omega \
\sigma_H (\omega, {\vec {\mathbf p}}) \
\frac{\cosh[\omega(\tau-\beta/2)]}{\sinh[\omega\beta/2]} \ . \label{corr_mom_1l}
\end{equation}
We note that in the analysis of lattice gauge theory, the Euclidean correlation function is
usually restricted to vanishing three momentum, $\vec{\mathbf p}=0$, 
and one can write 
$G_H(\tau T)=G_H(\tau,{\vec{\mathbf 0}})$. Because of the problem of analytic continuation
in the lattice gauge theory can not calculate spectral function $\sigma_H(\omega,{\vec{\mathbf p}})$
directly from Eq.~(\ref{spec_mom}), instead it uses Eq.~(\ref{corr_mom_1l}) 
to extract spectral function as discussed briefly below.

A finite temperature 
lattice gauge theory calculation is performed on
lattices with finite temporal extent $N_\tau$, which provides information
on the Euclidean correlation function, $G_H(\tau T)$, only for a discrete 
and finite set of Euclidean times 
$\tau =k/(N_\tau T), \ \ k=1,\cdots \ N_\tau$. 
The vector correlation function, $G_V(\tau T)$, had been
computed~\cite{Karsch:2001uw} within the quenched approximation of QCD using
non-perturbative improved clover fermions~\cite{Sheikholeslami:1985ij,Luscher:1996jn} for
temporal extent $N_\tau= 16$ and spatial extent $N_\sigma = 64$. Then by 
inverting the integral in (\ref{corr_mom_1l}) through a probabilistic
application based on the maximum entropy method (MEM)~\cite{Nakahara:1999vy,Asakawa:2000tr,Wetzorke:2000ez}, the spectral function was 
reconstructed~\cite{Karsch:2001uw} in lattice QCD.

\subsection{ Density Fluctuation and its Response}
Let \( {\cal O}_{\alpha } \) be a Heisenberg operator where $\alpha$ 
may be associated with a degree of freedom in the system.
In a static and uniform
external field \( {\cal F}_{\alpha } \), the (induced)
expectation value of the operator \( {\cal O}_\alpha \left( 0,
\overrightarrow{\mathbf x}\right) \) is 
written~\cite{forster,Callen:1951vq,Kubo:1957mj,Hatsuda:1994pi,Kunihiro:1991qu} as
\begin{equation}
\phi _{\alpha }\equiv \left\langle {\cal O} _{\alpha }\left
( 0,\overrightarrow{\mathbf x}\right) \right\rangle _{\cal F}=\frac{{\rm Tr}\left
[ {\cal O} _{\alpha }\left( 0,\overrightarrow{\mathbf x}\right) e^{-\beta \left
( {\cal H}+{\cal H}_{ex}\right) }\right] }{{\rm Tr}\left[ e^{-\beta
\left( {\cal H}+{\cal H}_{ex}\right) }
\right] }=\frac{1}{V}\int d^{3}x\, \left\langle {\cal O} _{\alpha }
\left( 0,\overrightarrow{\mathbf x}\right) \right\rangle \: , \label{eq1}
\end{equation}
where translational invariance is assumed, $V$ is the volume of the
system and \({\cal H}_{ex} \) is given by
\begin{equation}
{\cal H}_{ex}=-\sum _{\alpha }\int d^{3}x\, {\cal O} _{\alpha }\left( 0,
\overrightarrow{\mathbf x}\right) {\cal F}_{\alpha }\: .\label{eq2}
\end{equation}

The (static) susceptibility \( \chi _{\alpha \sigma } \) is defined as
the rate with which the expectation value changes
in response to that external field,
\begin{eqnarray}
\chi _{\alpha \sigma }(T) & = & \left. \frac{\partial \phi _{\alpha }}
{\partial {\cal F}_{\sigma }}\right| _{{\cal F}=0}
  = \beta \int d^{3}x\, \left\langle {\cal O} _{\alpha }\left
( 0,\overrightarrow{\mathbf x}\right) {\cal O} _{\sigma }
( 0,\overrightarrow{\mathbf 0})
 \right\rangle \: , \label{eq3}
\end{eqnarray}
where
$\langle {\cal O}_\alpha (0,{\vec {\mathbf x}})
{\cal O}_\sigma(0,{\vec {\mathbf 0}})\rangle $
is the two point correlation function with operators evaluated
at equal times. There is no broken symmetry as
\begin{equation}
\left.\left\langle {\cal O} _{\alpha }
\left ( 0,\overrightarrow{\mathbf x}\right ) \right\rangle
\right|_{{\mathcal F}\rightarrow 0}
 =\left. \left\langle {\cal O} _{\sigma } 
( 0,\overrightarrow{\mathbf 0}) \right\rangle 
\right |_{{\mathcal F}\rightarrow 0}=0  \ . \label{eq3i}
\end{equation}

\subsection{Thermodynamics functions and quark number susceptibility}
\label{qns_correlation}
The pressure of a given system is defined as
\be
{\cal P}=\frac{T}{V}\ln {\cal Z}
\label{press_lnz}
\ee
where $T$ is temperature, $V$ is the volume and ${\cal Z}$ is the partition function of a system containing quark-antiquark
gas. The entropy density is defined as
\be
S=\frac{\del {\cal P}}{\del T}
\ee
The number density for a given quark flavor can be written as
\be
\rho=\frac{\del {\cal P}}{\del\mu}= \frac{1}{V}\frac{\mbox{Tr}[{\cal N} e^{-\beta({\cal H}-\mu {\cal N})}]}
{\mbox{Tr}[ e^{-\beta({\cal H}-\mu {\cal N} )}]} =\frac{\langle{\cal N}\rangle}{V}
\label{rho_lnz}
\ee

with ${\cal N}$ is the quark number operator and $\mu$ is the chemical potential. If $\mu \rightarrow 0$, the quark
number density vanishes due to CP invariance.
Then the QNS for a given quark flavor follows can be written as
\begin{eqnarray}
\chi_q(T) &=& \left.\frac{\partial \rho}{\partial \mu}\right |_{\mu=0}
= \left.\frac{\partial^2 {\cal P}}{\partial \mu^2}\right |_{\mu=0}
\end{eqnarray}
\subsection{QNS and Temporal Euclidean
Correlation Function}
The QNS is a measure of the response of the quark number density with 
infinitesimal change in the quark chemical
potential, $\mu+\delta\mu $. Under such a situation the external field, 
${\cal F}_\alpha$, in Eq.~({\ref{eq2}}) can be identified as the  
quark chemical potential and the operator ${\cal O}_\alpha$ as
the temporal component ($J_0$) of the vector current, 
$J_\sigma(t,{\vec {\mathbf x}})= \overline{\psi} \Gamma_{ \sigma}\psi$, 
where $\Gamma^\sigma$ is in general a three point function.

Then the QNS for a given quark flavor can be written from
Eq.~(\ref{eq3}) as
\begin{eqnarray}
\chi_q(T) &=& \int \ d^4x \ \left \langle J_0(0,{\vec {\mathbf x}})J_0(0,{\vec {\mathbf 0}})
\right \rangle \ 
\nn
&=&- {\lim_{\vec{\mathbf p}\rightarrow 0}} 
{\mbox {Re}}\ G_{00}^R(0,{\vec{\mathbf p}}),
\label{eq4}
\end{eqnarray}
where $G_{00}^R$ is the retarded correlation function. To obtain 
(\ref{eq4}) in concise form, we have used the fluctuation-dissipation 
theorem given as
\begin{equation}
G_{00}(\omega,{\vec{\mathbf p}})=-\frac{2}{1-e^{-\omega/T}} {\mbox{Im}}
G_{00}^R(\omega,{\vec{\mathbf p}}), \label{eq4a}
\end{equation}

and the Kramers-Kronig dispersion relation
\begin{equation}
{\mbox{Re}}G_{00}^R(\omega,{\vec{\mathbf p}})=\int_{-\infty}^{\infty} 
\frac{d\omega^\prime}{2\pi}
 \frac{{\mbox{Im}}G_{00}^R(\omega',{\vec{\mathbf p}})}{\omega^\prime-\omega},
\label{eq4b}
\end{equation}
where $\lim_{\vec{\mathbf p}\rightarrow 0}{\mbox{Im}} G_{00}^R
(\omega,{\vec{\mathbf p}})$ is proportional to
$\delta(\omega)$ due to the quark number conservation~\cite{forster,Callen:1951vq,Kubo:1957mj,Hatsuda:1994pi,Kunihiro:1991qu}. 

Now, (\ref{eq3}) or (\ref{eq4}) indicates that the thermodynamic derivatives
with respect to the external source
are related to the temporal component of the static correlation function
associated with the number conservation of the system. This relation 
in (\ref{eq4}) is known as {\em the thermodynamic sum rule}~\cite{forster,Callen:1951vq,Kubo:1957mj}.

Owing to the quark number conservation the temporal spectral function 
$\sigma_{00}(\omega,{\vec{\mathbf 0}})$ in (\ref{spec_mom}) becomes
\begin{equation}
\sigma_{00}(\omega,{\vec{\mathbf 0}})=\frac{1}{\pi} {\mbox{Im}} G_{00}^R
(\omega,{\vec{\mathbf 0}})= - \omega \delta(\omega) \chi_q(T) \, . \label{eq6}
\end{equation} 

Using Eq.~(\ref{eq6}) in Eq.~(\ref{corr_mom_1l}), it is straight forward to obtain
the temporal correlation function as
\begin{equation}
G_{00}(\tau T)=-T\chi_q(T), \label{eq7}
\end{equation}
which is proportional to the QNS $\chi_q$ and $T$, but independent of $\tau$.

In the next section we present LO QNS in 
HTLpt for our purpose. 
\section{Leading order QNS in HTLpt}
\label{loqns}
The partition function including HTL improvement term can be written
as
\begin{eqnarray}
{\cal Z}[\mu]=\int {\cal D}[\psi] {\cal D}[{\bar \psi}] {\cal D}[{\cal A}]
e^{i\int d^4x(\mathcal{L}_{QCD}+\mathcal{L}_{HTL})(\psi,{\bar \psi},\mu)},
\label{i3}
\end{eqnarray}
where we have considered the external source as $\mu$, the quark chemical 
potential, and ${\cal A}$ is a background gauge field. The HTL improvement term
$\mathcal{L}_{HTL}$ is defined in Eq.~(\ref{htl_lag}) of Chapter~\ref{chapter:htl}.

Using the Eqs.~(\ref{press_lnz}),~(\ref{rho_lnz}) and (\ref{i3}), one could now write the number density as
\begin{eqnarray}
\rho(T,\mu)&=&N_cN_f\int \frac{d^4K}{(2\pi)^4}\ 
\mbox{Tr}\left [S(K)\Gamma_0(K,-K;0)\right ]
\nonumber\\
&=&iN_cN_fT\int \frac{d^3k}{(2\pi)^3}\sum_{k_0}\mbox{Tr}\left [S(K)\Gamma_0(K,-K;0)\right ], \label{i4}
\end{eqnarray}
where $`{\mbox{Tr}}$' is over Dirac indices. 
The quark Matsubara modes are given as $k_0=(2m+1)\pi i T+\mu$, 
where $m=0,1,2 \cdots$; $N_f$ is the number of massless flavors and
$N_c$ is the number of color. The quark propagator $S(K\equiv(k_0,k))$ and temporal component of three-point quark
gluon vertex can be found in~(\ref{electron_prop}) and (\ref{qqg}), respectively.
The zero momentum limit of the 3-point HTL function can also be obtained from 
the Ward identity as
\begin{equation}
\Gamma^0(K,-K;0)=\frac{\del}{\del k_0} \left (-i{S(k_0,k)}^{-1}\right )
=a\gamma^0+b\vec\gamma\cdot\hat k , \label{H2}
\end{equation}
where 
\begin{equation}
a\pm b=D_\pm'(k_0,k) , \label{H3}
\end{equation}
with $D_\pm(k_0,k) $ defined in Eq.~(\ref{Dpm}) and leads to
\begin{equation}
D'_\pm = \frac{D_\pm}{k_0\mp k} + \frac{2m_q^2}{k_0^2-k^2} \ .
\label{H3p}
\end{equation}

After performing the trace over Dirac matrices, the number density~(\ref{i4})
in terms of $D_\pm(k_0,k)$ can be written as 
\begin{eqnarray}
\rho^{HTL}(T,\mu)&=& 2N_cN_fT\int\frac{d^3k}{(2\pi)^3}
\sum_{k_0}\left[\frac{D_+'}{D_+}+
\frac{D_-'} {D_-}\right]
\nonumber\\
\!\!\!&=&\!\!\!2N_cN_fT\int\frac{d^3k}{(2\pi)^3}\sum_{k_0}\left[\frac{1}{k_0-k}
+\frac{1}{k_0+k}
 +\frac{2m_q^2}{k_0^2-k^2}\left(\frac{1}{D_+}+\frac{1}{D_-}
\right)\right]\!.\hspace{1.5cm} \label{H4}
\end{eqnarray}
Apart from the various poles due to quasiparticle (QP) in Eq.~(\ref{H4}) it has
landau damping (LD) part as $D_\pm(k_0,k)$ contain logarithmic terms which generate 
discontinuity for $k^2_0<k^2$. Eq.~(\ref{H4}) 
can be decomposed in individual contribution as
\begin{equation}
\rho^{HTL}(T,\mu)=\rho^{QP}(T,\mu)+\rho^{LD}(T,\mu) \ . \label{HT}
\end{equation}
%

\subsection{Quasiparticle part (QP)}
The pole part of the number density can be calculated using the contour integration method
introduce in Chapter~\ref{chapter:introduction} and 
calculating the residues of each term of Eq.~(\ref{H4}),
one can obtain the HTL quasiparticle  contributions to the quark number 
density as

\begin{eqnarray}
\rho^{QP}(T,\mu)&=&-N_cN_f\int \frac{d^3k}{(2\pi)^3}\left[
\tanh\frac{\beta (\omega_+- \mu)}{2}
+\tanh\frac{\beta(\omega_--\mu)}{2} 
- \tanh\frac{\beta(k-\mu)}{2}
\right.\nonumber \\
&&\left.\hspace*{2.7cm} 
- \tanh\frac{\beta (\omega_++ \mu)}{2}
-\tanh\frac{\beta(\omega_-+\mu)}{2}
+\tanh\frac{\beta(k+\mu)}{2}
\right]
\nonumber \\
&=&2N_cN_f\int \frac{d^3k} {(2\pi)^3}\left[
n(\omega_+-\mu)+n(\omega_--\mu)-n(k-\mu) -n(\omega_++\mu)
\right.\nonumber \\
&&\left. \hspace*{4cm}
-n(\omega_-+\mu)+n(k+\mu) 
\right ]   \ \ ,
\label{H6}
\end{eqnarray}
where $n(y)$ is the Fermi-Dirac distribution, $\omega_\pm$ is defined in Chapter~\ref{chapter:htl}.
Eq.~(H6) agrees with that of the two-loop approximately self-consistent 
$\Phi$-derivable HTL resummation of Ref.~\cite{Blaizot:2001vr} in leading order mass prescription.

Now, the pressure is obtained by integrating the first line of (\ref{H6}) 
w.r.t. $\mu$ as
\begin{eqnarray}
{\mathcal P}^{QP}(T,\mu) &=& 
2N_fN_c T \int \frac{d^3k}{(2\pi)^3} \left [ 
\ln\left(1+e^{-\beta(\omega_+-\mu)}\right) 
 +\ln\left( \frac{1+e^{-\beta(\omega_--\mu)}} 
 {1+e^{-\beta(k-\mu)}}\right ) 
\right. \nonumber \\
\!\!\!&+&\!\!\!\left.
\ln\left(1+e^{-\beta(\omega_++\mu)}\right) 
 +\ln\left( \frac{1+e^{-\beta(\omega_-+\mu)}} 
 {1+e^{-\beta(k+\mu)}}\right)
+\beta\omega_+ + \beta (\omega_--k)   \right]\!,\hspace{1.6cm}
\label{H7}
\end{eqnarray}
where both quasiparticles with energies $\omega_+$ and $\omega_-$ generate 
$T$ dependent ultra-violate (UV) divergences in LO HTL pressure due to 
quasiparticles. This also agrees with the {\em quasiparticle}
contribution as obtained in Ref.~\cite{Andersen:1999va}
for {$\mathbf \mu=0$} and adopting the same 
technique prescribed therein one can remove the UV divergences.
At very high temperature, Eq.~(\ref{H7}) reduces to free
case as 
\begin{eqnarray}
{\mathcal P}(T,\mu) &=& 
2N_fN_c T \int \frac{d^3k}{(2\pi)^3} \left [\beta k+ 
\ln\left(1+e^{-\beta(k-\mu)}\right) +
\ln\left(1+e^{-\beta(k+\mu)}\right) \right].\hspace{1.6cm}
\label{free_press}
\end{eqnarray}

The corresponding HTL QP entropy density in LO can be obtained as
\begin{eqnarray}
{\cal S}^{QP}(T,\mu)&=&\frac{\del {\cal P}^{QP}_I}{\del T}
=2N_cN_f\int\frac{d^3k}{(2\pi)^3}\left[\ln
\left(1+e^{-\beta(\omega_+-\mu)} \right)
+ \ln\left(\frac{1+e^{-\beta(\omega_--\mu)}}{1+e^{-\beta(k-\mu)}}
\right)\right.
\nonumber\\
\!\!\!&+&\!\!\!\ln\left(1+e^{-\beta(\omega_++\mu)}\right)
+ \ln\left(\frac{1+e^{-\beta(\omega_-+\mu)}}{1+e^{-\beta(k+\mu)}}\right)+
\frac{\beta (\omega_+-\mu)}{e^{\beta(\omega_+-\mu)}+1}
+\frac{\beta (\omega_--\mu)}{e^{\beta(\omega_--\mu)}+1}
\nonumber\\
&-&\left .
 \frac{\beta (k-\mu)}
{e^{\beta(k-\mu)}+1}+\frac{\beta (\omega_++\mu)}{e^{\beta(\omega_++\mu)}+1}
+\frac{\beta (\omega_-+\mu)}{e^{\beta(\omega_-+\mu)}+1}
-\frac{\beta (k+\mu)}{e^{\beta(k+\mu)}+1}\right] \ . \label{7s}
\end{eqnarray}

The  QNS in LO due to HTL QP can also be obtained 
from (\ref{H6}) as
\begin{eqnarray}
\chi^{QP}(T)&=&\left.\frac{\del}{\del\mu}\left[\rho^{QP}\right]
\right|_{\mu=0}\nonumber  \\
&=& 4N_cN_f\beta\int \frac{d^3k} {(2\pi)^3}\left [
n(\omega_+)\left(1-n(\omega_+)\right) 
\ + \ n(\omega_-)\left(1-n(\omega_-)\right) \right. \nonumber \\
&&\left. \ \ \ \ \ \ \ \ \ \ \ \ \ \ \ \ \ \ \ \ \ \ \ \ \
-\ n(k)\left(1-n(k)\right)
\right ] \ , \label{H8} 
\end{eqnarray}
where the $\mu$ derivative is performed only to the explicit $\mu$
dependence. Obviously (\ref{H8}) agrees well with that of 
the 2-loop approximately self-consistent  $\Phi$-derivable 
HTL resummation~\cite{Blaizot:2001vr}.
The above thermodynamical quantities in LO due to HTL
{\em quasiparticles} with excitation energies $\omega_\pm$ are similar 
in form to those of free case but the hard and soft contributions
are clearly separated out and one does not need an ad hoc separating scale 
as used in Ref.~\cite{Jiang:2010jz}.
\subsection{Landau Damping part (LD)}
The LD part of the quark number density in LO follows from (\ref{H4}) as
\begin{eqnarray}
\rho^{LD}(T,\mu) 
=  N_cN_f\int \frac{d^3k}{(2\pi)^3}\int \limits_{-k}^k 
{d\omega}
\left(\frac{2m_q^2}{\omega^2-k^2}\right)\beta_+(\omega,k)
\left[n(\omega-\mu)- n(\omega+\mu) \right ].\hspace{1.5cm}
\label{L3}
\end{eqnarray} 
where $\beta_\pm(\omega,k)$ is discontinuity for quark propagator  $D_\pm(k_0,k)$ for $k_0^2<k^2$ and can be obtained 
as
\begin{eqnarray}
\beta_\pm(\omega,k)\!\! &=&\!\! \frac{1}{\pi}\mbox{Disc}\frac{1}{D_\pm(k_0,k)} 
=\frac{1}{\pi} \mbox{Im}\left.\frac{1}{D_\pm(k_0,k)}
\right |_{\substack{k_0\rightarrow\omega+i\epsilon\\ \epsilon \rightarrow 0}}
\nonumber \\
&=&\!\! \frac{\frac{ m_q^2}{2k}\left(-1\pm\frac{\omega}{k}\right)
\Theta(k^2-\omega^2)}
{\left[\omega\mp 
k-\frac{m_q^2}{k}\left(\pm 1-\frac{\omega\mp k}{2k}\ln\frac{k+\omega}{k-\omega}
\right)\right]^2+\left[\pi\frac{m_q^2}{2k}\left(1\mp\frac{\omega}{ k}\right)
\right]^2} \, . \label{L2}
\end{eqnarray}
One can obtain the pressure due to LD contribution by integrating
(\ref{L3}) w.r.t. $\mu$, while considering $m_q$ is an implicit function 
of $T$ and $\mu$, as
\begin{eqnarray}
{\cal P}^{LD}(T,\mu)=N_cN_fT\int\frac{d^3 k}{(2\pi)^3}\int\limits_{-k}^k d\omega\left(
\frac{2m_q^2}{\omega^2-k^2}\right)\beta_+(\omega,k)\left[\ln\left(
1+e^{-\beta(\omega-\mu)}\right)\right.
\nonumber\\
\left.+\ln\left(1+e^{-\beta(\omega+\mu)}\right)+\beta\omega\right] \ ,
\label{L3p}
\end{eqnarray}
which has UV divergence like~\cite{Andersen:1999va} and can
be removed using the prescription therein.


The corresponding LD part of entropy density can be obtained as 
\begin{eqnarray}
{\cal S}^{LD}(T,\mu)&=&N_cN_f\int\frac{d^3 k}{(2\pi)^3}
\int\limits_{-k}^k d\omega\left(
\frac{2m_q^2}{\omega^2-k^2}\right)\beta_+(\omega,k)\left. \Big [\ln\left(
1+e^{-\beta(\omega-\mu)}\right)\right.
\nonumber\\
&&\left.+\ln\left(1+e^{-\beta(\omega+\mu)}\right)+\frac{\beta(\omega-\mu)}
{e^{\beta(\omega-\mu)}+1}+\frac{\beta(\omega+\mu)}{e^{\beta(\omega+\mu)}+1}
\right]  \ .
\label{L3s}
\end{eqnarray}

Also the LD part of the QNS becomes 
\begin{eqnarray}
\chi^{LD}(T)=\left.\frac{\del}{\del\mu}\left[\rho_I^{LD}(T,\mu)\right ]
\right|_{\mu=0}
&=&2N_cN_f\beta\int\frac{d^3k}{(2\pi)^3} 
\int\limits_{-k}^k d\omega\left(\frac{2m_q^2}{\omega^2-k^2}\right)\ 
\nonumber\\
&&\ \ \ \times \ \ 
\beta_+(\omega,k)\ n(\omega)\left(1-n(\omega)\right) \ , \label{L4}
\end{eqnarray}
where the $\mu$ derivative is again performed only to the explicit $\mu$
dependence. It is also to be noted that the LD contribution is of the
order of $m_q^4$. The LD contribution can not be compared with 
that of the 2-loop approximately self-consistent  $\Phi$-derivable HTL 
resummation~\cite{Blaizot:2001vr} as it does not 
have any closed form for the final expression. 

So total QNS can be written from Eqs.~(\ref{H8}) and  (\ref{L4}) as
\be
\chi_q^{HTL}(T)&=& \chi_{q}^{QP}+\chi_{q}^{LD}\nn
&=& 4N_cN_f\beta\int \frac{d^3k} {(2\pi)^3}\Big [
n(\omega_+)\left(1-n(\omega_+)\right)
\ + \ n(\omega_-)\left(1-n(\omega_-)\right)  \nonumber \\
&&
\hspace{2cm} - n(k)\left(1-n(k)\right)
\Big ]\nn
&+& 2N_cN_f\beta\int\frac{d^3k}{(2\pi)^3}
\int\limits_{-k}^k d\omega\left(\frac{2m_q^2}{\omega^2-k^2}\right)\
\beta_+(\omega,k)\ n(\omega)\left(1-n(\omega)\right)\hspace{1.5cm} 
\label{s5}
\ee
The QP part results in (\ref{H8}) 
is identical to that of the $2$-loop approximately self-consistent 
$\Phi$-derivable HTL resummation approach 
~\cite{Blaizot:2001vr,Blaizot:2002xz} within LO mass prescription. The LD part (\ref{L4}) cannot be compared directly to 
the LD part of Ref.~\cite{Blaizot:2001vr,Blaizot:2002xz} as no closed expression is given 
there. However, numerical results of the both QNS agree very well.
We also note that Ref.~\cite{Jiang:2010jz} used 
HTLpt but did not take into account properly the effect of the variation 
of the external field to the density fluctuation, which resulted in 
an over-counting in the  LO QNS. Moreover, it required an ad hoc
separation scale is required to distinguish between soft and hard momenta.
In the HTLpt approach in Ref.~\cite{Chakraborty:2001kx,Chakraborty:2003uw}
the HTL N-point functions were used uniformly for all momenta scale,
{\em i.e.}, both soft and hard momenta, which resulted in an over-counting
within the LO contribution~\cite{Blaizot:2001vr,Blaizot:2002xz}. The reason is that the HTL 
action is accurate only for soft momenta and for hard ones only in the vicinity of light cone.
\section{Results and Discussions:}\label{result_1loop}
\begin{figure}[h]
\begin{center}
\includegraphics[width=9cm]{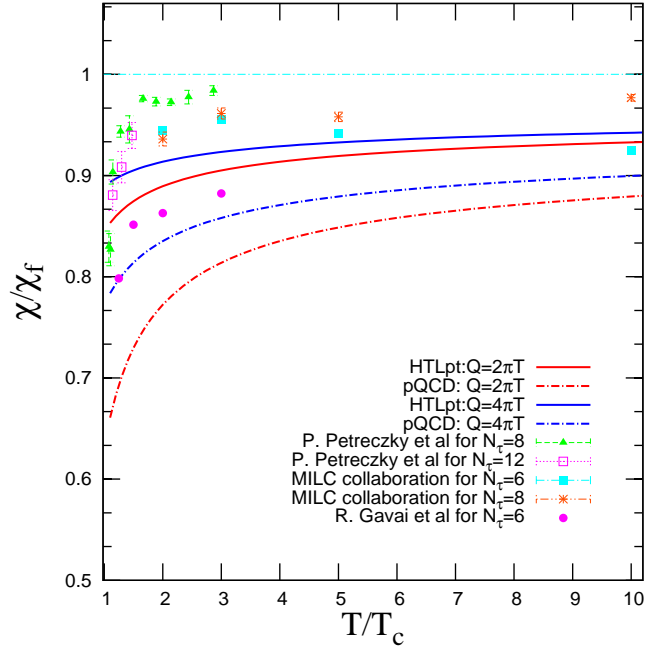}
\caption[The 2-flavor scaled QNS with that of free
one as a function of $T/T_C$.
The solid lines are for LO in HTLpt whereas the dashed lines are 
for LO (proportional to $g^2$) in pQCD. 
The symbols represent the
various lattice data.]{The 2-flavor scaled QNS with that of free
one as a function of $T/T_C$.
The solid lines are for LO in HTLpt whereas the dashed lines are 
for LO in pQCD~\cite{Blaizot:2001vr,Blaizot:2002xz,Toimela:1982hv,kapusta}. 
The different choices of the renormalization
scale are $Q =2\pi T \ {\mbox{(red)}}, \ \
{\mbox{and}} \ \ 4\pi T \ {\mbox{(blue)}}$. The symbols represent the
various lattice data~\cite{Allton:2005gk,Petreczky:2009cr,Bazavov:2009zn,
Bernard:2004je,Gavai:2001fr,Gavai:2002kq,Gavai:2001ie}. 
}
\label{qns_I}
\end{center}
\end{figure}

In Fig.~(\ref{qns_I}) we plot LO QNS in HTLpt for two-flavor system along-with perturbative and various 
lattice QCD data.

It is clear from Fig.~(\ref{qns_I}) that the dependence on renormalization scale in case of HTLpt is less than
that of pQCD~\cite{Blaizot:2001vr,Blaizot:2002xz,Toimela:1982hv,kapusta} results of order $g^2$. Also the 2 flavor LO QNS in HTLpt
is higher in all temperature range than pQCD results and HTLpt results are in good aggrement with lattice QCD data. Moreover,
it also shows the same trend as the available lattice
results for $(2+1)-$flavors~\cite{Allton:2005gk,Petreczky:2009cr,
Bazavov:2009zn,Bernard:2004je,Cheng:2009zi,Petreczky:2009at} as well as for $2-$flavors~\cite{Gavai:2001fr,
Gavai:2002kq,Gavai:2001ie}, though there
is a large variation among the various lattice results. In Fig.~(\ref{qns_I}), the green triangles (with $T_c=191$ MeV)
and purple squares (with $T_c=185$ MeV) represent
$p4$ lattice QCD data~\cite{Petreczky:2009cr,Bazavov:2009zn}, the squares
(cyan) and stars (saffron) are from asqtad lattice QCD
data~\cite{Bernard:2004je} for (2+1)-flavor, $m= 0.2m_s$ and
$T_c=186\pm 4$ MeV. 
The solid circles (purple) represent lattice QCD
data~\cite{Gavai:2001fr,Gavai:2002kq,Gavai:2001ie} for $T_c=0.49\Lambda_{\overline{MS}}$. 
The quark mass ranges between (0.1 to 0.2)$m_s$, 
where $m_s$ is the strange quark mass near its physical value.
Note that further lowering the quark
mass to its physical value seems to have a small effect~\cite{Cheng:2009zi} for
$T> 200$ MeV.  The details of these lattice results are also summarized 
in Ref.~\cite{Petreczky:2009at}. A detailed analysis on uncertainties of the
ingredients in the lattice QCD calculations is presented in Refs.~\cite{Bazavov:2009zn,Cheng:2009zi}.
This calls for further investigation both on the analytic side by improving the HTL resummation
schemes and on the lattice side by refining the various lattice ingredients. 
\begin{figure}[!tbh]
\subfigure{
{\includegraphics[height=7cm,width=7cm]{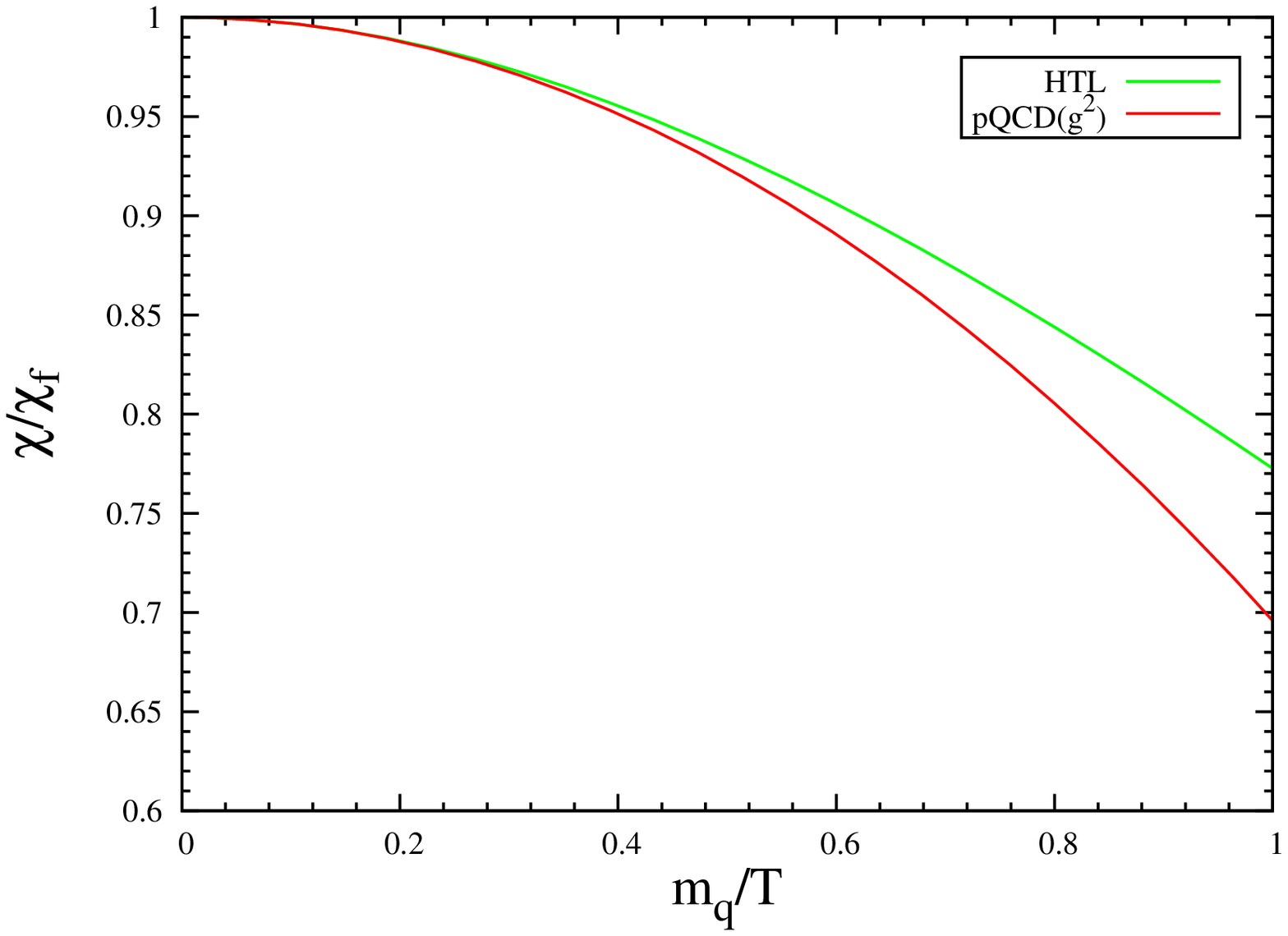}}
}
\subfigure{
{\includegraphics[height=7cm,width=7cm]{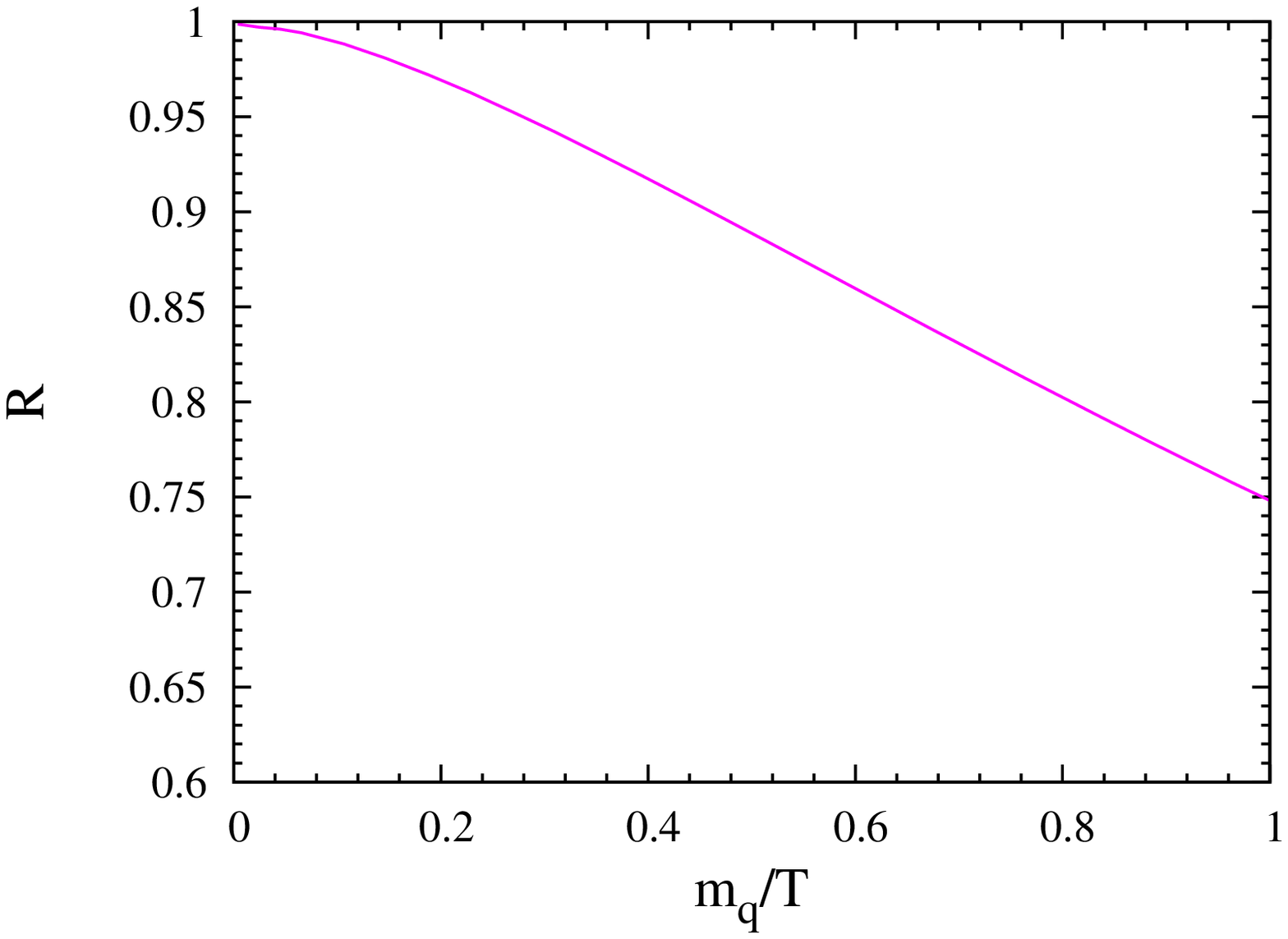}}
}
\vspace*{-0.22in}
\caption{(Color online) {\textit {Left panel:}}
The ratio of 2-flavour HTL to free quark
QNS and also that of LO  perturbative one  as a 
function of $m_q/T$. {\textit {Right panel:}} 
$R$ as a function of $m_q/T$.}
\label{pert_g2}
\end{figure}

In the left panel of Fig.~\ref{pert_g2} we display the LO HTL and
perturbative QNS~\cite{Blaizot:2001vr} scaled with free one vs $m_q/T$. In the weak 
coupling limit both approach unity whereas HTL has a little slower deviation 
from ideal gas value. Now, in the right panel we plot a ratio~\cite{Blaizot:2001vr,Blaizot:2002xz},
$R\equiv ({\chi_{htl}-\chi_f})/({\chi_{p(g^2)}-\chi_f})$, 
which measures the deviation of interaction of $\chi_{htl}$ from that of pQCD
in order $g^2$. It approaches unity in the weak coupling limit
implying the correct inclusion~\cite{Blaizot:2001vr,Blaizot:2002xz}
of order $g^2$ in our approach in a truly perturbative sense.

Recently, an improved lattice calculation~\cite{Ding:2010ga} has been 
performed within the quenched approximation of QCD where the temporal 
correlation function is determined. 
Using the LO HTLpt QNS in Eq.~(\ref{s5}) we now obtain the temporal
correlation function in Eq.~(\ref{eq7}) and compare with the recent
lattice data~\cite{Ding:2010ga}. 
\vspace{-.5cm}
\begin{figure}[ht]
\subfigure{
\includegraphics[height=0.48\textwidth, width=0.55\textwidth,
angle=270]{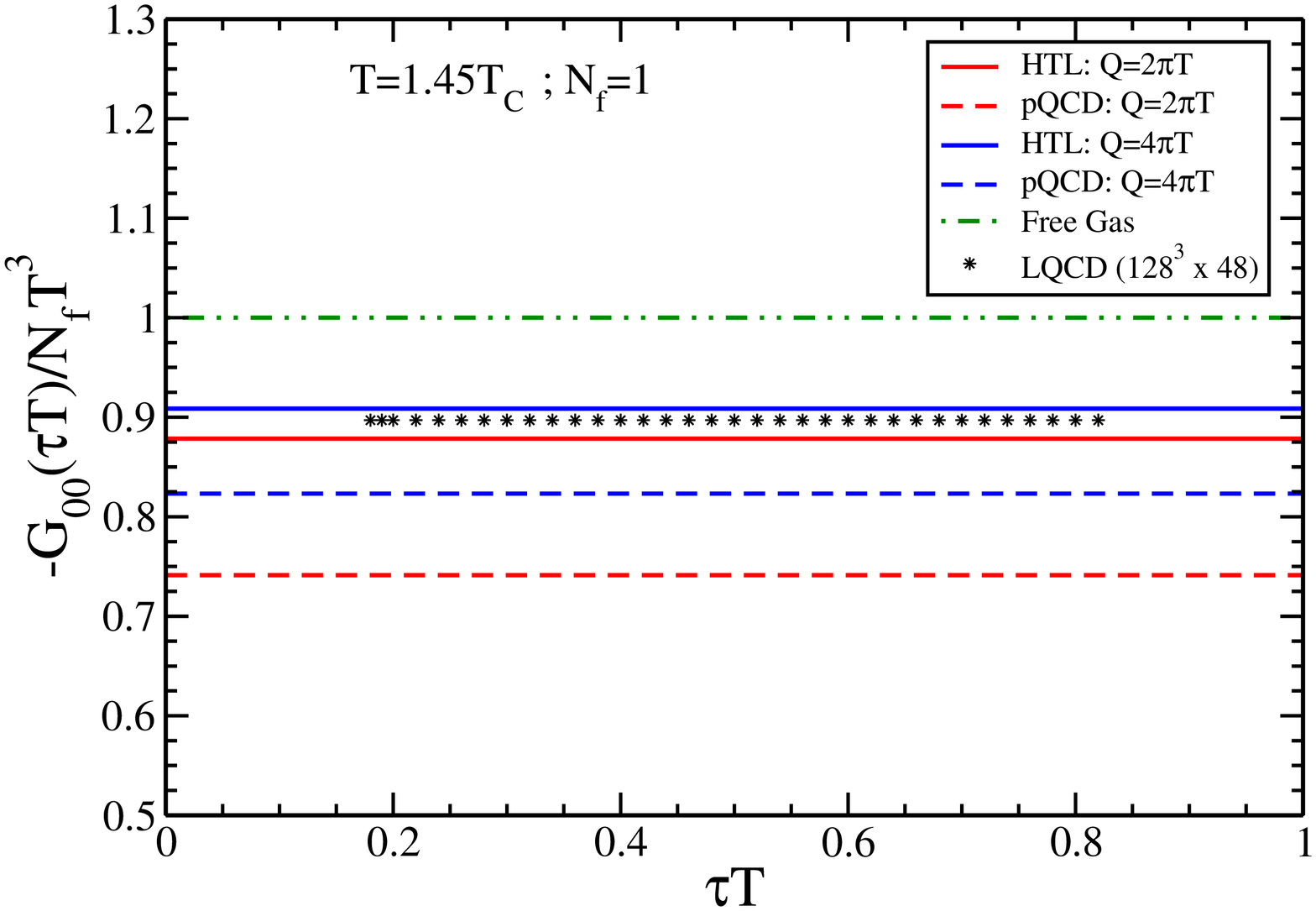}}
\subfigure{
\includegraphics[height=0.48\textwidth, width=0.55\textwidth,
angle=270]{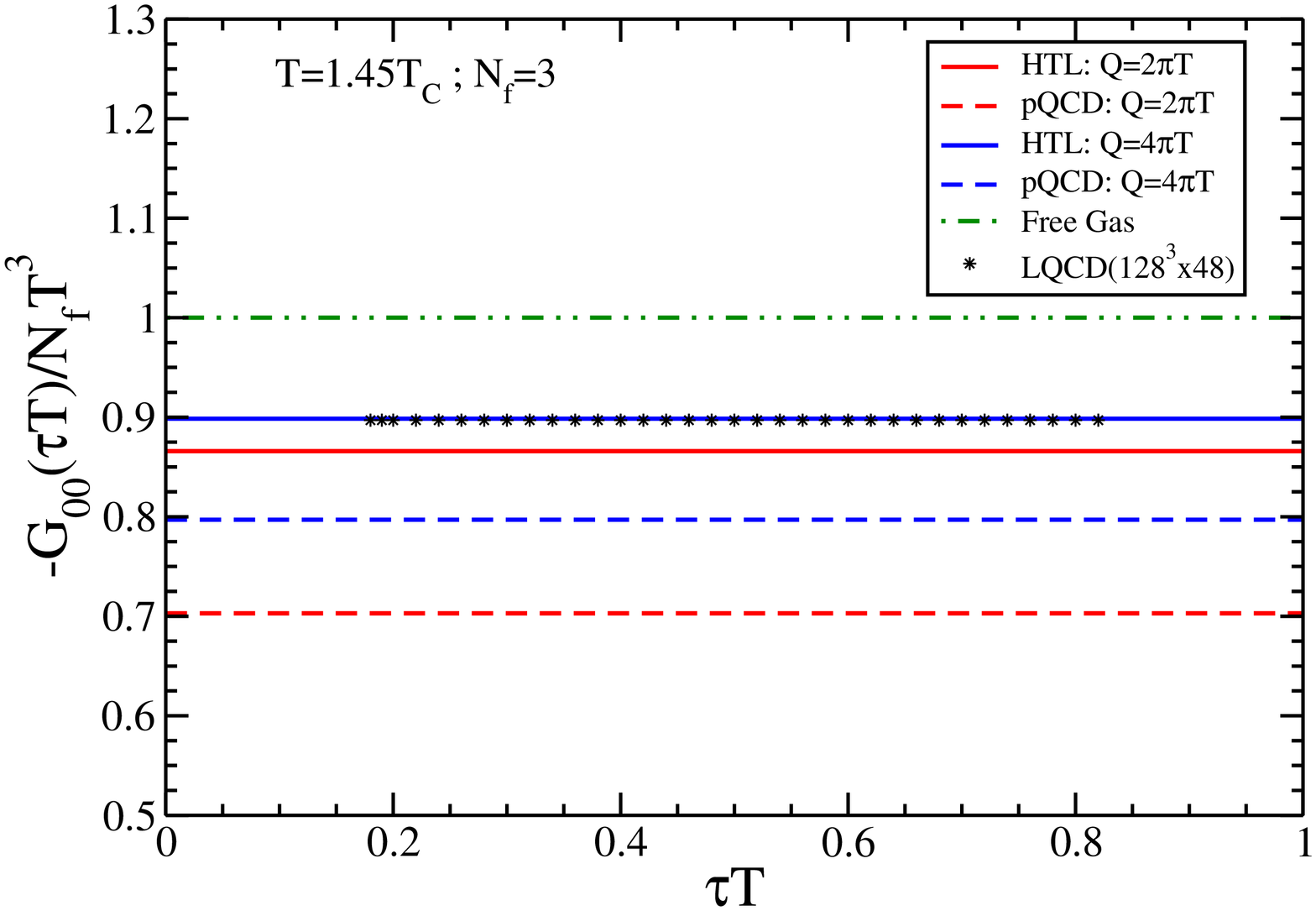}}
\caption[The scaled temporal
correlation function with $T^3$ for $N_f=1$ (left panel) and
$N_f=3$ (right panel) at $T=1.45T_C$ for $Q=2\pi T$ (red) and 
$4\pi T$ (blue) as a function of scaled Euclidean time. 
The symbols represent the recent lattice data.]{The scaled temporal
correlation function with $T^3$ for $N_f=1$ (left panel) and
$N_f=3$ (right panel) at $T=1.45T_C$ for $Q=2\pi T$ (red) and 
$4\pi T$ (blue) as a function of scaled Euclidean time, $\tau T$. 
The symbols represent the recent lattice data~\cite{Ding:2010ga} on 
lattices of size $128^3\times 48$ for quark mass $0.1T$ in quenched
QCD.}
\label{corr1}
\end{figure}
In Fig.~(\ref{corr1}) the scaled temporal correlation function with $T^3$
is shown for $N_f=1$ (left panel) and $N_f=3$ (right panel)
at $T=1.45T_c$. We first note that the correlation functions both in 
HTLpt and pQCD have weak flavor dependence due to the temperature dependent 
coupling, $\alpha_s$ as discussed before. The LO HTLpt result indicates 
an improvement over that of the pQCD one~\cite{Blaizot:2001vr,Blaizot:2002xz,
Toimela:1982hv,kapusta} for different choices of the renormalization scale as
shown in Fig.~(\ref{corr1}). Also, the HTLpt result shows a good agreement to
that of recent lattice gauge theory calculation~\cite{Ding:2010ga} performed 
on lattices up to size $128^3\times 48$ in quenched approximation for a quark 
mass $\sim 0.1T$. We also note that unlike the dynamical spatial part of the
correlation function in the vector channel the temporal part does not encounter
any infrared problem in the low energy part 
as it is related to the static quantity through the thermodynamic sum rule
associated with the corresponding symmetry, {\it viz.}, the number conservation of
the system. 

\begin{figure}[t]
\vspace{-.5cm}
\subfigure{
\includegraphics[height=0.48\textwidth, width=0.55\textwidth,
angle=270]{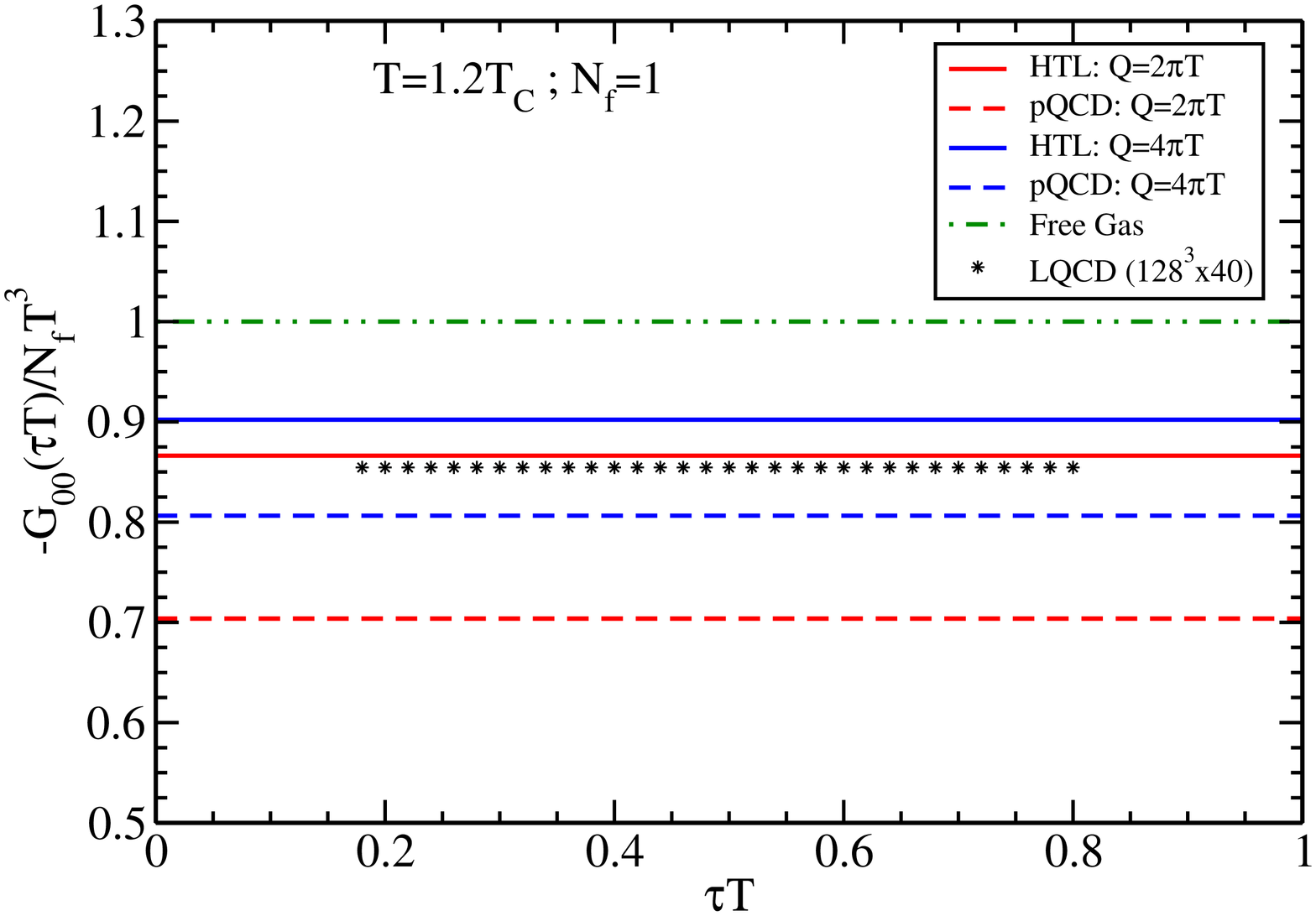}}
\subfigure{
\includegraphics[height=0.48\textwidth, width=0.55\textwidth,
angle=270]{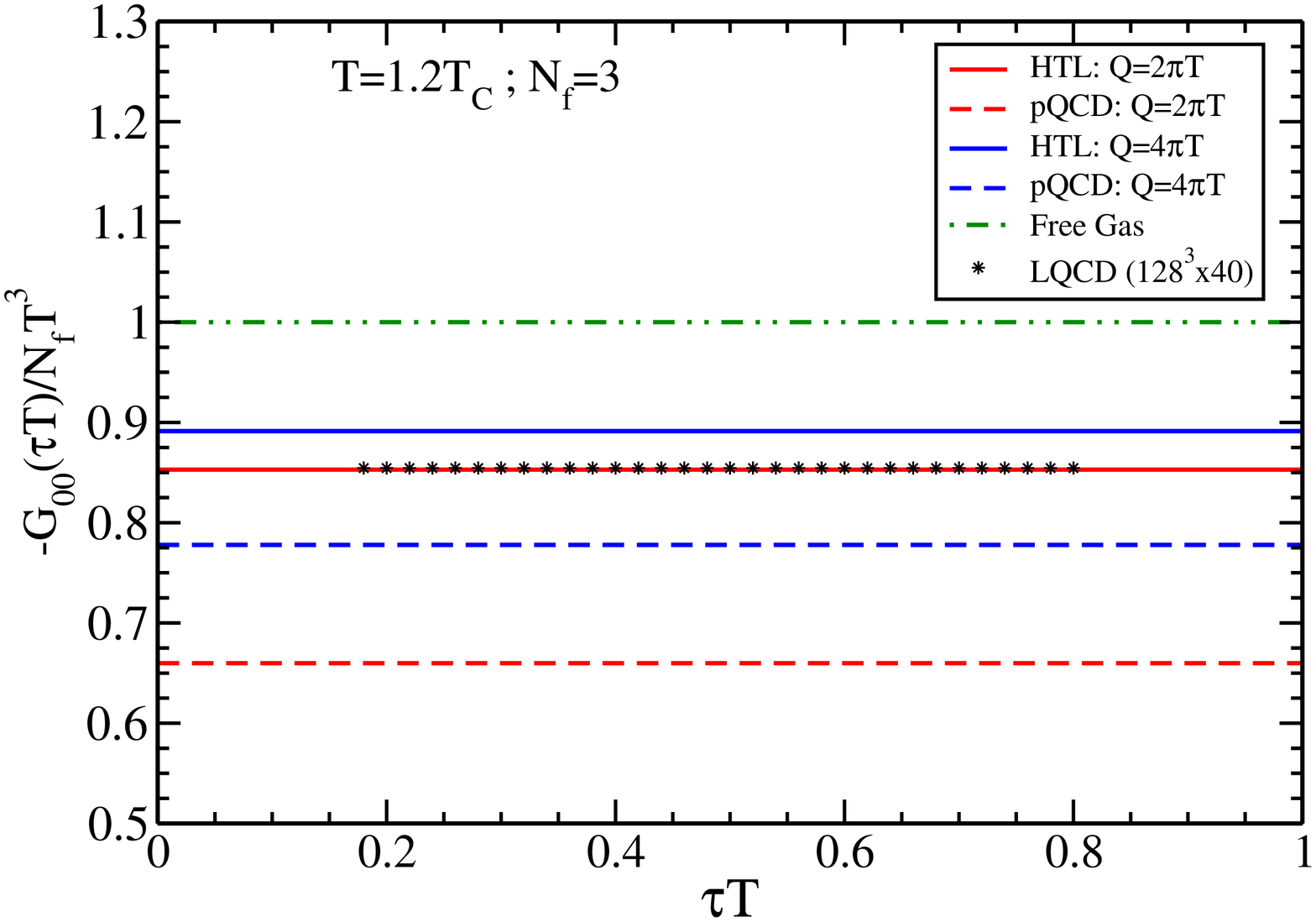}}
\caption{Same as Fig.~(\ref{corr1}) but at
$T=1.2T_C$ and the corresponding lattice data are preliminary~\cite{karsch_private}
 with
lattice size $148^3\times 40$.}
\label{corr2}
\end{figure}

\begin{figure}[ht]
\vspace{-.5cm}
\begin{center}
 \subfigure{
\includegraphics[height=0.48\textwidth, width=0.6\textwidth,
angle=270]{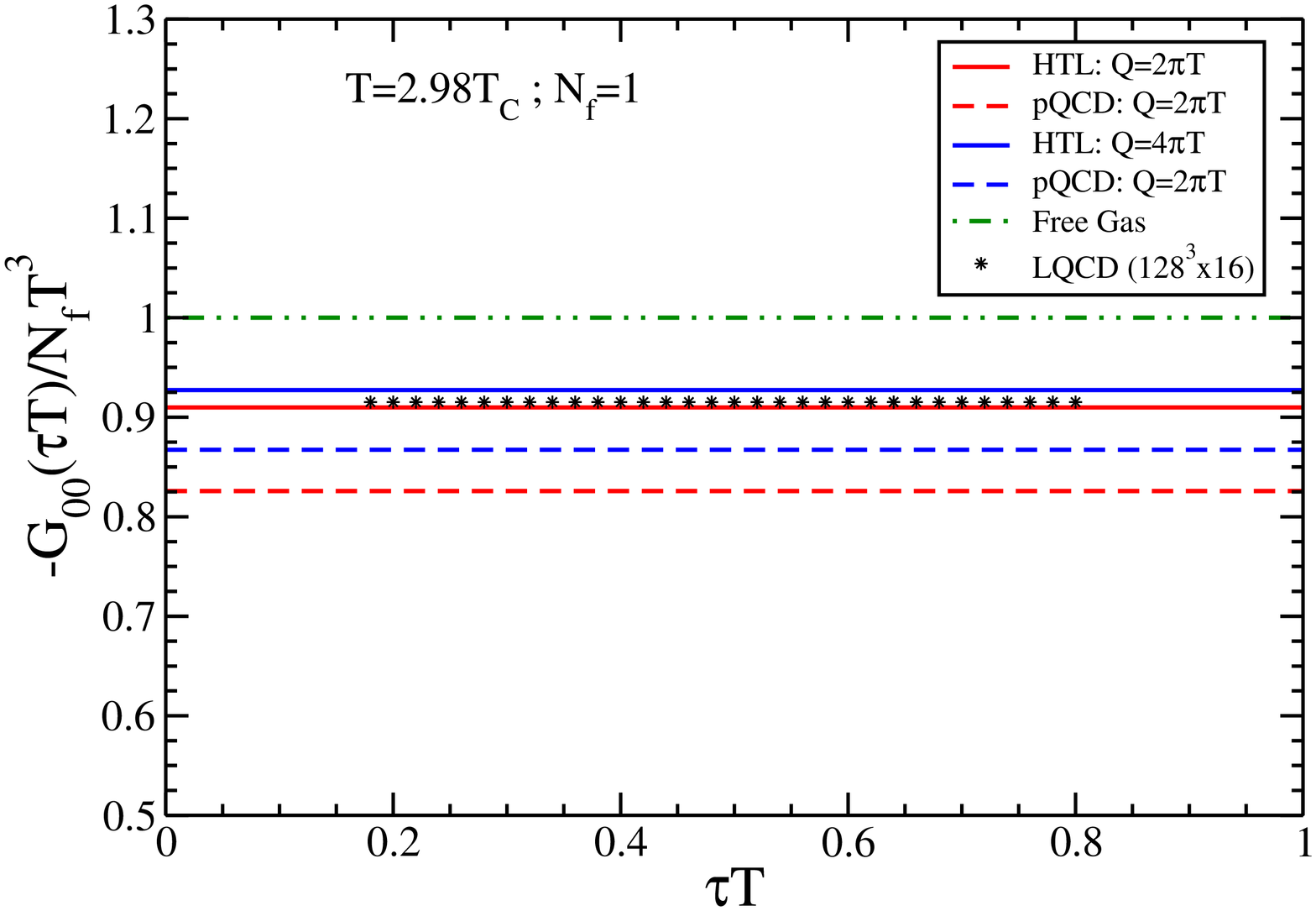}}
\subfigure{\includegraphics[height=0.48\textwidth, width=0.6\textwidth,
angle=270]{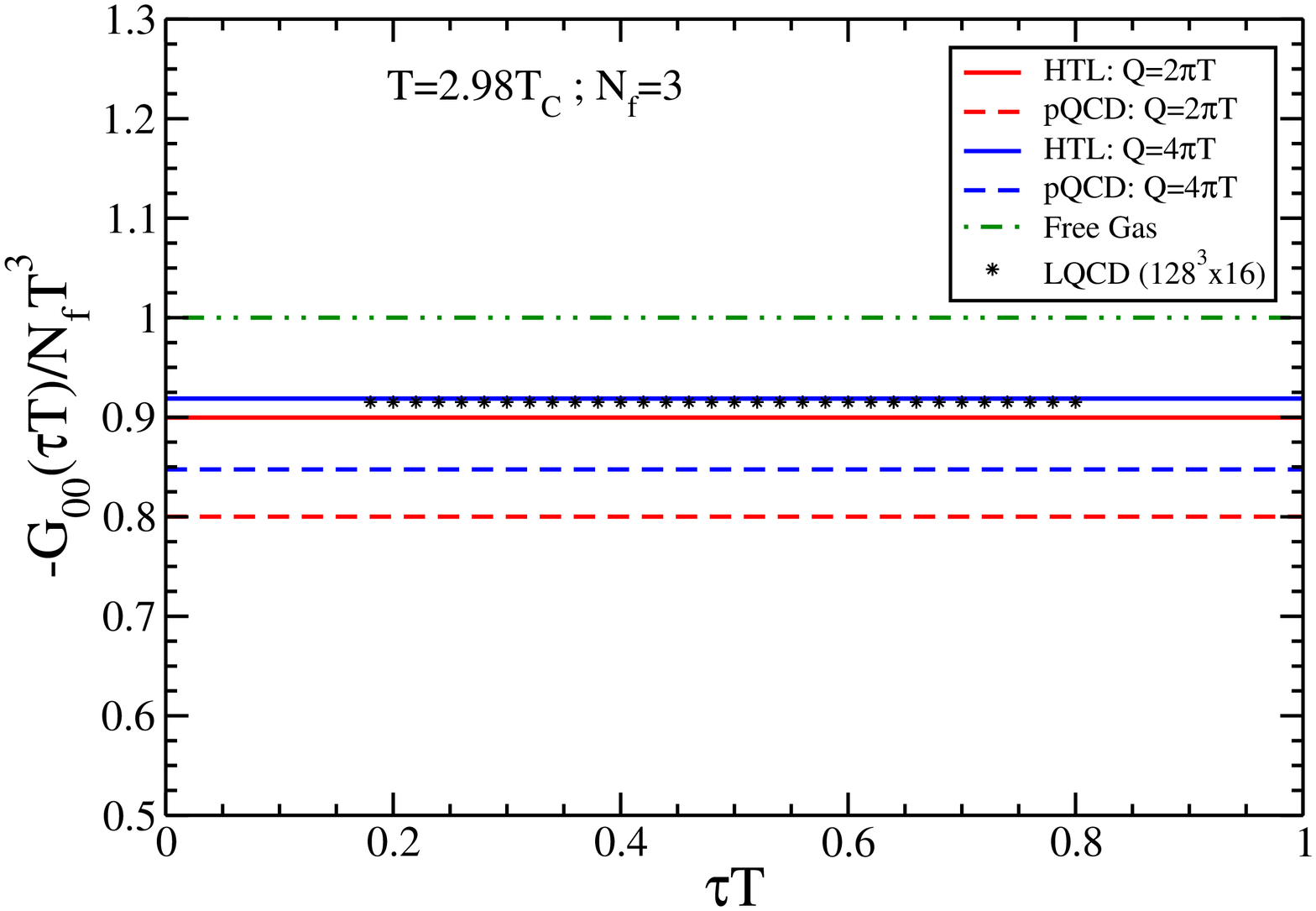}}
\vspace{-.4cm}
\caption{Same as Fig.~(\ref{corr1}) but at
$T=2.98T_C$ and the corresponding lattice data are preliminary~\cite{karsch_private}
 with lattice size $148^3\times 16$.}
\label{corr3}
\end{center}
\end{figure}

 We also presented two extreme cases of HTLpt temporal
correlation function at $T=1.2T_C$ in Fig.~(\ref{corr2}) and at $T\sim 3T_C$ 
in Fig.~(\ref{corr3}), respectively, for two different flavors and
compare with the corresponding preliminary lattice data~\cite{karsch_private}, which 
are also found to be in good agreement.
\section{Conclusion} 
\label{conclusion_1loop}
The LO QNS as a response of the conserved density fluctuation in HTLpt when 
compared with the available lattice data~\cite{Allton:2005gk,Petreczky:2009cr,Bazavov:2009zn,
Bernard:2004je,Gavai:2001fr,Gavai:2002kq,Gavai:2001ie}
in the literature within their wide variation shows the same trend 
but deviates from those in certain extent. The same HTL QNS is used to
compute the temporal part of the Euclidean correlation in vector current
which agrees quite well with that of improved lattice gauge theory 
calculations~\cite{Ding:2010ga,karsch_private} recently performed within quenched 
approximation on lattices up to size $128^3\times48$ for a quark mass $\sim 0.1T$.
Leaving aside the difference
in ingredients in various lattice calculations, one can expect that the HTLpt 
and lattice calculations are in close proximity for quantities associated 
with the conserved density fluctuation.
The leading order quark number susceptibility in HTL perturbation theory produces correct 
${\cal O}(g^2)$ perturbative results when expanded in a strict power
series in $g$. So make the results more reliable for heavy-ion-collisions experiments, we need to go beyond
leading order calculations. In the following two chapters, we will discuss about thermodynamic calculations beyond leading
order. 

%% file: text/2loop.tex
\chapter{Two-loop HTL Thermodynamics }
\label{chapter:2loop}
In this chapter, we study the two-loop pressure and also second and fourth order quark number susceptibility 
of QCD using the hard-thermal-loop perturbation theory. This chapter is based on:
{\it Two-loop HTL pressure at finite temperature and chemical potential},
N. Haque, M. G. Mustafa, and M. Strickland, {\bf Phys. Rev. D87 (2013) 105007} and 
{\it Quark Number Susceptibilities
from Two-Loop Hard Thermal Loop Perturbation Theory}, N. Haque, M. G. Mustafa, and M. Strickland, {\bf JHEP 1307
(2013) 184}.
\section{Introduction}
In Chapter~\ref{chapter:1loop}, HTL perturbation theory has been applied to calculate LO thermodynamics to the case of finite temperature 
and finite chemical potential and also to calculate LO QNS.
As discussed in Chapter~\ref{chapter:1loop}, LO QNS in HTLpt produced correct perturbative
order up to $g^2$ in strict perturbative expansion, it will be interesting to go beyond leading order at finite temperature
and finite chemical potential in HTLpt.


In HTL perturbation theory the next-to-leading order (NLO) thermodynamic potential was computed in~\cite{Andersen:2002ey,
Andersen:2003zk} at finite temperature and zero chemical potential. But in view of the ongoing RHIC beam
energy scan and planned FAIR experiments, one is motivated to reliably determine the thermodynamic 
functions at finite chemical potential. In this chapter we discuss the NLO pressure of quarks and
gluons at finite $T$ and $\mu$.  The computation utilizes a high temperature
expansion through fourth order in the ratio 
of the chemical potential to temperature. This allows us to reliably access the region of high 
temperature and small chemical potential. We compare our final result for the NLO HTLpt  pressure at finite
temperature and chemical potential with state-of-the-art perturbative quantum chromodynamics (QCD) 
calculations and available lattice QCD results.

Having the full thermodynamic potential as a function of chemical potential(s) and temperature allows us 
to compute the quark number susceptibilities. For massless quark flavors the QNS are usually defined as
\begin{eqnarray}
 \chi_{n}(T)\equiv \left. \frac{\partial^n\!{\cal P}}{\partial \mu^n} \right |_{\mu= 0} \ ,
\label{qns_def}
\end{eqnarray}
where $\cal P$ is the pressure of system, $\mu$ is the quark chemical potential and $T$ is the temperature of the system.

This chapter is organized as follows. In Sec.~(\ref{ingredients}) 
we discuss various quantities required to be calculated at finite chemical potential based on prior calculations of
the NLO thermodynamic at zero chemical potential~\cite{Andersen:2002ey,
Andersen:2003zk}. In Sec.~(\ref{scalarint}) we 
reduce the sum of various diagrams to scalar sum-integrals. A high temperature expansion is made in Sec.~(\ref{expand}) 
to obtain analytic expressions for both the LO and NLO thermodynamic potential. In Secs.~(\ref{asis}) and~(\ref{bsis}) 
we calculate the various sum-integrals and $d-$dimensional integrals that appear in Sec.~(\ref{expand}). We then use
the results of Secs.~(\ref{expand}),~(\ref{asis}) and~(\ref{bsis}) to compute the pressure in Sec.~(\ref{nlopress}).
We compute second and fourth order quark number susceptibilities in Sec.~(\ref{qns}). We finaly conclude in Sec.~(\ref{concl}).

\section{Ingredients for the NLO Thermodynamic potential in HTLpt}
\label{ingredients}
The LO HTLpt thermodynamic potential, $\Omega_{\rm LO}$, for an $SU(N_c)$ gauge theory with $N_f$ massless quarks 
in the fundamental representation can be written as ~\cite{Andersen:1999sf,Andersen:2002ey,
Andersen:2003zk}
\be
\Omega_{\rm LO}= d_A {\cal F}_{g}
+ d_F {\cal F}_q+\Delta_0{\cal E}_0\;,
\ee
where $d_F=N_f N_c$ and $d_A=N_c^2-1$ with $N_c$ is the number of
colors. ${\cal F}_q$ and ${\cal F}_g$ are the one loop contributions to quark and gluon free energies, respectively. 
The LO counter-term is the same as in the case of zero chemical potential~\cite{Andersen:1999sf}
\be
 \Delta_0{\cal E}_0 = \frac{d_A}{128\pi^2\epsilon}m_D^4\;.
\label{count0}
\ee
\begin{figure}[tbh]
\begin{center}
\includegraphics[width=7.5cm]{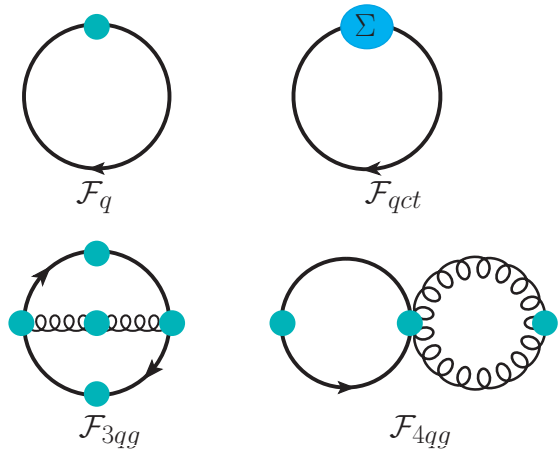}
\caption{Diagrams containing fermionic lines relevant for NLO thermodynamics potential in HTLpt with finite chemical potential.  
Shaded circles indicate HTL $n$-point functions.}
\label{diagramfig}
\end{center}
\end{figure}
At NLO one must consider the diagrams shown in Fig.~(\ref{diagramfig}).
The resulting NLO HTLpt thermodynamic potential can be written in the following general form~\cite{Andersen:2003zk} 
\be
\Omega_{\rm NLO}&=&\Omega_{\rm LO}+
d_A \left[{\cal F}_{3g}+{\cal F}_{4g}+{\cal F}_{gh}
+{\cal F}_{gct}
\right] 
+d_A s_F \left[{\cal F}_{3qg}+{\cal F}_{4qg}
\right] \nonumber \\
&&+ d_F {\cal F}_{qct}
+\Delta_1{\cal E}_0
+\Delta_1 m_D^2\frac{\partial}{\partial m_D^2}
\Omega_{\rm LO}+\Delta_1 m_q^2\frac{\partial}{\partial m_q^2}\Omega_{\rm LO}\;,
\label{OmegaNLO}
\ee
where $s_F=N_f/2$.
At NLO the terms that depend on the chemical potential 
are ${\cal F}_q$, ${\cal F}_{3qg}$, ${\cal F}_{4qg}$, ${\cal F}_{qct}$, 
$\Delta_1 m_q^2$, and $\Delta_1 m_D^2$ as displayed
in Fig.~(\ref{diagramfig}). The other terms, e.g. ${\cal F}_g$, ${\cal F}_{3g}$, ${\cal F}_{4g}$, 
${\cal F}_{gh}$ and ${\cal F}_{gct}$ coming from gluon and ghost loops remain the same as the $\mu=0$ case~\cite{Andersen:2002ey,
Andersen:2003zk}. 
We also add that the vacuum energy counter-term, $\Delta_1{\cal E}_0$, remains the 
same as the $\mu=0$ case whereas the 
mass counter-terms, $\Delta_1 m_D^2$ and $\Delta_1m_q^2$, have to be computed for $\mu\ne0$. These counter-terms are 
of order $\delta$.  This completes a general description of 
contributions one needs to compute in order to determine NLO HTLpt thermodynamic potential 
at finite chemical potential.  We now proceed to the scalarization of the necessary diagrams.

\section{Scalarization of the fermionic diagrams}
\label{scalarint}
\noindent
The one-loop quark contribution coming from the first diagram in Fig.~(\ref{diagramfig}) can be written as 
\be
{\cal F}_q =-\sumintf_{\{P\}}\ \log\det\left[P\!\!\!\!/-\Sigma(P)\right]\;
          = -2\sumintf_{\{P\}}\ \log P^2-2\sumintf_{\{P\}}\ 
\log\left[\frac{A_S^2-A_0^2}{P^2}\right],
\label{qlead}
\ee
where $A_0$ and $A_S$ are defined in Eq.~(\ref{A0AS_tauP}) of Chapter~\ref{chapter:htl} in Minkowski space.
We can write it in Euclidean space as 
\be
A_0(P)&=&iP_0-\frac{m_q^2}{iP_0}{\cal T}_P\;,
\label{aodef}
\\ 
A_S(P)&=&|{\bf p}|+\frac{m^2_q}{|{\bf p}|}\left[1-{\cal T}_P\right]\;,
\label{asdef}
\ee
and ${\cal T}_P$ is defined by the following integral~\cite{Andersen:2002ey,
Andersen:2003zk} in Euclidean space as
\be
{\cal T}_P&=&\left\langle \frac{P_0^2}{P_0^2+p^2c^2}\right\rangle_c 
= \frac{\omega(\epsilon)}{2}\int\limits_{-1}^1 dc \, (1-c^2)^{-\epsilon}\frac{iP_0}{iP_0-|{\bf p}|c} \, , 
\label{def-tf}
\ee
with $w(\epsilon) = 2^{2\epsilon}\,\Gamma(2-2\epsilon)/\Gamma^2(1-\epsilon)$.
In three dimensions {\em i.e} for $\epsilon \rightarrow 0$, Eq.~(\ref{def-tf}) reduces to 
\be
{\cal T}_P&=&\frac{iP_0}{2|{\bf p}|}\log \frac{iP_0+|{\bf p}|}{iP_0-|{\bf p}|} \, ,
\label{def-tf1}
\ee 
with $P\equiv (P_0,{\bf p})$.  
In practice, one must use the general form and only take the
limit $\epsilon \rightarrow 0$ after regularization/renormalization.

The HTL quark counter-term at one-loop order can be rewritten from the second diagram in Fig.~(\ref{diagramfig}) as
\be
\label{qct}
{\cal F}_{qct}=-4\sumintf_{\{P\}}\ 
\frac{P^2+m^2_q}{A_S^2-A_0^2}\;.
\ee

The two-loop contributions coming from the third and fourth diagrams in Fig.~(\ref{diagramfig}) are given, respectively, by
\be
{\cal F}_{3qg}&=&\frac{1}{2}g^2\sumintbf_{P\{Q\}}\mbox{Tr}\left[\Gamma^{\mu}(P,Q,R)S(Q)
\times \Gamma^{\nu}(P,Q,R)
S(R)\right]\Delta_{\mu\nu}(P) \, ,
\label{3qg}
\\  
{\cal F}_{4qg}&=&\frac{1}{2}g^2\sumintbf_{P\{Q\}}\mbox{Tr}\left[\Gamma^{\mu\nu}(P,-P,Q,Q)S(Q)\right]
\Delta_{\mu\nu}(P)\;,
\label{4qg}
\ee
where $S$ is the quark propagator which is given by $S = (\gamma^\mu {\cal A}_\mu)^{-1}$ with
${\cal A}_\mu = (A_0(P),A_S(P)\hat{\bf p})$ and $\Delta^{\mu\nu}$ is the gluon propagator.   
The general covariant gauge $\Delta^{\mu\nu}$ can be found in Eq.~(\ref{gprop-TC}) of Chapter~\ref{chapter:htl} in Minkowski space.
Also above $\Gamma^\mu$ and $\Gamma^{\mu\nu}$ are HTL-resummed $3$- and $4$-point functions can be
found in Chapter~\ref{chapter:htl}. Many more details
concerning the HTL $n$-point functions including the general Coulomb gauge propagator etc. can be found in 
Chapter~\ref{chapter:htl} and also in appendices of Refs.~\cite{Andersen:1999sf,Andersen:2002ey,Andersen:2003zk}.

In general covariant gauge, the sum of (\ref{3qg}) and (\ref{4qg})
reduces to 
\be
{\cal F}_{3qg+4qg}&=&\frac{1}{2}g^2\sumintbf_{P\{Q\}}\Bigg\{
\Delta_X(P)\mbox{Tr}\left[
\Gamma^{00}S(Q)
\right]
-\Delta_T(P)\mbox{Tr}\left[
\Gamma^{\mu}S(Q)\Gamma^{\mu}S(R')
\right]
\nonumber \\
&& \hspace{3cm}
\label{coll}
+\Delta_X(P)
\mbox{Tr}\left[
\Gamma^{0}S(Q)\Gamma^{0}S(R')
\right]\Bigg\}\;,
\ee
where $\Delta_T$ is the transverse gluon propagator,
$\Delta_X$ is a combination of the longitudinal and transverse gluon propagators~\cite{Andersen:2002ey,
Andersen:2003zk}, 
and $R' = Q - P$.  After performing the traces 
of the $\gamma$-matrices one obtains~\cite{Andersen:2002ey,
Andersen:2003zk}
\be
{\cal F}_{3qg+4qg}\!\!\!&=&\!\!\!
- g^2\sumintbf_{P\{Q\}}\frac{1}{A_S^2(Q)-A_0^2(Q)}
\Bigg[
2(d-1)\Delta_T(P)\frac{\hat{\bf q}\!\cdot\!\hat{\bf r}A_S(Q)A_S(R)-A_0(Q)A_0(R)}
{A_S^2(R)-A_0^2(R)} \nonumber \\
&-& 2\Delta_X(P)
\frac{A_0(Q)A_0(R)+A_S(Q)A_S(R)\hat{\bf q}\!\cdot\!\hat{\bf r}}
{A_S^2(R)-A_0^2(R) } 
\nonumber \\
&-&
4m_q^2
\Delta_X(P)
\Bigg\langle
\frac{A_0(Q)-A_s(Q)\hat{{\bf q}}\!\cdot\!\hat{{\bf y}}}{(P\!\cdot\!Y)^2
-(Q\!\cdot\!Y)^2}\frac{1}{(Q\!\cdot\!Y)}\Bigg\rangle_{\!\!\bf \hat y} \nonumber \\
&+&
\frac{8m_q^2\Delta_T(P)}{A_S^2(R)-A_0^2(R)}
\Bigg\langle
\frac{(A_0(Q)-A_S(Q)\hat{\bf q}\!\cdot\!\hat{\bf y})
(A_0(R)-A_S(R)\hat{\bf r}\!\cdot\!\hat{\bf y})}
{(Q\!\cdot\!Y)(R\!\cdot\!Y)}
\bigg\rangle_{\!\!\bf \hat y}
\nonumber \\
\!\!\!&+&\!\!\! 
 \frac{4 m_q^2\Delta_X(P)}{A_S^2(R)-A_0^2(R) }
\Bigg\langle
\frac{2A_0(R)A_S(Q)\hat{\bf q}\!\cdot\!\hat{\bf y}-A_0(Q)A_0(R)-A_S(Q)A_S(R)
\hat{\bf q}\!\cdot\!\hat{\bf r}
}{(Q\!\cdot\!Y)(R\!\cdot\!Y)}
\Bigg\rangle_{\!\!\bf \hat y} \Bigg] 
\nonumber
\\
&+& 
 O(g^2 m_q^4) \; ,
\label{2loop_si}
\ee
where $A_0$ and $A_S$ are defined in (\ref{aodef}) and (\ref{asdef}), respectively. We add that 
the exact evaluation of two-loop free energy could be performed numerically and would involve 
5-dimensional integrations; however, one would need to be able to identify all divergences and 
regulate the numerical integration appropriately.  Short of this, one can calculate the sum-integrals by expanding in
a power series in $m_D/T$, $m_q/T$, and $\mu/T$ in order to obtain semi-analytic expressions. 
%
\section{High temperature expansion}
\label{expand}
As discussed above, we make an expansion of two-loop free energies in a
power series of $m_D/T$ and $m_q/T$ to obtain a series which is nominally accurate to
order $g^5$.  
By ``nominally accurate'' we mean that we expand the scalar integrals treating $m_D$ and $m_q$
as ${\cal O}(g)$ keeping all terms which contribute through ${\cal O}(g^5)$; however, 
the resulting series is accurate to order $g^5$ in name
only. 

In practice, the HTL $n$-point functions can have both hard and soft momenta scales on each leg.
At one-loop order the contributions can be classified ``hard'' or ``soft'' depending on whether
the loop momenta are order $T$ or $gT$, respectively; however, since the lowest fermionic
Matsubara mode corresponds to $P_0 = \pi T$, fermion loops are always hard.
The two-loop contributions to the thermodynamic potential 
can be grouped into hard-hard (hh), hard-soft (hs), and soft-soft (ss) contributions.
However, we note that one of the momenta contributing is always hard since it corresponds to
a fermionic loop and therefore there will be no two-loop soft-soft contribution. Below we calculate 
the various contributions to the sum-integrals presented in Sec.~(\ref{scalarint}). 

\subsection{One-loop sum-integrals}
\label{1sis}

The one-loop sum-integrals (\ref{qlead}) and (\ref{qct})  
correspond to the first two diagrams in Fig.~(\ref{diagramfig}).  They represent
the leading-order quark contribution and order-$\delta$ HTL counterterm. We will
expand the sum-integrals through order $m_q^4$ taking $m_q$ to be of (leading) order $g$.
This gives a result which is nominally accurate (at one-loop) through order $g^5$.~\footnote{Of
course, this won't reproduce the full $g^5$ pQCD result in the limit $g\rightarrow0$.  
In order to reproduce all known coefficients through ${\cal O}(g^5)$, one would need to perform
a NNLO HTLpt calculation.}

\subsubsection*{Hard Contribution}
\label{hard}
The hard contribution to the one-loop quark self-energy in (\ref{qlead}) can be expanded in
powers of $m_q^2$ as
\be
{\cal F}^{(h)}_q=
-2\sumintf_{\{P\}}\ \log P^2-4m_q^2\sumintf_{\{P\}}\frac{1}{P^2}
+2m_q^4\sumintf_{\{P\}}\!
\left[\frac{2}{P^4}
-\frac{1}{p^2P^2}+\frac{2{\cal T}_P}{p^2P^2}
-\frac{\left({\cal T}_P\right)^2}{p^2P_0^2}
\right]\! .\hspace{5mm}
\label{1loop_exp}
\ee
Note that the function ${\cal T}_P$ does not appear in $m_q^2$ term.
The expressions for the sum-integrals in (\ref{1loop_exp}) are listed in Sec.(~\ref{asis}).
Using those expressions, the hard contribution to the quark free energy becomes
\be
{\cal F}_q^{(h)}\!\!\! &=&\!\!\! -\frac{7\pi^2}{180}T^4\left(1+\frac{120}{7}\hat\mu^2 
            +\frac{240}{7}\hat\mu^4\right) + \left(\frac{\Lambda}
            {4\pi T}\right)^{2\epsilon} \frac{m_q^2 T^2}{6}\biggr[\left(
            1+12\hat\mu^2\right)
\nn 
\!\!\!&+&\!\!\!\! \left.
            \epsilon\left(2-2\ln2+2\frac{\zeta'(-1)}{\zeta(-1)}+ 24(\gamma_E+2\ln2)
            \hat\mu^2 - 28\zeta(3)\hat\mu^4 + {\cal O}\left(\hat\mu^6\right)
            \right)\right] 
\nn &+&
           \frac{m_q^4}{12\pi^2}(\pi^2 - 6) \, .\hspace{5mm} 
\label{Quark1loop}
\ee
Expanding the HTL quark counter-term in~(\ref{qct}) one can write
\be
{\cal F}^{(h)}_{\rm qct}=
4m_q^2\sumintf_{\{P\}}\frac{1}{P^2}
-4m_q^4\sumintf_{\{P\}}\left[\frac{2}{P^4}
-\frac{1}{p^2P^2}
+\frac{2}{p^2P^2}{\cal T}_P
-\frac{1}{p^2P_0^2}\left({\cal T}_P\right)^2\right] ,
\label{1loop_ct}
\ee
where the expressions for various sum-integrals in (\ref{1loop_ct}) are listed in Sec.~(\ref{asis}).
Using those expressions, the hard contribution to the HTL quark counter-term becomes
\be
{\cal F}_{qct}^{(h)} = -\frac{m_q^2 T^2}{6}\left(1+12\hat\mu^2\right)
-   \frac{m_q^4}{6\pi^2}(\pi^2 - 6)  \, .
\label{count}
\ee
We note that the first term in ~(\ref{count}) 
cancels the order-$\epsilon^0$ term in the coefficient of $m_q^2$ in (\ref{Quark1loop}).
There are no soft contributions either from the leading-order quark 
term in (\ref{qlead}) or from the HTL quark counter-term in (\ref{qct}).

\subsection{Two-loop sum-integrals}
\label{2sis}

Since the two-loop sum-integrals given in~(\ref{coll}) 
contain an explicit factor of $g^2$, we only require an expansion
to order $m_q^2 m_D/T^3$ and $m_D^3/T^3$ in order to determine all terms contributing through order $g^5$.
We note that the soft scales are  given by $m_q$ and $m_D$ whereas the
hard scale is given by $T$, which leads to two different phase-space regions as discussed in Sec.~\ref{1sis}.
In the hard-hard region, all three momenta $P$, $Q$, and $R$ are hard whereas 
in the hard-soft region, two of the three momenta are hard and the other one is soft. 
\subsubsection*{The hh contribution}
The self-energies for hard momenta are suppressed~\cite{Braaten:1989kk,Braaten:1989mz,Braaten:1991gm,Andersen:2002ey,
Andersen:2003zk} by $m_D^2/T^2$
or $m_q^2/T^2$ relative to the propagators. For hard momenta, one just needs to expand in 
powers of gluon self-energies $\Pi_T$, $\Pi_L$, and quark self-energy $\Sigma$. So, the hard-hard 
contribution of $ {\cal F}_{3qg}$ 
and $ {\cal F}_{4qg}$ in (\ref{coll}) can be written as
\be
{\cal F}^{(hh)}_{3qg+4qg} &=&  (d-1)g^2\left[\ \sumintff_{\{PQ\}}\frac{1}{P^2Q^2} - 2 \sumintbf_{P\{Q\}}
\frac{1}{P^2Q^2}\right] 
\nn
&+&
 2m_D^2g^2\ \sumintbf_{P\{Q\}}\left[\frac{1}{p^2P^2Q^2}{\cal T}_P + \frac{1}
{P^4Q^2}- \frac{d-2}{d-1}\frac{1}{p^2P^2Q^2}\right]
\nn
 & + & m_D^2g^2\sumintff_{\{PQ\}}\left[\frac{d+1}{d-1}\frac{1}{P^2Q^2r^2} - \frac{4d}{d-1}
\frac{q^2}{P^2Q^2r^4} - \frac{2d}{d-1}\frac{P\cdot Q}{P^2Q^2r^4}\right]{\cal T}_R
\nn
 &+& m_D^2g^2\sumintff_{\{PQ\}}\Bigg[\frac{3-d}{d-1}\frac{1}{P^2Q^2R^2} + \frac{2d}{d-1}
\frac{P\cdot Q}{P^2Q^2r^4} - \frac{d+2}{d-1}\frac{1}{P^2Q^2r^2}
\nn
&& \hspace{2cm}+\frac{4d}{d-1}\frac{q^2}
{P^2Q^2r^4} - \frac{4}{d-1}\frac{q^2}{P^2Q^2r^2R^2}\Bigg]
\nn
&+& 2m_q^2g^2(d-1) \sumintff_{\{PQ\}}\left[\frac{1}{P^2Q_0^2Q^2} + \frac{p^2-r^2}
{q^2P^2Q_0^2R^2}\right]{\cal T}_Q\nonumber 
\ee
\be
&+& 2m_q^2g^2(d-1) \sumintbf_{P\{Q\}}\left[\frac{2}{P^2Q^4} - \frac{1}
{P^2Q_0^2Q^2}{\cal T}_Q\right]
\nn
&+& 2m_q^2g^2(d-1) \sumintff_{\{PQ\}}\left[\frac{d+3}{d-1}\frac{1}{P^2Q^2R^2} -\frac{2}{P^2Q^4}
- \frac{p^2-r^2}{q^2P^2Q^2R^2}\right] \, ,
\label{2loop_hh}
\ee
where the various sum-integrals and integrals are evaluated in Sec.~(\ref{asis}) and~(\ref{bsis}). Using those results, 
the hh contribution becomes
\be
{\cal F}^{(hh)}_{3qg+4qg} &=&  \frac{5\pi^2}{72}\frac{\alpha_s}{\pi}T^4\left[1 + \frac{72}{5}\ \hat\mu^2
+ \frac{144}{5}\ \hat\mu^4\right] 
\nn
 &-& \frac{1}{72}\frac{\alpha_s}{\pi}\left(\frac{\Lambda}
{4\pi T}\right)^{4\epsilon}\left[\frac{1+6(4-3\zeta(3))\ \hat\mu^2
-120(\zeta(3)-\zeta(5))\ \hat\mu^4 + {\cal O}
\left(\hat\mu^6\right)}{\epsilon} \right.\nn
&+& 1.3035 - 59.9055\ \hat\mu^2
- 75.4564\ \hat\mu^4 + {\cal O}\left(\hat\mu^6\right)\Big] m_D^2T^2
\nn&+& 
  \frac{1}{8}\frac{\alpha_s}{\pi}\left(\frac{\Lambda}{4\pi T}\right)^{4\epsilon}
\bigg[\frac{1-12\ \hat\mu^2}{\epsilon} + 8.9807 -152.793\ \hat\mu^2 
\nn
&&+115.826\ \hat\mu^4 + {\cal O}\left
(\hat\mu^6\right)\bigg] m_q^2T^2 \, . 
\label{2loop_hh_f}
\ee
\subsubsection*{The hs contribution}
Following Ref.~\cite{Andersen:2002ey,Andersen:2003zk} one can extract the hard-soft contribution from~(\ref{coll}) as the momentum 
$P$ is soft whereas momenta $Q$ and $R$ are always hard. The function associated with 
the soft propagator $\Delta_T(0,{\bf p})$ or $\Delta_X(0,{\bf p})$
can be expanded in powers of the soft momentum ${\bf p}$. For
$\Delta_T(0,{\bf p})$, the resulting integrals over ${\bf p}$\
are not associated with any scale and they vanish in dimensional regularization.
The integration measure $\int_{\bf p}$ scales like $m_D^3$,
the soft propagator $\Delta_X(0,{\bf p})$ scales like $1/m_D^2$,
and every power of $p$ in the numerator scales like $m_D$.

The contributions that survive only through order $g^2 m_D^3 T$ 
and $m_q^2m_Dg^3T$ from  $ {\cal F}_{3qg}$ and $ {\cal F}_{4qg}$ in (\ref{coll}) 
are
\be
{\cal F}_{3qg+4qg}^{(hs)}\!\!\!&=&\!\!\!g^2T\int\limits_{\bf p}\frac{1}{p^2+m^2_D}
        \sumintf_{\{Q\}}\left[\frac{2}{Q^2}-\frac{4q^2}{Q^4}\right]\nonumber
\ee
\be
\!\!\!&+&\!\!\!
2m_D^2g^2T\int\limits_{\bf p}\frac{1}{p^2+m_D^2}\sumintf_{\{Q\}}
     \left[\frac{1}{Q^4} - \frac{2(3+d)}{d}\frac{q^2}{Q^6}+\frac{8}{d}
      \frac{q^4}{Q^8}\right]
\nn
\!\!\!&-&\!\!\!
4m_q^2g^2T\int\limits_{\bf p}\frac{1}{p^2+m_D^2}
     \sumintf_{\{Q\}}\left[\frac{3}{Q^4}
     -\frac{4q^2}{Q^6} - \frac{4}{Q^4} {\cal T}_Q
      -\frac{2}{Q^2}\left\langle \frac{1}{(Q\!\cdot\!Y)^2} 
       \right\rangle_{\!\!\bf \hat y}\right]\!\!.\hspace{15mm}
\ee
Using the sum-integrals and integrals contained in Sec.~(\ref{asis}) and (\ref{bsis}), the hard-soft contribution 
becomes
\be
{\cal F}^{(hs)}_{3qg+4qg} \!\!\!&=&\!\!\! -\frac{1}{6}\alpha_s m_DT^3 (1+12\ \hat\mu^2)
      -\frac{\alpha_s}{2\pi^2}m_q^2 m_DT + \left(\frac{\Lambda}{4\pi T}\right)
^{2\epsilon}\left(\frac{\Lambda}{2 m_D}\right)^{2\epsilon} m_D^3 T
\nn
\!\!\!&\times&\!\!\! \frac{\alpha_s}{24\pi^2}\left[\frac{1}{\epsilon} + 1 + 2\gamma_E + 4\ln2 
- 14\zeta(3)\ \hat\mu^2 +62\zeta(5)
\ \hat\mu^4 + {\cal O}\left(\hat\mu^6\right)\right]\!\!.\hspace{15mm} 
\label{2loop_hs_f}
\ee
\subsubsection*{The ss contribution}
As discussed earlier in Sec.~(\ref{expand}) there is no soft-soft contribution from the diagrams in 
Fig.~(\ref{diagramfig}) since at least one of the loops is fermionic.
\subsection{Thermodynamic potential}
\label{subthpot}
Now we can obtain the thermodynamic potential through two-loop order
in HTL perturbation theory for which the contributions involving quark lines are computed here whereas the ghost and 
gluon contributions are computed in Refs.~\cite{Andersen:2002ey,
Andersen:2003zk}. We also follow the variational mass prescription
as discussed in Chapter~\ref{chapter:1loop} to determine the mass parameter $m_D$ and $m_q$ from respective gap equations.
 The details about the variational mass prescription will be discussed in Sec.~\ref{nlopress_variational_eqn}.
\subsubsection*{Leading order thermodynamic potential}
Using the expressions of ${\cal F}_q$ with finite quark chemical potential in (\ref{Quark1loop}) and 
${\cal F}_g$ from Ref.~\cite{Andersen:2002ey,
Andersen:2003zk}, the total contributions from the one-loop diagrams
including all terms through order $g^5$ becomes
\be
\Omega_{\rm one\;loop} &=& - d_A\frac{\pi^2T^4}{45}\Bigg\{1+\frac{7}{4}\frac{d_F}{d_A}
\left(1+\frac{120}{7}\hat\mu^2+\frac{240}{7}\hat\mu^4\right) 
\nn
&-& \frac{15}{2}
\left[1+\epsilon\left(2+2\frac{\zeta'(-1)}{\zeta(-1)}+
2\ln{\frac{\hat\Lambda}{2}}\right)\right]\hat m_D^2 
\nn
&-&
30\frac{d_F}{d_A}\bigg[\left(1+12\hat\mu^2\right)+\epsilon\Big(2-2\ln2 + 
2\frac{\zeta'(-1)}{\zeta(-1)}+2\ln{\frac{\hat\Lambda}{2}} + 24(\gamma_E+2\ln2)\hat\mu^2
\nn
&-&
28\zeta(3)\hat\mu^4+ {\cal O}\left(\hat\mu^6\right)\Bigg)\bigg]\hat m_q^2
+30\left(\frac{\Lambda}{2 m_D}\right)^{2\epsilon}\left[1+\frac{8}{3}\epsilon\right]\hat m_D^3
\nn
&+&
\frac{45}{8}\left(\frac{1}{\epsilon}+2\ln{\frac{\hat\Lambda}{2}}-7+2\gamma_E
+ \frac{2\pi^2}{3}\right)\hat m_D^4 - 60\frac{d_F}{d_A}(\pi^2-6)\hat m_q^4\Bigg\}
 \;,
\label{Omega-1loop}
\ee
where $\hat m_D$, $\hat m_q$, $\hat\Lambda$, and $\hat \mu$ are dimensionless variables:
\be
\hat m_D &=& \frac{m_D}{2 \pi T}  \;,
\\
\hat m_q &=& \frac{m_q}{2 \pi T}  \;,
\\
\hat \Lambda &=& \frac{\Lambda}{2 \pi T}  \;,
\\
\hat \mu &=& \frac{\mu}{2 \pi T}  \;. 
\ee
Adding the counterterm in (\ref{count0}),
we obtain the thermodynamic potential at leading order in the $\delta$-expansion:
\be
\Omega_{\rm LO}& =& - d_A\frac{\pi^2T^4}{45}\Bigg\{1+\frac{7}{4}\frac{d_F}{d_A}\left(
1+\frac{120}{7}\hat\mu^2+\frac{240}{7}\hat\mu^4\right)
\nn
&-& \frac{15}{2}\left[1+\epsilon
\left(2+2\frac{\zeta'(-1)}{\zeta(-1)}+2\ln{\frac{\hat\Lambda}{2}}\right)\right]\hat m_D^2 
\nn&-&
30\frac{d_F}{d_A}\left[\bigg(1+12\hat\mu^2\right)+\epsilon\Big(2-2\ln2+2\frac{\zeta'(-1)}
{\zeta(-1)}+2\ln{\frac{\hat\Lambda}{2}} + 24(\gamma_E+2\ln2)\hat\mu^2
\nn
&-&
28\zeta(3)\hat\mu^4 + {\cal O}\left(\hat\mu^6\right)\Big)\bigg]\hat m_q^2
+30\left(\frac{\Lambda}{2 m_D}\right)^{2\epsilon}\left[1+\frac{8}{3}\epsilon\right]
\ \hat m_D^3
\nn
&+&
\frac{45}{8}\left(2\ln{\frac{\hat\Lambda}{2}}-7+2\gamma_E+\frac{2\pi^2}{3}\right)
\hat m_D^4 - 60\frac{d_F}{d_A}(\pi^2-6)\hat m_q^4\Bigg\}
\;,
\label{Omega-LO}
\ee
where we have kept terms of ${\cal O}(\epsilon)$ since they will be needed for the
two-loop renormalization.
\subsubsection*{Next-to-leading order thermodynamic potential}
The complete expression for the next-to-leading order correction to the thermodynamic potential
is the sum of the contributions from all two-loop diagrams, the quark and gluon counter-terms,
and renormalization counterterms. Adding the contributions of the two-loop diagrams, ${\cal F}_{3qg+4qg}$, 
involving a quark line in (\ref{2loop_hh_f}) and (\ref{2loop_hs_f}) and the contributions of
${\cal F}_{3g+4g+gh}$ from Ref.~\cite{Andersen:2002ey,
Andersen:2003zk}, one obtains
\be
\Omega_{\rm two\;loop} &=& -d_A\frac{\pi^2T^4}{45}\frac{\alpha_s}{\pi}\Bigg\{-\frac{5}{4}
                       \left[c_A+\frac{5}{2}s_F\left(1+\frac{72}{5}\ \hat\mu^2+\frac{144}{5}
                       \ \hat\mu^4\right)\right]
\nn &+&               15\left(c_A + s_F\left(1+12\ \hat\mu^2\right)\right){\hat m_D}
-  \frac{55}{8}\bigg[\bigg(c_A-\frac{4}{11}s_F
\Big[1+6(4-3\zeta(3))\hat\mu^2
\nn
& -& 120(\zeta(3)-\zeta(5))\ \hat\mu^4 + {\cal O}\left(
                       \hat\mu^6\right)\Big]\bigg)\left(\frac{1}{\epsilon}+4\ln{\frac
                       {\hat\Lambda}{2}}\right)
\nn &-& 
                      s_F\left(0.471 - 34.876\ \hat\mu^2- 21.021\ \hat\mu^4 +{\cal O}
                      \left(\hat\mu^6\right)\right)
\nn\!\!\! &-&\!\!\! 
                     c_A\left(\frac{72}{11}\ln{\hat m_D}
                     -1.969\right) \bigg]{\hat m_D^2}- \frac{45}{2} s_F\bigg[\left(1-12\hat\mu^2\right)\left(\frac{1}
                     {\epsilon} + 4 \ln{\frac{\hat\Lambda}{2}}\right)
\nn
&+& 
                   8.981 -152.793\hat\mu^2 +115.826\hat\mu^4 + {\cal O}\left(\hat\mu^6\right)\bigg]
                     {\hat m_q^2} 
\nn &+&  
                   180 s_F\ {\hat m_D}{\hat m_q^2} + \frac{165}{4}\bigg[\left(c_A-\frac{4}
                   {11}s_F\right)\left(\frac{1}{\epsilon} + 4\ \ln{\frac{\hat\Lambda}{2}} 
                  - 2\ln{\hat m_D}\right) 
\nn
&+&               c_A\left(\frac{27}{11}+2\gamma_E\right)-\frac{4}{11}s_F\Big(1+2\gamma_E + 4\ln2 -14 \zeta(3)\ \hat\mu^2
\nn&+& 
                    62\zeta(5)
                  \ \hat\mu^4 + {\cal O}\left(\hat\mu^6\right)\Big)\bigg] {\hat m_D^3}
                  \Bigg\}\, ,
\label{Omega2loop}
\ee
where $c_A=N_c$ and $s_F=N_f/2$.

The HTL gluon counter-term is the same as obtained at zero chemical potential in Refs.~\cite{Andersen:2002ey,
Andersen:2003zk} and can be written as

\be
\Omega_{\rm gct} = - d_A\frac{\pi^2 T^4}{45}\left[\frac{15}{2}\ {\hat m_D^2} - 45 \hat m_D^3
                    -\frac{45}{4}\left(\frac{1}{\epsilon} + 2\ln{\frac{\hat\Lambda}{2}} - 7 
                   + 2\gamma_E + \frac{2\pi^2}{3}\right) {\hat m_D^4}\right]\!.\hspace{1cm}
\label{OmegaGct}
\ee
The HTL quark counterterm as given by~(\ref{count}) is
\be
\Omega_{qct} = -d_F\frac{\pi^2 T^4}{45}\left[30(1 + 12\ \hat\mu^2)\ {\hat m_q^2} +
                     120(\pi^2-6)\ {\hat m_q^4}\right] \, .
\label{OmegaQct}
\ee
The ultraviolet divergences that remain after adding (\ref{Omega2loop}), (\ref{OmegaGct}), and (\ref{OmegaQct})
can be removed by renormalization of the vacuum energy density ${\cal E}_0$
and the HTL mass parameter $m_D$ and $m_q$.
The renormalization contributions~\cite{Andersen:2002ey,
Andersen:2003zk} at first order in $\delta$ are
\be
\Delta\Omega = \Delta_1{\cal E}_0 + \Delta_1 m_D^2\frac{\partial}{\partial m_D^2}\Omega_{LO}
+ \Delta_1 m_q^2\frac{\partial}{\partial m_q^2}\Omega_{LO} \, .
\label{del_tot}
\ee
The counterterm $\Delta_1{\cal E}_0$ at first order in $\delta$ will be same as the zero chemical
potential counterterm
\be
\Delta_1{\cal E}_0 = - \frac{d_A}{64\pi^2\epsilon} m_D^4 \, .
\label{del1e0}
\ee
The mass counterterms necessary at first order in $\delta$ are found to be
\be
\Delta_1 \hat m_D^2 &=& -\frac{\alpha_s}{3\pi\epsilon}\Bigg[\frac{11}{4}c_A-s_F-s_F\left(1+6\hat m_D
                  \right)
\nn
&\times&\bigg[(24-18\zeta(3))\hat\mu^2 + 120(\zeta(5)-\zeta(3))\hat\mu^4 + 
                  {\cal O}\left(\hat\mu^6\right)\bigg]\Bigg]\ \hat m_D^2
\label{delmd}
\ee
and
\be
\Delta_1 \hat m_q^2 \;&=& -\frac{\alpha_s}{3 \pi \epsilon}\left[\frac{9}{8}\frac{d_A}{c_A}\right] 
\frac{1-12\ \hat\mu^2}{1+12\ \hat\mu^2}\ \hat m_q^2
\;.
\label{delmq}
\ee
Using the above counterterms, the complete contribution from the counterterms in (\ref{del_tot}) at first order
in $\delta$ at finite chemical potential becomes
\be
\Delta\Omega &=& - d_A\frac{\pi^2T^4}{45}\Bigg\{\frac{45}{4\epsilon}\hat m_D^4+\frac{\alpha_s}
                 {\pi}\bigg[\frac{55}{8}\bigg(c_A-\frac{4}{11}s_F\Big[1+(24-18\zeta(3))
                 \hat\mu^2 
\nn
&+& 120(\zeta(5)-\zeta(3))\hat\mu^4 + {\cal O}\left(\hat\mu^6\right)
                 \Big]\bigg) \left(\frac{1}{\epsilon} + 2 + 2\frac{\zeta'(-1)}{\zeta(-1)}+2\ln{\frac{\hat
                 \Lambda}{2}}\right)\hat m_D^2
\nn&-&
\frac{165}{4}\left(c_A-\frac{4}{11}s_F\right)\left(\frac{1}{\epsilon}+2+
2\ln{\frac{\hat\Lambda}{2}} - 2\ln{\hat m_D}\right)\hat m_D^3
\nn
&-&
\frac{165}{4}\frac{4}{11}s_F\left[(24-18\zeta(3))\hat\mu^2 + 120(\zeta(5)-
\zeta(3))\hat\mu^4 + {\cal O}\left(\hat\mu^6\right)\right]
\nn
&\times& \left(2\frac{\zeta'(-1)}{\zeta(-1)}
 +2\ln{\hat m_D} \right)\hat m_D^3
\nn
&+&
\frac{45}{2} s_F \frac{1-12\ \hat\mu^2}{1+12\ \hat\mu^2}\Bigg (\frac{1+12\ \hat\mu^2}{\epsilon}
+ 2 + 2\ln{\frac{\hat\Lambda}{2}} -2\ln 2
 +2\frac{\zeta'(-1)}{\zeta(-1)}  \nn 
&+&
 24(\gamma_E+2\ln2)\ \hat\mu^2-28\zeta(3)\ \hat\mu^4 + {\cal O}
\left(\hat\mu^6\right)
 \Bigg ) \hat m_q^2
\bigg]\Bigg\} .
\label{OmegaVMct}
\ee

Adding the contributions from the two-loop diagrams in~(\ref{Omega2loop}), the
HTL gluon and quark counterterms in~(\ref{OmegaGct}) and~(\ref{OmegaQct}), the
contribution from vacuum and mass renormalizations in~(\ref{OmegaVMct}), and
the leading-order thermodynamic potential in~(\ref{Omega-LO}) we
obtain the complete expression for the QCD thermodynamic potential 
at next-to-leading order in HTLpt:

\be
\Omega_{\rm NLO}\!\!\!&=&\!\!\!
           - d_A \frac{\pi^2 T^4}{45} \Bigg\{ 
	   1 + \frac{7}{4} \frac{d_F}{d_A}\left(1+\frac{120}{7}
           \hat\mu^2+\frac{240}{7}\hat\mu^4\right) - 15 \hat m_D^3 
\nn \!\!\!&-&\!\!\!
           \frac{45}{4}\left(\log\frac{\hat\Lambda}{2}-\frac{7}{2}+\gamma_E+\frac{\pi^2}{3}\right)\hat m_D^4
           + 60 \frac{d_F}{d_A}\left(\pi^2-6\right)\hat m_q^4	
\nn
\!\!\!&+&\!\!\! \frac{\alpha_s}{\pi} \Bigg[ -\frac{5}{4}\left(c_A + \frac{5s_F}{2}\!\left(1+\frac{72}{5}
           \hat\mu^2+\frac{144}{5}\hat\mu^4\right)\!\right) 
	   + 15 \left(c_A+s_F(1+12\hat\mu^2)\right)\!\hat m_D
\nn \!\!\!&-&\!\!\! 
	 \frac{55}{4}\Bigg\{ c_A\left(\log\frac{\hat\Lambda}{2}- \frac{36}{11}\log\hat m_D - 2.001\right)
		- \frac{4}{11} s_F \left[\left(\log\frac{\hat\Lambda}{2}-2.337\right)\right.
\nn \!\!\!&+&\!\!\! 
	\left. (24-18\zeta(3))\left(\log\frac{\hat\Lambda}{2} -15.662\right)\hat\mu^2
	+ 120\left(\zeta(5)-\zeta(3)\right)
\right.\nn
\!\!\!&\times&\!\!\!\left.\left(\log\frac{\hat\Lambda}{2} -1.0811\right)\hat\mu^4 + 
         {\cal O}\left(\hat\mu^6\right)\right] \!\!\Bigg\} \hat m_D^2-45 \, s_F \left\{\log\frac{\hat\Lambda}{2}
        + 2.198  -44.953\hat\mu^2\right. 
\nn \!\!\!&-&\!\!\! 
	\left.\left(288 \ln{\frac{\hat\Lambda}{2}} 
       +19.836\right)\hat\mu^4 + {\cal O}\left(\hat\mu^6\right)\right\} \hat m_q^2
       + \frac{165}{2}\left\{ c_A\left(\log\frac{\hat\Lambda}{2}+\frac{5}{22}+\gamma_E\right)\right.
\nn \!\!\!&-&\!\!\!
	\left. \frac{4}{11} s_F \left(\log\frac{\hat\Lambda}{2}-\frac{1}{2}+\gamma_E+2\ln2 -7\zeta(3)\hat\mu^2+
         31\zeta(5)\hat\mu^4 + {\cal O}\left(\hat\mu^6\right) \right)\right\}\hat m_D^3
\nn \!\!\!&+&\!\!\!
         15 s_F \left(2\frac{\zeta'(-1)}{\zeta(-1)}
         +2\ln \hat m_D\right)\Big[(24-18\zeta(3))\hat\mu^2 
\nn
 \!\!\!&+&\!\!\! 120(\zeta(5)-\zeta(3))\hat\mu^4 + 
           {\cal O}\left(\hat\mu^6\right)\Big] \hat m_D^3
	+ 180\,s_F\hat m_D \hat m_q^2 \Bigg]
\Bigg\} .
\label{Omega-NLO}
\ee
%
\section{The necessary Sum-Integrals}
\label{asis}
In the imaginary-time (Euclidean time) formalism for the field theory of a hot and dense medium, 
the 4-momentum $P=(P_0,{\bf p})$ is Euclidean with $P^2=P_0^2+{\bf p}^2$. 
The Euclidean energy $P_0$ has discrete values:
$P_0=2n\pi T$ for bosons and $P_0=(2n+1)\pi T-i\mu$ for fermions,
where $n$ is an integer running from $-\infty$ to $\infty$, $\mu$ is the quark chemical potential,
and $T=1/\beta$ is the temperature of the medium. Loop diagrams usually then involve sums over $P_0$ and integrals 
over ${\bf p}$. In dimensional regularization, the integral over spatial momentum  is generalized
to $d = 3-2 \epsilon$ spatial dimensions. The dimensionally regularized bosonic and fermionic sum-integrals are defined in
Eq.~(\ref{sumint-def}). All other frequency sums that appear in this chapter can be evaluated using the contour integration technique
as discussed in Chapter~\ref{chapter:htl} and are listed below:
\subsection{Simple one loop sum-integrals}
\label{a1sis}
The specific fermionic one-loop sum-integrals needed are
\be 
\sumintf_{\{P\}} \ \ln P^2 = \frac{7\pi^2}{360}T^4 \left(1 + \frac{120\ {\hat\mu^2}}{7}
+\frac{240\ {\hat\mu^4}}{7}\right) .
\ee
\vspace{-.5cm}
\be
\sumintf_{\{P\}} \ \frac{1}{P^2}\!\!\!&=&\!\!\!-\frac{T^2}{24}\left(\frac{\Lambda}
{4\pi T}\right)^{2\epsilon}\Bigg[1 + 12\hat\mu^2 + \epsilon\bigg(
2-2\ln2+2\frac{\zeta'(-1)}{\zeta(-1)} 
\nn
&+&  24(\gamma_E + 2\ln2)\hat\mu^2 -28\zeta(3)
\ \hat\mu^4 +{\cal O}\left(\hat\mu^6\right)\bigg)
\nn
&+& \epsilon^2\bigg(4 + \frac{\pi^2}{4}-4\ln2
 + 4(1-\ln2)\frac{\zeta'(-1)}{\zeta(-1)}-2\ln^22
\nn
&+&
2\frac{\zeta''(-1)}{\zeta(-1)} + 94.5749\ \hat\mu^2 - 143.203\ \hat\mu^4 
+ {\cal O}\left(\hat\mu^6\right)\bigg )\Bigg]. \hspace{1cm} 
\ee
\vspace{-.5cm}
\be
\sumintf_{\{P\}} \ \frac{1}{P^4}&=&\frac{1}{(4\pi)^2}\left(\frac{\Lambda}
{4\pi T}\right)^{2\epsilon}\Bigg[\frac{1}{\epsilon} +  2\gamma_E + 4\ln2 
 -14\ \zeta(3)\ \hat\mu^2 
\nn
&+& 62\ \zeta(5)\ \hat\mu^4 + {\cal O}\left(\hat\mu^6\right)+\epsilon\bigg( 4(2\gamma_E + \ln2)\ln 2 
\nn
&-&  4\gamma_1 + \frac{\pi^2}{4}
 - 71.6013\ \hat\mu^2 + 356.329\ \hat\mu^4 +{\cal O}\left(\hat\mu^6\right)\bigg)
\Bigg].
\label{1byP4}
\ee
\vspace{-.5cm}
\be
\sumintf_{\{P\}} \ \frac{p^2}{P^4}=-\frac{T^2}{16}\left(\frac{\Lambda}
{4\pi T}\right)^{2\epsilon}\left[1 +12\ \hat\mu^2 + \epsilon\left(\frac{4}{3} - 
2\ln2 +2\frac{\zeta'(-1)}{\zeta(-1)} 
\right.\right.\nn\left.\left. 
+ 8(3\gamma_E+6\ln2-1) \ \hat\mu^2- \ 28\ \zeta(3) \ \hat\mu^4 +{\cal O}
\left(\hat\mu^6\right)\right)\right] .
\ee
\vspace{-.5cm}
\be
\sumintf_{\{P\}} \frac{p^2}{P^6}&=&\frac{3}{4(4\pi)^2}\left(\frac{\Lambda}
{4\pi T}\right)^{2\epsilon}\Bigg[\frac{1}{\epsilon} + 2\gamma_E 
-\frac{2}{3} + 4\ln2 
\nn
&& -\ 14\zeta(3)\hat\mu^2 + 62 \zeta(5)\ \hat\mu^4 +
{\cal O}\left(\hat\mu^6\right) 
\Bigg].\hspace{.7cm}
\ee
\vspace{-.5cm}
\be
\sumintf_{\{P\}} \ \frac{p^4}{P^6}=-\frac{5T^2}{64}\left(\frac{\Lambda}
{4\pi T}\right)^{2\epsilon}\left[1 +12\ \hat\mu^2 + \epsilon\left(\frac{14}{15} - 
2\ln2 +2\frac{\zeta'(-1)}{\zeta(-1)}
\right.\right.\nn\left.\left.
 + 8\left(-\frac{8}{5}+3\gamma_E+6\ln2\right)\ \hat\mu^2 -\  28\ \zeta(3)\ \hat\mu^4 
+{\cal O}\left(\hat\mu^6\right)\right)\right] .
\ee
\vspace{-.5cm}
\be
\sumintf_{\{P\}} \ \frac{p^4}{P^8}=\frac{1}{(4\pi)^2}\left(\frac{\Lambda}
{4\pi T}\right)^{2\epsilon}\frac{5}{8}\Bigg[\frac{1}{\epsilon} + \bigg( 2\gamma_E 
-\frac{16}{15} + 4\ln2 
 -14\ \zeta(3)\ \hat\mu^2
\nn
 +\ 62\ \zeta(5)\ \hat\mu^4 +
{\cal O}\left(\hat\mu^6\right) \bigg) 
\Bigg] .
\ee
\vspace{-.5cm}
\be
\sumintf_{\{P\}} \ \frac{1}{p^2P^2}\!\!\!&=&\!\!\!\frac{1}{(4\pi)^2}\left(\frac{\Lambda}
{4\pi T}\right)^{2\epsilon}2\Bigg[\frac{1}{\epsilon} + \Big(2+ 2\gamma_E 
+ 4\ln2  -14\ \zeta(3)\ \hat\mu^2
\nn
\!\!\!&+&\!\!\! 62\ \zeta(5)\ \hat\mu^4 +
{\cal O}\left(\hat\mu^6\right) \Big)
+\epsilon\Big(4+8\ln2+4\ln^2 2+4\gamma_E
\nn
\!\!\!&+&\!\!\! 8\gamma_E\ln2+\frac{\pi^2}{4}-4\gamma_1
-105.259\ \hat\mu^2 + 484.908\ \hat\mu^4 + {\cal O}\left(\hat\mu^6\right)\Big)
\Bigg]\!.\hspace{10mm}
\label{1byp2P2}
\ee
where $\gamma_1$, appearing in Eqs.~(\ref{1byP4}) and (\ref{1byp2P2}), is the first Stieltjes 
gamma constant defined by the equation
\be
\zeta(1+z)=\frac{1}{z}+\gamma_E-\gamma_1z+{\cal O}(z^2).
\ee
and the numerical value of $\gamma_1\approx-0.0728158$.
\subsection{HTL one loop sum-integrals}
We also need some more difficult one-loop sum-integrals 
that involve the HTL function 
defined in (\ref{def-tf}).
The specific fermionic sum-integrals needed are
\be
\sumintf_{\{P\}} \ \frac{1}{P^4}{\cal T}_P &=& \frac{1}{(4\pi)^2}\left(\frac{\Lambda}
{4\pi T}\right)^{2\epsilon}\frac{1}{2}\Bigg[\frac{1}{\epsilon} + \bigg(1 + 2\gamma_E + 
4\ln2 - 14\ \zeta(3)\ \hat\mu^2 
\nn
&&\hspace{5.2cm} +\ 62\ \zeta(5)\ \hat\mu^4 
+{\cal O}\left(\hat\mu^6\right)\bigg)\Bigg]\!.\hspace{5mm}
\ee
\vspace{-1cm}
\be
\sumintf_{\{P\}} \ \frac{1}{p^2P^2}{\cal T}_P \!\!\!&=&\!\!\! \frac{2}{(4\pi)^2}\left(\frac{\Lambda}
{4\pi T}\right)^{2\epsilon}  \Bigg[\frac{\ln2}{\epsilon} + \frac{\pi^2}{6}+2\gamma_E\ln2
\nn
 &+&
\ln2 \Big( 5\ln2   -14\ \zeta(3)\hat\mu^2+ 62\ \zeta(5)\hat\mu^4 
+
{\cal O}\left(\hat\mu^6\right)\Big)
\nn
&+&\epsilon\left(17.5137 - 85.398\ \hat\mu^2 + 383.629\ \hat\mu^4 +
{\cal O}\left(\hat\mu^6\right)\right)
\Bigg]. \;\;\;
\ee
\vspace{-1cm}
\be
\sumintf_{\{P\}}\frac{1}{P^2P_0^2}{\cal T}_P \!\!\!&=&\!\!\! \frac{1}{(4\pi)^2}\left(\frac{\Lambda}
{4\pi T}\right)^{2\epsilon}  \Bigg[\frac{1}{\epsilon^2} + \frac{2}{\epsilon}\Big(
\gamma_E+2\ln2- 7\ \zeta(3)\hat\mu^2 
\nn
&+& 31\ \zeta(5)\hat\mu^4 +{\cal O}\left(\hat\mu^6\right)\Big)+\frac{\pi^2}{4} + 4\ln^22
\nn
\!\!\!&+&\!\!\!8\gamma_E\ln2 -4\gamma_1 -71.601\hat\mu^2 
+ 356.329\hat\mu^4 + {\cal O}\left(\hat\mu^6\right)\Bigg]\!.\hspace{15mm}
\ee
\vspace{-1cm}
\be
\sumintf_{\{P\}} \ \frac{1}{p^2P_0^2}\left({\cal T}_P\right)^2 = \frac{4}{(4\pi)^2}\left(\frac
{\Lambda}{4\pi T}\right)^{2\epsilon} \!\!\!&\ln2&\!\!\! \Bigg[\frac{1}{\epsilon} + \left(
 2\gamma_E + 5\ln2\right)- 14\ \zeta(3)\ \hat\mu^2
\nn&&\ \ \ \ +\ 62\ \zeta(5)\ \hat\mu^4 
+{\cal O}\left(\hat\mu^6\right)\Bigg] .
\ee
\vspace{-1cm}
\be
\sumintf_{\{P\}} \ \frac{1}{P^2}\left\langle\frac{1}{(P\cdot Y)^2}\right\rangle_{\bf\hat y} =
- \frac{1}{(4\pi)^2}\left(\frac
{\Lambda}{4\pi T}\right)^{2\epsilon}\!\!\!\!\!\!  &&\Bigg[\frac{1}{\epsilon} - 1+  2\gamma_E + 4\ln2
 - 14\ \zeta(3)\ \hat\mu^2
\nn&&\ \ \ \ \ \ +\  62\ \zeta(5)\ \hat\mu^4 
+{\cal O}\left(\hat\mu^6\right)\Bigg] .
\ee
\subsection{Simple two loop sum-integrals}
\be
\sumintff_{\{PQ\}} \frac{1}{P^2Q^2R^2}=\frac{T^2}{(4\pi)^2}\left(\frac{\Lambda}
{4\pi T}\right)^{4\epsilon}\left[\frac{\hat\mu^2}{\epsilon} + 2 (4\ln2+2\gamma_E+1)
\hat\mu^2-\frac{28}{3}\zeta(3)\hat\mu^4+{\cal O}\left(\hat\mu^6\right)\right]\!.\hspace{.5cm}
\label{2s1}
\ee
\be
 \sumintff_{\{PQ\}} \ \frac{1}{P^2Q^2r^2}&=&-\frac{T^2}{(4\pi)^2}\left(\frac{\Lambda}
{4\pi T}\right)^{4\epsilon}\frac{1}{6}\Bigg[\frac{1}{\epsilon}
\left(1+12\hat\mu^2\right) + 4-2\ln2+ 4 \frac{\zeta'(-1)}{\zeta(-1)}
\nn
&&\hspace{2cm}+\ 48\left(1+\gamma_E+\ln2\right)\ \hat\mu^2
- 76\zeta(3)\ \hat\mu^4+{\cal O}\left(\hat\mu^6\right)\Bigg]\!.\hspace{15mm}
\label{2s2}
\ee
\be
 \sumintff_{\{PQ\}} \ \frac{p^2}{P^2Q^2r^4}&=&\frac{T^2}{(4\pi)^2}\left(\frac{\Lambda}
{4\pi T}\right)^{4\epsilon}\left(-\frac{1}{12}\right)\Bigg[\frac{1}{\epsilon}
\left(1+12\hat\mu^2\right)+\frac{11}{3}+2\gamma_E
\nn
&-&2\ln2+2\frac{\zeta'(-1)}{\zeta(-1)}
+ 4\left(7 + 12\gamma_E + 12\ln2 - 3\zeta(3)\right)\ \hat\mu^2
\nn
&-& 4\left(27\zeta(3)-20\zeta(5)\right)\ \hat\mu^4+{\cal O}\left(\hat\mu^6\right)\Bigg] .
\label{2s3}
\ee
\be
 \sumintff_{\{PQ\}} \ \frac{P\cdot Q}{P^2Q^2r^4}\!\!\!&=&\!\!\!-\frac{T^2}{(4\pi)^2}\left(\frac{\Lambda}
{4\pi T}\right)^{4\epsilon}\frac{1}{36}\Bigg[1 - 6\gamma_E + 6 \frac{\zeta'(-1)}{\zeta(-1)}
+24\left\{2 +  3\zeta(3)\right\}
\ \hat\mu^2
\nn
&&\hspace{5cm} + \ 48(7\zeta(3)-10\zeta(5))\ \hat\mu^4+{\cal O}(\hat\mu^6)\Bigg]\!.\hspace{15mm}
\label{2s4}
\ee
\be
\sumintff_{\{PQ\}} \frac{p^2}{r^2P^2Q^2R^2}&=&-\frac{T^2}{(4\pi)^2}\left(\frac{\Lambda}
{4\pi T}\right)^{4\epsilon}\frac{1}{72}\Bigg[\frac{1}{\epsilon}
\Big(1 -12(1-3\zeta(3))\ \hat\mu^2 
\nn &+& 240(\zeta(3)-\zeta(5))\ \hat\mu^4+{\cal O}
\left(\hat\mu^6\right)\Big) 
\nn
&-& \left( 7.001 - 108.218\ \hat\mu^2
-304.034\ \hat\mu^4+{\cal O}(\hat\mu^6)\right)\Bigg] .
\label{2s5}
\ee
\be
 \sumintff_{\{PQ\}} \ \frac{p^2}{q^2P^2Q^2R^2}&=&\frac{T^2}{(4\pi)^2}\left(\frac{\Lambda}
{4\pi T}\right)^{4\epsilon}\frac{5}{72}\Bigg[\frac{1}{\epsilon}\bigg(1 - 
\frac{12}{5}\left(1+7\zeta(3)\right)\ 
\hat\mu^2 
\nn
&-&\frac{24}{5}\left(14\zeta(3)-31\zeta(5)\right)\ \hat\mu^4+{\cal O}
\left(\hat\mu^6\right)\bigg)
\nn
&+& \left(9.5424 - 185.706\ \hat\mu^2
+916.268\ \hat\mu^4+{\cal O}(\hat\mu^6)\right)\Bigg] .
\label{2s6}
\ee
\be
 \sumintff_{\{PQ\}} \ \frac{r^2}{q^2P^2Q^2R^2}\!\!\!&=&\!\!\! -\frac{T^2}{(4\pi)^2}\left(\frac{\Lambda}
{4\pi T}\right)^{4\epsilon}\frac{1}{18}\Bigg[\frac{1}{\epsilon}\bigg(1 + 
3(-2+7\zeta(3))\hat\mu^2 
\nn
&+& 6(14\zeta(3)-31\zeta(5))\hat\mu^4 + \ {\cal O}(\hat\mu^6)\bigg)
\nn 
 &+&\left(8.143 + 96.935 \ \hat\mu^2
- 974.609\ \hat\mu^4+{\cal O}(\hat\mu^6)\right)\Bigg] .
\label{2s7}
\ee
The generalized two loop sum-integrals can be written from \cite{Andersen:2003zk} as
{\small
\be
 \sumintff_{\{PQ\}}\!\!\! &&\!\!\!|F(P) G(Q) H(R) =
\int\limits_{PQ} F(P) G(Q) H(R)
\nn &&
- \int\limits_{p_0,{\bf p}} \epsilon(p_0) n_F(|p_0|) \, 2 \, {\rm Im} F(-i p_0+ \varepsilon,{\bf p}) 
	\, {\rm Re} \int\limits_Q G(Q) H(R)\bigg|_{P_0 = -ip_0 + \varepsilon}
\nn && 
- \int\limits_{p_0,{\bf p}} \epsilon(p_0) n_F(|p_0|) \, 2 \, {\rm Im} G(-i p_0+ \varepsilon,{\bf p}) 
	\, {\rm Re} \int\limits_Q H(Q) F(R)\bigg|_{P_0 = -ip_0 + \varepsilon}
\nn && 
+ \int\limits_{p_0,{\bf p}} \epsilon(p_0) n_B(|p_0|) \, 2 \, {\rm Im} H(-i p_0+ \varepsilon,{\bf p}) 
	\, {\rm Re} \int\limits_Q F(Q) G(R)\bigg|_{P_0 = -ip_0 + \varepsilon}
\nn &&  
+ \int\limits_{p_0,{\bf p}}\!\!\! \epsilon(p_0) n_F(|p_0|) \, 2 \, {\rm Im} F(-i p_0+ \varepsilon,{\bf p}) \,
        \!\!\!\int\limits_{q_0,{\bf q}}\!\!\! \epsilon(q_0) n_F(|q_0|) \, 2 \, {\rm Im} G(-i q_0+ \varepsilon,{\bf q}) 
	\, {\rm Re} H(R)\bigg|_{R_0 = i (p_0 + q_0)+ \varepsilon}
\nn && 
- \int\limits_{p_0,{\bf p}}\!\!\! \epsilon(p_0) n_F(|p_0|) \, 2 \, {\rm Im} G(-i p_0+ \varepsilon,{\bf p}) \,
        \!\!\!\int\limits_{q_0,{\bf q}}\!\!\! \epsilon(q_0) n_B(|q_0|) \, 2 \, {\rm Im} H(-i q_0+ \varepsilon,{\bf q}) 
	\, {\rm Re} F(R)\bigg|_{R_0 = i (p_0 + q_0)+ \varepsilon}
\nn &&  
- \int\limits_{p_0,{\bf p}}\!\!\! \epsilon(p_0) n_B(|p_0|) \, 2 \, {\rm Im} H(-i p_0+ \varepsilon,{\bf p}) \,
        \!\!\!\int\limits_{q_0,{\bf q}}\!\!\! \epsilon(q_0) n_F(|q_0|) \, 2 \, {\rm Im} F(-i q_0+ \varepsilon,{\bf q}) 
	\, {\rm Re} G(R)\bigg|_{R_0 = i (p_0 + q_0)+ \varepsilon}
\;.\nn
\label{int-2loop}
\ee
}
After applying Eq.~(\ref{int-2loop}) and using the delta function to calculate the $P_0$ and $Q_0$
integrations, the sum-integral (\ref{2s5}) reduces to 
\be
 \sumintff_{\{PQ\}} \ \frac{1}{P^2Q^2R^2} =\int\limits_{\bf pq}\frac{n_F^{-}(p)-n_F^{+}(p)}{2p}\frac{n_F^{-}(q)
-n_F^{+}(q)}{2q}\frac{2p\ q}{\Delta(p,q,r)} \, ,
\ee
where 
\be
n_F^{\pm}(p) = \frac{1}{e^{\beta(p\pm\mu)}+1}
\ee
and
\be
 \Delta(p,q,r)=p^4+q^4+r^4 -2(p^2q^2+q^2r^2+p^2r^2) = -4p^2q^2(1-x^2) \, ,
\ee
and using the result of Eq.~(\ref{th-2s1}), we get sum-integral (\ref{2s1}) and agree with \cite{Vuorinen:2003fs}.


After applying Eq.~(\ref{int-2loop}), the sum-integral (\ref{2s2}) reduces to

\be
 \sumintff_{\{PQ\}} \ \frac{1}{P^2Q^2r^2} = -2 \int\limits_{\bf p}\frac{n_F(p)}{2 p}\int\limits_Q
\frac{1}{Q^2r^2} + \int\limits_{\bf pq}\frac{n_F(p)n_F(q)}{4pq}\frac{1}{r^2} \, ,
\ee
where $n_F(p)=n^-_F(p)+n^+_F(p)$ . Now using the result of 4-dimensional integrals from
\cite{Andersen:2003zk} and applying Eq.~(\ref{th-f}) and Eq.~(\ref{th-2s2}), we can calculate sum-integral
Eq.~(\ref{2s2}). The sum-integrals (\ref{2s3}) can be calculated in same way:
 \be
 \sumintff_{\{PQ\}} \ \frac{p^2}{P^2Q^2r^4} = -2 \int\limits_{\bf p}\frac{n_F(p)}{2 p}\int\limits_Q
\frac{p^2}{Q^2r^4} + \int\limits_{\bf pq}\frac{n_F(p)n_F(q)}{4pq}\frac{p^2}{r^4}.
\ee
The sum-integral (\ref{2s4}) can be written as
\be
 \sumintff_{\{PQ\}} \ \frac{P\cdot Q}{P^2Q^2r^4}= \sumintff_{\{PQ\}} \ \frac{P_0 Q_0}{P^2Q^2r^4}+
\frac{1}{2} \sumintff_{\{PQ\}} \ \frac{1}{P^2Q^2r^2} -  \sumintff_{\{PQ\}} \ \frac{p^2}{P^2Q^2r^4}
\ee
Using Eq.~(\ref{int-2loop}) and after doing $P_0$ and $Q_0$ integrations, first  
sum-integral above reduces to
\be
 \sumintff_{\{PQ\}} \ \frac{P_0 Q_0}{P^2Q^2r^4} =  \int\limits_{\bf pq}\frac{n_F^{-}(p)-n_F^{+}(p)}{2\ p}
\ \frac{n_F^{-}(q)-n_F^{+}(q)}{2\ q}\ \frac{p\ q}{r^4} \, ,
\ee 
and the result is given in Eq.~(\ref{th-2s4}). 
The second term and third terms sum-integrals above are linear combinations of Eq.~(\ref{2s2})
and Eq.~(\ref{2s3}). Adding all of them, we get required sum-integral.


Similarly after applying Eq.~(\ref{int-2loop}), the sum-integral (\ref{2s5}) reduces to 
\be
 \sumintff_{\{PQ\}}\ \frac{p^2}{r^2P^2Q^2R^2} &=& \left.\int\limits_{\bf p}\frac{n_B(p)}{p}\int\limits_Q
\frac{r^2}{p^2Q^2R^2}\right|_{P_0=-i p} 
\nn
&-& \left.\int\limits_{\bf p}\frac{n_F(p)}{2p}
\int\limits_Q\frac{1}{Q^2R^2}\left(\frac{q^2}{r^2}+\frac{p^2}{q^2}\right)\right|_{P_0=-i p} 
\nn
&+& \int\limits_{\bf pq}\frac{n_F(p)n_F(q)}{4pq}\ \frac{q^2}{r^2}\ \frac{r^2-p^2-q^2}{
\Delta(p,q,r)} 
\nn
&-&  \int\limits_{\bf pq}\frac{n_F(p)n_B(q)}{4pq}\ \frac{p^2+r^2}{q^2}
\ \frac{r^2-p^2-q^2}{\Delta(p,q,r)} \, .
\label{2s5_1}
\ee 
So
\be
\left\langle\frac{p^2+r^2}{q^2}\ \frac{r^2-p^2-q^2}{\Delta(p,q,r)}\right\rangle_{\hat p
\cdot\hat q} = \frac{1}{2 q^2\ \epsilon} \, ,
\ee
and
\be
\left\langle\frac{q^2}{r^2}\ \frac{r^2-p^2-q^2}{\Delta(p,q,r)}\right\rangle_{\hat p
\cdot\hat q} &=&\left\langle\frac{q^2}{\Delta(p,q,r)}\right\rangle_x - \left\langle
\frac{q^2(p^2+q^2)}{\Delta(p,q,r)}\right\rangle_x \, ,
\nn
&=&\frac{1-2\epsilon}{8\epsilon}\frac{1}{p^2} - \frac{1}{2\epsilon}\left\langle
\frac{q^2}{r^4}\right\rangle_x-\frac{1-2\epsilon}{8\epsilon}\frac{1}{p^2}
\nn 
&=& 
- \frac{1}{2\epsilon}\left\langle
\frac{q^2}{r^4}\right\rangle_x \, .
\ee
Using the above angular integration, Eq.~(\ref{2s5_1}) becomes
\be
 \sumintff_{\{PQ\}}\ \frac{p^2}{r^2P^2Q^2R^2} &=& \left.\int\limits_{\bf p}\frac{n_B(p)}{p}\int\limits_Q
\frac{r^2}{p^2Q^2R^2}\right|_{P_0=-i p}
\nn
&-& \left.\int\limits_{\bf p}\frac{n_F(p)}{2p}
\int\limits_Q\frac{1}{Q^2R^2}\left(\frac{q^2}{r^2}+\frac{p^2}{q^2}\right)\right|_{P_0=-i p} 
\nn
&-&\frac{1}{2\epsilon} \ \int\limits_{\bf pq}\frac{n_F(p)n_F(q)}{4pq}\  
\frac{p^2}{r^4}  - \frac{1}{2 \epsilon}\ \int\limits_{\bf pq}\frac{n_F(p)n_B(q)}{4pq}\ 
 \frac{1}{ q^2} \, .
\label{2s5_2}
\ee 
Using the 4-dimensional integrals from \cite{Andersen:2003zk} and Eqs.~(\ref{th-b}), ({\ref{th-f}}), 
({\ref{th-f1}}) and ({\ref{th-ff}}), 
we obtain the sum-integral (\ref{2s5}).

Similarly after  applying Eq.~(\ref{int-2loop}), the sum-integral (\ref{2s6}) reduces to 
\be
 \sumintff_{\{PQ\}}\ \frac{p^2}{q^2P^2Q^2R^2} &=& \left.\int\limits_{\bf p}\frac{n_B(p)}{p}\int\limits_Q
\frac{q^2}{Q^2r^2R^2}\right|_{P_0=-i p} 
\nn
&-& \left.\int\limits_{\bf p}\frac{n_F(p)}{2p}
\int\limits_Q\frac{1}{Q^2R^2}\left(\frac{p^2}{q^2}+\frac{q^2}{p^2}\right)\right|_{P_0=-i p} 
\nn
&+& \int\limits_{\bf pq}\frac{n_F(p)n_F(q)}{4pq}\ \frac{p^2}{q^2}\ \frac{r^2-p^2-q^2}{
\Delta(p,q,r)} 
\nn
&-&  \int\limits_{\bf pq}\frac{n_F(p)n_B(q)}{4pq}\ \left(\frac{p^2}{r^2}
+\frac{r^2}{p^2}\right)
\ \frac{r^2-p^2-q^2}{\Delta(p,q,r)} \, .
\label{2s6_1}
\ee 
Now
\be
\left\langle\frac{p^2}{q^2}\ \frac{r^2-p^2-q^2}{\Delta(p,q,r)}\right\rangle_{\hat p
\cdot\hat q} = 0 \, ,
\ee
and
\be
\left\langle\left(\frac{p^2}{r^2}+\frac{r^2}{p^2}\right)\ \frac{r^2-p^2-q^2}{\Delta(p,q,r)}\right\rangle_{\hat p
\cdot\hat q} = 
\frac{1}{2\epsilon}\frac{1}{p^2}- \frac{1}{2\epsilon}\left\langle
\frac{p^2}{r^4}\right\rangle_x \, .
\ee
Using the above angular average, we find
\be
 \sumintff_{\{PQ\}}\ \frac{p^2}{q^2P^2Q^2R^2} &=& \left.\int\limits_{\bf p}\frac{n_B(p)}{p}\int\limits_Q
\frac{q^2}{Q^2r^2R^2}\right|_{P_0=-i p}
\nn
&-& \left.\int\limits_{\bf p}\frac{n_F(p)}{2p}
\int\limits_Q\frac{1}{Q^2R^2}\left(\frac{p^2}{q^2}+\frac{q^2}{p^2}\right)\right|_{P_0=-i p} 
\nn
 &-&\frac{1}{2\epsilon}  \int\limits_{\bf pq}\frac{n_F(p)n_B(q)}{2pq}\ \frac{1}{p^2}
+ \frac{1}{2\epsilon}  \int\limits_{\bf pq}\frac{n_F(p)n_B(q)}{2pq}\ \frac{p^2}{r^4}.
\label{2s6_1a}
\ee
Using the 4-dimensional integrals from \cite{Andersen:2003zk} and Eqs.~(\ref{th-b}), ({\ref{th-f}}),
({\ref{th-f1}}) and ({\ref{th-fb}}), we obtain the sum-integral (\ref{2s6}).

Similarly after  applying Eq.~(\ref{int-2loop}), the sum-integral (\ref{2s7}) reduces to 
\be
 \sumintff_{\{PQ\}}\ \frac{r^2}{p^2P^2Q^2R^2} &=& \left.\int\limits_{\bf p}\frac{n_B(p)}{p}\int\limits_Q
\frac{p^2}{Q^2r^2R^2}\right|_{P_0=-i p} 
\nn
&-& \left.\int\limits_{\bf p}\frac{n_F(p)}{2p}
\int\limits_Q\frac{1}{Q^2R^2}\left(\frac{r^2}{p^2}+\frac{r^2}{q^2}\right)\right|_{P_0=-i p} 
\nn
&+& \int\limits_{\bf pq}\frac{n_F(p)n_F(q)}{4pq}\ \frac{r^2}{p^2}\ \frac{r^2-p^2-q^2}{
\Delta(p,q,r)} 
\nn
&-&  \int\limits_{\bf pq}\frac{n_F(p)n_B(q)}{4pq}\ \left(\frac{q^2}{r^2}
+\frac{q^2}{p^2}\right)
\ \frac{r^2-p^2-q^2}{\Delta(p,q,r)} \, .
\label{2s7_1}
\ee 
Now
\be
\left\langle\frac{r^2}{p^2}\ \frac{r^2-p^2-q^2}{\Delta(p,q,r)}\right\rangle_{\hat p
\cdot\hat q} = \frac{1}{2p^2\epsilon}\, ,
\ee
and
\be
\left\langle\left(\frac{q^2}{r^2}+\frac{q^2}{p^2}\right)\ \frac{r^2-p^2-q^2}{\Delta(p,q,r)}\right\rangle_{\hat p
\cdot\hat q} = - \frac{1}{2\epsilon}\left\langle
\frac{q^2}{r^4}\right\rangle_x \, .
\ee
Using the above angular average, we have
\be
 \sumintff_{\{PQ\}}\ \frac{p^2}{q^2P^2Q^2R^2} &=& \left.\int\limits_{\bf p}\frac{n_B(p)}{p}\int\limits_Q
\frac{q^2}{Q^2r^2R^2}\right|_{P_0=-i p}
\nn
&-& \left.\int\limits_{\bf p}\frac{n_F(p)}{2p}
\int\limits_Q\frac{1}{Q^2R^2}\left(\frac{p^2}{q^2}+\frac{q^2}{p^2}\right)\right|_{P_0=-i p} 
\nn
 &+& \frac{1}{2\epsilon}  \int\limits_{\bf pq}\frac{n_F(p)n_B(q)}{2pq}\ \frac{1}{p^2}
+ \frac{1}{2\epsilon}  \int\limits_{\bf pq}\frac{n_F(p)n_B(q)}{2pq}\ \frac{q^2}{r^4} \, .
\label{2s7_2}
\ee
Using the 4-dimensional integrals from \cite{Andersen:2003zk} and Eqs.~(\ref{th-b}), 
({\ref{th-f}}), ({\ref{th-f1}}) and ({\ref{th-bf}}), 
we obtain the sum-integral (\ref{2s6}).
\subsection{HTL two loop sum-integrals}
\be
 \sumintff_{\{PQ\}} \ \frac{1}{P^2Q^2r^2}{\cal T}_R \!\!\!&=&\!\!\! -\frac{T^2}{(4\pi)^2}\left(\frac{\Lambda}
{4\pi T}\right)^{4\epsilon} \frac{1}{48} \Bigg[\frac{1}{\epsilon^2}
\nn
&+&\left(2+12(1+8\ \hat\mu^2)\ln2 + 4\frac{\zeta'(-1)}{\zeta(-1)}\right)\frac{1}{\epsilon}
\nn
\!\!\!&+&\!\!\!\left(136.362 + 460.23 \ \hat\mu^2 - 273.046\ \hat\mu^4+{\cal O}\left(\hat\mu^6\right)\right)\Bigg] .
\ee
\vspace{-1cm}
\be
\sumintff_{\{PQ\}}\frac{p^2}{P^2Q^2r^4}{\cal T}_R\!\!\!&=&\!\!\!-\frac{1}{576}\frac{T^2}{(4\pi)^2}\left(\frac{\Lambda}
{4\pi T}\right)^{4\epsilon}\Bigg[\frac{1}{\epsilon^2}
+\left(\frac{26}{3}+4(13 + 144\hat\mu^2)\ln2 + 4\frac{\zeta'(-1)}{\zeta(-1)}
\right)\!\frac{1}{\epsilon}
\nn
&+&\left(446.397 + 2717.86\ \hat\mu^2 - 1735.61\ \hat\mu^4+{\cal O}(\hat\mu^6)\right)\Bigg] .
\ee
\vspace{-1cm}
\be
 \sumintff_{\{PQ\}} \ \frac{P\cdot Q}{P^2Q^2r^4}{\cal T}_R\!\!\!&=&\!\!\!\frac{T^2}{(4\pi)^2}\left(\frac{\Lambda}
{4\pi T}\right)^{4\epsilon}\left(-\frac{1}{96}\right)\Bigg[\frac{1}{\epsilon^2}
+\left(4\ln2 + 4\frac{\zeta'(-1)}{\zeta'(-1)}\right)\frac{1}{\epsilon}
\nn
\!\!\!&+&\!\!\!\left(69.1737 + 118.244\ \hat\mu^2
+ 136.688\ \hat\mu^4+{\cal O}\left(\hat\mu^6\right)\right)\Bigg] .
\ee

\be
 \sumintff_{\{PQ\}} \ \frac{r^2-p^2}{P^2q^2Q_0^2R^2}{\cal T}_Q &=&-\frac{T^2}{(4\pi)^2}\left(
\frac{\Lambda}{4\pi T}\right)^{4\epsilon}\frac{1}{8}\Bigg[\frac{1}{\epsilon^2}\left(1+4
\ \hat\mu^2\right)
\nn
&+&\frac{1}{\epsilon}\bigg(2+2\gamma_E+\frac{10}{3}\ln2+2\frac{\zeta'(-1)}{\zeta(-1)}
\nn
 &+&
2\ (8\gamma_E + 16\ln2 -7\zeta(3)) \ \hat\mu^2
\nn
&-&\frac{2}{3}\left(98\zeta(3)-93\zeta(5)\right)
\ \hat\mu^4 + {\cal O}\left(\hat\mu^6\right)\bigg)
\nn
&+&
46.8757 - 41.1192  \ \hat\mu^2 + 
64.0841 \ \hat\mu^4+{\cal O}\left(\hat\mu^6\right)\Bigg]\!.\ \
\ee
\section{Integrals}
\label{bsis}
\subsection{Three dimensional integrals}
We require one integral that does not involve the 
Bose-Einstein distribution function.
The momentum scale in these integrals is set by the mass
$m=m_D$.
The one-loop integral is
\be
\int_{\bf p} \frac{1}{p^2+m^2} & = & 
- \frac{m}{4\pi} \left( \frac{\Lambda}{2 m} \right)^{2 \epsilon}
\left[1 + 2 \epsilon  \right] \,.
\label{bi3}
\ee

\subsection{Thermal Integrals}
\be
\frac{{\Lambda}^{2\epsilon}}{(4\pi)^2}\int\limits_{\bf p}\frac{n_B(p)}{p}p^{-2\epsilon}\!\!\!&=&\!\!\!
\frac{T^2}{(4\pi)^2}\left(\frac{\Lambda}{4\pi T}\right)^{4\epsilon}\left(\frac{1}{12}
     \right)\Bigg[1 + \epsilon\left(2-2\ln2+ 4 \frac{\zeta'(-1)}{\zeta(-1)}\right) \nn
\!\!\!&+&\!\!\! 
     2\epsilon^2\ \Bigg(\frac{7\pi^2}{8} - 2 + \ln^2 2 -2 \ln2 +4(1+\ln2)\nn
&&\hspace{1cm} +4(1+\ln2)\frac
     {\zeta'(-1)}{\zeta(-1)} + 4 \frac{\zeta''(-1)}{\zeta(-1)}\Bigg)\Bigg].
\label{th-b}
\ee
\be
\frac{{\Lambda}^{2\epsilon}}{(4\pi)^2}\int\limits_{\bf p}\frac{n_F(p)}{2p}p^{-2\epsilon}
&=&
    \frac{T^2}{(4\pi)^2}\left(\frac{\Lambda}{4\pi T}\right)^{4\epsilon}\left(\frac{1}{24}
     \right)\Big[\left(1+12\hat\mu^2\right)
\nn &+& 
    \epsilon\Big\{2-2\ln2+ 4 \frac{\zeta'(-1)}{\zeta(-1)}+24\left(2\gamma_E+5\ln2-1\right)\ \hat\mu^2 
\nn &-& 56\zeta(3)\ \hat\mu^4+{\cal O}\left(\hat\mu^6\right)\Big\}\Big].
\label{th-f}
\ee
\be
\frac{{\Lambda}^{2\epsilon}}{(4\pi)^2}\int\limits_{\bf p}\frac{n_F(p)}{2p}
\ \frac{1}{p^2}\ p^{-2\epsilon}
\!\!\!&=&\!\!\!
    -\frac{T^2}{(4\pi)^2}\left(\frac{\Lambda}{4\pi T}\right)^{4\epsilon}\Bigg[\frac{1}
    {\epsilon} + 2 + 2\gamma_E + 10\ln2 
\nn
&&\hspace{.5cm} -\ 28\zeta(3)\ \hat\mu^2 + 124\zeta(5)\ \hat\mu^4 
    + {\cal O}\left(\hat\mu^6\right)\Bigg].
\label{th-f1}
\ee
\be
\int\limits_{\bf pq}\frac{n_F(p)n_F(q)}{4pq}\frac{1}{r^2} = 
           \frac{T^2}{(4\pi)^2}\left[\frac{1}{3}(1-\ln2) + 4(2\ln2-1)\hat\mu^2+\frac{10}{3}
           \zeta(3)\ \hat\mu^4+{\cal O}\left(\hat\mu^6\right)\right]\!.\hspace{10mm}
\label{th-2s2}
\ee
\be
\int\limits_{\bf pq}\frac{n_F(p)n_F(q)}{4pq}\frac{p^2}{r^4} &=&
             \frac{T^2}{(4\pi)^2}\left(-\frac{1}{36}\right)\Bigg[\Bigg(5 + 6\gamma_E + 
             6\ln2 - 6 \frac{\zeta'(-1)}{\zeta(-1)} 
\nn
&-& 12(12 \ln2 -13 + 3 \zeta(3))\hat\mu^2 
\nn
&+& 12 \left(
       20 \zeta(5)-13 \zeta(3) \right)\hat\mu^4+{\cal O}\left(\hat\mu^6\right)\Bigg)
\nn
&+& \epsilon\left(3.0747 + 31.2624 \ \hat\mu^2 +262.387 \ \hat\mu^4
             + {\cal O}\left(\hat\mu^6\right) \right)\Bigg].\hspace{2cm}
\label{th-ff}
\ee
\be
\int\limits_{\bf pq}\frac{n_B(p)n_F(q)}{2pq}\frac{p^2}{r^4}\!\!\!&=&\!\!\!-\frac{1}{36}\frac{T^2}{(4\pi)^2}
\Bigg[\Bigg\{7 - 6\gamma_E -18\ln2 + 6 \frac{\zeta'(-1)}{\zeta(-1)}+ 6 ( 21 \zeta(3)-21)
\hat\mu^2
\nn
 &+& 
6\left(126 \zeta(3) - 155 \zeta(5)\right)\ \hat\mu^4+{\cal O}\left(\hat\mu^6\right)\Bigg\} 
\nn
&+& \epsilon\left (29.5113 +
158.176 \ \hat\mu^2 - 557.189\ \hat\mu^4 + {\cal O}\left(\hat\mu^6\right)\right)\Bigg].
\label{th-fb}
\ee
\be
\int\limits_{\bf pq}\frac{n_B(p)n_F(q)}{2pq}\frac{q^2}{r^4}
\!\!\!&=&\!\!\!\frac{T^2}{(4\pi)^2}\left(
\frac{1}{18}\right)\Bigg[\bigg(1 -
6\gamma_E - 12\ln2 + 6 \frac{\zeta'(-1)}{\zeta(-1)} + 12 \hat\mu^2
\nn
\!\!\!&-&\!\!\!6 \left(28 \zeta(3) 
-31 \zeta(5)\right)\hat\mu^4 
+ {\cal O}\left(\hat\mu^6\right)\bigg)
\nn
\!\!\!&+&\!\!\! \epsilon\Big(31.0735 + 222.294\ \hat\mu^2 -\ 416.474\ \hat\mu^4 + {\cal O}\left(
\hat\mu^6\right)\Big)\Bigg]\!.\hspace{1.5cm}
\label{th-bf}
\ee
 \be
\int\limits_{\bf pq}\frac{n_F^{-}(p)-n_F^{+}(p)}{2\ p}\!\!\!\!\!\!&&\!\!\!\!\!\! \frac{n_F^{-}(q)-n_F^{+}(q)}{2\ q}
\ \frac{p\ q}{r^4} 
\nn
\!\!\!&=&\!\!\! \frac{T^2}{(4\pi)^2}\frac{1}{3}\left[(1-3\zeta(3))\ \hat\mu^2 -20(\zeta(3)-
\zeta(5))\ \hat\mu^4 + {\cal O}\left(\hat\mu^6\right)\right]\!.\hspace{2cm}
\label{th-2s4}
\ee
Thermal integrals containing the triangle function:

\be
\int_{\bf pq}\!\!\!&&\!\!\!\frac{n_F^{-}(p)-n_F^{+}(p)}{2p}\frac{n_F^{-}(q)
-n_F^{+}(q)}{2q}\frac{2p\ q}{\Delta(p,q,r)}
\nn
&&= \frac{T^2}{(4\pi)^2}\left(\frac{\Lambda}
{4\pi T}\right)^{4\epsilon}\left[\frac{\hat\mu^2}{\epsilon}+ 2(4\ln2+2\gamma_E + 1)
\ \hat\mu^2-\frac{28}{3}\zeta(3)\ \hat\mu^4+{\cal O}\left(\hat\mu^6\right)\right]\!.\hspace{2cm}
\label{th-2s1}
\ee


Thermal integrals containing both the triangle function and HTL average are listed below:
\be
\int\limits_{\bf pq}\frac{n_F(p)n_F(q)}{4pq}\!\!\!&{\mbox Re}&\!\!\!\left\langle c^2
\frac{r^2c^2-p^2-q^2}{\Delta(p+i\varepsilon,q,rc)}\right\rangle_c 
\nn
   \!\!\!&=&\!\!\! \frac{T^2}{(4\pi)^2}\left[0.01458+0.23807\ \hat\mu^2 + 0.82516
      \ \hat\mu^4+{\cal O}\left(\hat\mu^6\right)\right]\!.\hspace{2.cm}
\label{f1}
\ee
\be
\int\limits_{\bf pq}\frac{n_F(p)n_F(q)}{4pq}\!\!\!&{\mbox Re}&\!\!\!\left\langle c^4
\frac{r^2c^2-p^2-q^2}{\Delta(p+i\varepsilon,q,rc)}\right\rangle_c
\nn
   \!\!\!&=&\!\!\! \frac{T^2}{(4\pi)^2}\left[0.017715 + 0.28015\ \hat\mu^2 + 
       0.87321\ \hat\mu^4+{\cal O}\left(\hat\mu^6\right)\right]\!.\hspace{2.cm}
\label{f2}
\ee
\be
\int\limits_{\bf pq}\frac{n_F(p)n_F(q)}{4pq}\!\!\!&{\mbox Re}&\!\!\!\left\langle \frac{q^2}
{r^2}c^2\frac{r^2c^2-p^2-q^2}{\Delta(p+i\varepsilon,q,rc)}\right\rangle_c
 \nn
   \!\!\!&=&\!\!\! -\frac{T^2}{(4\pi)^2}\left[0.01158 + 0.17449\ \hat\mu^2 + 0.45566\ 
      \hat\mu^4 + {\cal O}\left(\hat\mu^6\right)\right]\!.\hspace{2.cm}
\label{f3}
\ee
\be
\int\limits_{\bf pq}\frac{n_B(p)n_F(q)}{2p q}\!\!\!&{\mbox Re}&\!\!\!\left\langle \frac{p^2-q^2}
 {r^2}\frac{r^2c^2-p^2-q^2}{\Delta(p+i\varepsilon,q,rc)}\right\rangle_c
\nn
   \!\!\!&=&\!\!\! \frac{T^2}{(4\pi)^2}\left[0.17811 + 1.43775\ \hat\mu^2 - 2.45413\ 
     \hat\mu^4 + {\cal O}\left(\hat\mu^6\right)\right]\!.\hspace{2.cm}
\label{f4}
\ee
%
Second set of integrals involve the variables $r_c=|{\bf p+q}/c|$:
 \be
\int\limits_{\bf pq}\frac{n_F(p)n_B(q)}{2p q}\!\!\!&{\mbox Re}&\!\!\!\left\langle c^{-1+2\epsilon}
\frac{r_c^2-p^2-q^2}{\Delta(p+i\varepsilon,q,r_c)}\right\rangle_c
\nn
   \!\!\!&=&\!\!\! \frac{T^2}{(4\pi)^2}\left[0.19678 + 1.07745\ \hat\mu^2 - 2.63486\ \hat\mu^4
     + {\cal O}\left(\ \hat\mu^6\right)\right]\!.\hspace{2.cm}
\label{f5}
\ee
\be
\int\limits_{\bf pq}\frac{n_F(p)n_B(q)}{2p q}\!\!\!&{\mbox Re}&\!\!\!\left\langle c^{1+2\epsilon}
\frac{r_c^2-p^2-q^2}{\Delta(p+i\varepsilon,q,r_c)}\right\rangle_c
\nn
   \!\!\!&=&\!\!\! \frac{T^2}{(4\pi)^2}\left[0.048368 + 0.23298\ \hat\mu^2 - 0.65074\ \hat\mu^4
   + {\cal O}\left(\ \hat\mu^6\right)\right]\!.\hspace{2.cm}
\label{f6}
\ee
\be
\int\limits_{\bf pq}\frac{n_F(p)n_B(q)}{2p q}\hspace{-10mm}&&\frac{p^2}{q^2}{\mbox Re}\left\langle 
c^{1+2\epsilon}\frac{r_c^2-p^2-q^2}{\Delta(p+i\varepsilon,q,r_c)}\right\rangle_c
   = \frac{1}{96}\frac{T^2}{(4\pi)^2}\left(\frac{\Lambda}{4\pi T}\right)^{4\epsilon}
\nn
\!\!\!&\times&\!\!\!\Bigg[\frac{\left(1+12\ \hat\mu^2\right)}{\epsilon}
+\left(7.7724+81.1057\ \hat\mu^2-48.5858\ \hat\mu^4+{\cal O}
   \left(\hat\mu^6\right)\right)\Bigg]\!.\hspace{2.cm}
\label{f7}
\ee
\be
\int\limits_{\bf pq}\frac{n_F(p)n_B(q)}{2p q}\!\!\!&{\mbox Re}&\!\!\!\left\langle c^{1+2\epsilon}
\frac{r_c^2}{q^2}\frac{r_c^2-p^2-q^2}{\Delta(p+i\varepsilon,q,r_c)}\right\rangle_c
    = \frac{T^2}{(4\pi)^2}\left(\frac
    {\Lambda}{4\pi T}\right)^{4\epsilon}\frac{11-8\ln2}{288}
\nn
\!\!\!&\times&\!\!\!\!\left[\frac{1}{\epsilon}
    \left(1+12\ \hat\mu^2\right)
   +\left(7.799 + 70.516\hat\mu^2 - 57.928\hat\mu^4 + {\cal O}
    \left(\hat\mu^6\right)\right)\right]\!.\hspace{2cm}
\label{f8}
\ee
\be
\int\limits_{\bf pq}\frac{n_F(p)n_F(q)}{4p q}\!\!\!&{\mbox Re}\!\!\!&\frac{1}{24}\left\langle c^{-1+2\epsilon}
\frac{r_c^2-p^2}{q^2}\frac{r_c^2-p^2-q^2}{\Delta(p+i\varepsilon,q,r_c)}\right\rangle_c
    = -\frac{T^2}{(4\pi)^2}\left(\frac{\Lambda}{4\pi T}\right)^{4\epsilon}
\nn    
\!\!\!&\times&\!\!\!\Bigg[\left(1+12\ \hat\mu^2\right)\frac{1}{\epsilon^2}
    +\frac{2}{\epsilon}\bigg(1+\gamma_E + \ln2+ \frac{\zeta'(-1)}{\zeta(-1)} 
\nn
&+&\!\!\!(24\gamma_E
    + 48\ln2-7\zeta(3))\hat\mu^2 +(31\zeta(5)-98\zeta(3))\hat\mu^4 +
     {\cal O}\left(\hat\mu^6\right)\!\bigg)
\nn
    &+& \left(40.3158 + 261.822\ \hat\mu^2 - 1310.69\ \hat\mu^4 + {\cal O}\left(\hat
    \mu^6\right)\right)\Bigg].
\label{f9}
\ee
\be
\int\limits_{\bf pq}\frac{n_B(p)n_F(q)}{2p q}\!\!\!&{\mbox Re}&\!\!\!\left\langle c^{-1+2\epsilon}
\frac{r_c^2-p^2}{q^2}\frac{r_c^2-p^2-q^2}{\Delta(p+i\varepsilon,q,r_c)}\right\rangle_c 
     = -\frac{T^2}{(4\pi)^2}\left(\frac{\Lambda}{4\pi T}\right)^{4\epsilon}\frac{1}{12}
    \left[\frac{1}{\epsilon^2}
\right.\nn\!\!\!&+&\!\!\!\left.
     \frac{1}{\epsilon}\left(2+2\gamma_E+4\ln2+2\frac{\zeta'(-1)}{\zeta(-1)}-14\zeta(3)
     \hat\mu^2 + 62\zeta(5)\hat\mu^4 + {\cal O}\left(\hat\mu^6\right)\!\right)
\right.\nn \!\!\!&+&\!\!\!\left.
     \left(52.953 - 190.103\ \hat\mu^2 + 780.921\ \hat\mu^4 +
     {\cal O}\left(\hat\mu^6\right)\right)\right]. \hspace{4cm}
\label{f10}
\ee
The integral (\ref{f2}) can be evaluated directly in three dimensions at finite chemical 
potential. The other  integrals Eqs.~(\ref{f3})--(\ref{f10}) can be evaluated following the same
procedure as discussed in \cite{Andersen:2003zk} at finite chemical potential.
%
\section{Pressure}
\label{nlopress}
In the previous section we have computed both LO and NLO thermodynamic potential in presence of quark chemical potential
and temperature. All other thermodynamic quantities can be calculated using standard thermodynamic relations. The pressure
is defined as 
\be
P = -{\Omega}(T,\mu,m_q,m_D) \, ,
\label{pressure-potential}
\ee
\subsection{LO Pressure}
The LO HTLpt pressure through ${\cal O}(g^4)$ at any $\mu$ can be written in leading order of $\epsilon$
from eqn.~(\ref{Omega-LO}) as
\be
{\cal P}_{\rm LO}& =&  d_A\frac{\pi^2T^4}{45}\Bigg[1+\frac{7}{4}\frac{d_F}{d_A}\left(
1+\frac{120}{7}\hat\mu^2+\frac{240}{7}\hat\mu^4\right)
\nn
 &-& \frac{15}{2} \hat m_D^2 
     -30\frac{d_F}{d_A}\left(1+12\hat\mu^2\right)\hat m_q^2+30\hat m_D^3
\nn
&+&  \frac{45}{4}\left(\ln{\frac{\hat\Lambda}{2}}-\frac{7}{2}+\gamma_E+\frac{\pi^2}{3}\right)
     \hat m_D^4 - 60\frac{d_F}{d_A}(\pi^2-6)\hat m_q^4\Bigg]
 \;.
\label{pressure-LO}
\ee

At leading order, the weak coupling expressions for the mass parameters are 
\be 
m_D^2 &=& \frac{g^2T^2}{3} \left [ c_A +s_F\left(1+12\hat\mu^2\right)  \right ]\, ;  \hspace*{0.2in}
m_q^2 = \frac{g^2T^2}{4} \frac{c_F}{2} \left (1 + 4\hat\mu^2\right ) \, . \label{mass_lo}
\ee
\vspace{1cm}
\subsection{NLO HTLpt Pressure and Variational Mass Gap Equations}\label{nlopress_variational_eqn}
The NLO HTLpt pressure through ${\cal O}[(\mu/T)^4]$ can be obtain from Eq.~(\ref{Omega-NLO}) 
using (\ref{pressure-potential}) as

\be
{\cal P}_{\rm NLO}\!\!\!&=&\!\!\!
           d_A \frac{\pi^2 T^4}{45} \Bigg\{ 
	   1 + \frac{7}{4} \frac{d_F}{d_A}\left(1+\frac{120}{7}
           \hat\mu^2+\frac{240}{7}\hat\mu^4\right) - 15 \hat m_D^3 
\nn \!\!\!&-&\!\!\!
           \frac{45}{4}\left(\log\frac{\hat\Lambda}{2}-\frac{7}{2}+\gamma_E + \frac{\pi^2}{3}\right)\hat m_D^4
          + 60 \frac{d_F}{d_A}\left(\pi^2-6\right)\hat m_q^4	
\nn
\!\!\!&+&\!\!\! \frac{\alpha_s}{\pi} \Bigg[ -\frac{5}{4}\left(c_A + \frac{5s_F}{2}\!\left(1+\frac{72}{5}
           \hat\mu^2+\frac{144}{5}\hat\mu^4\right)\!\right) 
	   + 15 \left(c_A+s_F(1+12\hat\mu^2)\right)\!\hat m_D
\nn \!\!\!&-&\!\!\! 
	 \frac{55}{4}\Bigg\{ c_A\left(\log\frac{\hat\Lambda}{2}- \frac{36}{11}\log\hat m_D - 2.001\right)
		- \frac{4}{11} s_F \left[\left(\log\frac{\hat\Lambda}{2}-2.337\right)\right.
\nn \!\!\!&+&\!\!\! 
	\left. (24-18\zeta(3))\left(\log\frac{\hat\Lambda}{2} -15.662\right)\hat\mu^2
	+ 120\left(\zeta(5)-\zeta(3)\right)
\right.\nn
\!\!\!&\times&\!\!\!\left.\left(\log\frac{\hat\Lambda}{2} -1.0811\right)\hat\mu^4 + 
         {\cal O}\left(\hat\mu^6\right)\right] \!\!\Bigg\} \hat m_D^2-45 \, s_F \left\{\log\frac{\hat\Lambda}{2}
        + 2.198  -44.953\hat\mu^2\right. 
\nn \!\!\!&-&\!\!\! 
	\left.\left(288 \ln{\frac{\hat\Lambda}{2}} 
       +19.836\right)\hat\mu^4 + {\cal O}\left(\hat\mu^6\right)\right\} \hat m_q^2
       + \frac{165}{2}\left\{ c_A\left(\log\frac{\hat\Lambda}{2}+\frac{5}{22}+\gamma_E \right)\right.
\nn \!\!\!&-&\!\!\!
	\left. \frac{4}{11} s_F \left(\log\frac{\hat\Lambda}{2}-\frac{1}{2}+\gamma_E + 2\ln2 -7\zeta(3)\hat\mu^2+
         31\zeta(5)\hat\mu^4 + {\cal O}\left(\hat\mu^6\right) \right)\right\}\hat m_D^3
\nn \!\!\!&+&\!\!\!
         15 s_F \left(2\frac{\zeta'(-1)}{\zeta(-1)}
         +2\ln \hat m_D\right)\Big[(24-18\zeta(3))\hat\mu^2 
\nn
 \!\!\!&+&\!\!\! 120(\zeta(5)-\zeta(3))\hat\mu^4 + 
           {\cal O}\left(\hat\mu^6\right)\Big] \hat m_D^3
	+ 180\,s_F\hat m_D \hat m_q^2 \Bigg]
\Bigg\} .
\label{pressure-NLO}
\ee
which is accurate up to ${\cal O}(g^3)$ and nominally accurate to ${\cal O}(g^5)$ since it was obtained from an expansion of two-loop 
thermodynamic potential in a power series in $m_D/T$ and $m_q/T$ treating both $m_D$ and $m_q$ having leading terms proportional to $g$.  
Using the result above, the mass parameters $m_D$ and $m_q$ can be obtained by solving the two variational equations:
\be
\left.\frac{\partial \Omega_{\rm NLO}}{\partial \hat{m}_D}\right|_{m_q=\text{constant}} &=& 0 \, ,
\nonumber \\
\left.\frac{\partial \Omega_{\rm NLO}}{\partial \hat{m}_q}\right|_{m_D=\text{constant}} &=& 0 \, .
\ee
This leads to the following two gap equations which will be solved numerically
\begin{eqnarray}
 && \hspace{-7mm} 
 45\hat m_D^2\left[1+\left(\ln\frac{\hat\Lambda}{2}-\frac{7}{2}+\gamma_E+
\frac{\pi^2}{3}\right)\hat m_D\right]
\nonumber\\&=&
\frac{\alpha_s}{\pi}\Bigg\{15(c_A+s_F(1+12\hat\mu^2))
-\frac{55}{2}\left[c_A\left(\ln\frac{\hat\Lambda}{2}-\frac{36}{11}\ln{\hat m_D}-3.637\right)
\right.
\nonumber\\
&-&\left.\frac{4}{11}s_F\left\{\ln\frac{\hat\Lambda }{2}-2.333+(24-18\zeta(3))
\left(\ln\frac{\hat\Lambda }{2}-15.662\right)\hat\mu^2
\right.\right.
\nonumber\\
&+&\left.\left.120(\zeta(5)-\zeta(3))\left(\ln\frac{\hat\Lambda }{2}-1.0811\right)\hat\mu^4
\right\}\right]\hat m_D
+\frac{495}{2}\left[c_A\left(\ln\frac{\hat\Lambda }{2}
+\frac{5}{22}+\gamma_E\right)
\right.
\nonumber\\
&-&\left.
\frac{4}{11}s_F\left\{\ln\frac{\hat\Lambda }{2}-\frac{1}{2}
+\gamma_E + 2\ln2-7\zeta(3)\hat\mu^2+31\zeta(5)\hat\mu^4
\right.\right.
\nonumber\\
&-&\left.\left.
\left(\frac{\zeta'(-1)}{\zeta(-1)}+\ln \hat m_D+
\frac{1}{3}\right)\!\left((24-18\zeta(3))\hat\mu ^2+120(\zeta(5)-\zeta(3))\hat\mu^4\right)\right\}\right] \hat m_D^2
\nn
&+&180 s_F\hat m_q^2\Bigg\}, \nonumber \\
\label{gap_md}
\end{eqnarray}
and
\begin{eqnarray}
 \hat m_q^2&=&\frac{d_A}{8d_F\left(\pi ^2-6\right)}\frac{\alpha_s s_F}{\pi }\left[3\left(\ln\frac{\hat\Lambda }{2}
+2.198-44.953\ \hat\mu^2 \right. \right.
\nonumber \\
&& \left. \left.
\hspace{4cm}
-\left(288\ln\frac{\hat\Lambda }{2}+19.836\right)\hat\mu^4\right)-12\hat m_D\right]. 
\nonumber \\
\label{gap_mq}
\end{eqnarray}

\begin{figure}
\subfigure{\includegraphics[width=7cm,height=8cm]{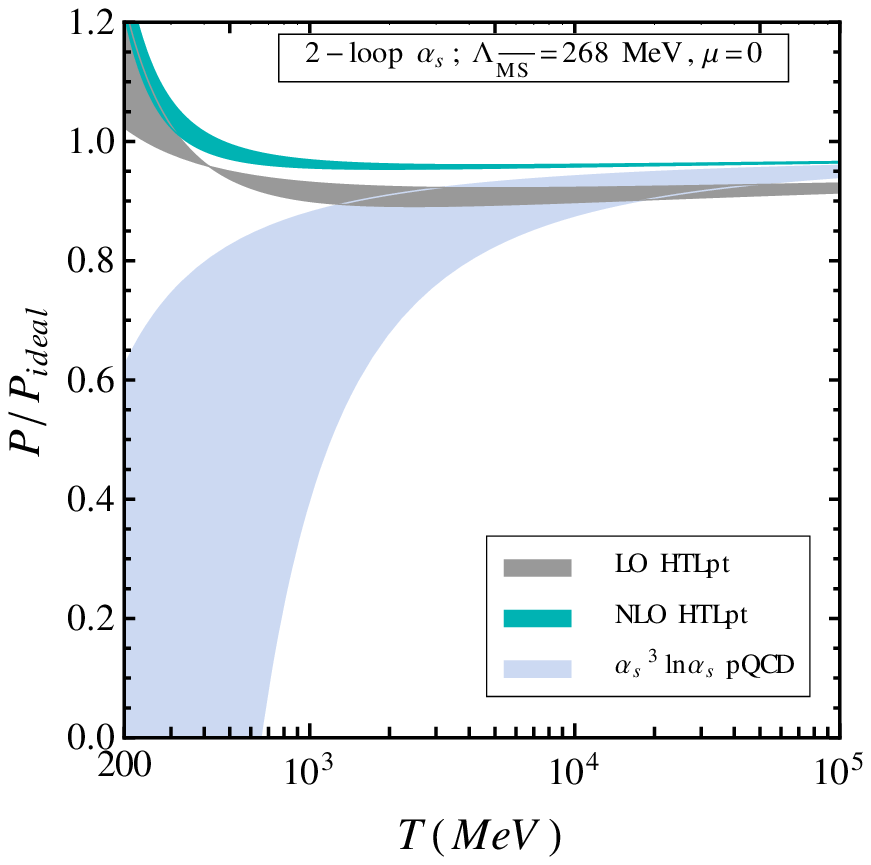}}
\subfigure{\includegraphics[width=7cm,height=8cm]{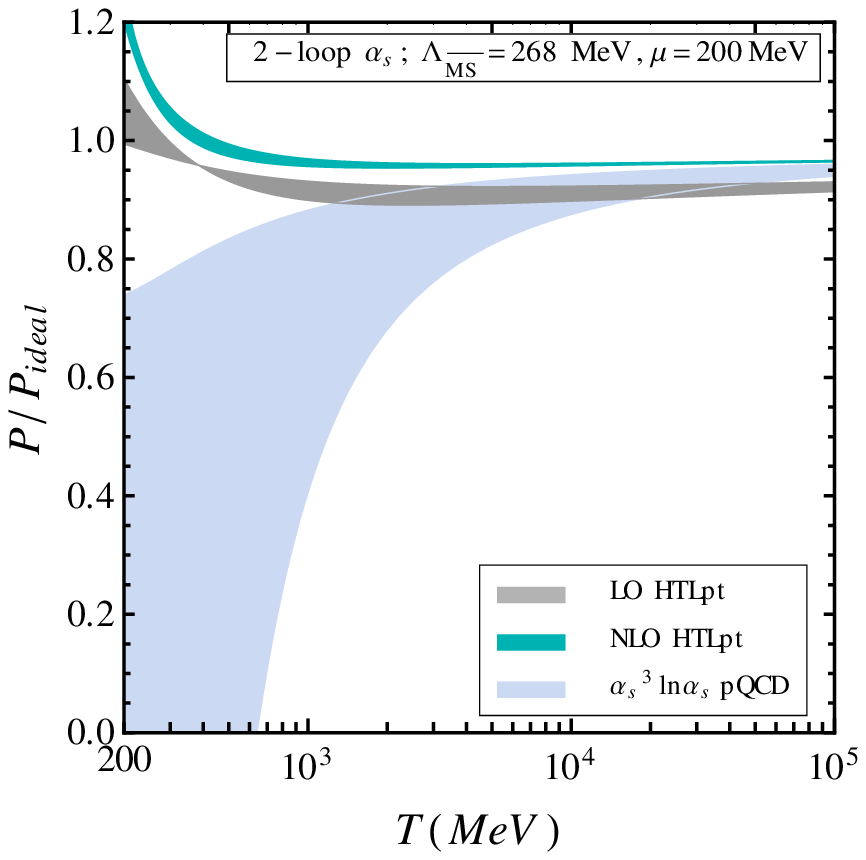}}
\caption[The NLO HTLpt pressure scaled with ideal gas pressure plotted along with four-loop
pQCD pressure for two different values of chemical potential with $N_f=3$  and 2-loop running
coupling constant $\alpha_s$.]{The NLO HTLpt pressure scaled with ideal gas pressure plotted along with four-loop
pQCD pressure~\cite{Vuorinen:2002ue,Vuorinen:2003fs} for two different values of chemical potential with $N_f=3$  and 2-loop running
coupling constant $\alpha_s$. The bands are obtained by varying the renormalization scale by a factor of 
2 around its central value $\Lambda=2\pi \sqrt{T^2+\mu^2/\pi^2}$ \cite{Vuorinen:2002ue,Vuorinen:2003fs,Rebhan:2003wn,
Cassing:2007nb,Gardim:2009mt,Kurkela:2009gj}. 
We use $\Lambda_{\overline{\rm MS}}=290$ MeV based on recent lattice calculations~\cite{Bazavov:2012ka} 
of the three-loop running of $\alpha_s$.}
\label{press_htl_nlo}
\end{figure}

\begin{figure}
\subfigure{\includegraphics[width=7cm,height=8cm]{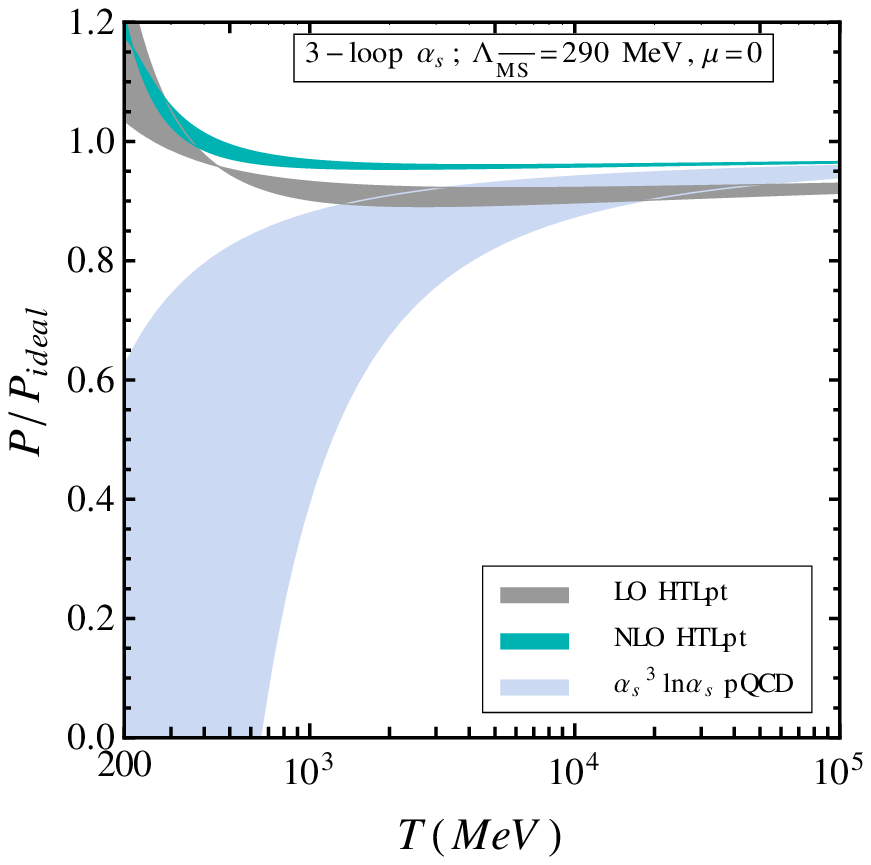}}
\subfigure{\includegraphics[width=7cm,height=8cm]{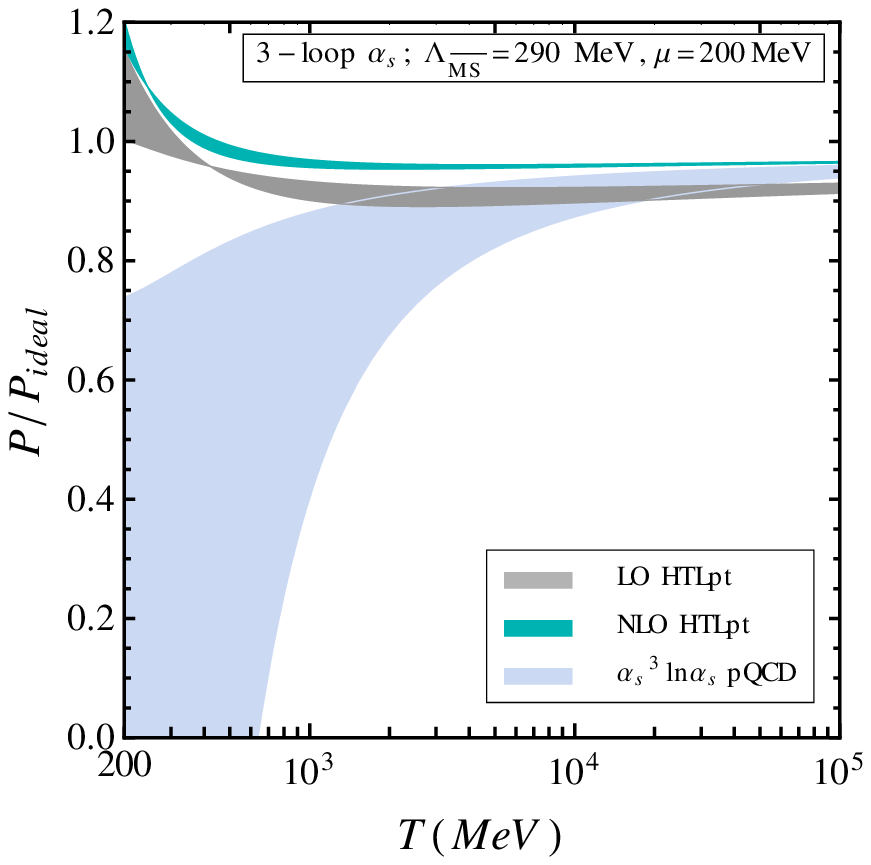}}
\caption{Same as Fig.~\ref{press_htl_nlo} but for 3-loop  $\alpha_s$.
}
\label{press_htl_nlo1}
\end{figure}
\noindent
For convenience and comparison with lattice data \cite{Borsanyi:2012cr}, we define the pressure difference
\begin{eqnarray}
 \Delta P(T,\mu)=P(T,\mu)-P(T,0) \, .
\end{eqnarray}

\begin{figure}
 \subfigure{
\includegraphics[width=7.3cm,height=7cm]{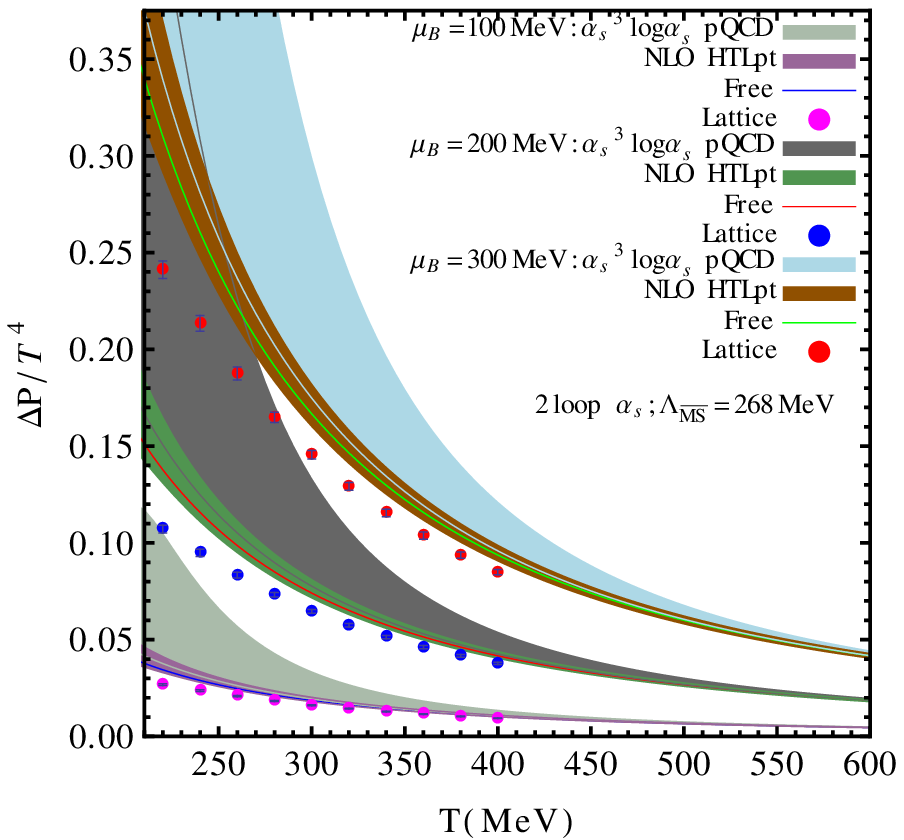}
}
\hspace{-.5cm}
\subfigure{
\includegraphics[width=7.3cm,height=7cm]{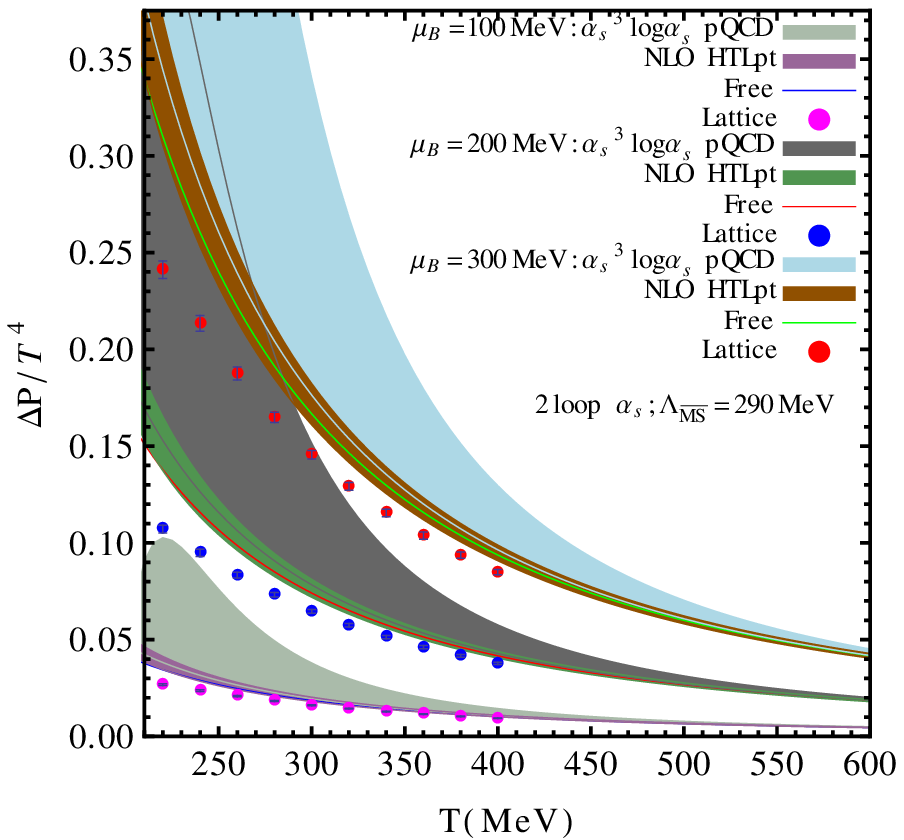}
}
\caption[{\it Left panel}: $\Delta P$ for $N_f=3$ is plotted as a function of $T$ for two-loop
HTLpt result along with those of four-loop pQCD  and 
lattice QCD data using 2-loop running coupling constant $\alpha_s$.  
{\it Right panel}: Same as left panel but using 3-loop running coupling.]{(Left panel) $\Delta P$ for $N_f=3$ is 
plotted as a function of $T$ for two-loop
HTLpt result along with those of four-loop pQCD up to $\alpha_s^3\ln\alpha_s$ \cite{Vuorinen:2002ue,Vuorinen:2003fs} and 
lattice QCD~\cite{Borsanyi:2012cr} up to ${\cal O}\left(\mu^2\right)$ using 2-loop running coupling constant $\alpha_s$.  
(Right panel) Same as left panel but using 3-loop running coupling. 
In both cases three different values of $\mu$ are shown as specified in the legend. The bands in both HTLpt and pQCD are
obtained by varying the renormalization scale by a factor of 2 around its  central value $\Lambda=2\pi\sqrt{T^2+\mu^2/\pi^2}$
\cite{Vuorinen:2002ue,Vuorinen:2003fs,Rebhan:2003wn,Cassing:2007nb,Gardim:2009mt,Kurkela:2009gj}.
}
\label{diff_press}
\end{figure}

\noindent
where we note that we have discarded terms of ${\cal O}(\hat\mu^6)$ and higher above.  In figs.~(\ref{press_htl_nlo})
and~(\ref{press_htl_nlo1}) we present a comparison of NLO HTLpt pressure with that of four-loop 
pQCD~\cite{Vuorinen:2002ue,Vuorinen:2003fs} as a function of the temperature for two and three loop running of $\alpha_s$.  
The only difference 
between Figs.~(\ref{press_htl_nlo}) and~(\ref{press_htl_nlo1}) is the choice of order of the running coupling used. 
As can be seen from these figures, the dependence on the order of the running coupling is quite small. However, 
we note that in both figures even at extremely large
temperatures there is a sizable correction when going from LO to NLO.  This was already seen in the $\mu=0$
results of Ref.~\cite{Andersen:2002ey,
Andersen:2003zk} where it was found that due the logarithmic running of the coupling,
it was necessary to go to very large temperatures in order for the LO and NLO predictions to overlap.  This
is due to over-counting problems at LO which lead to an order-$g^2$ perturbative coefficient which is twice
as large as it should be \cite{Andersen:2002ey,
Andersen:2003zk,Haque:2013qta}.  This problem is corrected at NLO, but the end result is that there is a reasonably
large correction ($\sim 5\%$) at the temperatures shown.

The NLO HTLpt result differs from 
the pQCD result through order $\alpha_s^3\ln\alpha_s$ at low temperatures.
A NNLO HTLpt calculation at finite $\mu$ would
agree better with pQCD $\alpha_s^3\ln\alpha_s$ as found in $\mu=0$ case~\cite{Andersen:2011sf}. The  HTLpt result 
clearly indicates a modest improvement over pQCD in respect of convergence and sensitivity of the renormalization 
scale. In Fig.~\ref{diff_press} the pressure difference, $\Delta P$,  is also compared with the same quantity computed using
pQCD~\cite{Vuorinen:2002ue,Vuorinen:2003fs} and lattice QCD~\cite{Borsanyi:2012cr}. Both LO and NLO HTLpt results are less sensitive to 
the choice of the renormalization scale than the weak coupling results with the inclusion of successive orders of approximation. 
Comparison 
with available lattice QCD data~\cite{Borsanyi:2012cr} suggests that HTLpt and pQCD cannot accurately account for the lattice QCD results
below approximately $3\,T_c$; however, the results are in very good qualitative agreement with the lattice QCD results without
any fine tuning.

\section{Quark Number Susceptibility}
\label{qns}

We are now in a position to obtain the second and fourth-order HTLpt QNS following Eq.~(\ref{qns_def}).  
We note that the pure gluonic loops at any order do not contribute to QNS, however, gluons contribute through 
the dynamical fermions through fermionic loops. This makes QNS proportional 
to only quark degrees of freedom.  Below we present \mbox{(semi-)}analytic expressions for both LO and NLO QNS.

To obtain the second and fourth-order quark number susceptibilities in HTLpt, one requires 
expressions for $m_D$,  $\frac{\partial^2}{\partial\mu^2}m_D$,  $m_q$, and $\frac{\partial^2}{\partial\mu^2}m_q$ 
at $\mu=0$ from Eqs.~(\ref{gap_md}) and (\ref{gap_mq}).\footnote{Note that odd derivatives 
with respect to $\mu$ vanish at $\mu=0$. Fourth-order derivatives at $\mu=0$
are nonzero, however, they appear as multiplicative factors of the gap equations and are therefore not required, as we will see below.}
We list these here for completeness.  The result
for the limit of the $m_D$ gap equation necessary is
\begin{eqnarray}
 && \hspace{-1cm} 
 45\hat m_D^2(0)\left[1+\left(\ln\frac{\hat\Lambda}{2}-\frac{7}{2}+\gamma_E+
\frac{\pi^2}{3}\right)\hat m_D(0)\right]
\nonumber\\&=&
\frac{\alpha_s}{\pi}\Bigg\{15(c_A+s_F)
-\frac{55}{2}\left[c_A\left(\ln\frac{\hat\Lambda}{2}-\frac{36}{11}\ln{\hat m_D(0)}-3.637\right)
\right.
\nonumber\\
&-&\left.\frac{4}{11}s_F\left\{\ln\frac{\hat\Lambda }{2}-2.333
\right\}\right]\hat m_D(0)
+\frac{495}{2}\left[c_A\left(\ln\frac{\hat\Lambda }{2}
+\frac{5}{22}+\gamma_E\right)
\right.
\nonumber\\
&-&\left.
\frac{4}{11}s_F\left(\ln\frac{\hat\Lambda }{2}-\frac{1}{2}
+\gamma_E+2\ln2
\right)\right] {\hat m}_D^2(0)
+180 s_F\hat m_q^2(0)\Bigg\} . \label{gap_md0}
\end{eqnarray}
For $m_q$ one obtains
\begin{eqnarray}
 \hat m_q^2(0)=\frac{d_A}{8d_F\left(\pi ^2-6\right)}\frac{\alpha_s s_F}{\pi }\left[3\left(\ln\frac{\hat\Lambda }{2}
+2.198\right)-12\hat m_D(0)\right] . \label{gap_mq0}
\end{eqnarray}
For $\frac{\partial^2}{\partial\mu^2}m_D$ one obtains
\begin{eqnarray}
 && \hspace{-5mm} 45
 \left[2+3\left(\ln\frac{\hat\Lambda }{2}-\frac{7}{2}+\gamma_E+\frac{\pi^2}{3}\right)
 \hat m_D(0)\right] \hat m_D(0) \hat m_D''(0)
\nonumber\\
&=&\frac{\alpha_s}{\pi } \Bigg\{360 s_F-\frac{55}{2} 
\hat m_D''(0) \left[c_A \left(
\ln\frac{\hat\Lambda }{2}-\frac{36}{11} \ln\hat m_D(0)-6.9097\right)
-\frac{4}{11} s_F \left(\ln\frac{\hat\Lambda }{2}-2.333\right)\right]
\nonumber\\
&& + 495 
\hat m_D''(0) \left[c_A \left(\frac{5}{22}+\gamma_E+\ln\frac{\hat\Lambda }{2}\right)-\frac{4}{11}
 s_F \left(\ln\frac{\hat\Lambda }{2}-\frac{1}{2}+\gamma_E+2 \ln2\right)\right]
\hat m_D(0)
\nonumber\\
&& + 20  s_F \left(\ln\frac{\hat\Lambda}
{2}-15.662\right) (24-18 \zeta(3)) \hat m_D(0)
\nonumber\\
&& + 180 s_F \hat m_D(0)^2 \left[7 \zeta(3)+ \left(\frac{\zeta'(-1)}{\zeta(-1)}+\ln\hat m_D(0)+\frac{1}{3}\right) 
(24-18 \zeta(3))\right]
\nonumber\\
&& +360s_F\hat m_q(0)\hat m_q''(0)\Bigg\} . 
\nonumber \\
\label{gap_mdpp0}
\end{eqnarray}
For $\frac{\partial^2}{\partial\mu^2}m_q$ one obtains
\begin{eqnarray}
 \hat m_q(0)\hat m_q''(0)=-\frac{3d_A}{8d_F\left(\pi ^2-6\right)}\frac{\alpha_s s_F}{\pi }\left[44.953+2\hat m_D''(0)\right] .
\label{gap_mqpp0}
\end{eqnarray}
In the expressions above, $m_D(0)\equiv m_D(T,\Lambda,\mu=0)$,  $m_D''(0)\equiv \left.\frac{\partial^2}{\partial\mu^2}m_D(T,\Lambda,\mu)
 \right|_{\mu=0}$ 
and similarly for $m_q$. 
%
\subsection{LO HTLpt second-order QNS}
An analytic expression for the LO HTLpt second-order QNS can be obtained using Eq.~(\ref{pressure-LO})
\begin{eqnarray}
\chi_2^{\rm LO}(T)&=&\left.\frac{\partial^2 }{\partial\mu^2}{\cal P}_{\rm LO}(T,\Lambda,\mu)\right|_{\mu=0}=\frac{1}{(2\pi T)^2}
\left.\frac{\partial^2 }{\partial\hat\mu^2}{\cal P}_{\rm LO}(T,\Lambda,\hat\mu)\right|_{\hat\mu=0}
\nonumber\\
&=&\frac{d_F T^2}{3}\Bigg[1-\frac{3c_F}{4}\left(\frac{g}{\pi}\right)^2
                   +\frac{c_F}{4}\sqrt{3(c_A+s_F)}\left(\frac{g}{\pi }\right)^3                  -
\frac{c_F^2}{64}\left(\pi^2-6\right)\left(\frac{g}{\pi}\right)^4 
\nonumber\\
&&+\ \frac{c_F}{16}(c_A+s_F)\left(\log\frac{\hat\Lambda
}{2}-\frac{7}{2}+\gamma_E + \frac{\pi ^2}{3}
\right)\left(\frac{g}{\pi }\right)^4\Bigg] \ , \label{chi2_lo}
\end{eqnarray}
where the LO Debye and quark masses listed in Eqs.~(\ref{mass_lo}) and their $\mu$ derivatives have been used. 


\subsection{LO HTLpt fourth-order QNS}

An analytic expression for the LO HTLpt fourth-order QNS can also be obtained using Eq.~(\ref{pressure-LO}) 
\begin{eqnarray}
\chi_4^{\rm LO}(T)&=&\left.\frac{\partial^4 }{\partial\mu^4}{\cal P}_{\rm LO}(T,\Lambda,\mu)\right|_{\mu=0}=\frac{1}{(2\pi T)^4}
\left.\frac{\partial^4 }{\partial\hat\mu^4}{\cal P}_{\rm LO}(T,\Lambda,\hat\mu)\right|_{\hat\mu=0}
\nonumber\\
&=&\frac{2d_F}{\pi^2}\Bigg[1-\frac{3}{4}c_F
\left(\frac{g}{\pi }\right)^2
                +\frac{3}{8}c_F s_F\sqrt{\frac{3}{c_A+s_F}}\left(\frac{g}{\pi
}\right)^3
                -\frac{c_F^2\left(\pi^2-6\right)}{64}\left(\frac{g}{\pi}\right)^4
\nonumber\\
&&              +\ \frac{3}{16}c_F
s_F\left(\log\frac{\hat\Lambda}{2}-\frac{7}{2}+\gamma_E
              +\frac{\pi^2}{3}\right)\left(\frac{g}{\pi}\right)^4\Bigg] \ , \label{chi4_lo}
\end{eqnarray}
where, once again, the LO Debye and quark masses listed in Eqs.~(\ref{mass_lo}) and their $\mu$ derivatives 
have been used. We note that both $\chi_2^{\rm LO}$ in (\ref{chi2_lo}) and $\chi_4^{\rm LO}$ in (\ref{chi4_lo})
are the same as those recently obtained by Andersen et al.~\cite{Andersen:2012wr}; however, the closed-form expressions obtained
here have not been explicitly listed therein.
\subsection{NLO HTLpt second-order QNS}
A semi-analytic expression for the NLO  HTLpt second-order QNS can be obtained from Eq.~(\ref{pressure-NLO})
\begin{eqnarray}
 \chi_2^{\rm NLO}(T)&=&\left.\frac{\partial^2 }{\partial\mu^2}{\cal P}_{\rm NLO}(T,\Lambda,\mu)\right|_{\mu=0}=\frac{1}{(2\pi T)^2}
\left.\frac{\partial^2 }{\partial\hat\mu^2}{\cal P}_{\rm NLO}(T,\Lambda,\hat\mu)\right|_{\hat\mu=0}
\nonumber\\
\!\!\!&=&\!\!\! \frac{d_AT^2}{2}\Bigg[\frac{2}{3}\frac{d_F}{d_A} + \frac{\alpha_s}{\pi}s_F\Bigg\{-1
        +4\ \hat m_D(0)
        +\frac{2}{3}\left(\ln{\frac{\hat\Lambda}{2}-15.662}\right)
\nonumber\\
\!\!\!&\times&\!\!\!(4-3\zeta(3))\,\hat m_D^2(0)+ 44.953\ \hat m_q^2(0)
\nonumber\\     
\!\!\!&+&\!\!\!\left[\frac{14}{3}\zeta(3)+\left(\frac{\zeta'(-1)}{\zeta(-1)}
      + \ln\hat m_D(0)\right)(16-12\zeta(3))\right]\hat m_D^3(0)\Bigg\}\Bigg].\hspace{1cm}
\label{chi2_nlo}
\end{eqnarray}
\begin{figure}[tbh]
\subfigure{
\includegraphics[width=0.48\textwidth,height=0.5\textwidth]{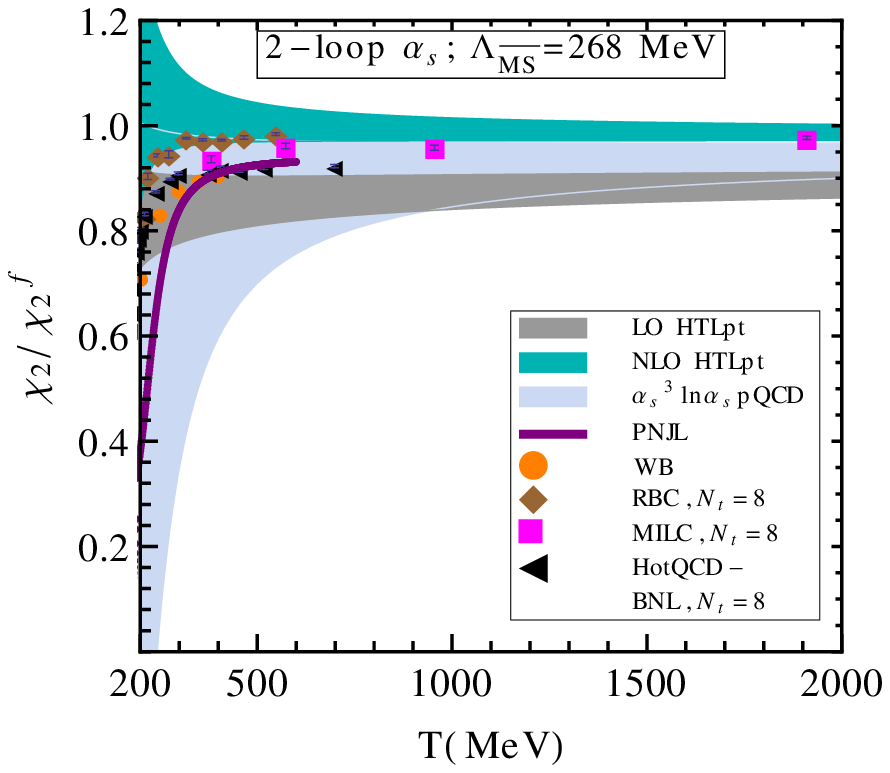}
}
\subfigure{
\includegraphics[width=0.48\textwidth,height=0.5\textwidth]{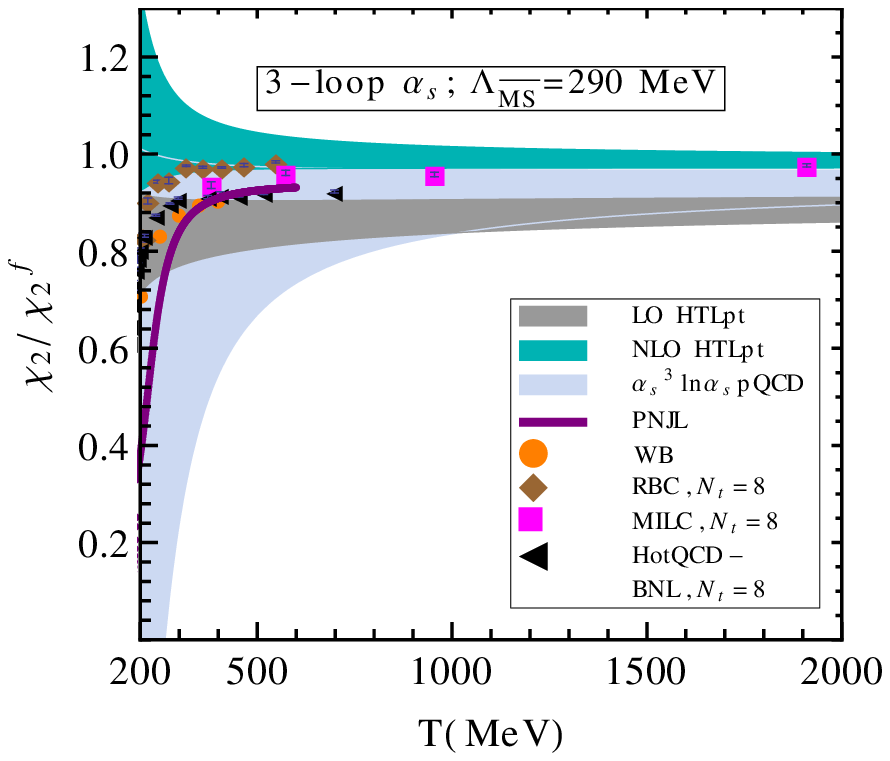}
}
\caption[{\em Left panel} : $\chi_2$  scaled by the free field value for LO (grey band) and NLO (sea green band) in 2-loop HTLpt, 
4-loop pQCD (sky blue band) along-with various LQCD data using 2-loop $\alpha_s$.{\em Right panel} : Same as left panel but using 
3-loop $\alpha_s$.]{{\em Left panel} : $\chi_2$  scaled by the free field value for LO (grey band) and NLO (sea green band) in 
2-loop HTLpt, 
4-loop pQCD (sky blue band)~\cite{Vuorinen:2003fs}, LQCD (various symbols)~\cite{Borsanyi:2012cr,Petreczky:2012rq,Petreczky:2009cr,
Bernard:2004je},and PNJL model 
(thick purple line)~\cite{Bhattacharyya:2010jd,Bhattacharyya:2010ef} are plotted as a function of the temperature.
The bands in HTLpt and pQCD are obtained by varying the ${\overline{\rm MS}}$ renormalization scale ($\Lambda$) around its 
central value by a factor of two. We also used  $\Lambda_{\overline{\rm MS}}=268$ MeV  and 2-loop $\alpha_s$ for HTLpt and pQCD.
{\em Right panel} : Same as left panel but using 3-loop $\alpha_s$ and $\Lambda_{\overline{\rm MS}}=290$ MeV.}
\label{fig_chi2}
 \end{figure}
We note that no $\mu$ derivatives of the mass parameters appear in (\ref{chi2_nlo}) and, as a result, $\chi_2^{\rm NLO}(T)$
reduces to such a simple and compact form. This is because the second derivatives of the mass parameters with respect to
$\mu$ always appear as multiplicative factors of the gap equations (\ref{gap_md0}) and (\ref{gap_mq0}) 
and hence these contributions vanish.  Numerically solving for the variational masses using Eq.~(\ref{gap_md0})
and (\ref{gap_mq0}) one can directly compute $\chi_2^{\rm NLO}(T)$ from (\ref{chi2_nlo}).  Alternatively, 
we have also computed $\chi_2^{\rm NLO}(T)$ by performing numerical differentiation of the pressure in (\ref{pressure-NLO})
which leads to the same result within numerical errors.

In Fig.~(\ref{fig_chi2}) we have plotted the $N_f = 3$ second-order QNS scaled by the corresponding
free gas limit as a function of the temperature. As discussed above, the bands shown
for the HTLpt and pQCD~\cite{Vuorinen:2003fs} results indicate the sensitivity of $\chi_2$ to the choice of the
renormalization scale $\Lambda$. However, $\chi_2$ in both HTLpt and pQCD depends only weakly on
the chosen order of the running of the strong coupling and in turn only depends weakly
on $\Lambda_{MS}$ , as can be seen clearly from both panels of Fig.~(\ref{fig_chi2}). The LO HTLpt 
prediction for $\chi_2$ seems to agree reasonably well with the available Wuppertal-Budapest LQCD data which are obtained
using the tree-level improved Symanzik action and a stout smeared staggered fermionic
action with light quark masses $\sim 0.035 \, m_s$, with $m_s$ being the strange quark mass near its physical value;
however, there is a sizable variation among different lattice computations~\cite{Borsanyi:2012cr,Petreczky:2009cr,Bernard:2004je}
considering improved lattice actions and a range of quark masses. However,
lowering the quark mass $(∼ 0.035ms , ms$ is the strange quark mass) nearer to its physical
value~\cite{Borsanyi:2012cr} seems to have a very small effect in the temperature range, as seen from the
LQCD data. Note that for the Wuppertal-Budapest (WB) lattice data shown in Fig.~(\ref{fig_chi2}),
Ref.~\cite{Borsanyi:2012cr} provided a parameterization of their $\chi_2$ data
\be
\chi_2(T ) = e^{-(h_3 /t + h_4 /t^2)} f_3\ [\tanh(f_4 t + f_5 ) + 1] ,
\ee
where $t = T /(200 {\mbox MeV}),\ h_3 = −0.5022,\ h_4 = 0.5950,\ f_3 = 0.1359,\ f_4 = 6.3290,\ \mbox{and}\
f5 = −4.8303.$ The authors of Ref.~\cite{Borsanyi:2012cr} performed the fit for data \cite{Borsanyi:2011sw} in the temperature
range $125$ MeV $ < T \le 400$ MeV. Using the parameterization above, we display their
data up to $400$ MeV with a step size of $50$ MeV. The  
RBC-Bielefeld collaboration~\cite{Petreczky:2012rq} data for $\chi_2$ shown in Fig.~(\ref{fig_chi2}) 
used  a p4 action whereas the MILC collaboration~\cite{Bernard:2004je} used an asqtad action. 
In both cases the light quark mass ranges from (0.1-0.2)$\,m_s$. The results for $\chi_2$ obtained using a
nonperturbative PNJL model \cite{Bhattacharyya:2010jd,Bhattacharyya:2010ef} which includes an six-quark interaction are only
available very close to the phase transition temperature.
We see in Fig.~(\ref{fig_chi2}) that NLO HTLpt~(\ref{chi2_nlo}) exhibits a modest improvement over the
pQCD calculation shown, which is accurate to ${\cal O}(\alpha_s^3\ln \alpha_s)$. However, the NLO $\chi_2$ is higher
than the LO one at higher temperature and it goes beyond the free gas value at lower
temperatures. It should be mentioned that, although the 2-loop calculation improves upon
the LO results by rectifying over-counting which causes incorrect coefficients in the weak
coupling limit, it does so by pushing the problem to higher order in $g$. The reason can
be understood in the following way: in HTLpt the loop and coupling expansion are not
symmetrical, therefore at a given loop order there are contributions from higher orders in
coupling. Since the NLO HTL pressure and thus QNS is only strictly accurate to order ${\cal O}(g^3)$
there is over-counting occurring at higher orders in $g$, namely at ${\cal O}(g^4)$ and ${\cal O}(g^5)$.
A next-to-next-to-leading order (NNLO) HTLpt calculation would fix the problem 
through ${\cal O}(g^5)$ thereby guaranteeing that, when expanded in a strict power series in $g$, the HTLpt result would reproduce the 
perturbative result order-by-order through ${\cal O}(g^5)$.

\subsection{NLO HTLpt fourth-order QNS}
A semi-analytic expression for the NLO HTLpt fourth-order QNS can also be obtained from Eq.~(\ref{pressure-NLO})

\begin{eqnarray}
 \chi_4^{\rm NLO}(T)
                &=&\left.\frac{\partial^4 }{\partial\mu^4}{\cal P}_{\rm NLO}(T,\Lambda,\mu)\right|_{\mu=0}
                 =\frac{1}{(2\pi T)^4}\left.\frac{\partial^4 }
                  {\partial\hat\mu^4}{\cal P}_{\rm NLO}(T,\Lambda,\hat\mu)\right|_{\hat\mu=0}
\nonumber\\
     &=&\!\!\!\!\frac{d_A}{4\pi^2}\Bigg[ 8\frac{ d_F}{d_A}
     +\frac{\alpha_s}{\pi }s_F 
      \Bigg\{-12 +6  \hat m_D''(0)
\nonumber\\
     &+&\!\!\!\! 3 \hat m_D^2(0) 
    \left[ \left(\frac{\zeta'(-1)}{\zeta(-1)}+\ln\hat m_D(0)+\frac{1}{3}\right)
     (24-18 \zeta(3)) +7\zeta(3)\right] \hat m_D''(0)
\nonumber\\
    &+&\!\!\!\! \hat m_D(0)\hat m_D''(0)  \left(\ln\frac{\hat\Lambda} {2}-15.662\right) 
    (8-6 \zeta(3)) 
\nonumber\\
    &-&\!\!\!\!4 \hat m_D^3(0) \left[31 \zeta(5)-120 \left(
    \frac{\zeta'(-1)}{\zeta(-1)}+\ln\hat m_D(0)\right) (\zeta(5)-\zeta(3))
    \right]
\nonumber\\
    &+&\!\!\!\! 80\hat m_D^2(0)  
        \left(\ln\frac{\hat\Lambda} {2}-1.0811\right) (\zeta(5)-\zeta(3))
   + 134.859\ \hat m_q(0)\hat m_q''(0)\Bigg\}\Bigg], \hspace{2cm}
\label{chi4_nlo}
\end{eqnarray}
where the double derivatives of the mass parameters with respect to $\mu$ survive, but the fourth derivatives 
of the mass parameters disappear as discussed earlier. One can now directly 
compute the fourth-order susceptibility by using numerical solutions of the gap equations in (\ref{gap_md0}) 
and (\ref{gap_mqpp0}). Alternatively, we have also computed  $\chi_4^{\rm NLO}(T)$ by performing numerical 
differentiation of the pressure in (\ref{pressure-NLO}) which leads to the same result within numerical errors.
\begin{figure}
 \subfigure{
\includegraphics[width=0.48\textwidth]{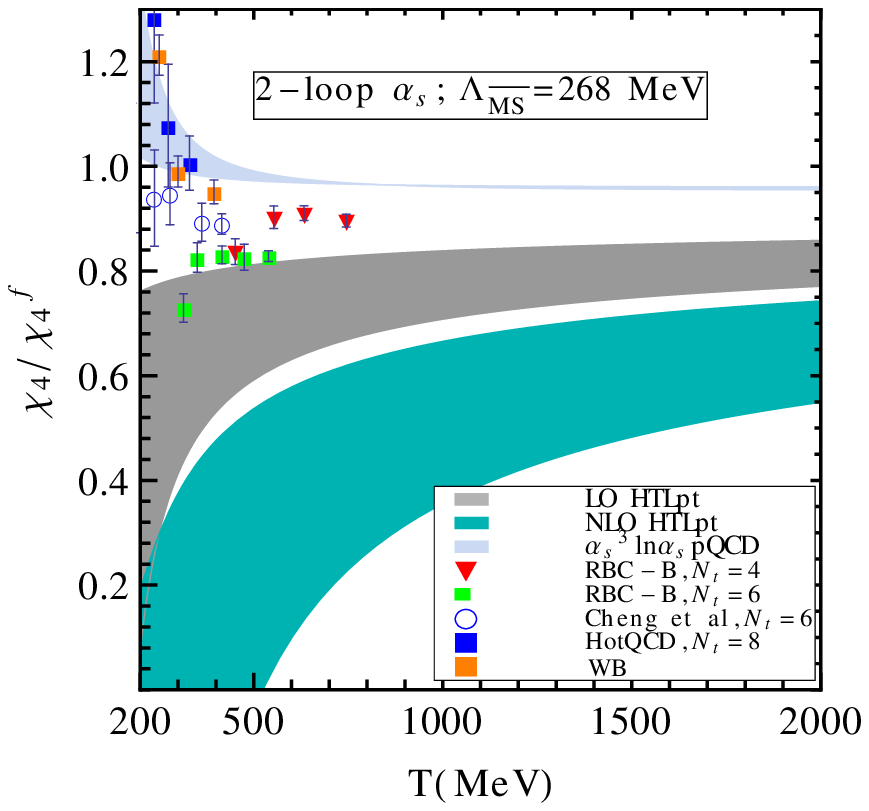}
}
\subfigure{
\includegraphics[width=0.48\textwidth]{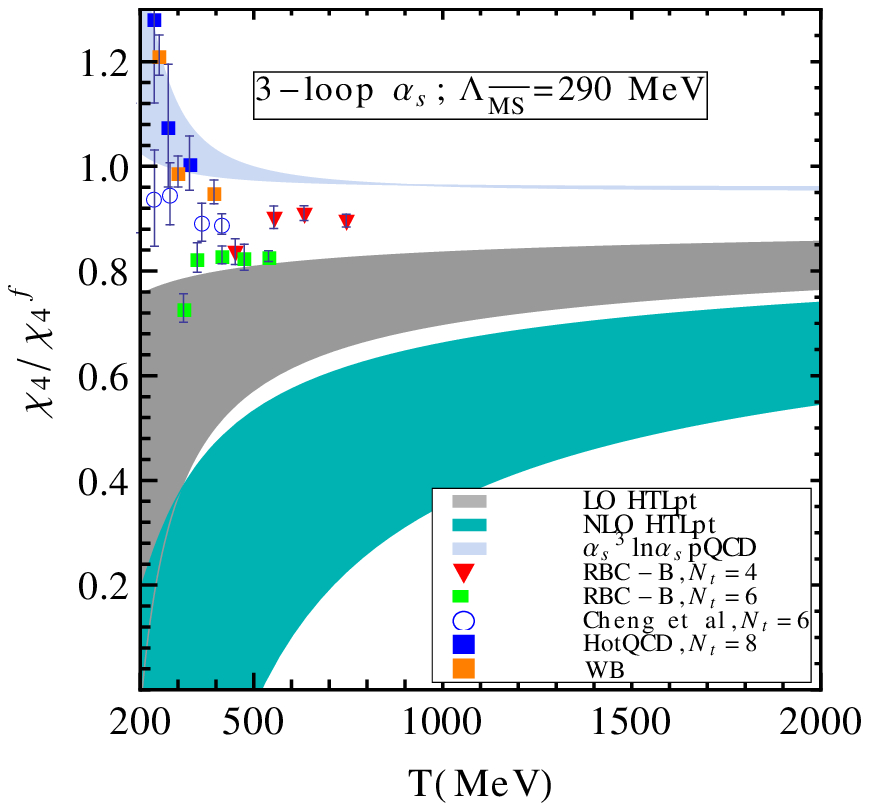}
}
\caption[$\chi_4$ scaled by the free field value for LO and NLO HTLpt, 4-loop pQCD, and
LQCD are plotted as a function of the temperature]{{\rm Left panel}: $\chi_4$ scaled by the free field value for LO and NLO HTLpt as given, respectively,
in (\ref{chi4_lo}) and (\ref{chi4_nlo}), 4-loop pQCD~\cite{Vuorinen:2003fs}, and
LQCD~\cite{Petreczky:2009cr,Cheng:2008zh} are plotted as a function of the temperature. The bands in HTLpt 
and pQCD are obtained by varying the 
${\overline{\rm MS}}$ renormalization scale ($\Lambda$) around its central value by a factor of two.  We used  
$\Lambda_{\overline{\rm MS}}=268$ MeV  and 2-loop $\alpha_s$ for HTLpt and pQCD.
Lattice QCD results~\cite{Petreczky:2009cr,Cheng:2008zh} are represented by symbols.  The Wuppertal-Budapest (WB) lattice
data are taken from Ref.~\cite{Borsanyi:2012rr}.
{\rm Right panel}: Same as left panel but using 3-loop $\alpha_s$ and $\Lambda_{\overline{\rm MS}}=290$ MeV.}
\label{fig_chi4}
\end{figure}

In Fig.~(\ref{fig_chi4}) we plot the fourth-order QNS ($\chi_4$) scaled by the corresponding free gas value
for HTLpt as given in (\ref{chi4_lo}) and (\ref{chi4_nlo}), pQCD,  and LQCD. Both the HTLpt and pQCD results
exhibit a very weak dependence on the choice of order of the running of 
$\alpha_s$ and thus $\Lambda_{\overline{\rm MS}}$. Nevertheless, the HTLpt results are found to be far below
the pQCD result~\cite{Vuorinen:2003fs} which is accurate to ${\cal O}(\alpha_s^3\ln(\alpha_s))$ and 
the LQCD results~\cite{Petreczky:2009cr,Cheng:2008zh}. Also, the correction to $\chi_4$ when going from LO to NLO is quite large.
This is due to the fact that the fourth order susceptibility is highly sensitive to the erroneous 
${\cal O}(g^4)$ and ${\cal O}(g^5)$ terms which appear at NLO.  It is expected that carrying the HTLpt calculation
to NNLO would improve this situation; however, only explicit calculation can prove this.  We note additionally 
that although the pQCD result is very close to the Stefan-Boltzmann limit, the dimensional-reduction resummation
method yields a fourth-order QNS which is approximately 20\% below the Stefan-Boltzmann limit
\cite{Andersen:2012wr} which places it slightly higher than the LO HTLpt result shown in Fig.~(\ref{fig_chi4}).



\section{Conclusions and Outlook}
\label{concl} 
In this chapter we have generalized the zero chemical potential NLO HTLpt calculation of the QCD thermodynamic
potential \cite{Andersen:2002ey,Andersen:2003zk} to finite chemical potential. We have obtained (semi-)analytic
expressions for the thermodynamic potential at both LO and NLO in HTL perturbation theory. We have also obatained
second and fourth order quark number susceptibilities at both LO and NLO from that (semi-)analytic expressions 
for the thermodynamic potential. The results for thermodynamic potential obtained here are trustworthy at high
temperatures and small chemical potential since we performed an expansion in the ratio of the chemical potential
over the temperature.

This NLO thermodynamic potential will be useful for the study of finite temperature and chemical potential QCD matter. This is important
in view of the ongoing RHIC beam energy scan and proposed heavy-ion experiments at FAIR. Using the NLO HTLpt thermodynamic potential, 
we have obtained a variational solution for both mass parameters, $m_q$ and $m_D$, and we have used this to obtain the pressure 
at finite temperature and chemical potential. When compared with the weak coupling expansion 
of QCD, the HTLpt pressure helps somewhat with the problem of oscillation of successive approximations found in pQCD. Furthermore, the 
scale variation of the NLO HTLpt result for pressure is smaller than that obtained with the weak coupling result. 

The LO result for $\chi_2$ obtained here shows reasonable agreement with available LQCD data; however, at this
point in time there is still a fairly sizable variation of this quantity between the different
lattice groups. Moving forward it would seem that a detailed analysis of the uncertainties
in the various LQCD calculations is necessary before detailed conclusions can be drawn. Unlike the LO results, our NLO
calculation takes into account dynamical quark contributions by including two-loop graphs
which involve fermion loops; however, they suffer from the same problem that NLO HTLpt
calculations at zero chemical potential faced: the NLO $\chi_2$ in Eq.~(\ref{chi2_nlo}) gets ${\cal O}(g^3)$ correct but
the ${\cal O}(g^4)$ and ${\cal O}(g^5)$ contributions are incorrect if they are expanded out in a strict power
series in g. As a result, our NLO result for $\chi_2$ scaled to the free limit is closer to unity
than the corresponding LO result and only shows a weak dependence on the chosen value of the renormalization scale. 
Our NLO result for $\chi_4$ (Eq.~(\ref{chi4_nlo})) in which $\mu$ derivatives of
the variational mass parameters survive is significantly below the pQCD and lattice data.

As was the case with the pressure at zero chemical potential, it seems that fixing
this problem will require going to NNLO. In the case of the zero chemical potential
pressure, performing such a calculation resulted in much improved agreement between
HTL perturbation theory and LQCD calculations above $\sim 2T_c$. At the very least a NNLO calculation will
fix the over-counting problems through ${\cal O}(g^5)$
and NNLO calculation at finite temperature and chemical potential will be discuss in the next Chapter~\ref{chapter:3loop}.

%% file: text/3loop.tex

\chapter{Three-loop HTLpt thermodynamics}
\label{chapter:3loop}

In this chapter, we study the three loop thermodynamics of QCD using the hard 
thermal loop perturbation theory. We show that at three loop order hard thermal
loop perturbation theory results are compatible with 
lattice results for the pressure, energy density, entropy density, various
order diagonal and off-diagonal susceptibilities, speed of sound down to 
temperatures $T\sim 250$ MeV. This chapter is based on:
{\it Three-loop HTLpt Pressure and Susceptibilities at Finite Temperature and
Density},  N. Haque, J. O. Andersen, M. G. Mustafa, M. Strickland, and N. Su, 
{\bf Phys. Rev.  D89 (2014) 061701} and {\it Three-loop HTLpt thermodynamics at finite temperature
and chemical potential}, N. Haque, A. Bandyopadhyay, J. O. Andersen, M. G. Mustafa,
M. Strickland, and N. Su, {\bf JHEP 1405 (2014) 027}.
As we discussed in Chapter~\ref{chapter:2loop}, at leading order thermodynamic potential is  only correct $g^0$ and $g^3$
terms when one expands in a strict power series in $g$. Similarly,  at next-to-leading order thermodynamic potential one obtains
correct $g^0$, $g^2$ and $g^3$ terms when one expands in a strict power series in $g$. So to make the results more reliable at moderate
coupling constant $g$, one needs to calculate thermodynamic functions beyond next-to-leading order in HTL perturbation theory.
In this chapter we calculate the thermodynamic potential at finite temperature $(T)$ and chemical potential $(\mu)$
to three-loop order in HTLpt. The three-loop thermodynamic potential is renormalized using only known vacuum, mass,
and coupling constant counterterms and 
the final result is completely analytic.  The resulting analytic thermodynamic potential is then used to obtain expressions for 
the pressure, energy density, entropy density, trace anomaly, speed of sound, and various quark number susceptibilities.  We find 
that there is good agreement between our NNLO HTLpt results and lattice data down to temperatures on the order of 250 MeV.


This chapter is organized as follows. In Sec.~(\ref{feyn_diag}) we discuss the diagrams that contribute to the HTL 
perturbation theory thermodynamic potential through NNLO. In Sec.~(\ref{expansion}) the necessary diagrams are reduced
to scalar sum-integrals and expanded in powers of $m_D/T$ and $m_q/T$.  We list the necessary non-trivial sum-integrals
and $d-$ dimensional integrals in Secs.~(\ref{app:sum-integrals}) and (\ref{app:threedints}) respectively.
In Sec.~({\ref{nnloomega}}) we present our final results for the NNLO thermodynamic potential. In Sec.~(\ref{pres})
we discuss the mass prescription for the in-medium masses $m_D$ and $m_q$. We present our results for the thermodynamic
functions and compare them with results from lattice gauge simulations in Sec.~(\ref{thermof}). In Sec.~(\ref{sec:qns})
we present our results for the \mbox{second-,} fourth-, and sixth-order baryon and quark number susceptibilities and 
compare them with results from lattice QCD data. In Sec.~(\ref{outlook}) we summarize and conclude.
\section{Contributions to the HTLpt thermodynamic potential through NNLO}
\label{feyn_diag}
\begin{figure}[tbh]
\begin{center}
\includegraphics[width=13cm]{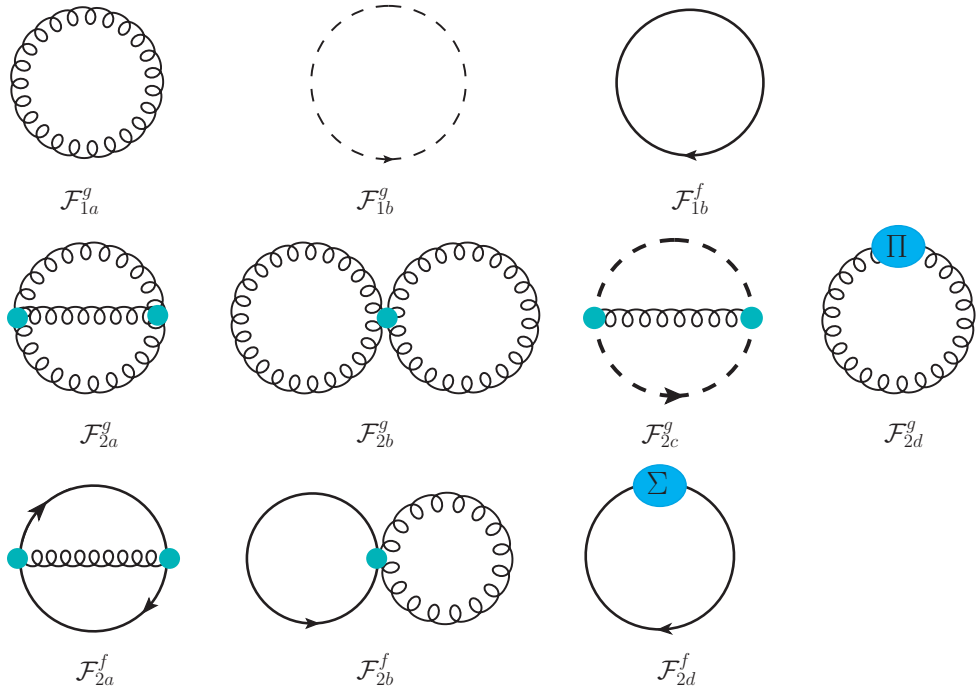}
\caption{One and two loop Feynman diagrams that will contribute to the thermodynamic potential.}
\label{feyn_diag12}
\end{center}
\end{figure}

\begin{figure}
 \vspace{-1cm}\includegraphics[width=15cm,height=18.5cm]{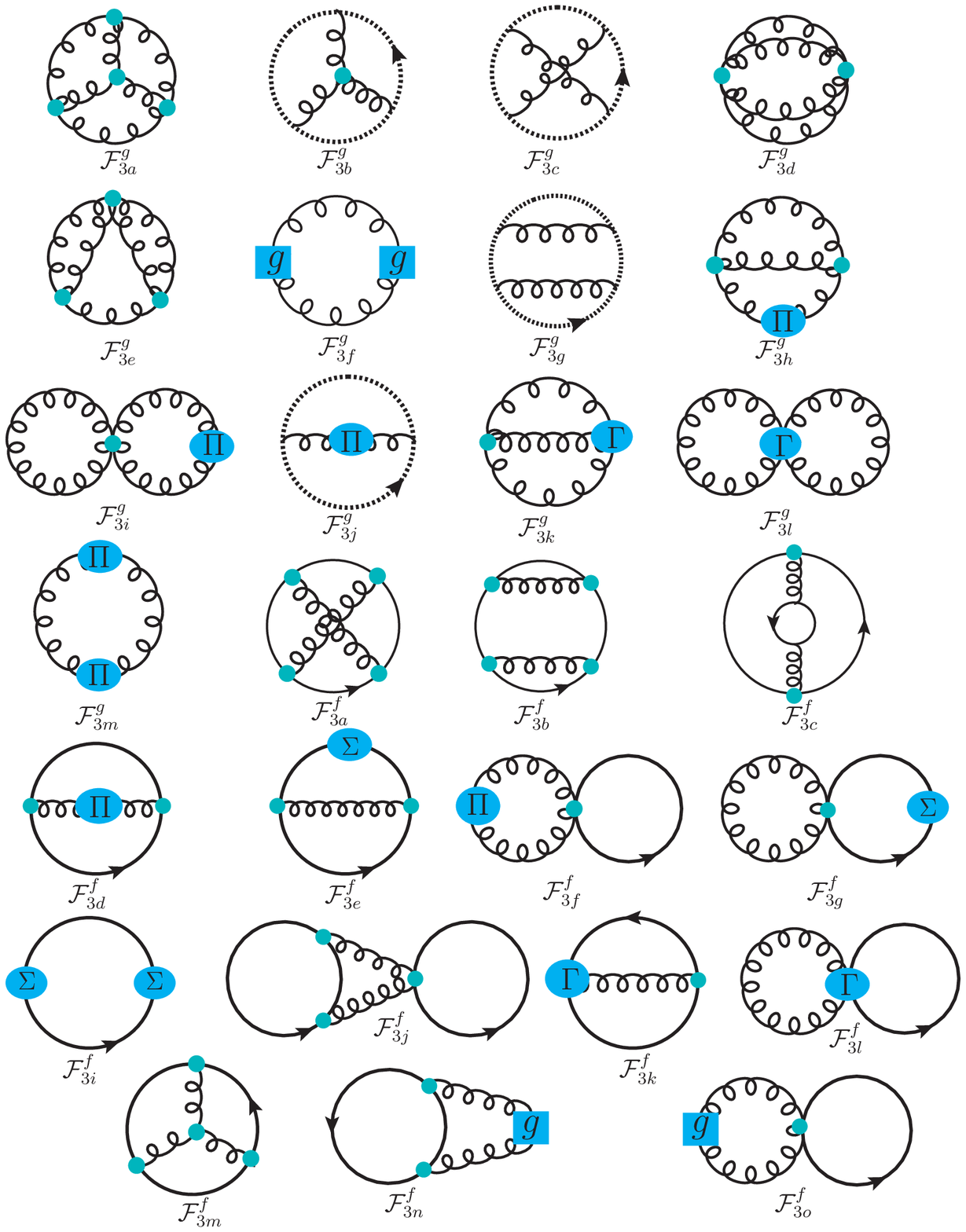}
\caption{Three loop HTL Feynman diagrams that will contribute to the thermodynamic potential.}
\label{feyn_diag3}
\vspace{-2cm}
\end{figure}

\begin{figure}
\includegraphics[width=15cm]{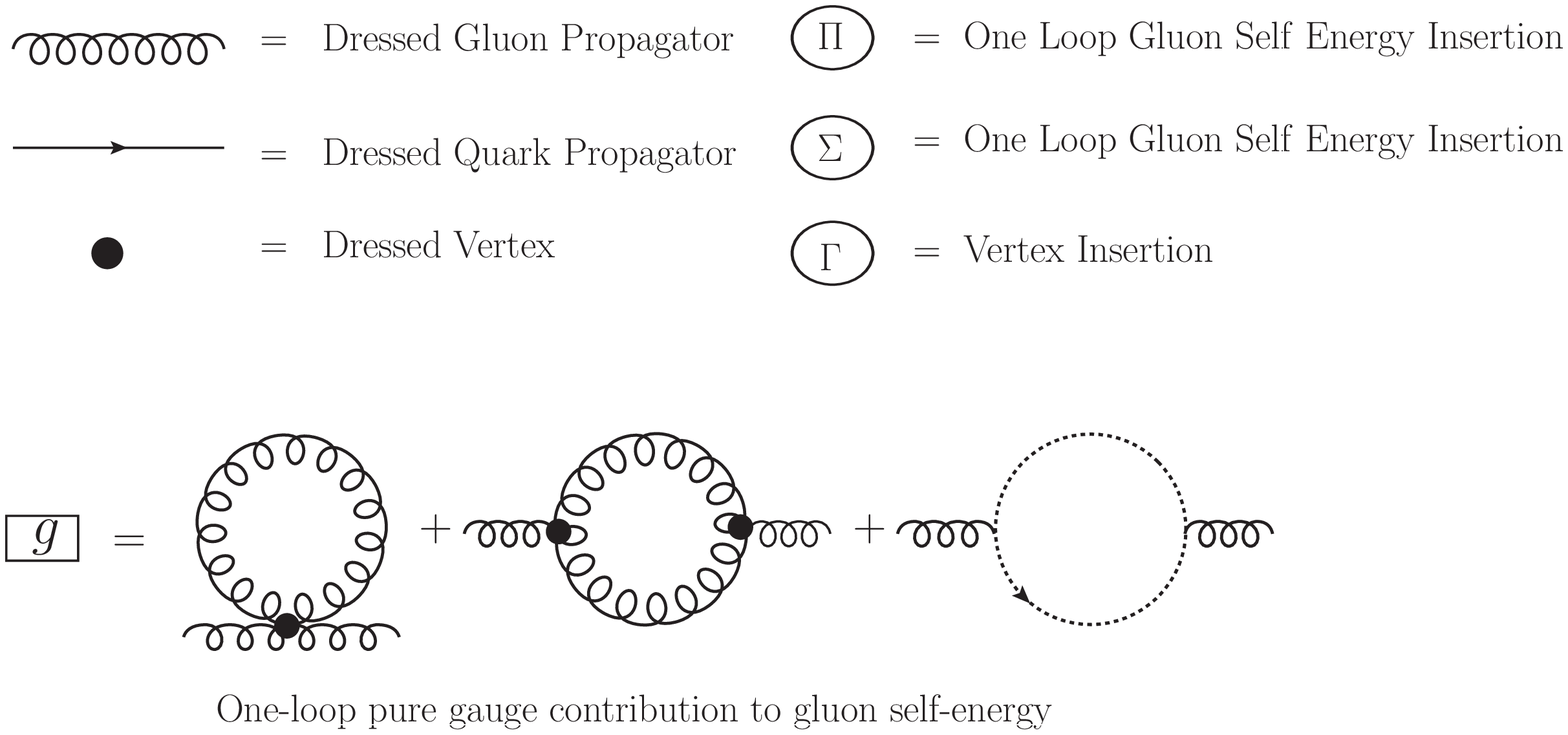}
\caption{The shorthand notations used in Fig.~(\ref{feyn_diag3}).}
\label{shorthand}
\end{figure}

The diagrams needed for the computation of the HTLpt thermodynamic potential through NNLO are listed in 
Figs.~(\ref{feyn_diag12}) and (\ref{feyn_diag3}). The shorthand notations used in Fig.~(\ref{feyn_diag3})
have been explained in Fig.~(\ref{shorthand}).

In Ref.~\cite{Andersen:2011sf} the authors computed the NNLO thermodynamic
potential at zero chemical potential.  Here we extend the NNLO calculation to finite chemical potential. For this purpose,
one needs to only consider diagrams which contain at least one quark propagator; however,
for completeness we also list the purely gluonic contributions below.  In the results we will express thermodynamic quantities in
terms of dimensionless variables:  $\hat{m}_D = m_D/(2\pi T)$, $\hat{m}_q = m_q/(2\pi T)$, $\hat{\mu} = \mu/(2\pi T)$, 
$\hat{\Lambda} = \Lambda/(2\pi T)$ and $\hat{\Lambda}_g = \Lambda_g/(2\pi T)$ where $\Lambda$ and $\Lambda_g$ are renormalization scale 
scales for gluon and fermion respectively as discussed in Sec.~(\ref{sec:scales}).

\newpage

\newpage
The complete NNLO HTLpt thermodynamic potential can be expressed in terms of each diagrams of Figs.~(\ref{feyn_diag12})
 and (\ref{feyn_diag3}) as
\begin{eqnarray}
 \Omega_{\rm NNLO}&=&d_A\left[{\cal F}_{1a}^g+{\cal F}_{1b}^g+{\cal F}_{2d}^g+{\cal F}_{3m}^g\right]+d_F\[{\cal F}_{1b}^f
                 +{\cal F}_{2d}^f+{\cal F}_{3i}^f\]\nonumber\\
&&\hspace{-.5cm}+d_Ac_A\Big[{\cal F}_{2a}^g+{\cal F}_{2b}^g+{\cal F}_{2c}^g+{\cal F}_{3h}^g+{\cal F}_{3i}^g+{\cal F}_{3j}^g
           +{\cal F}_{3k}^g+{\cal F}_{3l}^g\Big]   \nonumber\\
&&\hspace{-.5cm}+d_As_F\Big[{\cal F}_{2a}^f+{\cal F}_{2b}^f+{\cal F}_{3d}^f+{\cal F}_{3e}^f+{\cal F}_{3f}^f+{\cal F}_{3g}^f+
       {\cal F}_{3k}^f+{\cal F}_{3l}^f\Big]      \nonumber\\
&&\hspace{-.5cm}  +d_Ac_A^2\Big[{\cal F}_{3a}^g+{\cal F}_{3b}^g+{\cal F}_{3c}^g+{\cal F}_{3d}^g
                  +{\cal F}_{3e}^g+{\cal F}_{3f}^g+{\cal F}_{3g}^g\Big]+d_As_{2F}
         \Big[{\cal F}_{3a}^f+{\cal F}_{3b}^f\Big]   \nonumber\\
&&\hspace{-.5cm}+d_Ac_As_{F}
         \Big[-\frac{1}{2}{\cal F}_{3a}^f+{\cal F}_{3m}^f+{\cal F}_{3n}^f+{\cal F}_{3o}^f\Big]
         +d_As_F^2\Big[{\cal F}_{3c}^f+{\cal F}_{3j}^f\Big]   \nonumber\\
&&\hspace{-.5cm} + \Delta_0{\cal E}_0+\Delta_1{\cal E}_0+ \Delta_2{\cal E}_0+\Delta_1m_D^2\frac{\partial}
       {\partial m_D^2}\Omega_{\rm LO}+\Delta_1m_q^2\frac{\partial}{\partial m_q^2}\Omega_{\rm LO}
       \nonumber\\
&&\hspace{-.5cm}+\Delta_2m_D^2\frac{\partial}
       {\partial m_D^2}\Omega_{\rm LO} +\Delta_2m_q^2\frac{\partial}{\partial m_q^2}\Omega_{\rm LO} + 
       \Delta_1m_D^2\frac{\partial}{\partial m_D^2}
       \Omega_{\rm NLO}
\nonumber \\
&& \hspace{-.5cm} + \Delta_1m_q^2\frac{\partial}
       {\partial m_q^2}\Omega_{\rm NLO}+\frac{1}{2}\left[\frac{\partial^2}{(\partial m_D^2)^2}\Omega_{\rm LO}
       \right]\left(\Delta_1 m_D^2\right)^2 
        \nonumber\\
&&\hspace{-.5cm}+\frac{1}{2}\left[\frac{\partial^2}{(\partial m_q^2)^2}
       \Omega_{\rm LO}\right]\left(\Delta_1 m_q^2\right)^2
       +d_A\left[\frac{c_A{\cal F}_{2a+2b+2c}^g+s_F{\cal F}_{2a+2b}^f}{\alpha_s}\right]\Delta_1\alpha_s,
\label{omega_initial}
\end{eqnarray}
where the necessary counterterms at any order in $\delta$ can be calculated using the following counter terms
\be
\Delta{\cal E}_0&=&\frac{d_A}{128\pi^2\epsilon}(1-\delta)^2m_D^4\ ,
\label{ctE}\\
\Delta m_D^2&=&\frac{11c_A-4s_F}{12\pi\epsilon}\alpha_s\delta(1-\delta)m_D^2\ ,
\label{ctmD}
\\
\Delta m_q^2&=&\frac{3}{8\pi\epsilon}\frac{d_A}{c_A}\alpha_s \delta(1-\delta)m_q^2\ ,
\label{ctmq}
\\
\delta\Delta\alpha_s&=&-\frac{11c_A-4s_F}{12\pi\epsilon}\alpha_s^2\delta^2\ ,
\label{ctalpha}
\ee.

The expressions for the one- and two-loop diagrams above can be found in Refs.~\cite{Andersen:2002ey,Andersen:2003zk}.
The expressions for the three-loop bosonic diagrams ${\cal F}_{3a}^g$--${\cal F}_{3m}^g$ are presented in section 
3 of Ref.~\cite{Andersen:2010ct}. The three-loop diagrams specific to QCD, i.e. the 
non-Abelian diagrams involving quarks, are given by
\be
{\cal F}_{\rm 3a}^f
\!\!\!&=&\!\!\!\frac{1}{4}g^4
\sumintbff_{P\{QR\}}{\rm Tr}\left[\Gamma^{\mu}(-P,Q-P,Q)S(Q)
\Gamma^{\alpha}(Q-R,Q,R)S(R)\Gamma^{\nu}(P,R,R-P)
\right.\nn \!\!\!&\times&\!\!\!\left.
 S(R-P)\Gamma^{\beta}(-Q+R,R-P,Q-P)S(Q-P)\right]
\Delta_{\mu\nu}(P)\Delta_{\alpha\beta}(Q-R)\ ,\hspace{1.5cm}
\label{f3a}
\\
{\cal F}_{\rm 3b}^f\!\!\!&=&\!\!\!
\frac{1}{2}g^4\sumintbff_{P\{QR\}}{\rm Tr}
\left[\Gamma^{\mu}(P,P+Q,Q)S(Q)\Gamma^{\beta}(-R+Q,Q,R)S(R)
\Gamma^{\alpha}(R-Q,R,Q)
\right.\nn \!\!\!\!&\times&\!\!\!\left.
 S(Q)\Gamma^{\nu}(-P,Q,P+Q)S(P+Q)\right]
\Delta_{\mu\nu}(P)\Delta_{\alpha\beta}(R-Q)\ ,\hspace{1.5cm}
\label{f3b}
\\
{\cal F}_{\rm 3c}^f\!\!\!&=&\!\!\!-\frac{1}{4}
g^4\sumintbff_{P\{QR\}}
{\rm Tr}\left[
\Gamma^{\mu}(P,P+Q,Q)S(Q)\Gamma^{\beta}(-P,Q,P+Q)S(P+Q)
\right]
 \nn \!\!\!&\times&\!\!\!
{\rm Tr}\left[
\Gamma^{\nu}(-P,R,P+R)S(P+R)
\Gamma^{\alpha}(P,P+R,R)S(R)
\right]\Delta^{\mu\nu}(P)
\Delta^{\alpha\beta}(P)\ ,\hspace{1.5cm}
\label{f3c} 
\\
{\cal F}_{\rm 3j}^f&=&-\frac{1}{2}
g^4\sumintbff_{P\{QR\}}
{\rm Tr}\left[
\Gamma^{\alpha\beta}(P,-P,R,R)
S(R)\right]\Delta^{\alpha\mu}(P)\Delta^{\beta\nu}(P)
\nn
\!\!\!&\times&\!\!\!
{\rm Tr}\left[
\Gamma^{\mu}(P,P+Q,Q)S(Q)\Gamma^{\nu}(-P,Q,P+Q)S(P+Q)\right]\ ,\hspace{1.5cm}
\\
{\cal F}_{\rm 3m}^f
\!\!\!&=&\!\!\! \frac{1}{6}
\sumintfff_{\{PQR\}}{\rm Tr}\left[
\Gamma^{\alpha}(R-P,R,P)S(P)\Gamma^{\beta}(P-Q,P,Q)S(Q)
\Gamma^{\gamma}(Q-R,Q,R)S(R)\right]
\nn \!\!\!&\times&\!\!\! 
\Gamma^{\mu\nu\delta}(P-R,Q-P,R-Q)
\Delta^{\alpha\mu}(P-R)\Delta^{\beta\nu}(Q-P)\Delta^{\gamma\delta}(R-Q)\ ,\hspace{1.5cm}
\\
{\cal F}^f_{\rm 3n}\!\!\!&=&\!\!\!
{-}\sumintb_{P}\ \bar{\Pi}_g^{\mu\nu}(P)\Delta^{\nu\alpha}(P)
\bar{\Pi}_f^{\alpha\beta}(P)\Delta^{\beta\mu}(P)\ ,\hspace{1.5cm}
\\
{\cal F}^f_{\rm 3o}\!\!\!&=&\!\!\!
-\frac{1}{2}
g^2\sumintbf_{P\{Q\}}
{\rm Tr}\left[
\Gamma^{\alpha\beta}(P,-P,Q,Q)
S(Q)\right]\Delta^{\alpha\mu}(P)\Delta^{\beta\nu}(P)
\bar{\Pi}_g^{\mu\nu}(P)\ ,\hspace{1.5cm}
\ee
where
\be
\bar{\Pi}^{\mu\nu}_g(P)
&=&\frac{1}{2}g^2\sumintb_Q\ 
\Gamma^{\mu\nu,\alpha\beta}(P,-P,Q,-Q)\Delta^{\alpha\beta}(Q)
\nn
 &+&\frac{1}{2}g^2\sumintb_Q\ \Gamma^{\mu\alpha\beta}(P,Q,-P-Q)\Delta^{\alpha\beta}(Q)
\Gamma^{\nu\gamma\delta}(P,Q,-P-Q)\Delta^{\gamma\delta}(-P-Q)
\nn
&+&g^2\sumintb_Q\ \frac{Q^{\mu}(P+Q)^{\nu}}{Q^2(P+Q)^2}\;,
\\ 
\bar{\Pi}^{\mu\nu}_f(P)
&=&
{-}g^2\sumintf_{\{Q\}}\ {\rm Tr}\left[\Gamma^{\mu}(P,Q,Q-P)
S(Q)\Gamma^{\nu}(P,Q,Q-P)S(Q-P)\right]
\;.
\ee
Thus $\bar{\Pi}^{\mu\nu}(P)$ is the one-loop gluon self-energy with
HTL-resummed propagators and vertices:
\be
\bar{\Pi}^{\mu\nu}(P) = 
c_A\bar{\Pi}^{\mu\nu}_g(P)+s_F\bar{\Pi}^{\mu\nu}_f(P)\;.
\ee

\section{Expansion in mass parameters}
\label{expansion}
In Refs.~\cite{Andersen:2002ey,Andersen:2003zk} the NLO HTLpt thermodynamic potential was reduced to scalar sum-integrals.
Evaluating these scalar sum-integrals exactly seems intractable, however, the sum-integrals can be calculated approximately
by expanding them in powers of $m_D/T$ and $m_q/T$ following the method developed in Ref.~\cite{Andersen:2001ez}. We will adopt the same
strategy in this chapter and include all terms through order $g^5$ assuming that $m_D$ and $m_q$ are ${\cal O}(g)$ at leading order.
At each loop order, the contributions can be divided into those coming from hard and soft momenta, which are the momenta
proportional to the scales $T$ and $gT$ respectively. In the one-loop diagrams, the contributions are either hard $(h)$ or
soft $(s)$, while at the two-loop level, there are hard-hard $(hh)$, hard-soft $(hs)$, and soft-soft $(ss)$ contributions.
At three loops there are hard-hard-hard $(hhh)$, hard-hard-soft $(hhs)$, hard-soft-soft $(hss)$, and soft-soft-soft $(sss)$
contributions.
\subsection{One-loop sum-integrals}
We now review the mass expansion of the necessary one-loop sum-integrals considering separately
the contributions from hard and soft momenta.  We list the purely gluonic contributions
when they are necessary for simpler exposition of the final result.  Note that in order to simplify
the results, when possible, it is best to add the corresponding iterated polarization and self-energy 
insertions that appear at higher order in $\delta$, e.g. below we will also include ${\cal F}^g_{\rm 2d}$,
${\cal F}^f_{\rm 2d}$, ${\cal F}^g_{\rm 3m}$, and ${\cal F}^f_{\rm 3i}$ as ``one-loop'' contributions.
\vspace{-1cm}
\subsubsection*{Hard contributions}
For one loop gluon $({\cal F}_{1a}^g)$ and one loop ghost $({\cal F}_{1b}^g)$ diagrams, we need to expand in order $m_D^2$:
\be
{\cal F}_{1a+1b}^{g(h)}&=&\frac{1}{2}(d-1)\sumintb_P\ \ln P^2+\frac{1}{2}
m_D^2\sumintb_P\frac{1}{P^2}
\nn
&-&
\frac{1}{4(d-1)}m_D^4\sumintb_P\left[\frac{1}{P^4}
-\frac{2}{p^2P^2}-\frac{2d}{p^4}{\cal T}_P
+\frac{2}{p^2P^2}{\cal T}_P
+\frac{d}{p^4}{\cal T}_P^2
\right]\!.\ 
\label{Flo-h}
\ee
The one-loop graph with a gluon self-energy insertion $({\cal F}_{2d}^g)$ has an explicit factor of $m_D^2$ and, therefore, 
we only need to expand the sum-integral to first order in $m_D^2$:
\be
{\cal F}_{2d}^{g(h)}&=&-\frac{1}{2}
m_D^2\sumintb_P\frac{1}{P^2}
\nn
&+&\frac{1}{2(d-1)}m_D^4\sumintb_P\left[
\frac{1}{P^4}-\frac{2}{p^2P^2}-\frac{2d}{p^4}{\cal T}_P
+\frac{2}{p^2P^2}{\cal T}_P
+\frac{d}{p^4}{\cal T}_P^2
\right]\!. \hspace{6mm}
\label{ct1}
\ee
The one-loop graph with two gluon self-energy insertions $({\cal F}_{3m}^g)$ 
must be expanded to zeroth order in $m_D^2$
\be
{\cal F}_{3m}^{g(h)}=
-\frac{1}{4(d-1)}m_D^4\sumintb_P\left[
\frac{1}{P^4}-\frac{2}{p^2P^2}-\frac{2d}{p^4}{\cal T}_P
+\frac{2}{p^2P^2}{\cal T}_P
+\frac{d}{p^4}{\cal T}_P^2
\right]\!. \hspace{6mm}
\label{ct2}
\ee

The sum of Eqs.~(\ref{Flo-h})-(\ref{ct2}) is very simple
\be
{\cal F}_{1a+1b+2d+3m}^{g(h)} =
\frac{1}{2}(d-1)\sumintb_P\ \ln\left(P^2\right) = -\frac{\pi^2}{45}T^4
\;.
\ee
This is the free energy of an ideal gas consisting of a single massless boson.

The one-loop fermionic graph ${\cal F}_{1b}^f$ needs to expanded to second order in $m^2_q$:
\be
{\cal F}_{1b}^{f(h)}&=&-2\sumintf_{\{P\}}\ln P^2-4m_q^2
\sumintf_{\{P\}}\frac{1}{P^2}
\nn
&&+2m_q^4\sumintf_{\{P\}}\left[
\frac{2}{P^4}
-\frac{1}{p^2P^2}+
\frac{2}{p^2P^2}{\cal T}_P
-\frac{1}{p^2P_0^2}{\cal T}_P^2
\right] \! . \hspace{9mm}
\label{f1b}
\ee
The one-loop fermion loop with a fermion self-energy insertion ${\cal F}_{2d}^f$
must be expanded to first order in $m_q^2$:
\be
{\cal F}_{2d}^{f(h)}=4m_q^2\sumintf_{\{P\}}\frac{1}{P^2}
-4m_q^4\sumintf_{\{P\}}\left[
\frac{2}{P^4}-\frac{1}{p^2P^2}
+
\frac{2}{p^2P^2}{\cal T}_P
-\frac{1}{p^2P_0^2}{\cal T}_P^2
\right]\!.
\label{f2d}
\ee
The one-loop fermion loop with two self-energy insertions ${\cal F}_{3i}^f$ must be
expanded to zeroth order in $m_q^2$:
\be
{\cal F}_{3i}^{f(h)}&=&2m_q^4\sumintf_{\{P\}}\left[
\frac{2}{P^4}-\frac{1}{p^2P^2}
+
\frac{2}{p^2P^2}{\cal T}_P
-\frac{1}{p^2P_0^2}{\cal T}_P^2
\right].
\label{f3i}
\ee
The sum of Eqs.~(\ref{f1b})-(\ref{f3i}) is particularly simple
\be
{\cal F}_{1b+2d+3i}^{f(h)}&=&-2\sumintf_{\{P\}}\ \ln P^2 \nn
&=&
-\frac{7\pi^2}{180}T^4\[1+\frac{120}{7}\hmu^2+\frac{240}{7}\hmu^4\]
\;.
\ee
This is the free energy of an ideal gas consisting of a single massless fermion.
\vspace{-.5cm}
\subsubsection*{Soft contributions}
\vspace{-.5cm}
The soft contributions in the diagrams ${\cal F}^g_{1a+1b}$, ${\cal F}^g_{2d}$, and ${\cal F}^g_{3m}$
arise from the $P_0=0$ term in the sum-integral.  At soft momentum $P=(0,{\bf p})$, the HTL self-energy
functions reduce to $\Pi_T(P) = 0$ and $\Pi_L(P) = m_D^2$.  The transverse term vanishes in dimensional
regularization because there is no momentum scale in the integral over ${\bf p}$.  Thus the soft contributions
come from the longitudinal term only and read
\be
{\cal F}^{g(s)}_{\rm 1a+1b}
&=&\frac{1}{2}T\int\limits_p\ln\left(p^2+m_D^2\right) = 
- \frac{m_D^3T}{12\pi}
\left( \frac{\Lambda_g}{2 m_D} \right)^{2 \epsilon}\left[
1+\frac{8}{3}\epsilon 
\right] , 
\label{f1as}
\ee
\vspace{-1cm}
\be
{\cal F}^{g(s)}_{\rm 2d}&=&
-\frac{1}{2}m_D^2T\int\limits_p\frac{1}{p^2+m_D^2} =
\frac{m^3_DT}{8\pi} \left( \frac{\Lambda_g}{2 m_D} \right)^{2 \epsilon}
\left[1 + 2 \epsilon  
 \right],
\label{f2ds}
 \ee
\vspace{-1cm}
\be
{\cal F}^{g(s)}_{\rm 3m}&=& - \frac{1}{4}m_D^4T\int\limits_p\frac{1}{(p^2+m_D^2)^2} 
= - \frac{m^3_DT}{32\pi}
\label{f3ms}
\;.
\ee
The total soft contribution from Eqs.~(\ref{f1as})-(\ref{f3ms}) is
\be
{\cal F}_{\rm 1a+1b+2d+3m}^{g(s)}&=&-\frac{m_D^3T}{96\pi}\left( \frac{\Lambda_g}{2 m_D} \right)^{2 \epsilon}\left[
1+\frac{8}{3}\epsilon
\right] .
\ee 
There are no soft contributions from the leading-order fermion diagrams or HTL counterterms (polarization and self-energy insertions).

\subsection{Two-loop sum-integrals}
For hard momenta, the self-energies are suppressed by $m_D/T$ and $m_q/T$ relative to the inverse free propagators,
so we can expand in powers of $\Pi_T$, $\Pi_L$, and $\Sigma$.  As was the case for the one-loop contributions, 
we once again treat the polarization and self-energy insertion NNLO diagrams as two-loop graphs in order to simplify
the resulting expressions.
\vspace{-1cm}
\subsubsection*{Hard-hard (hh) contribution}
\vspace{-1cm}
We first consider the contribution from fermionic diagrams.  The $(hh)$ contribution from ${\cal F}_{\rm 2a}^f$ 
and ${\cal F}_{\rm 2b}^f$ reads
\be
{\cal F}_{\rm 2a+2b}^{f(hh)}
\!\!\!&=&\!\!\!(d-1)g^2\left[\ \sumintff_{\{PQ\}}\frac{1}{P^2Q^2}
      -\sumintbf_{P\{Q\}}\frac{2}{P^2Q^2}\right] 
\nonumber \\ 
\!\!\!&+&\!\!\! 2m_D^2g^2\sumintbf_{P\{Q\}}\left[\frac{1}{p^2P^2Q^2}
     {\cal T}_P+\frac{1}{P^4Q^2}
     - \frac{d-2}{d-1}\frac{1}{p^2P^2Q^2}\right]
\nonumber \\ 
\!\!\!&+&\!\!\! m_D^2g^2\sumintff_{\{PQ\}}
    \left[ \frac{d+1}{d-1}\frac{1}{P^2Q^2r^2}
    -\frac{4d}{d-1}\frac{q^2}{P^2Q^2r^4}-\frac{2d}{d-1}
    \frac{P\!\cdot\!Q}{P^2Q^2r^4}\right]{\cal T}_R  
\nonumber \\ 
\!\!\!&+&\!\!\! m_D^2g^2\sumintff_{\{PQ\}}\left[ \frac{3-d}{d-1}\frac{1}{P^2Q^2R^2}+
    \frac{2d}{d-1}\frac{P\!\cdot\! Q}{P^2Q^2r^4}
    -\frac{d+2}{d-1}
    \frac{1}{P^2Q^2r^2} \right.
\nonumber \\
\!\!\!&+&\!\!\! \left.\frac{4d}{d-1}\frac{q^2}{P^2Q^2r^4} - \frac{4}{d-1}\frac{q^2}{P^2Q^2r^2R^2} 
    \right] +2m_q^2g^2(d-1)\sumintff_{\{PQ\}}\left[ \frac{d+3}{d-1}\frac{1}{P^2Q^2R^2}\right.
\nn   
 \!\!\!&-&\!\!\! \left. \frac{2}{P^2Q^4} 
     + \frac{r^2-p^2}{q^2P^2Q^2R^2}\right]
     + 2m_q^2g^2(d-1)\sumintff_{\{PQ\}}\left[ \frac{1}{P^2Q_0^2Q^2}
     + \frac{p^2-r^2}{P^2q^2Q_0^2R^2}
     \right] {\cal T}_Q
\nonumber \\
\!\!\!&+&\!\!\! 2m_q^2g^2(d-1)\sumintbf_{P\{Q\}} \left[\frac{2}{P^2Q^4}
    -\frac{1}{P^2Q_0^2Q^2}{\cal T}_Q\right] .
\label{f2ab}
\ee
\vspace{0cm}
We consider next the $(hh)$ contributions from ${\cal F}_{\rm 3d}^f$ and ${\cal F}_{\rm 3f}^f$.
The easiest way to calculate these term, is to expand the two-loop
diagrams ${\cal F}_{\rm 2a}^f$ and ${\cal F}_{\rm 2b}^f$ to first order in $m_D^2$. This yields
\be
{\cal F}_{\rm 3d+3f}^{f(hh)}&=&
- 2m_D^2g^2\sumintbf_{P\{Q\}}
\left[\frac{1}{p^2P^2Q^2}{\cal T}_P+\frac{1}{P^4Q^2}
- \frac{d-2}{d-1}\frac{1}{p^2P^2Q^2}
\right] 
\nonumber \\ 
&-& m_D^2g^2\sumintff_{\{PQ\}}
\left[ \frac{d+1}{d-1}\frac{1}{P^2Q^2r^2}
-\frac{4d}{d-1}\frac{q^2}{P^2Q^2r^4}-\frac{2d}{d-1}
\frac{P\!\cdot\!Q}{P^2Q^2r^4}\right]{\cal T}_R  
\nonumber \\ 
&-&
 m_D^2g^2\sumintff_{\{PQ\}}\left[ \frac{3-d}{d-1}\frac{1}{P^2Q^2R^2}
\frac{2d}{d-1}\frac{P\!\cdot\! Q}{P^2Q^2r^4}
-\frac{d+2}{d-1}
\frac{1}{P^2Q^2r^2} \right.
\nonumber \\ && \left.
\hspace{4cm}
+\frac{4d}{d-1}\frac{q^2}{P^2Q^2r^4}
-\frac{4}{d-1}\frac{q^2}{P^2Q^2r^2R^2} 
\right] .
\label{f3df}
\ee
Next we consider the $(hh)$ contribution from the diagrams ${\cal F}_{3e}^f,{\cal F}_{3g}^f, {\cal F}_{3k}^f$ and ${\cal F}_{3l}^f$
\be
{\cal F}_{3e+3g+3k+3l}^{f(hh)}&=&
- 2m_q^2g^2(d-1)\;\; \sumintff_{\{PQ\}}\left[ \frac{1}{P^2Q_0^2Q^2}
+\frac{p^2-r^2}{P^2q^2Q_0^2R^2}
\right] {\cal T}_Q \nn
\!\!\!&-&\!\!\! 
2m_q^2g^2(d-1) \; \sumintbf_{P\{Q\}} \left[\frac{2}{P^2Q^4}
+\frac{1}{P^2Q_0^2Q^2}{\cal T}_Q\right] 
\nn
\!\!\!&-&\!\!\!\! 
 2m_q^2g^2(d-1) \sumintff_{\{PQ\}}\!\left[ \frac{d+3}{d-1}\frac{1}{P^2Q^2R^2}
- \frac{2}{P^2Q^4} 
+\frac{r^2-p^2}{q^2P^2Q^2R^2}\right]\!.\hspace{1cm} 
\label{f3egkl}
\ee
The sum of Eqs.(\ref{f2ab})-(\ref{f3egkl}) is
\be
{\cal F}_{2a+2b+3d+3e+3f+3g+3k+3l}^{f(hh)}&=&(d-1)g^2\left[\ \sumintff_{\{PQ\}}\frac{1}{P^2Q^2}
-\sumintbf_{P\{Q\}}\frac{2}{P^2Q^2}\right] 
\nn
&=&\frac{\pi^2}{72}\frac{\alpha_s}{\pi}T^4\lb1+12\hmu^2\rb\lb5+12\hmu^2\rb \, .
\ee
For completeness, the (hh) contribution coming from two-loop pure-glue diagrams is \cite{Andersen:2010ct}
\vspace{-.5cm}
\be
{\cal F}_{2a+2b+2c+3h+3i+3j+3k+3l}^{g(hh)}=\frac{1}{4}(d-1)^2g^2 \sumintbb_{PQ}\frac{1}{P^2Q^2}
=\frac{\pi^2}{36}\frac{\alpha_s}{\pi}T^4
\ee
\subsubsection*{Hard-soft (hs) contribution}
\vspace{-.5cm}
In the $(hs)$ region, one gluon momentum is soft but the fermionic momentum is always hard.  The terms that contribute
through order $g^2 m_D^3 T$ and $g^2m_q^2m_DT$ from ${\cal F}_{2a}^f$ and ${\cal F}_{2b}^f$ were calculated in 
Ref.~\cite{Andersen:2002ey,Andersen:2003zk,Haque:2012my} and read
\vspace{-.3cm}
\be
{\cal F}_{2a+2b}^{f(hs)}\!\!\!&=&\!\!\!2g^2T\int\limits_p\frac{1}{p^2+m^2_D}
\sumintf_{\{Q\}}\left[
\frac{1}{Q^2}-\frac{2q^2}{Q^4}\right]
\nn
\!\!\!&+&\!\!\! 2m_D^2g^2T\int\limits_p\frac{1}{p^2+m_D^2}
\sumintf_{\{Q\}}
\left[\frac{1}{Q^4}
-\frac{2}{d}(3+d)\frac{q^2}{Q^6}+\frac{8}{d}\frac{q^4}{Q^8}
\right]
\nn 
\!\!\!&-&\!\!\! 4m_q^2g^2T\int\limits_p\frac{1}{p^2+m_D^2}
\sumintf_{\{Q\}}\left[\frac{3}{Q^4}
-\frac{4q^2}{Q^6} -\frac{4}{Q^4} {\cal T}_Q
-\frac{2}{Q^2}\bigg\langle \frac{1}{(Q\!\cdot\!Y)^2} \bigg\rangle_{\!\!\bf \hat y}
\right]\!.\ \ \ \ \ 
\label{2a2bhs}
\ee

The $(hs)$ contribution from diagrams ${\cal F}_{\rm 3d}^f$ and ${\cal F}_{\rm 3f}^f$ can again be calculated from the diagrams 
${\cal F}^f_{2a}$ and ${\cal F}^f_{2b}$ by Taylor expanding their contribution to first order in $m_D^2$. This yields
\vspace{-.5cm}
\be
{\cal F}_{\rm 3d+3f}^{f(hs)}\!\!\!&=&\!\!\!
2m_D^2g^2T\int\limits_p\frac{1}{(p^2+m_D^2)^2}
\sumintf_{\{Q\}}\left[\frac{1}{Q^2}-\frac{2q^2}{Q^4}\right]
\nn
\!\!\!&-&\!\!\! 2m_D^2g^2T\int\limits_p\frac{p^2}{(p^2+m^2_D)^2}\sumintf_{\{Q\}}
               \left[\frac{1}{Q^4}-\frac{2}{d}(3+d)\frac{q^2}{Q^6}+\frac{8}{d}\frac{q^4}{Q^8}\right]
\nn
 \!\!\!&-&\!\!\! 4m_D^2m_q^2g^2T\int\limits_p\frac{1}{(p^2+m_D^2)^2}\sumintf_{\{Q\}}\left[\frac{3}{Q^4}
               -\frac{4q^2}{Q^6} -\frac{4}{Q^4}{\cal T}_{Q}-\frac{2}{Q^2}\bigg\langle
               \frac{1}{(Q\!\cdot\!Y)^2} \bigg\rangle_{\!\!\bf \hat y}
               \right] \! . \hspace{13mm}
\label{3d3fhs}
\ee
We also need the $(hs)$ contributions from the diagrams ${\cal F}_{3e}^f,{\cal F}_{3g}^f, {\cal F}_{3k}^f$ and ${\cal F}_{3l}^f$. 
Again we calculate these contributions by expanding the two-loop diagrams ${\cal F}_{2a}^f$ and ${\cal F}_{2b}^f$ to first order 
in $m_q^2$. This yields
\be
{\cal F}_{3e+3g+3k+3l}^{f(hs)}=4m_q^2g^2T\int_{p}\frac{1}{p^2+m_D^2}
\sumintf_{\{Q\}}\left[\frac{3}{Q^4}
-\frac{4q^2}{Q^6} - \frac{4}{Q^4} {\cal T}_Q
-\frac{2}{Q^2}\bigg\langle \frac{1}{(Q\!\cdot\!Y)^2} \bigg\rangle_{\!\!\bf \hat y}
\right] \! . \hspace{9mm}
\label{3egklhs}
\ee
The sum of Eqs.~(\ref{2a2bhs})-(\ref{3egklhs}) is
\be
\hspace{6mm} &\;& \hspace{-1cm} {\cal F}_{2a+2b+3d+3e+3f+3g+3k+3l}^{f(hs)}
\nn
&=&2g^2 T\left[\int_p\frac{1}{p^2+m_D^2}+m_D^2\int_p\frac{1}{\lb p^2+m_D^2\rb^2}
\right]\sumintf_{\{Q\}}\left[\frac{1}{Q^2}-\frac{2q^2}{Q^4}\right]\nn
&+&2g^2m_D^2T \[\int_p\frac{1}{p^2+m_D^2}-\int_p\frac{p^2}{\lb p^2+m_D^2\rb^2}\]
\sumintf_{\{Q\}}
\left[\frac{1}{Q^4}-\frac{2}{d}(3+d)\frac{q^2}{Q^6}+\frac{8}{d}\frac{q^4}{Q^8}
\right]
\nn
&-& 4m_q^2m_D^2T\int_p\frac{1}{\lb p^2+m_D^2\rb^2}
\sumintf_{\{Q\}}\left[\frac{3}{Q^4}
-\frac{4q^2}{Q^6} -\frac{4}{Q^4} {\cal T}_Q
-\frac{2}{Q^2}\bigg\langle \frac{1}{(Q\!\cdot\!Y)^2} \bigg\rangle_{\!\!\bf \hat y}
\right]
\nn
&=&-2(d-1)g^2 T\left[\int_p\frac{1}{p^2+m_D^2}+m_D^2\int_p\frac{1}{\lb p^2+m_D^2\rb^2}
\right]\sumintf_{\{Q\}}\frac{1}{Q^2}
\nn
\!\!\!&-&\!\!\!\frac{d-1}{3}g^2m_D^4T\!\int\limits_p\!\frac{1}{\lb p^2+m_D^2\rb^2}\sumintf_{\{Q\}}\frac{1}{Q^4}-8\frac{d-3}{d-1}
g^2m_q^2m_D^2T\!\int\limits_p\!\frac{1}{\lb p^2+m_D^2\rb^2}\sumintf_{\{Q\}}\frac{1}{Q^4}.\hspace{1.5cm} 
\label{F2a-3l}
\ee
Using the results from Secs.~(\ref{app:sum-integrals}) and (\ref{app:threedints}), Eq.~(\ref{F2a-3l}) can be written as
\be
\nn
{\cal F}_{2a+2b+3d+3e+3f+3g+3k+3l}^{f(hs)}&=&-\frac{1}{12}\alpha_s\lb1+12\hmu^2\rb m_DT^3
-\frac{\alpha_s}{4\pi^2}m_q^2m_D^2T
\nonumber \\
&& 
-\frac{\alpha_s}{48\pi^2}\[\frac{1}{\epsilon}-1-\aleph(z)\]
\left(\frac{\Lambda}{4\pi T}\right)^{2\epsilon}\left(\frac{\Lambda}{2m_D}\right)^{2\epsilon}\, .
\ee
where $\aleph(z)=\Psi(z)+\Psi(z^*)$ with $z=1/2-i\hmu$ and $\Psi$ being the digamma function
(see app.~\ref{app:aleph} for more details and useful properties of $\aleph(z)$).

\subsection{Three-loop sum-integrals}

We now list the mass-expanded sum-integrals necessary at three-loops.  As before we organize the
contributions according to whether the momentum flowing in a given propagator is hard or soft.

\subsubsection*{Hard-hard-hard (hhh) contribution}
The $(hhh)$ contributions from diagrams ${\cal F}_{3a}^{f}$ and ${\cal F}_{3b}^{f}$ which are defind in Eqs.~(\ref{f3a})
and~(\ref{f3b}) can be written as
\begin{eqnarray}
 {\cal F}_{3a}^{f(hhh)}&=&g^4(d-1)\Biggl[4\Big(\sumintb_{P}\ \frac{1}{P^2}
   -2\sumintf_{\{P\}}\frac{1}{P^2}\Big)\sumintff_{\{QR\}}\frac{1}{Q^2R^2(Q+R)^2}
   \nonumber \\
&&\hspace{-1.5cm}
-\frac{1}{2}(d-7)\sumintfff_{\{PQR\}}\frac{1}{P^2Q^2R^2(P+Q+R)^2}
\nonumber\\
&&\hspace{-1.5cm}
+(d-3)\sumintfff_{\{PQR\}}\frac{1}{P^2Q^2(P-R)^2
(Q-R)^2}  + 2 \sumintbff_{\{PQ\}R}\frac{(P-Q)^2}{P^2Q^2R^2(P-R)^2(Q-R)^2}\Bigg] \nn
&&\hspace{-1.5cm} = g^4(d-1)\Big\{4\lb{\cal I}_1^0-2\tilde{\cal I}_1^0\rb\tilde\tau -\frac{1}{2}\lb d-7\rb{\cal N}_{0,0}
 +(d-3)\widetilde M_{0,0} + 2{\cal N}_{1,-1}\Big\}.
\label{f3a_hhh}
\end{eqnarray}
\begin{eqnarray}
 {\cal F}_{3b}^{f(hhh)}&=&-g^4(d-1)^2\Bigg[\Big(\sumintb_{P}\ \frac{1}{P^2}
   -\sumintf_{\{P\}}\frac{1}{P^2}\Big)^2\sumintf_{\{P\}}\frac{1}{P^4}-2\sumintfff_{\{PQR\}}
    \frac{1}{P^2Q^2R^2(Q+R)^2}   \nonumber\\
&& +\sumintfff_{\{PQR\}}\frac{1}{P^2Q^2(P-R)^2(Q-R)^2}+ \sumintfff_{\{PQR\}}\frac{(P-Q)^2}{P^2Q^2R^2(P-R)^2(Q-R)^2}
\Biggr] \nn
&=&-g^4(d-1)^2 \Big\{\lb{\cal I}_1^0-\tilde{\cal I}_1^0\rb^2\tilde{\cal I}_2^0-2\tilde{\cal I}_1^0\tilde{\tau}
+\widetilde{\cal M}_{0,0}+\widetilde{\cal M}_{1,-1}\Big\}.
\label{f3b_hhh}
\end{eqnarray}
Using the results of sum-integrals from Sec.~(\ref{app:sum-integrals}), the hhh contribution of the sum
of  ${\cal F}_{3a}^f$ and ${\cal F}_{3b}^f$ from Eqs.~(\ref{f3a_hhh}) and~(\ref{f3b_hhh}) can be written as
\begin{eqnarray}
{\cal F}_{3a}^{f(hhh)}+{\cal F}_{3b}^{f(hhh)} && \nonumber \\
&& \hspace{-2.5cm} = 
\frac{\alpha_s^2T^4}{ 192}\Bigg[35-32\frac{\zeta'(-1)}{\zeta(-1)}+472 \hat\mu^2
+384  \frac{\zeta'(-1)}{\zeta(-1)} \hat\mu ^2+1328  \hat\mu^4
    \nonumber\\
&& \hspace{-2.5cm}
-64\Big(36i\hat\mu\aleph(2,z)-6(1+8\hat\mu^2)\aleph(1,z)-3i\hat\mu(1+4\hat\mu^2)\aleph(0,z)\Big)\Bigg]\!.\ \ 
\end{eqnarray}
where $\aleph(n,z)$ is defined in appendix (\ref{app:aleph}).

Hard-hard-hard contribution from the term  ${\cal F}_{3c}^f$ that has been defined in Eq.~(\ref{f3c}) can be written as
\begin{eqnarray}
 {\cal F}_{3c}^{f(hhh)} &=& -\frac{ 5\alpha_s^2 }{216}T^4\left(\frac{\Lambda}{4\pi T}\right)^{6\epsilon}\Bigg[ \left(1 +
\frac{72}{5}\hat\mu^2+\frac{144}{5}\hat\mu^4\right)\frac{1}{\epsilon}
+ \frac{31}{10}+\frac{6}{5}\gamma_E\nn
&-& \frac{68}{25}\frac{\zeta'(-3)}{\zeta(-3)}+\frac{12}{5}(25+12\gamma_E)\hat\mu^2 + 120\hat\mu^4
 - \frac{8}{5}(1+12\hat\mu^2)\frac{\zeta'(-1)}{\zeta(-1)} 
\nn
 &-&  \frac{144}{5}\Big[8\aleph(3,z)+3\aleph(3,2z) +12 i \hat\mu\,(\aleph(2,z)+\aleph(2,2z)) 
\nn
 &-& (3+20\hat\mu^2)\aleph(1,z)-
i \hat\mu(1+12\hat\mu^2)\,\aleph(0,z)  
 -12\hat\mu^2\aleph(1,2z) \Big]\Bigg] \, ,
\end{eqnarray}
The $(hhh)$ contribution of the term, which is proportional to $c_A s_F$ is 
\begin{eqnarray}
&-&\frac{1}{2}{\cal F}_{3a}^{f(hhh)}+{\cal F}_{3m+3n}^{f(hhh)}\nonumber\\
&=&          g^2(d-1)\Bigg\{2(d-5)\sumintbbf_{PQ\{R\}}\frac{1}{P^4Q^2R^2}+\frac{1}{2}(d-3)
             \sumintfff_{\{PQR\}}\frac{1}{P^2Q^2(P-Q)^2(Q-R)^2}\nonumber\\
&&\qquad  -\frac{1}{4}(d-7)\sumintbff_{\{PQ\}R}\frac{1}{P^2Q^2(P-R)^2(Q-R)^2}+\sumintbff_{\{PQ\}R}
\frac{(P-Q)^2}{P^2Q^2R^2(P-R)^2(Q-R)^2}\nn
&&\qquad -(d-3)\sumintbff_{P\{QR\}}\frac{1}{P^2Q^2R^2(Q-R)^2}-2\sumintfff_{\{PQR\}}\frac{1}{P^2Q^2R^2(Q-R)^2}
\nonumber\\ 
&&\qquad +2\sumintfff_{\{PQR\}}
      \frac{R^4}{P^2Q^2(P-Q)^4(Q-R)^4}\Bigg\} 
\nn
&=& g^2(d-1)\Bigg[2(d-5){\cal I}_2^0{\cal I}_1^0\tilde {\cal I}_1^0+\frac{1}{2}(d-3)\widetilde {\cal M}_{00}-\frac{1}{4}
(d-7){\cal N}_{00}\nn
&&\hspace{3cm}+{\cal N}_{1,-1}-(d-3){\cal I}_1^0\tilde\tau-2\tilde {\cal I}_1^0\tilde\tau\Bigg]\nn
&=&         -\frac{25\alpha_s^2T^4}{864}\Bigg[\left(1+\frac{72}{25}\hat\mu^2-\frac{1584}{25}
             \hat\mu^4\right)\lb\frac{1}{\epsilon}
           +6\ln\frac{\hat\Lambda}{2}\rb
\nn
&&  -\ \frac{369}{250}\left(1+\frac{2840}{123}\hat\mu^2+\frac{28720}{123}\hat\mu^4\right)
       +\frac{48}{25}\left(1+12\hat\mu^2\right)\gamma_E
            \nn
&&   +\ \frac{536}{125}\frac{\zeta'(-3)}{\zeta(-3)}
          +\frac{32}{25}\lb1+6\hat\mu^2\rb\frac{\zeta'(-1)}{\zeta(-1)}+\frac{288}{25}\Big[26\aleph(3,z)
\nn           
&& +\ \left(3-68\hat\mu^2\right)\aleph(1,z)+72i\hmu\aleph(2,z)+2i\hmu\aleph(0,z)\Big]\Bigg] ,          
\end{eqnarray}
where the integrals appearing above are evaluated in Sec.~(\ref{app:threeloopsumints}).
Finally, we note that there is no $(hhh)$ contribution from ${\cal F}_{3o}^f$ since this is a purely HTL diagram.
\subsubsection*{hhs contribution}
\vspace{-.7cm}
The $(hhs)$ contribution to the ${\cal F}_{\rm 3a}^f$, ${\cal F}_{\rm 3b}^f$, and ${\cal F}_{\rm 3c+3j}^f$ are
\begin{eqnarray}
 {\cal F}_{\rm 3a}^{f(hhs)}&=&
2(d-1)g^4T\int\limits_p\frac{1}{p^2+m_D^2}\Bigg[\ \ 
\sumintbf_{Q\{R\}}\frac{4Q_0R_0}{Q^2R^4(Q+R)^2}
\nonumber \\
&& \hspace{4cm}
-\sumintff_{\{QR\}}\frac{1}{Q^2R^2(Q-R)^2}
+\sumintff_{\{QR\}}\frac{2Q_0R_0}{Q^4R^4}\Bigg] \!,\hspace{1cm}
\label{3ahhs} 
\end{eqnarray}
\vspace{-1cm}
\begin{eqnarray}
 {\cal F}_{\rm 3b}^{f(hhs)}&=&
2(d-1)g^4T\int\limits_p\frac{1}{p^2+m_D^2}\left[\ \ 
\sumintbf_{Q\{R\}}-\frac{4Q_0R_0}{Q^2R^4(Q+R)^2}+\sumintff_{\{QR\}}\frac{1}{Q^2R^2(Q-R)^2}\right.
\nonumber\\
&&
\hspace{2cm}
+ \left.(d-3)\sumintf_{\{Q\}}\frac{1}{Q^4}
\left(\sumintb_{R}\frac{1}{R^2}-\sumintf_{\{R\}}\frac{1}{R^2}\right)\right] , 
\end{eqnarray}
\vspace{-1cm}
\begin{eqnarray}
{\cal F}_{3c+3j}^{f(hhs)}&=&
-4g^4T\int\limits_p\frac{1}{(p^2+m_D^2)^2}
\left[\ \sumintf_{\{Q\}}\left(
\frac{1}{Q^2}-\frac{2q^2}{Q^4}\right)
\right]^2
\nn && 
\hspace{-1.5cm}
+8g^4T\int\limits_p\frac{p^2}{(p^2+m_D^2)^2}
\sumintf_{\{Q\}}\left[\frac{1}{Q^2}-\frac{2q^2}{Q^4}\right]
\sumintf_{\{R\}}\left[
\frac{1}{R^4}-\frac{2}{d}(3+d)
\frac{r^2}{R^6}+\frac{8}{d}\frac{r^4}{R^8}
\right]\nn
&&
\hspace{-1.5cm}
-16m_q^2g^4T\int\limits_{p}\!\!\frac{1}{(p^2+m_D^2)^2}
\sumintf_{\{Q\}}\left[\frac{1}{Q^2}-\frac{2q^2}{Q^4}\right]
\sumintf_{\{R\}}\left[
\frac{3}{R^4}-\frac{4r^2}{R^6}
-\frac{4}{R^4}{\cal T}_R-\frac{2}{R^2}
\bigg\langle\frac{1}{(R\!\cdot\!Y)^2} \bigg\rangle_{\!\!\bf \hat y}
\right]\;
\nn
&& \hspace{-1.5cm}
 = -4g^4T(d-1)^2\int\limits_p\!\!\frac{1}{(p^2+m_D^2)^2}
\sumintff_{\{QR\}}
\frac{1}{Q^2R^2}
+\frac{4}{3}g^4T(d-1)^2\int\limits_p\!\!\frac{p^2}{(p^2+m_D^2)^2}
\sumintff_{\{QR\}}\frac{1}{Q^2R^4}
\nn
&&+32m_q^2g^4T(d-3)
\int\limits_{p}\frac{1}{(p^2+m_D^2)^2}
\sumintff_{\{QR\}}\frac{1}{Q^2R^4}\;.
\end{eqnarray}
Computing the necessary sum-integrals one finds
\begin{eqnarray}
 {\cal F}^{f(hhs)}_{3a+3b}=\frac{\alpha_s^2 m_DT^3}{4\pi}(1+12\hat\mu^2) \, ,
\end{eqnarray}
\vspace{-1cm}
\begin{eqnarray}
 {\cal F}^{f(hhs)}_{3c+3j}&=&
\frac{\alpha_s^2m_DT^3}{12\pi}\left[\frac{1+12\hat\mu^2}{\epsilon}+(1+12\hat\mu^2)\left(\frac{4}{3}
-\aleph(z)\right)+24\aleph(1,z)\right]
\nonumber\\
&-&\frac{\pi\alpha_s^2T^5}{18m_D}\left(1+12\hat\mu^2\right)^2-\frac{\alpha_s^2m_q^2T^3}{3\pi m_D}\left(1+12\hat\mu^2\right) .
\end{eqnarray}

Similarly, one obtains
\begin{eqnarray}
&& -\frac{1}{2}{\cal F}_{3a}^{f(hhs)}+{\cal F}_{3m+3n+3o}^{f(hhs)} 
\nonumber \\ && \hspace{.5cm} 
= g^2T(d-1)\Bigg\{2(d-1)^2\int\limits_p\frac{1}{(p^2+m_D^2)^2}\sumintbf_{Q\{R\}}\frac{1}{Q^2R^2}
\nonumber \\ && \hspace{1cm} 
+\frac{1}{2}(d-3)\int\limits_p\frac{1}{(p^2+m_D^2)}\sumintff_{\{QR\}}\frac{1}{Q^2R^2(Q-R)^2}
\nonumber\\
&&\ \hspace{1cm} - \int\limits_p\frac{1}{(p^2+m_D^2)}\sumintff_{\{QR\}}\frac{2Q_0R_0}{Q^4R^4}  
-\frac{1}{3}\left(d^2-11d+46\right)\int\limits_p\frac{p^2}{(p^2+m_D^2)^2}\sumintbf_{Q\{R\}}\frac{1}{Q^4R^2}
\nn                     
&& \hspace{1cm} -\frac{1}{3}(d-1)^2\int\limits_p\frac{p^2}{(p^2+m_D^2)^2}\sumintbf_{Q\{R\}}\frac{1}{Q^2R^4}
\nonumber\\
&& \hspace{1cm}  +4m_q^2(d-1)\int\limits_p\frac{1}{(p^2+m_D^2)^2}\sumintb_{Q}\frac{1}{Q^2}\sumintf_{\{R\}}\left[\frac{3}{R^4}
                     -\frac{4r^2}{R^6}-\frac{4}{R^4}{\cal T}_R-\frac{2}{R^2}\left\langle\frac{1}{(R.Y)^2}\right\rangle_
                     {\hat{\bf y}}\right]\Bigg\} \, ,
\nonumber\\ && \hspace{.5cm} 
= g^2T(d-1)\Bigg\{2(d-1)^2\int\limits_p\frac{1}{(p^2+m_D^2)^2}\sumintbf_{Q\{R\}}\frac{1}{Q^2R^2}
\nonumber\\ && \hspace{1cm}
-\frac{1}{3}\left(d^2-11d+46\right)\int\limits_p\frac{p^2}{(p^2+m_D^2)}\sumintbf_{Q\{R\}}\frac{1}{Q^4R^2}
\nonumber\\ && \hspace{1cm}  
                     -\frac{1}{3}(d-1)^2\int\limits_p\frac{p^2}{(p^2+m_D^2)^2}\sumintbf_{Q\{R\}}\frac{1}{Q^2R^4}
          +  8m_q^2(d-3)\int\limits_p\frac{1}{(p^2+m_D^2)^2}\sumintbf_{Q\{R\}}\frac{1}{Q^2R^4}
                     \Bigg\}
\nonumber\\ && \hspace{.5cm}
= -\frac{\alpha_s m_DT^3}{48\pi}\left(\frac{\Lambda}{2m_D}\right)^{2\epsilon}\left(\frac{\Lambda}{4\pi T}\right)^{4\epsilon}
         \Bigg[\frac{1}{\epsilon} \left(7+132\hat\mu^2\right)+\frac{88}{3}+440\hat\mu^2+22\left(1+12\hat\mu^2\right)\gamma_E
\nonumber\\   
&&\hspace{1.5cm} -8\frac{\zeta'(-1)}{\zeta(-1)} +\ 4\aleph(z)+264\aleph(1,z)\Bigg]
-\frac{\pi\alpha_s^2T^5}{9m_D}\left(1+12\hat\mu^2\right)-\frac{\alpha_s^2}{3\pi m_D}m_q^2T^3 \, .
\end{eqnarray}

\subsubsection*{hss contribution}
\vspace{-1cm}
The only three loop diagram involving a fermionic line that has a $(hss)$ contribution is 
${\cal F}_{3n}^f$ which can be written as
\begin{eqnarray}
 {\cal F}_{3n}^{f(hss)}&=&g^4T^2\int\limits_{pq}\left[\frac{4}{\left(p^2+m_D^2\right)
       \left(q^2+m_D^2\right)\lb{\bf p}+{\bf q}\rb^2}-\frac{2}
      {\left(p^2+m_D^2\right) \left(q^2+m_D^2\right)^2}\right.
\nonumber\\
&&
\left.\hspace{1cm}-\frac{8 m_D^2}
      {\lb{\bf p}+{\bf q}\rb^2\left(p^2+m_D^2\right)
       \left(q^2+m_D^2\right)^2}\right]\sumintf_{\{R\}}\lb\frac{1}{R^2}-\frac{2r^2}{R^4}\rb
\nonumber\\
&=&-g^4T^2(d-1)\int\limits_{pq}\left[\frac{4}{\left(p^2+m_D^2\right)
       \left(q^2+m_D^2\right)\lb{\bf p}+{\bf q}\rb^2}-\frac{2}
      {\left(p^2+m_D^2\right) \left(q^2+m_D^2\right)^2}\right.
\nonumber\\
&&
\left.\hspace{5cm}-\frac{8 m_D^2}
      {\lb{\bf p}+{\bf q}\rb^2\left(p^2+m_D^2\right)
       \left(q^2+m_D^2\right)^2}\right]\sumintf_{\{R\}}\frac{1}{R^2}
\nonumber\\
&=&  \frac{\alpha_s^2T^4}{12}\left[\frac{1+12\hat\mu^2}{\epsilon}+2\left(1+12\hat\mu^2+12\aleph(1,z)\right)\right]
      \left(\frac{\Lambda}{2m_D}\right)^{4\epsilon}\left(\frac{\Lambda}{4\pi T}\right)^{2\epsilon}
\end{eqnarray}

\section{Sum-Integrals}
\label{app:sum-integrals}
In Sec~(\ref{asis}) of Chapter~\ref{chapter:2loop} we have listed all the one and two loop sum-integrals
at small chemical potential up to $\mathcal{O}(\mu/T)^4$. In this chapter we are calculating thermodynamic
potential at general value of chemical potential. So we are enlisting here all the necessary one, two and
three loop sum-integrals at any value of temperature and chemical potential.

We can define a set of ``master" sum integral as in~\cite{Vuorinen:2002ue,Vuorinen:2003fs}
\be
{\cal I}_n^m &=& \sumintb_{P}\frac{P_0^m}{P^{2n}}\ ,\\
{\widetilde{\cal I}}_n^m &=& \sumintf_{\{P\}}\frac{P_0^m}{P^{2n}}\ ,\\
\tilde{\tau}&=&\sumintff_{\{PQ\}}\frac{1}{P^2Q^2(P+Q)^2}\ ,
\ee
\be
{\cal M}_{m,n}&=&\sumintbbb_{PQR}\frac{1}{P^2Q^2(R^2)^m[(P-Q)^2]^n(P-R)^2(Q-R)^2}\ ,\\
\widetilde{\cal M}_{m,n}&=&\sumintfff_{\{PQR\}}\frac{1}{P^2Q^2(R^2)^m[(P-Q)^2]^n(P-R)^2(Q-R)^2}\ ,\\
 {\cal N}_{m,n}&=&\sumintbff_{\{PQ\}R}\frac{1}{P^2Q^2(R^2)^m[(P-Q)^2]^n(P-R)^2(Q-R)^2} \ .
\ee

\subsection{One loop sum-integrals}
%
The specific bosonic sun-integrals needed are
\begin{eqnarray}
{\cal I}_1^0 = \sumintb_{P}\ \frac{1}{P^2}=\frac{T^2}{12}\left(\frac{\Lambda_g}{4\pi T}\right)^{2\epsilon}
                              \left[1+2\epsilon\left(1+\frac{\zeta'(-1)}{\zeta(-1)}\right)\right] ,
\end{eqnarray}
\begin{eqnarray}
{\cal I}_2^0 = \sumintb_{P}\ \frac{1}{P^4}=\frac{1}{\left(4\pi\right)^2}\left(\frac{\Lambda_g}{4\pi T}\right)^{2\epsilon}\left[\frac{1}
{\epsilon}+2\gamma_E \right] .
\end{eqnarray}
The specific fermionic sun-integrals needed are
\begin{eqnarray}
\widetilde{\cal I}_1^0=\sumintf_{\{P\}}\frac{1}{P^2}=-\frac{T^2}{24}\left(\frac{\Lambda}{4\pi T}\right)^{2\epsilon}\left[1+12\hat\mu^2+2\epsilon
\left(1+12\hat\mu^2
       +12\aleph(1,z)\right)\right] ,
\end{eqnarray}
and
\begin{eqnarray}
\widetilde{\cal I}_2^0=\sumintf_{\{ P\} }\frac{1}{P^4}=\frac{1}{\left(4\pi\right)^2}\left(\frac{\Lambda}{4\pi T}\right)^{2\epsilon}\left[\frac{1}
{\epsilon}-\aleph(z)\right] .
\end{eqnarray}

Using the two basic one-loop sum-integrals above, we can construct other one-loop sum-integrals that will be necessary here as follows:
\begin{eqnarray}
 \sumintf_{\{P\}}\frac{1}{P^2}=\frac{2}{d}\sumintf_{\{P\}}\frac{p^2}{P^4}
\end{eqnarray}
\begin{eqnarray}
 \sumintf_{\{P\}}\frac{1}{P^4} &=& \frac{4}{d}\sumintf_{\{P\}}\frac{p^2}{P^6}=\frac{24}{d(d+2)}\sumintf_{\{P\}}\frac{p^4}{P^8} \, , \nn
&=& (d-1)\sumintf_{\{P\}}\frac{1}{P^4}{\cal T}_P=-\frac{2}{d-1}\sumintf_{\{P\}}\frac{1}{P^2}\left\langle\frac{1}{(P.Y)^2}\right\rangle_{\hat{\bf y}} \, .
\end{eqnarray}
This allows us to compute the following sum-integrals
\begin{eqnarray}
\sumintf_{\{P\}}\frac{p^2}{P^4}=-\frac{T^2}{16}\left(\frac{\Lambda}{4\pi T}\right)^{2\epsilon}\left[1+12\hat\mu^2
        +\frac{4}{3}\epsilon\left(1+12\hat\mu^2
       +18\aleph(1,z)\right)\right] ,
\end{eqnarray}
\begin{eqnarray}
\sumintf_{\{P\}}\frac{1}{P^4}{\cal T}_P=\frac{1}{2}\frac{1}{\left(4\pi\right)^2}\left(\frac{\Lambda}{4\pi T}\right)^{2\epsilon}
\left[\frac{1}{\epsilon}+1-\aleph(z)\right] ,
\end{eqnarray}
\begin{eqnarray}
\sumintf_{\{P\}}\frac{p^2}{P^6}=\frac{3}{4}\frac{1}{\left(4\pi\right)^2}\left(\frac{\Lambda}{4\pi T}\right)^{2\epsilon}
\left[\frac{1}{\epsilon}-\frac{2}{3}
            -\aleph(z)\right] ,
\end{eqnarray}
\begin{eqnarray}
\sumintf_{\{P\}}\frac{p^4}{P^8}=\frac{5}{8}\frac{1}{\left(4\pi\right)^2}\left(\frac{\Lambda}{4\pi T}\right)^{2\epsilon}
\left[\frac{1}{\epsilon}-\frac{16}{15}-\aleph(z)\right] ,
\end{eqnarray}
\begin{eqnarray}
\sumintf_{\{P\}}\frac{1}{P^2}\left\langle\frac{1}{(P.Y)^2}\right\rangle_{\hat{\bf y}}=-\frac{1}{\left(4\pi\right)^2}
      \left(\frac{\Lambda}{4\pi T}
      \right)^{2\epsilon}\left[\frac{1}{\epsilon}-1-\aleph(z)\right] ,
\end{eqnarray}
\begin{eqnarray}
 \sumintf_{\{P\}}\frac{P_0}{P^4}=\frac{1}{\left(4\pi\right)}
      \left(\frac{\Lambda}{4\pi T}\right)^{2\epsilon}\left[i\hat\mu+\ \aleph(0,z)\epsilon\right] .
\end{eqnarray}

\subsection{Two loop sum-integrals}
\label{app:twoloopsumints}

For the purposes of this chapter we only need one new two-loop sum-integral
\be
\tilde{\tau}=\sumintff_{\{PQ\}}\frac{1}{P^2Q^2(P+Q)^2}=-\frac{T^2}{\lb4\pi\rb^2}\lb\frac{\Lambda}{4\pi T}\rb^{4\epsilon}
\[\frac{\hmu^2}{\epsilon}+2\hmu^2-2i\hmu\aleph[0,z]\] \, .
\ee

\subsection{Three loop sum-integrals}
\label{app:threeloopsumints}
%
%
The three-loop sum-integrals necessary are
\begin{eqnarray}
{\cal M}_{00}&=&\sumintbbb_{PQR}\frac{1}{P^2Q^2R^2\left(P+Q+R\right)^2}
\nn
&=&\frac{1}{(4\pi)^2}\left(\frac{T^2}{12}\right)^2\left(\frac{\Lambda_g}
         {4\pi T}\right)^{6\epsilon}\left[\frac{6}{\epsilon}+\frac{182}{5}-12\frac{\zeta'(-3)}{\zeta(-3)}+48\frac{\zeta'(-1)}
         {\zeta(-1)}\right] .
\end{eqnarray}
\vspace{-1cm}
%
\begin{eqnarray}
{\cal N}_{00}&=&\sumintfff_{\{PQR\}}\frac{1}{P^2Q^2R^2\left(P+Q+R\right)^2}\nn
             &=&\frac{1}{(4\pi)^2}\left(\frac{T^2}{12}\right)^2\left(\frac{\Lambda}
         {4\pi T}\right)^{6\epsilon}\Bigg[\frac{3}{2\epsilon}\left(1+12\hat\mu^2\right)^2+\frac{173}
          {20}+210\hat\mu^2
\nn
&&   +1284\hat\mu^4-\frac{24}{5}\Zc + 144\Big((1+8\hat\mu^2)\aleph(1,z)+4
          \hat\mu^2\aleph(1,2z)
\nn
&-&  4i\hat\mu\big[\aleph(2,z)+\aleph(2,2z)\big]-2\aleph(3,z)-\aleph(3,2z)
\Big)\Bigg].
\end{eqnarray}
\vspace{-1cm}
%
\begin{eqnarray}
 \widetilde{\cal M}_{00}&=&\sumintbbf_{PQ\{R\}}\frac{1}{P^2Q^2R^2\left(P+Q+R\right)^2}\nn
                        &=&-\frac{1}{(4\pi)^2}\left(\frac{T^2}{12}\right)^2\left(\frac{\Lambda}
         {4\pi T}\right)^{6\epsilon}\Bigg[\frac{3}{4\epsilon}\left(1+24\hat\mu^2-48\hat\mu^4\right)+\frac{179}
          {40}
\nn
     &+&  111\hat\mu^2-210\hat\mu^4+48\frac{\zeta'(-1)}{\zeta(-1)}\hat\mu^2+\frac{24}{5}\frac{\zeta'(-3)}{\zeta(-3)}
\nn 
     &+&72\Big((1-8\hat\mu^2)\aleph(1,z)+6\aleph(3,z)+12i\hat\mu\aleph(2,z)\Big)\Bigg] .
\end{eqnarray}
\vspace{-1cm}
%
\begin{eqnarray}
{\cal N}_{1,-1} &=&
-\frac{1}{2\lb4\pi\rb^2}\lb\frac{T^2}{12}\rb^2\left(\frac{\Lambda}
         {4\pi T}\right)^{6\epsilon}\Bigg[\frac{3}{2\epsilon}\lb1+12\hmu^2\rb\lb1-4\hmu^2\rb
\nn
&+& \frac{173}{20}+114\hmu^2+132\hmu^4
- \frac{12}{5}\Zc - 96\hmu^2\Za
\nn
&-& 144\Big[2\aleph(3,z) + 2\aleph(3,2z) - 4i \hmu\,\aleph(2,z) + 8i\hmu\,\aleph(2,2z)
\nn
&-& \lb1-4\hmu^2\rb\aleph(1,z)
- 8\hmu^2\aleph(1,2z) -\frac{1}{3}i\hmu\lb1+12\hmu^2\rb\aleph(0,z)\Big] \Bigg].
\end{eqnarray}
%

\begin{eqnarray}
H_3&=& \sumintbbf_{\{P\}QR}\frac{Q\cdot R}{P^2Q^2R^2\left(P+Q\right)^2\left(P+R\right)^2}
\nn
&=&\frac{1}{(4\pi)^2}\left(\frac{T^2}{12}\right)^2
                \left(\frac{\Lambda}{4\pi T}\right)^{6\epsilon}\Bigg[\frac{3}{8\epsilon}\left(1+12\hat\mu^2\right)^2
                +\frac{361}{160}-\frac{3}{5}\frac{\zeta'(-3)}{\zeta(-3)}
\nonumber\\
\!\!\!&+&\!\!\!\frac{141}{4}\hat\mu^2+\frac{501}{2}\hat\mu^4
         -9\Bigg\{\left(\frac{1}{8}+\hat\mu^2+2\hat\mu^4\right)\aleph(z)+2i\hat\mu\left(1+4\hat\mu^2\right)\aleph(0,z) 
\nonumber\\
&+& 2\left(1-12\hat\mu^2\right)\aleph(1,z)+24i\hat\mu\aleph(2,z)+16\aleph(3,z)\Bigg\}\Bigg] \, .
\end{eqnarray}

\begin{eqnarray}
\widetilde{\cal M}_{-2,2}&=&\sumintfff_{\{PQR\}}\frac{R^4}{P^2Q^2(P-Q)^4(Q-R)^2(R-P)^2}\nn
\!\!\!&=&\!\!\!-\frac{1}{(4\pi)^2}\left(\frac{T^2}{12}\right)^2
         \left(\frac{\Lambda}{4\pi T}\right)^{6\epsilon}\Bigg[\frac{1}{12\epsilon}\left(29+288\hat\mu^2-144
         \hat\mu^4\right) +\frac{89}{12}+4\gamma_E \nonumber\\ 
\!\!\!&+&\!\!\!     2(43+24\gamma_E)\hat\mu^2   - 68\hat\mu^4+\frac{10}{3}\left(1+\frac{84}
        {5}\hat\mu^2\right)\frac{\zeta'(-1)}{\zeta(-1)} + \frac{8}{3}\frac{\zeta'(-3)}{\zeta(-3)}
        \nonumber\\
&+&   24\left[10\aleph(3,z)+18i\hat\mu\aleph(2,z)+2(2-5\hat\mu^2)\aleph(1,z)
      +i\hat\mu\aleph(0,z)\right]\Bigg] .\hspace{1cm}
\end{eqnarray}

\section{Three-dimensional integrals}
\label{app:threedints}
%
Dimensional regularization can be used to regularize both the ultraviolet divergences and infrared divergences in 3-dimensional integrals over momenta. The spatial dimension is generalized to  $d = 3-2\epsilon$ dimensions. Integrals are evaluated at a value of $d$ for which they converge and then analytically continued to $d=3$. We use the integration measure 
\begin{equation}
 \int\limits_p\;\equiv\;
  \left(\frac{e^{\gamma_E}\Lambda^2}{4\pi}\right)^\epsilon\;
\:\int \frac{d^{3-2\epsilon}p}{(2 \pi)^{3-2\epsilon}}\;.
\label{int-def}
\end{equation}

\subsection{One-loop integrals}

The general one-loop integral is given by
\be\nonumber
I_n&\equiv&\int\limits_p\frac{1}{(p^2+m^2)^n}\\
&=&\frac{1}{8\pi}(e^{\gamma_E}\Lambda^2)^{\epsilon}
\frac{\Gamma(n-\mbox{$\frac{3}{2}$}+\epsilon)}
{\Gamma(\mbox{$\frac{1}{2}$})
\Gamma(n)}m^{3-2n-2\epsilon}
\;.
\ee
Specifically, we need
\be\nonumber
I_0^{\prime}&\equiv&
\int_p\ln(p^2+m^2)\\
&=&
-\frac{m^3}{6\pi}\left(\frac{\Lambda}{2m}\right)^{2\epsilon}
\left[
1+\frac{8}{3}
\epsilon
+{\cal O}\left(\epsilon^2\right)
\right]\;,\\ 
I_1&=&-\frac{m}{4\pi}\left(\frac{\Lambda}{2m}\right)^{2\epsilon}
\left[
1+2\epsilon+{\cal O}\left(\epsilon^2\right)
\right]\;,\\
\label{i2}
I_2&=&\frac{1}{8\pi m}\left(\frac{\Lambda}{2m}\right)^{2\epsilon}
\left[1+{\cal O}\left(\epsilon\right)
\right]
\;.
\ee

\subsection{Two-loop integrals}

We also need a few two-loop integrals on the form
\be
J_n&=&\int\limits_{pq}\frac{1}{p^2+m^2}\frac{1}{(q^2+m^2)^n}
\frac{1}{({\bf p}+{\bf q})^2} \;.
\ee
Specifically, we need $J_1$ and $J_2$ which were calculated in Ref.~\cite{Andersen:2010ct}:
\be
J_1&=&
\frac{1}{4(4\pi)^2}\left(\frac{\Lambda}{2m}\right)^{4\epsilon}
\left[
\frac{1}{\epsilon}+2
+{\cal O}(\epsilon)
\right]\;,\\
J_2&=&
\frac{1}{4(4\pi)^2m^2}\left(\frac{\Lambda}{2m}\right)^{4\epsilon}
\left[1+{\cal O}(\epsilon)
\right]\;.
\ee


\section{NNLO HTLpt thermodynamic potential}
\label{nnloomega}
In Sec.~(\ref{expansion}) we expanded each term of Eq.~(\ref{omega_initial}) form small $m_D/T$ and $m_q/T$.
In Sec.~(\ref{app:sum-integrals}) and Sec.~(\ref{app:threedints}) we calculated necessary sum-integrals and 
three dimensional integrals that arise in Sec.~(\ref{expansion}). In this section we will summarized the results from 
Secs.~(\ref{expansion}), (\ref{app:sum-integrals}) and (\ref{app:threedints}) to get final thermodynamic
potential.

We consider first  the case that all quarks have the same chemical potential $\mu_f = \mu = \mu_B/N_f$ where $f$
is a flavor index with $f \in \{ \mu_u, \mu_d, \mu_s, \cdots, \mu_{N_f} \}$. Here we are considering $N_f=3$, 
so $\mu_f = \mu = \mu_B/3$.  After presenting the steps
needed for this case, we give the general result with separate chemical potentials for each flavor.
\subsection{NNLO result for equal chemical potentials}
When all quarks have the same chemical potential $\mu_i = \mu = \mu_B/3$ we can straightforwardly combine
the results for the various sum-integrals.  In this case, the unrenormalized three-loop HTLpt thermodynamic potential is
\newpage
\vspace*{-2cm}
\begin{eqnarray}
\frac{\Omega_{\rm 3 loop}}{\Omega_0}
     &=& \frac{7}{4}\frac{d_F}{d_A}\lb1+\frac{120}{7}\hmu^2+\frac{240}{7}\hmu^4\rb
        +\frac{s_F\alpha_s}{\pi}\Bigg[-\frac{5}{8}\left(1+12\hat\mu^2\right)\left(5+12\hat\mu^2\right)
\nn
    &+& \frac{15}{2}\left(1+12\hat\mu^2\right)\hat m_D+\frac{15}{2}\bigg(\frac{1}{\epsilon}-1
    -\aleph(z)+4\ln{\frac{\hat\Lambda}{2}}-2\ln\hat m_D\Big)\hat m_D^3
\nn
    &-& 90\hat m_q^2 \hat m_D\Bigg] + s_{2F}\left(\frac{\alpha_s}{\pi}\right)^2\Bigg[\frac{15}{64}\bigg\{35-32\frac{\zeta'(-1)}
      {\zeta(-1)}+472 \hat\mu^2 
\nn      
    &+& 384  \Za \hat\mu ^2+1328  \hat\mu^4 + 64\Big(-36i\hat\mu\aleph(2,z)+6(1+8\hat\mu^2)\aleph(1,z)
\nn 
     &+&3i\hat\mu(1+4\hat\mu^2)\aleph(0,z)\Big)\bigg\} - \frac{45}{2}\hat m_D\left(1+12\hat\mu^2\right)\Bigg]
\nn
   &+& \left(\frac{s_F\alpha_s}{\pi}\right)^2\Bigg[\frac{5}{4\hat m_D}\left(1+12\hat\mu^2\right)^2+30\left(1+12\hat\mu^2
      \right)\frac{\hat m_q^2}{\hat m_D}
\nn
   &+& \frac{25}{24}\Bigg\{ \left(1 +\frac{72}{5}\hat\mu^2+\frac{144}{5}\hat\mu^4\right)\lb\frac{1}{\epsilon}
        +6\ln\frac{\hat\Lambda}{2}\rb
        +\frac{31}{10}+\frac{6}{5}\gamma_E - \frac{68}{25}\frac{\zeta'(-3)}{\zeta(-3)}
\nn
   &+&\frac{12}{5}(25+12\gamma_E)\hat\mu^2 + \frac{24}{5}(61+36\gamma_E)\hat\mu^4
        - \frac{8}{5}(1+12\hat\mu^2)\frac{\zeta'(-1)}{\zeta(-1)} 
\nn
   &-& \frac{144}{5}\Big[8\aleph(3,z)+3\aleph(3,2z) +12 i \hat\mu\,(\aleph(2,z)+\aleph(2,2z)) -12\hat\mu^2\aleph(1,2z)
\nn
   &-&i \hat\mu(1+12\hat\mu^2)\,\aleph(0,z) - (3+20\hat\mu^2)\aleph(1,z)\Big]\Bigg\}
\nn
  &-&\frac{15}{2}\Bigg\{\lb1+12\hat\mu^2\rb\lb\frac{1}{\epsilon}+4\L-2\ln\hat m_D\rb +  (1+12\hat\mu^2)\left(\frac{4}{3}
       -\aleph(z)\right)
\nn
&+& 24\aleph(1,z)\Bigg\}\hat m_D\Bigg] + \left(\frac{c_A\alpha_s}{3\pi}\right)\left(\frac{s_F\alpha_s}{\pi}\right)
\Bigg[\frac{15}{2\hat m_D}\lb1+12\hmu^2\rb
\nn     
&-& \frac{235}{32}\Bigg\{\bigg(1+\frac{792}{47}\hat\mu^2+\frac{1584}{47}\hat\mu^4\bigg)\lb\frac{1}{\epsilon}
     +6\ln\frac{\hat\Lambda}{2}\rb-\frac{48 \gamma_E }{47}\lb1+12\mu^2\rb
     \nonumber\\
    &&+\frac{1809}{470}\left(1+\frac{8600}{603}\hat\mu^2+\frac{28720}{603}\hat\mu^4\right)
   -\frac{32}{47}\lb1+6\mu^2\rb\Za
    \nonumber\\
 &-&\frac{464}{235}\frac{\zeta'(-3)}{\zeta(-3)}-\frac{288 }{47}\lb1+12\hmu^2\rb\ln\hat m_D
   -\frac{288}{47}\Big[2i\hat\mu\aleph(0,z)
 \nonumber\\
&-& \left(3+68\hat\mu^2\right)\aleph(1,z)
   +72i\hmu \aleph(2,z)+26\aleph(3,z)\Big]\Bigg\}
\nn
   &+&\frac{315}{8}\Bigg\{\lb1+\frac{132}{7}\mu^2\rb\lb\frac{1}{\epsilon}+6\L-2\ln\hat m_D\rb+\frac{88}{21}
   +\frac{22}{7}\lb1+12\hmu^2\rb\gamma_E
\nn
   &+&\frac{440}{7}\hmu^2-\frac{8}{7}\frac{\zeta'(-1)}{\zeta(-1)}+\frac{4}{7}\aleph(z)+\frac{264}{7}\aleph(1,z)\Bigg\}\hat m_D
     +90\frac{\hat m_q}{\hat m_D}\Bigg] ,
\label{threeloopunrenorm}
\end{eqnarray}
where $\Omega_0=-d_A\pi^2T^4/45$. 

The sum of all counterterms through order $\delta^2$ is
\begin{eqnarray}
\frac{\Delta\Omega}{\Omega_0} &=& \frac{\Delta\Omega_1+\Delta\Omega_2}{\Omega_0} \nn
&=& \frac{s_F\alpha_s}{\pi}\Bigg[- \frac{15}{2}
 \left(\frac{1}{\epsilon}+2\L - 
  2 \ln \hat m_D  \right) \hat m_D^3 \Bigg]
 \nonumber \\ 
&+& \left(\frac{c_A\alpha_s}{3\pi}\right)\left(\frac{s_F\alpha_s}{\pi}\right)\Bigg[
   \frac{235}{32}\Bigg\{\left(1+\frac{792}{47}\hat\mu^2+\frac{1584}{47}\hmu^4\right)\lb\frac{1}{\epsilon}
   +4\ln\frac{\hat\Lambda}{2}\rb+\frac{56}{47}\Za 
   \nonumber\\
   &+& \frac{149}{47}\lb1+\frac{2376}{149}\hmu^2+\frac{4752}{149}\hmu^4\rb  +\frac{1584}{47}\lb1
    +4\hmu^2\rb\aleph(1,z)+\frac{1056}{47}\Za\hmu^2\Bigg\}
   \nonumber\\
&-&\frac{315}{8}\Bigg\{\lb1+\frac{132}{7}\hmu^2\rb\lb\frac{1}{\epsilon}+4\ln\frac{\hat\Lambda}{2}
    -2\ln{\hat m_D}\rb
   -\frac{8}{7}\frac{\zeta'(-1)}{\zeta(-1)}+\frac{61}{21}+44\hmu^2
\nonumber\\
&+& \frac{264}{7}\aleph(1,z)\Bigg\}\hat m_D
\Bigg]+\left(\frac{s_F\alpha_s}{\pi}\right)^2 \Bigg[ - \frac{25}{24}
   \Bigg\{\lb1+\frac{72}{5}\hmu^2+\frac{144}{5}\hmu^4\rb\lb\frac{1}{\epsilon} + 4\L+ 3\rb 
\nonumber\\
&+&\frac{144}{5}\lb1+4\hmu^2\rb\aleph(1,z) + 
   \frac{8}{5}\lb1+12\hmu^2\rb\frac{\zeta'(-1)}{\zeta(-1)}\Bigg\}  
\nn
&+& \frac{15}{2}\Bigg\{\lb1+12\hmu^2\rb\lb\frac{1}{\epsilon}+4\L - 
   2 \ln \hat m_D + \frac{7}{3}\rb+24\aleph(1,z)\Bigg\}\hat m_D
\Bigg]
+ \frac{\Delta\Omega^{\rm YM}}{\Omega_0} \ ,
\nonumber \\
\label{deltaomega}
\end{eqnarray}
where $\Delta\Omega^{\rm YM}$ is the pure-glue three-loop HTLpt counterterm~\cite{Andersen:2011sf}
\be
\frac{\Delta\Omega^{\rm YM}}{\Omega_0} &=& \frac{45}{8\epsilon}\hat m_D^4+ \frac{495}{8} \lb\frac{c_A\alpha_s}{3 \pi}\rb\lb
\frac{1}{\epsilon}+2\Lg-2\ln\hat m_D\rb\hat m_D^3
\nn
&&+\lb\frac{c_A\alpha_s}{3\pi}\rb^2\Bigg[\frac{165}{16}\lb\frac{1}{\epsilon}+4\Lg+2+4\Za\rb
\nn
&&-\frac{1485}{8}\lb\frac{1}{\epsilon}+4\Lg-2\ln\hat m_D+\frac{4}{3}+2\Za\rb\hat m_D\Bigg] \, .
\ee
Adding the total three-loop HTLpt counterterm~(\ref{deltaomega}) to the unrenormalized three-loop HTLpt thermodynamic potential
(\ref{threeloopunrenorm}) we obtain our final result for the NNLO HTLpt thermodynamic potential in the case that
all quarks have the same chemical potential 
\begin{eqnarray}
\frac{\Omega_{\rm NNLO}}{\Omega_0}
&=& \frac{7}{4}\frac{d_F}{d_A}\lb1+\frac{120}{7}\hmu^2+\frac{240}{7}\hmu^4\rb
    -\frac{s_F\alpha_s}{\pi}\bigg[\frac{5}{8}\left(1+12\hat\mu^2\right)\left(5+12\hat\mu^2\right)
    \nn
    &&-\frac{15}{2}\left(1+12\hat\mu^2\right)\hat m_D-\frac{15}{2}\bigg(2\ln{\frac{\hat\Lambda}{2}-1
   -\aleph(z)}\Big)\hat m_D^3
      +90\hat m_q^2 \hat m_D\bigg]
\nn
&+& s_{2F}\left(\frac{\alpha_s}{\pi}\right)^2\bigg[\frac{15}{64}\bigg\{35-32\lb1-12\hmu^2\rb\frac{\zeta'(-1)}
      {\zeta(-1)}+472 \hat\mu^2+1328  \hat\mu^4\nn
&+& 64\Big(-36i\hat\mu\aleph(2,z)+6(1+8\hat\mu^2)\aleph(1,z)+3i\hat\mu(1+4\hat\mu^2)\aleph(0,z)\Big)\bigg\}
\nn
&-& \frac{45}{2}\hat m_D\left(1+12\hat\mu^2\right)\bigg] +\left(\frac{s_F\alpha_s}{\pi}\right)^2\left[\frac{5}{4\hat m_D}\left(1+12\hat\mu^2\right)^2+30\left(1+12\hat\mu^2
        \right)\frac{\hat m_q^2}{\hat m_D}\right.\nn
&+&   \left.\frac{25}{12}\Bigg\{ \left(1 +\frac{72}{5}\hat\mu^2+\frac{144}{5}\hat\mu^4\right)\ln\frac{\hat\Lambda}{2}
       + \frac{1}{20}\lb1+168\hmu^2+2064\hmu^4\rb\right.\nn
&+& \left.
         \frac{3}{5}\lb1+12\hmu^2\rb^2\gamma_E -\frac{8}{5}(1+12\hat\mu^2)\frac{\zeta'(-1)}{\zeta(-1)}
        - \frac{34}{25}\frac{\zeta'(-3)}{\zeta(-3)}\right.
\nn
&-&       
  \frac{72}{5}\Big[8\aleph(3,z)+3\aleph(3,2z)+ 12 i \hat\mu\,(\aleph(2,z)+\aleph(2,2z))\nn  
&-& 12\hat\mu^2\aleph(1,2z)
       -\left.i \hat\mu(1+12\hat\mu^2)\,\aleph(0,z)  
       - 2(1+8\hat\mu^2)\aleph(1,z)\Big]\Bigg\}\right.\nn
&-&\left.\frac{15}{2}\lb1+12\hat\mu^2\rb\lb2\L-1-\aleph(z)\rb\hat m_D\right]
\nn
&+& \left(\frac{c_A\alpha_s}{3\pi}\right)\left(\frac{s_F\alpha_s}{\pi}\right)\Bigg[\frac{15}{2\hat m_D}\lb1+12\hmu^2\rb
     -\frac{235}{16}\Bigg\{\bigg(1+\frac{792}{47}\hat\mu^2+\frac{1584}{47}\hat\mu^4\bigg)\ln\frac{\hat\Lambda}{2}
     \nonumber\\
&-&\frac{144}{47}\lb1+12\hmu^2\rb\ln\hat m_D+\frac{319}{940}\left(1+\frac{2040}{319}\hat\mu^2+\frac{38640}{319}\hat\mu^4\right)
   -\frac{24 \gamma_E }{47}\lb1+12\mu^2\rb \nn
&-&
   \frac{44}{47}\lb1+\frac{156}{11}\hmu^2\rb\frac{\zeta'(-1)}{\zeta(-1)}
    -\frac{268}{235}\frac{\zeta'(-3)}{\zeta(-3)}
   -\frac{72}{47}\Big[4i\hat\mu\aleph(0,z)
    \nonumber\\
&+&\left(5-92\hat\mu^2\right)\aleph(1,z)+144i\hmu\aleph(2,z)
   +52\aleph(3,z)\Big]\Bigg\}+90\frac{\hat m_q^2}{\hat m_D}
\nonumber\\
   &+&\frac{315}{4}\Bigg\{\lb1+\frac{132}{7}\hmu^2\rb\L+\frac{11}{7}\lb1+12\hmu^2\rb\gamma_E+\frac{9}{14}\lb1+\frac{132}{9}\hmu^2\rb
\nn  
 &+&\frac{2}{7}\aleph(z)\Bigg\}\hat m_D 
\Bigg]
+ \frac{\Omega_{\rm NNLO}^{\rm YM}}{\Omega_0} \, .
\label{finalomega1}
\end{eqnarray}
where $\Omega_{\rm NNLO}^{\rm YM}$ is the NNLO pure glue thermodynamic potential~\cite{Andersen:2010ct}
{\small
\be
\frac{\Omega_{\rm NNLO}^{\rm YM}}{\Omega_0} \!\!\!&=&\!\!\! 1-\frac{15}{4}\hat m_D^3+\frac{c_A\alpha_s}{3\pi}\Bigg[-\frac{15}{4}
+\frac{45}{2}\hat m_D-\frac{135}{2}\hat m_D^2-\frac{495}{4}\lb\Lg+\frac{5}{22}+\gamma_E\rb  \hat m_D^3 \Bigg]
\nn
&-&\lb\frac{c_A\alpha_s}{3\pi}\rb^2\Bigg[\frac{165}{8}\lb\Lg-\frac{72}{11}\ln\hat m_D-\frac{84}{55}-\frac{6}{11}
\gamma_E-\frac{74}{11}\Za+\frac{19}{11}\Zc\rb
\nn
&-&\frac{1485}{4}\lb\Lg-\frac{79}{44}+\gamma_E+\ln2-\frac{\pi^2}{11}\rb\hat m_D-\frac{45}{4\hat m_D}\Bigg] \, ,
\label{nnloym}
\ee
}
It is worth to mention here that the NNLO pure glue thermodynamic potential in Eq.~(\ref{nnloym}) looks like chemical
potential independent as there are no explicit chemical potential dependence, but the chemical potential implicitly
present within Debye mass $m_D$.
It appears in pure glue diagrams from the internal quark loop in effective gluon propagators and effective vertices.

Note that the full thermodynamic potential (\ref{finalomega1}) reduces to thermodynamic potential of Ref.~\cite{Andersen:2011sf}
in the limit $\mu\rightarrow 0$. In addition, the above thermodynamic potential produces the correct ${\cal O}(g^5)$ 
perturbative result when expanded in a strict power series in $g$~\cite{Vuorinen:2002ue,Vuorinen:2003fs}.

\subsection{NNLO result -- General case}
It is relatively straightforward to generalize the previously obtained result (\ref{finalomega1}) to the case that each quark
 has a separate chemical potential $\mu_f$.  The final result is
\begin{eqnarray}
\frac{\Omega_{\rm NNLO}}{\Omega_0}
&=& \frac{7}{4}\frac{d_F}{d_A}\frac{1}{N_f}\sum\limits_f\lb1+\frac{120}{7}\hmu_f^2+\frac{240}{7}\hmu_f^4\rb
    -\frac{s_F\alpha_s}{\pi}\frac{1}{N_f}\!\sum\limits_f\bigg[\frac{5}{8}\left(5+72\hat\mu_f^2+144\hat\mu_f^4\right)
    \nn
    &-&\frac{15}{2}\left(1+12\hat\mu_f^2\right)\hat m_D-\frac{15}{2}\bigg(2\ln{\frac{\hat\Lambda}{2}-1
   -\aleph(z_f)}\Big)\hat m_D^3
      +90\hat m_q^2 \hat m_D\bigg]\nonumber
\ee
\be
&+& \frac{s_{2F}}{N_f}\left(\frac{\alpha_s}{\pi}\right)^2\sum\limits_f\bigg[\frac{15}{64}\bigg\{35-32\lb1-12\hmu_f^2\rb\frac{\zeta'(-1)}
      {\zeta(-1)}+472 \hat\mu_f^2+1328  \hat\mu_f^4\nn
      &+& 64\Big(-36i\hat\mu_f\aleph(2,z_f)+6(1+8\hat\mu_f^2)\aleph(1,z_f)+3i\hat\mu_f(1+4\hat\mu_f^2)\aleph(0,z_f)\Big)\bigg\}\nn
      &-& \frac{45}{2}\hat m_D\left(1+12\hat\mu_f^2\right)\bigg] \nn
&+& \left(\frac{s_F\alpha_s}{\pi}\right)^2
      \frac{1}{N_f}\sum\limits_{f}\frac{5}{16}\Bigg[96\left(1+12\hat\mu_f^2\right)\frac{\hat m_q^2}{\hat m_D}
     +\frac{4}{3}\lb1+12\hmu_f^2\rb\lb5+12\hat\mu_f^2\rb
      \ln\frac{\hat{\Lambda}}{2}\nn
    &+&\frac{1}{3}+4\gamma_E+8(7+12\gamma_E)\hat\mu_f^2+112\mu_f^4-\frac{64}{15}\frac{\zeta^{\prime}(-3)}{\zeta(-3)}-
   \frac{32}{3}(1+12\hat\mu_f^2)\frac{\zeta^{\prime}(-1)}{\zeta(-1)}\nn
   &-&    96\Big\{8\aleph(3,z_f)+12i\hat\mu_f\aleph(2,z_f)-2(1+2\hat\mu_f^2)\aleph(1,z_f)-i\hat\mu_f\aleph(0,z_f)\Big\}\Bigg] \nn
&+& \left(\frac{s_F\alpha_s}{\pi}\right)^2
      \frac{1}{N_f^2}\sum\limits_{f,g}\Bigg[\frac{5}{4\hat m_D}\left(1+12\hat\mu_f^2\right)\left(1+12\hat\mu_g^2\right)
     +90\Bigg\{ 2\left(1 +\gamma_E\right)\hat\mu_f^2\hat\mu_g^2
      \nn
&-&\Big\{\aleph(3,z_f+z_g)+\aleph(3,z_f+z_g^*)+ 4i\hat\mu_f\left[\aleph(2,z_f+z_g)+\aleph(2,z_f+z_g^*)\right]
         \nn
&-& 4\hat\mu_g^2\aleph(1,z_f)-(\hat\mu_f+\hat\mu_g)^2\aleph(1,z_f+z_g)- (\hat\mu_f-\hat\mu_g)^2\aleph(1,z_f+z_g^*)
     \nn
&-& 4i\hat\mu_f\hat\mu_g^2\aleph(0,z_f)\Big\}\Bigg\}-\frac{15}{2}\lb1+12\hat\mu_f^2\rb\lb2\L-1-\aleph(z_g)\rb  \hat m_D\Bigg]
\nn
&+& \left(\frac{c_A\alpha_s}{3\pi}\right)\left(\frac{s_F\alpha_s}{\pi N_f}\right)\sum\limits_f\Bigg[
     -\frac{235}{16}\Bigg\{\bigg(1+\frac{792}{47}\hat\mu_f^2+\frac{1584}{47}\hat\mu_f^4\bigg)\ln\frac{\hat\Lambda}{2}
     \nonumber\\
&-&\frac{144}{47}\lb1+12\hmu_f^2\rb\ln\hat m_D+\frac{319}{940}\left(1+\frac{2040}{319}\hat\mu_f^2+\frac{38640}{319}\hat\mu_f^4\right)   
\nn
&-&\frac{24 \gamma_E }{47}\lb1+12\hat\mu_f^2\rb-\frac{44}{47}\lb1+\frac{156}{11}\hmu_f^2\rb\frac{\zeta'(-1)}{\zeta(-1)}
    -\frac{268}{235}\frac{\zeta'(-3)}{\zeta(-3)}
    \nonumber
\ee
\be
    &-&\frac{72}{47}\Big[4i\hat\mu_f\aleph(0,z_f)+\left(5-92\hat\mu_f^2\right)\aleph(1,z_f)+144i\hmu_f\aleph(2,z_f)
   +52\aleph(3,z_f)\Big]\Bigg\}
\nn
&+& \frac{15}{2\hat m_D}\lb1+12\hmu_f^2\rb+90\frac{\hat m_q^2}{\hat m_D}
   +\frac{315}{4}\Bigg\{\lb1+\frac{132}{7}\hmu_f^2\rb\L
\nonumber\\
   &+&\frac{11}{7}\lb1+12\hmu_f^2\rb\gamma_E+\frac{9}{14}\lb1+\frac{132}{9}\hmu_f^2\rb
+\frac{2}{7}\aleph(z_f)\Bigg\}\hat m_D 
\Bigg]
+ \frac{\Omega_{\rm NNLO}^{\rm YM}}{\Omega_0} \, ,
\label{finalomega}
\end{eqnarray}
where the sums over $f$ and $g$ include all quark flavors, $z_f = 1/2 - i \hat{\mu}_f$, and $\Omega_{\rm NNLO}^{\rm YM}$
is the pure-glue contribution as before.
\section{Mass prescription}
\label{pres}

As discussed in Sec.~(\ref{htl_lag_sec}) of Chapter~\ref{chapter:htl} and also in Ref.~\cite{Andersen:2011sf},
the two-loop perturbative electric gluon mass, first introduced by Braaten and
Nieto in~\cite{Braaten:1995cm,Braaten:1995jr} is the most suitable for three-loop HTLpt calculations. We use the Braaten-Nieto (BN)
mass prescription for $m_D$ in the remainder of the chapter.  Originally, the two-loop perturbative mass was calculated in
Refs.~\cite{Braaten:1995cm,Braaten:1995jr} for zero chemical potential, however, Vuorinen has generalized it to finite
chemical potential. The resulting expression for $m_D^2$ is \cite{Vuorinen:2002ue,Vuorinen:2003fs}
\begin{eqnarray}
\hat m_D^2&=&\frac{\alpha_s}{3\pi} \Biggl\{c_A
+\frac{c_A^2\alpha_s}{12\pi}\lb5+22\gamma_E+22\Lg\rb +
\frac{1}{N_f} \sum\limits_{f}
\Biggl[ s_F\lb1+12\hmu_f^2\rb
\nonumber\\
  &+&\frac{c_As_F\alpha_s}{12\pi}\lb\lb9+132\hmu_f^2\rb+22\lb1+12\hmu_f^2\rb\gamma_E+2\lb7+132\hmu_f^2\rb\L+4\aleph(z_f)\rb
\nonumber\\  
&+&\frac{s_F^2\alpha_s}{3\pi}\lb1+12\hmu_f^2\rb\lb1-2\L+\aleph(z_f)\rb
 -\frac{3}{2}\frac{s_{2F}\alpha_s}{\pi}\lb1+12\hmu_f^2\rb \Biggr] \Biggr\} \, .
\end{eqnarray}
The effect of the in-medium quark mass parameter $m_q$ in thermodynamic functions is small and following
Ref.~\cite{Andersen:2011sf} we take $m_q=0$ which is the three loop variational solution.
The maximal effect on the susceptibilities comparing the perturbative quark mass, 
$\hat{m}_q^2 =c_F \alpha_s(T^2+\mu^2/\pi^2)/8\pi$, with the variational solution, $m_q=0$, is approximately 0.2\% at $T=200$ MeV.
At higher temperatures, the effect is much smaller, e.g. 0.02\% at $T=1$ GeV.

\section{Thermodynamic functions} 
\label{thermof}

In this section we present our final results for the NNLO HTLpt pressure, energy density, entropy density,
trace anomaly, and speed of sound. We will plotted our NNLO result both using one loop(\ref{1loop_running}) and three loop
(\ref{3loop_running}) running coupling. For both one- and three-loop running we fix the scale 
$\Lambda_{\overline{\rm MS}}$ by requiring that $\alpha_s({\rm 1.5\;GeV}) = 0.326$ which is obtained
from lattice measurements \cite{Bazavov:2012ka}.  For one-loop running, this procedure gives
$\Lambda_{\overline{\rm MS}} = 176$ MeV, and for three-loop running, one obtains $\Lambda_{\overline{\rm MS}} = 316$ MeV.

\subsection{Scales} 
\label{sec:scales}
For the renormalization scale we use separate scales, $\Lambda_g$ and $\Lambda$, for purely-gluonic and 
fermionic graphs, respectively.  We take the central values of these renormalization scales to be 
$\Lambda_g = 2\pi T$ and $\Lambda=2\pi \sqrt{T^2+\mu^2/\pi^2}$.  In all plots the thick lines 
indicate the result obtained using these central values and the light-blue band indicates the variation of 
the result under variation of both of these scales by a factor of two, e.g. $\pi T \leq \Lambda _g \leq 4 \pi T$.
For all numerical results below we use $c_A = N_c=3$ and $N_f=3$.

\subsection{Pressure} 
The QGP pressure can be obtained directly from the thermodynamic potential (\ref{finalomega1})
\be
{\cal P}(T,\Lambda,\mu)=-\Omega_{\rm NNLO}(T,\Lambda,\mu) \, ,
\ee
where $\Lambda$ above is understood to include both scales $\Lambda_g$ and $\Lambda$.

We note that in the ideal gas limit, the pressure becomes
\be
{\cal P}_{\rm ideal}(T,\mu)=\frac{d_A\pi^2T^4}{45}\left[1+\frac{7}{4}\frac{d_F}{d_A}\left(1+\frac{120}{7}\hmu^2
+\frac{240}{7}\hmu^4\right)\right] .
\ee

\begin{figure}[tbh]
\subfigure{
\hspace{-2mm}\includegraphics[width=7.5cm]{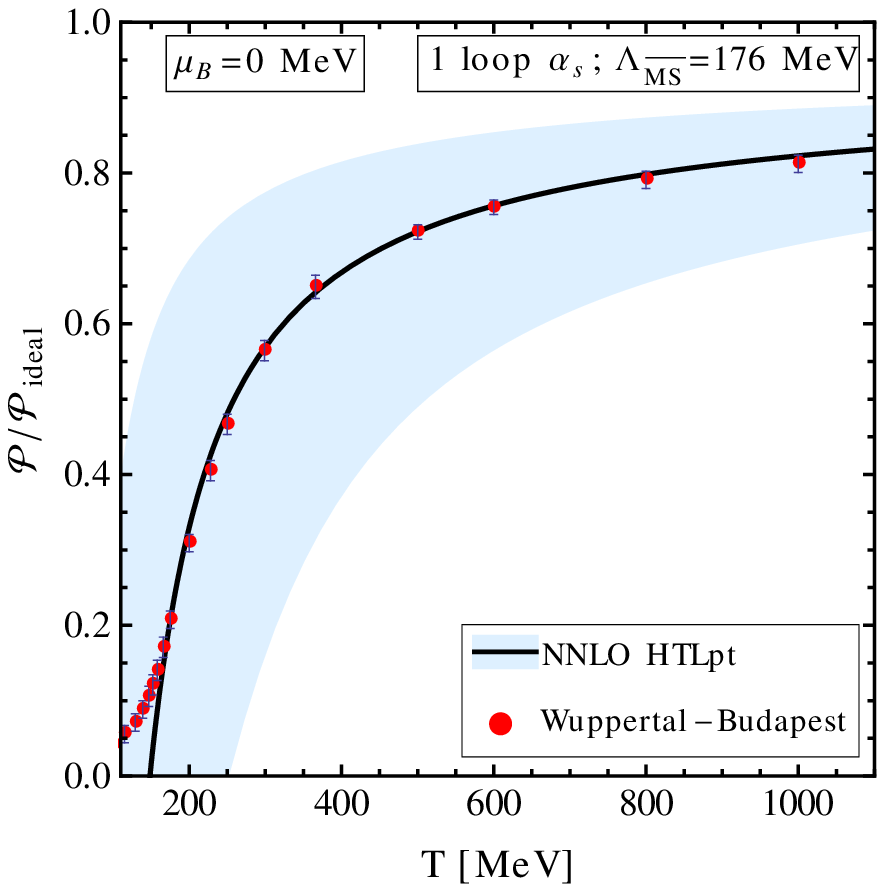}} 
\subfigure{
\includegraphics[width=7.5cm]{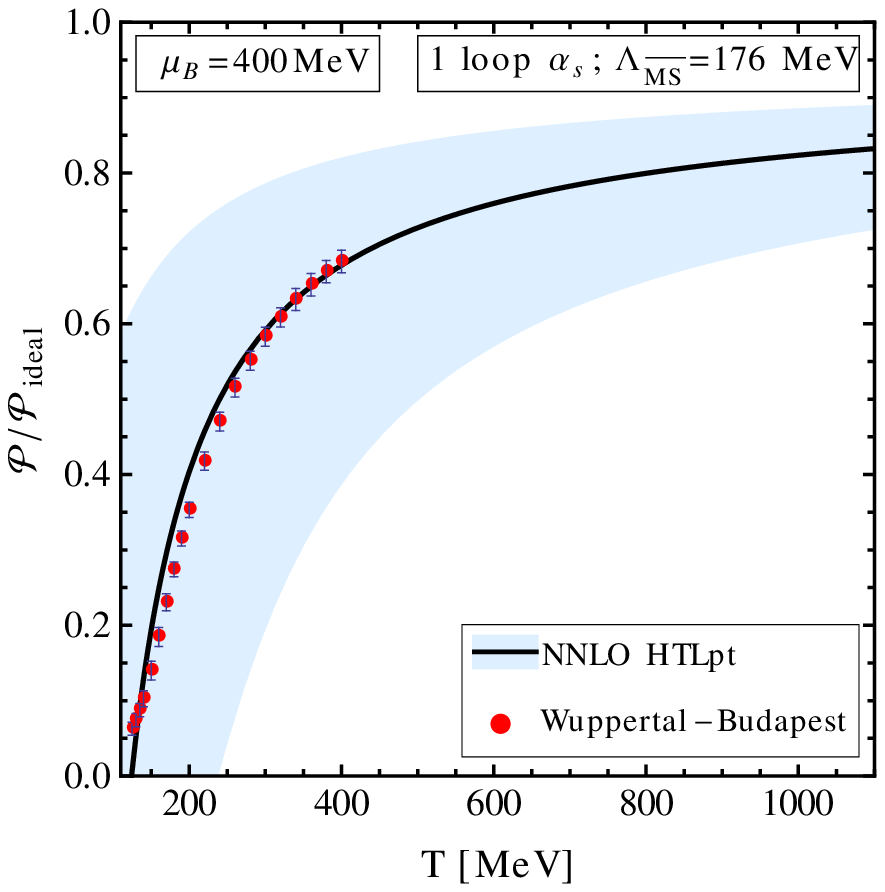}}
\caption[Comparison of the $N_f=2+1$, $\mu_B=0$ (left) and $\mu_B=400$ MeV (right) NNLO HTLpt 
pressure using one-loop running coupling constant with lattice data.]{
Comparison of the $N_f=2+1$, $\mu_B=0$ (left) and $\mu_B=400$ MeV (right) NNLO HTLpt 
pressure with lattice data from Borsanyi et al. \cite{Borsanyi:2010cj,Borsanyi:2012uq}.
For the HTLpt results a one-loop running coupling constant was used.
}
\label{pres_1l}
\end{figure}
\begin{figure}[tbh]
\subfigure{
\hspace{-2mm}\includegraphics[width=7.5cm]{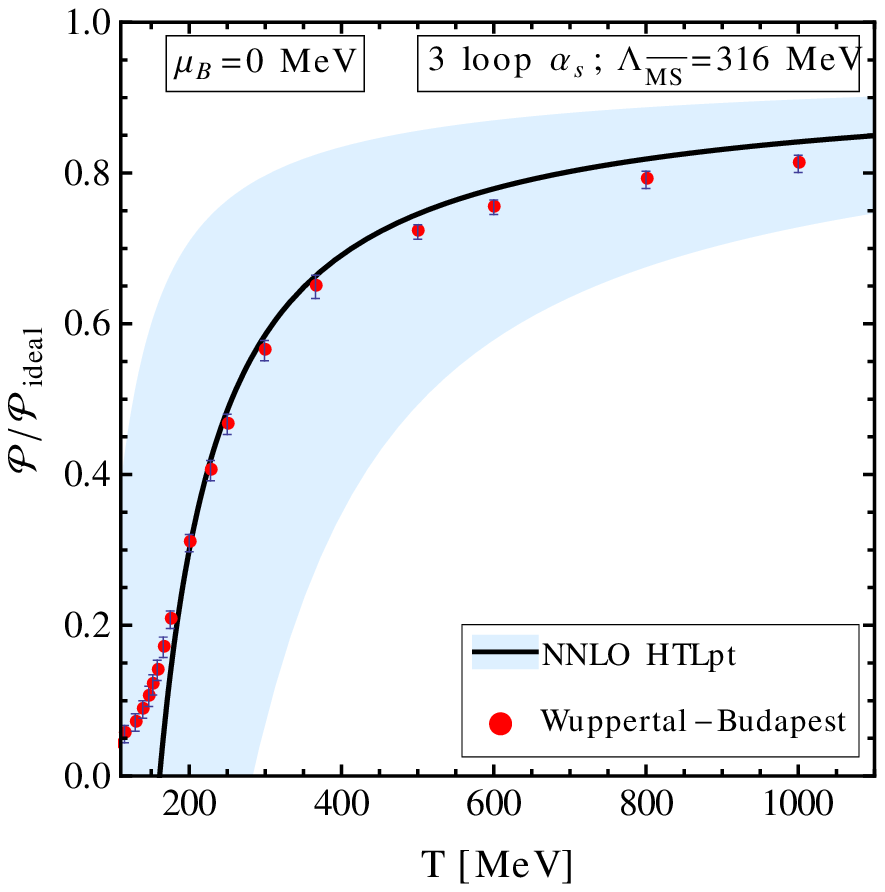}} 
\subfigure{
\includegraphics[width=7.5cm]{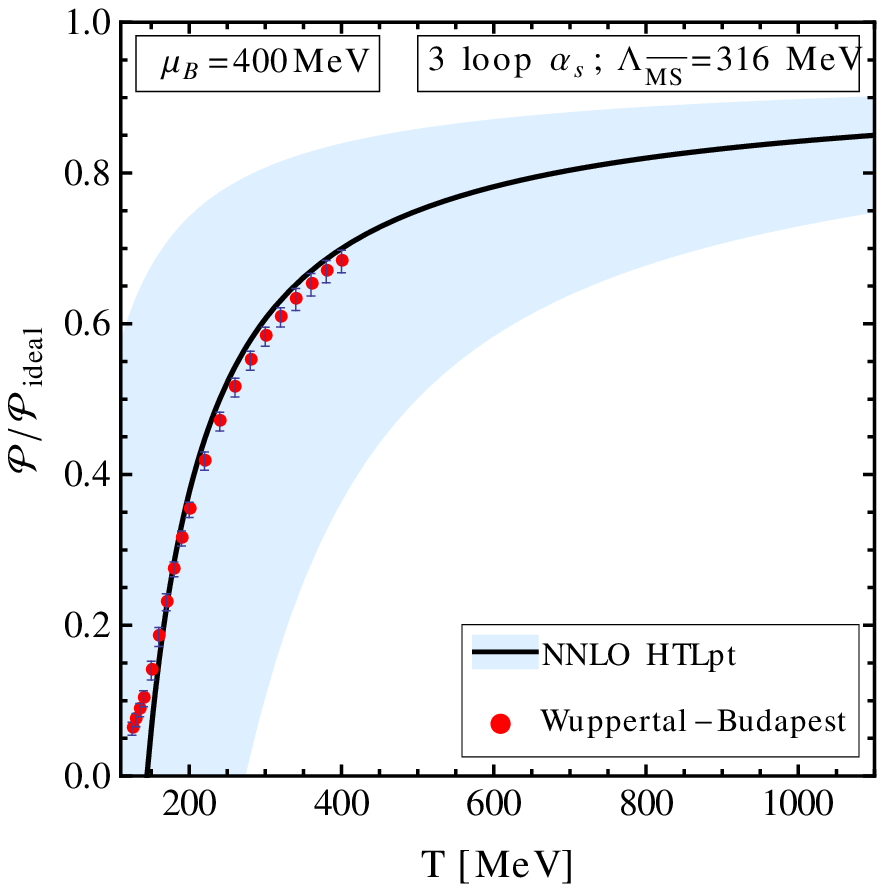}} 
\caption[Comparison of the $N_f=2+1$, $\mu_B=0$ (left) and $\mu_B=400$ MeV (right) NNLO HTLpt 
pressure using three-loop running coupling constant with lattice data.]
{Same as Fig.~(\ref{pres_1l}) except with a three-loop running coupling constant.
}
\label{pres_3l}
\end{figure}

In Figs.~(\ref{pres_1l}) and (\ref{pres_3l}) we compare the scaled NNLO HTLpt pressure for $\mu_B=0$ (left) and $\mu_B=400$ MeV
(right) with lattice data from Refs.~\cite{Bazavov:2009zn,Borsanyi:2010cj,Borsanyi:2012uq}. The
deviations below $T\sim 200$ MeV are due to the fact that our calculation does not include hadronic degrees of freedom
which dominate at low temperatures (see e.g. fits in~\cite{Huovinen:2009yb}) or
nonperturbative effects~\cite{KorthalsAltes:1999xb,Pisarski:2000eq,KorthalsAltes:2000gs,Zwanziger:2004np,Vuorinen:2006nz,
deForcrand:2008aw,Fukushima:2013xsa}.
Further, in order to gauge the sensitivity
of the results to the order of the running coupling, in Fig.~(\ref{pres_1l}) we show the results obtained using a  one-loop
running and in Fig.~(\ref{pres_3l}) the results obtained using a three-loop running.  As can be seen by comparing these two sets,
the sensitivity of the results to the order of the  running coupling is small for $T \gtrsim 250$ MeV.  As a result, unless 
the order of the running coupling turns out to have a significant effect on a given observable (see e.g. the fourth-order 
baryon number susceptibility), we will show the results obtained using a one-loop running coupling consistent with the 
counterterms necessary to renormalize the NNLO thermodynamic potential (\ref{ctalpha}).

\begin{figure}[tbh]
\centerline{\includegraphics[width=9cm]{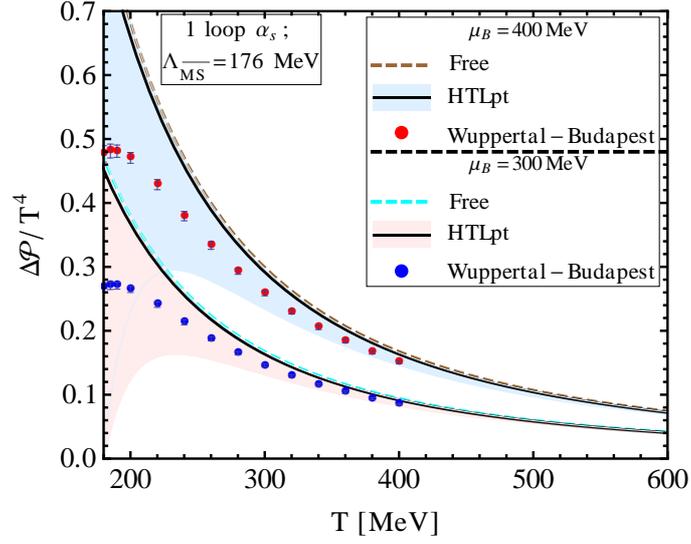}}
\caption[Comparison of the Stefan-Boltzmann limit (dashed lines) and NNLO HTLpt (solid lines) results for
the scaled pressure difference with lattice data.]{Comparison of the Stefan-Boltzmann limit (dashed lines)
 and NNLO HTLpt (solid lines) results for
the scaled pressure difference with lattice data from Borsanyi et al. \cite{Borsanyi:2012uq}.}
\label{dPfig}
\end{figure}

For an additional comparison we can compute the change in the pressure
\be
\Delta {\cal P} = {\cal P}(T,\Lambda,\mu)-{\cal P}(T,\Lambda,0) \, .
\ee
In Fig.~(\ref{dPfig}) we plot $\Delta {\cal P}$ as a function of the temperature for $\mu_B = 300$ MeV and $\mu_B = 400$ MeV.
The solid lines are the NNLO HTLpt result and the dashed lines are the result obtained in the Stefan-Boltzmann limit.
We note that  in Fig.~(\ref{dPfig}) the lattice data from the Wuppertal-Budapest group  \cite{Borsanyi:2012uq} is 
computed up to ${\cal O}(\mu_B^2)$, whereas the HTLpt result includes all orders in $\mu_B$. As can be seen from this figure,
the NNLO HTLpt result is quite close to the result obtained in the Stefan-Boltzmann limit.  Note that the small correction
in going from the Stefan-Boltzmann limit to NNLO HTLpt indicates that the fermionic sector is, to good approximation,
weakly coupled for $T \gtrsim 300$ MeV.

\subsection{Energy density}
Once the pressure is known, it is straightforward to compute other thermodynamic functions such
as the energy density by computing derivatives of the pressure with respect to the temperature
and chemical potential.  The energy density can be obtained via
\begin{eqnarray}
{\cal E}=T\frac{\partial{\cal P}}{\partial T}+\mu\frac{\partial{\cal P}}{\partial \mu}-{\cal P} \, .
\end{eqnarray}
%
We note that in the ideal gas limit, the entropy density becomes
\be
{\cal E}_{\rm ideal}(T,\mu)=\frac{d_A\pi^2T^4}{15}\left[1+\frac{7}{4}\frac{d_F}{d_A}\left(1+\frac{120}{7}\hmu^2
+\frac{240}{7}\hmu^4\right)\right].
\ee

\begin{figure}[tbh]
\subfigure{
\hspace{-2mm}\includegraphics[width=7.5cm]{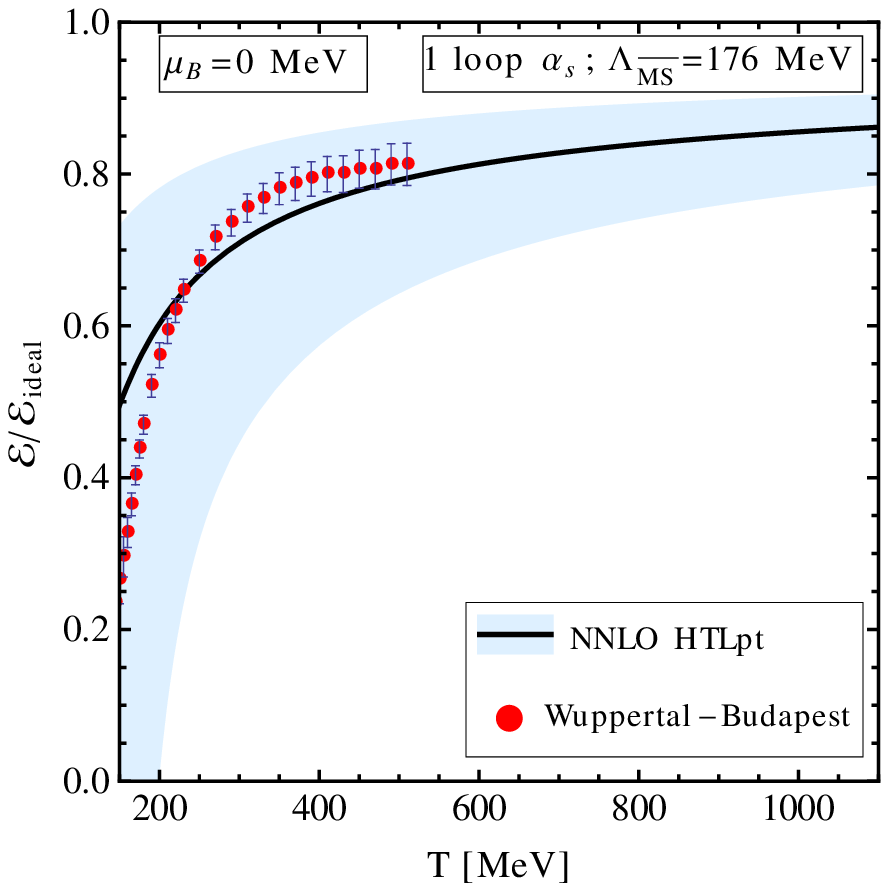}} 
\subfigure{
\includegraphics[width=7.5cm]{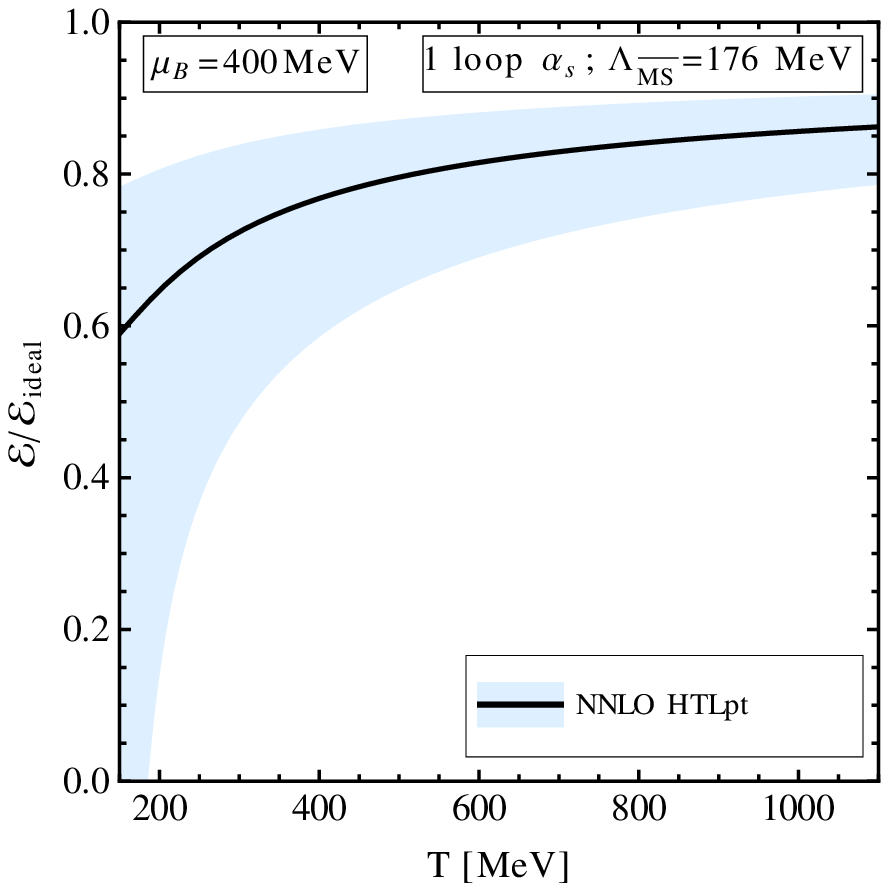}} 
\caption[Comparison of the $N_f=2+1$, $\mu_B=0$ (left) and $\mu_B=400$ MeV (right) NNLO HTLpt 
energy density with lattice data.]{
Comparison of the $N_f=2+1$, $\mu_B=0$ (left) and $\mu_B=400$ MeV (right) NNLO HTLpt 
energy density with lattice data.
The $\mu_B=0$ lattice data are from \cite{Borsanyi:2010cj}. 
For the HTLpt results a one-loop running coupling constant was used.}
\label{ed_1l}
\end{figure}

In Fig.~(\ref{ed_1l}) we plot the scaled NNLO HTLpt energy density for $\mu_B=0$ (left) and $\mu_B=400$ 
MeV (right) together with $\mu_B=0$ lattice data from Ref.~\cite{Borsanyi:2010cj}. 
As we can see from this figure, there is reasonable agreement between the NNLO HTLpt energy density
and the lattice data when the central value of the scale is used.

\subsection{Entropy density}
Similarly, we can compute the entropy density
\be
{\cal S}(T,\mu) = \frac{\partial{\cal P}}{\partial T} \, .
\ee
We note that in the ideal gas limit, the entropy density becomes
\be
{\cal S}_{\rm ideal}(T,\mu)=\frac{4d_A\pi^2T^3}{45}\left[1+\frac{7}{4}\frac{d_F}{d_A}\left(1+\frac{60}{7}\hmu^2\right)\right] .
\ee
\begin{figure}[tbh]
\subfigure{
\hspace{-2mm}\includegraphics[width=7.5cm]{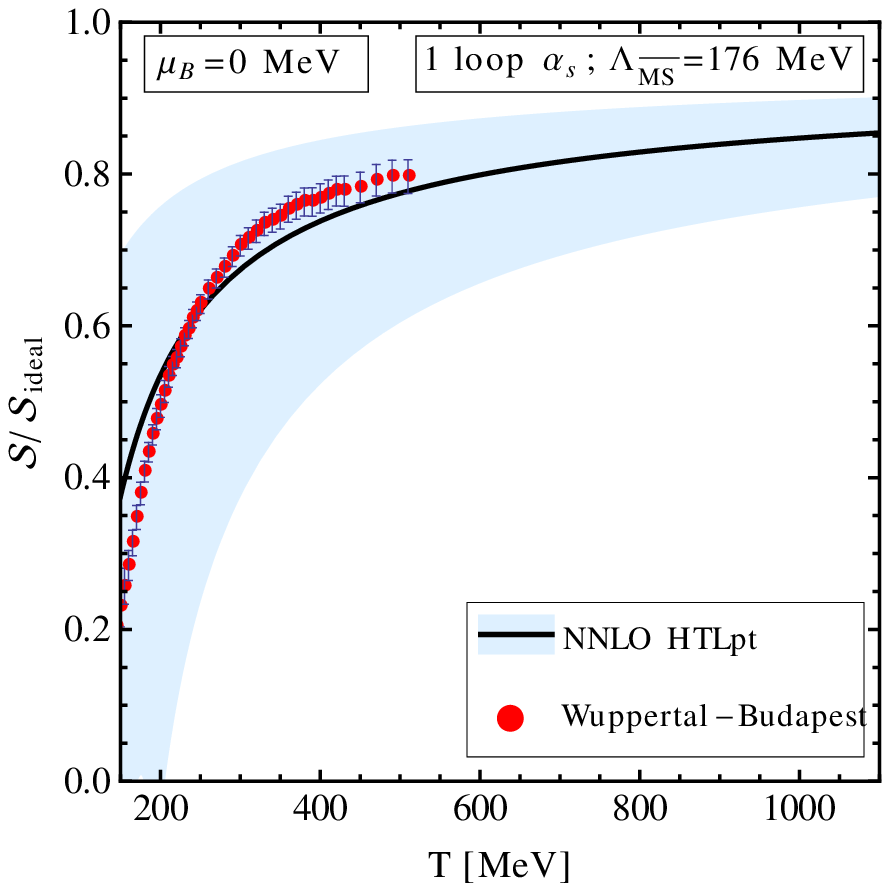}} 
\subfigure{
\includegraphics[width=7.5cm]{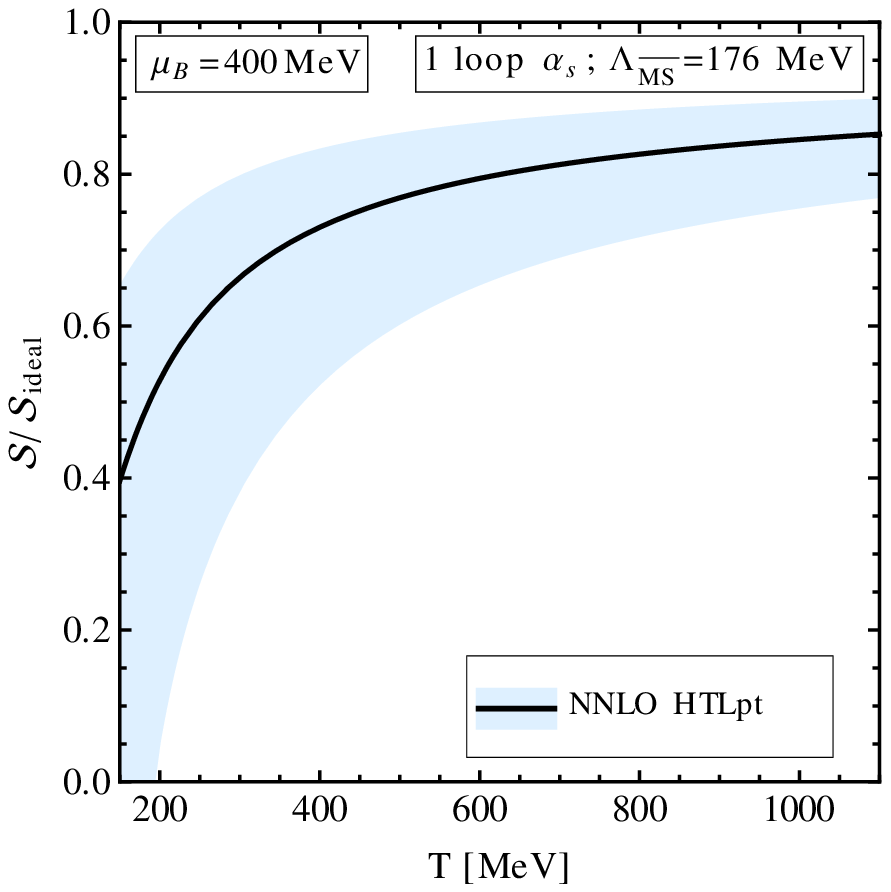}} 
\caption[Comparison of the $N_f=2+1$, $\mu_B=0$ (left) and $\mu_B=400$ MeV (right) NNLO HTLpt 
entropy density with lattice data.]{
Comparison of the $N_f=2+1$, $\mu_B=0$ (left) and $\mu_B=400$ MeV (right) NNLO HTLpt 
entropy density with lattice data.
The $\mu_B=0$ lattice data are from \cite{Borsanyi:2010cj}. 
For the HTLpt results a one-loop running coupling constant was used.}
\label{en_1l}
\end{figure}

%
In Fig~(\ref{en_1l}) we plot the scaled NNLO HTLpt entropy density for $\mu_B=0$ (left) and $\mu_B=400$ MeV (right)
together with $\mu_B=0$ lattice data from Ref.~\cite{Borsanyi:2010cj}.  As we can see from this figure, 
there is quite good agreement between the NNLO HTLpt entropy density and the lattice data when the central value of the scale is used.

\subsection{Trace anomaly}
Since it is typically the trace anomaly itself which is computed on the lattice and then integrated to obtain the other
thermodynamic functions,it is interesting to compare directly with lattice data for the trace anomaly. The trace
anomaly is simply ${\cal I} = {\cal E}-3 {\cal P}$.  In the ideal gas limit, the trace anomaly goes to 
zero since ${\cal E}=3{\cal P}$. When interactions are included, however, the trace anomaly (interaction measure) becomes non-zero.
\begin{figure}[tbh]
\subfigure{
\includegraphics[width=7.5cm]{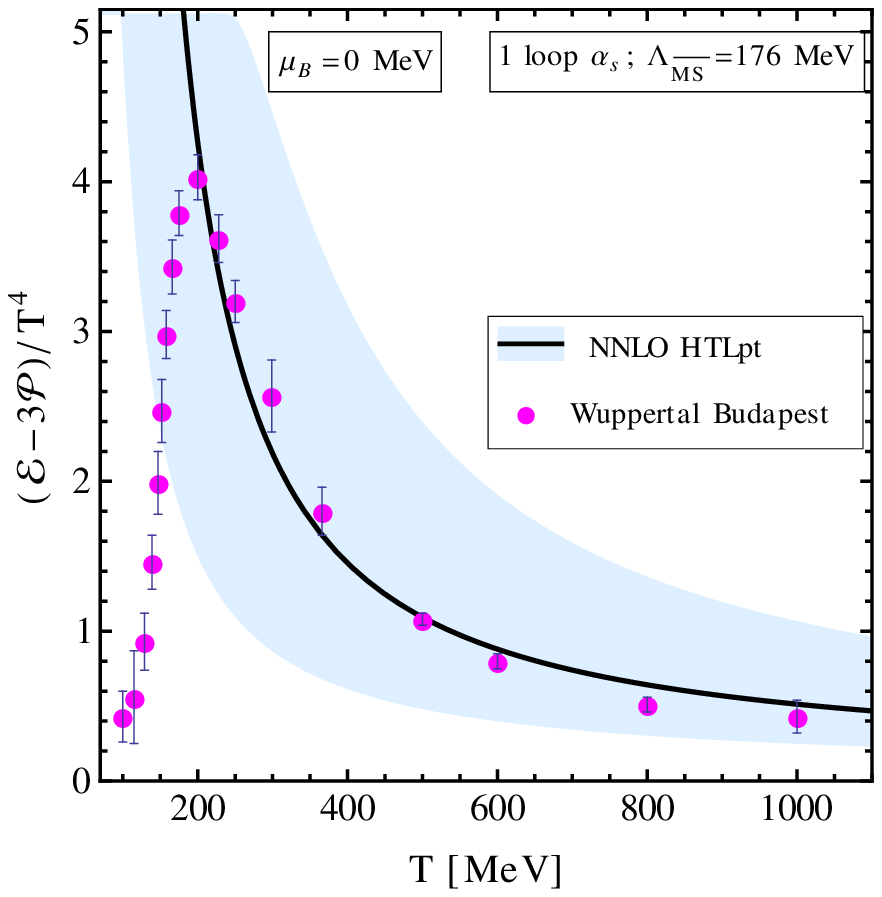}} 
\subfigure{
\includegraphics[width=7.5cm]{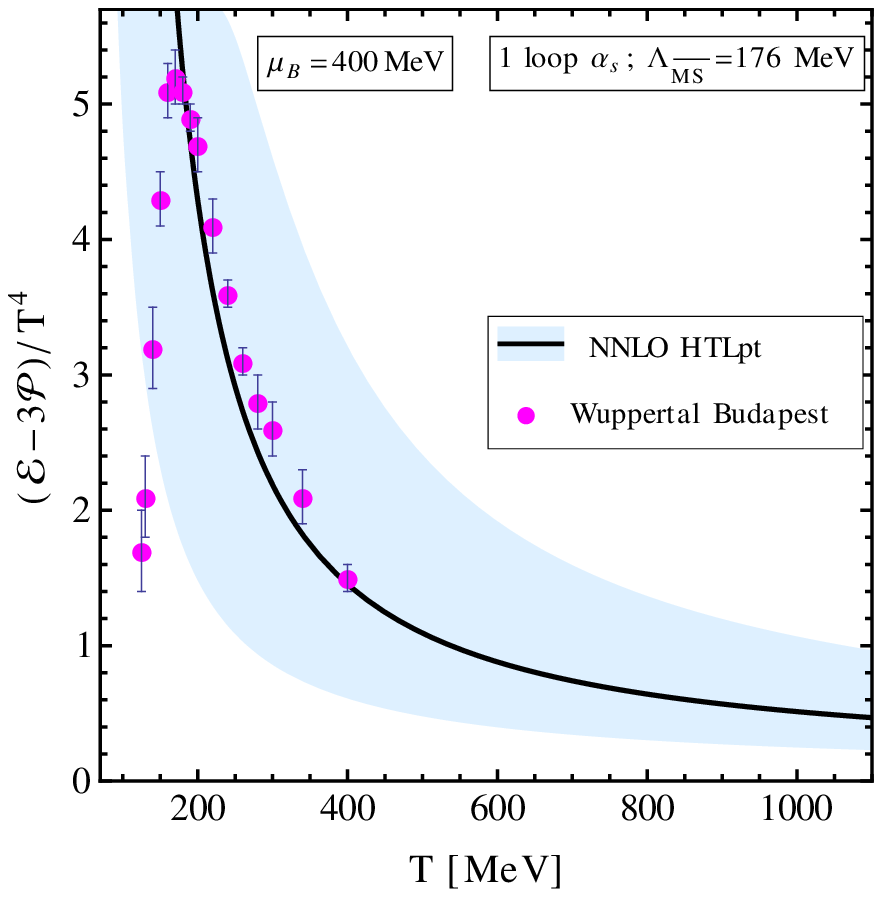}} 
\caption[Comparison of the $N_f=2+1$, $\mu_B=0$ (left) and $\mu_B=400$ MeV (right) NNLO HTLpt 
trace anomaly with lattice data.]{
Comparison of the $N_f=2+1$, $\mu_B=0$ (left) and $\mu_B=400$ MeV (right) NNLO HTLpt 
trace anomaly with lattice data.
The $\mu_B=0$ lattice data are from \cite{Borsanyi:2010cj} and the $\mu_B=400$ MeV lattice data are from \cite{Borsanyi:2012cr}. 
For the HTLpt results a one-loop running coupling constant was used.
}
\label{ta_1l}
\end{figure}

In Fig.~(\ref{ta_1l}) we plot the scaled NNLO HTLpt trace anomaly for $\mu_B=0$ (left) and $\mu_B=400$ MeV (right)
together with lattice data from Refs.~\cite{Borsanyi:2010cj} and \cite{Borsanyi:2012cr}.
As we can see from this figure, there is quite good agreement between the NNLO HTLpt trace anomaly and
the lattice data for $T \gtrsim 220$ MeV when the central value of the scale is used.

\subsection{Speed of sound}
Another quantity which is phenomenologically interesting is the speed of sound.  The speed of sound is defined as
\be
c_s^2=\frac{\del{\cal P}}{\del{\cal E}} \, .
\ee
%
\begin{figure}[tbh]
\subfigure{
\hspace{-3mm}
\includegraphics[width=7.2cm]{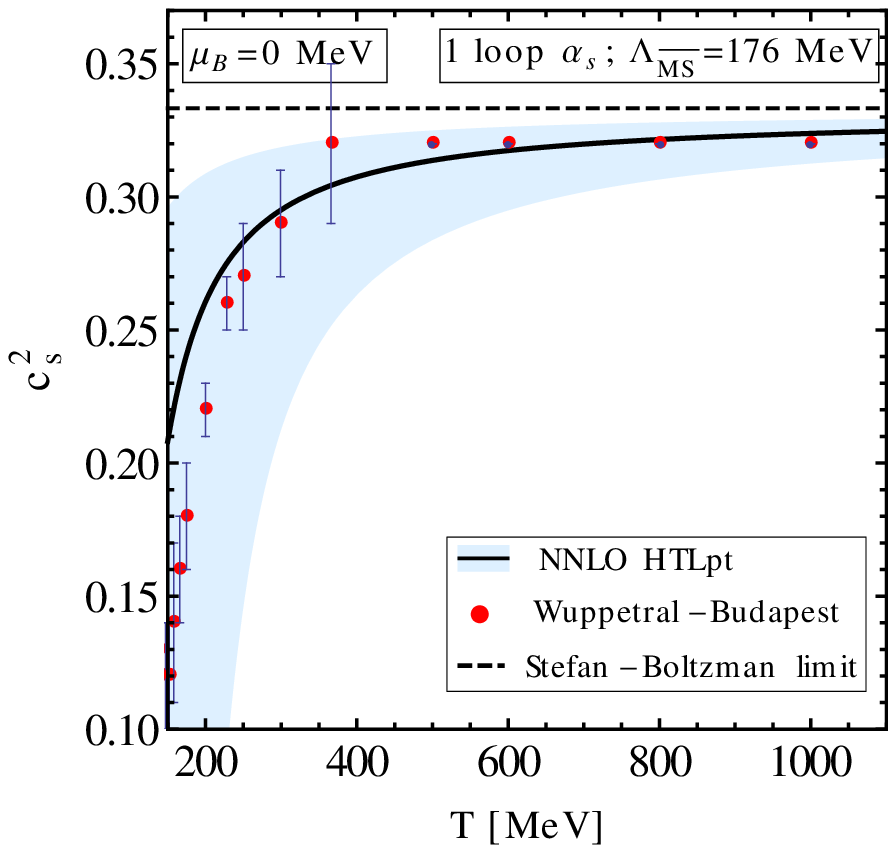}} 
\hspace{-3mm}
\subfigure{
\includegraphics[width=7.2cm]{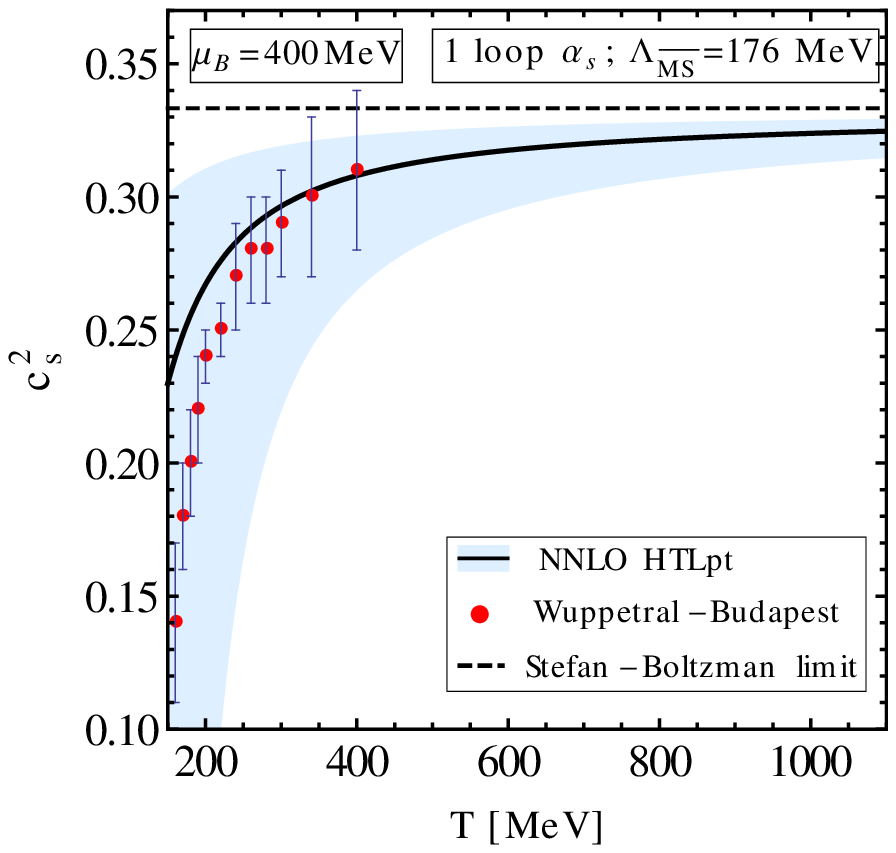}} 
\caption[Comparison of the $N_f=2+1$, $\mu_B=0$ (left) and $\mu_B=400$ MeV (right) NNLO HTLpt 
speed of sound squared with lattice data.]{
Comparison of the $N_f=2+1$, $\mu_B=0$ (left) and $\mu_B=400$ MeV (right) NNLO HTLpt 
speed of sound squared with lattice data.
The $\mu_B=0$ lattice data are from \cite{Borsanyi:2010cj} and the $\mu_B=400$ MeV lattice data are
from \cite{Borsanyi:2012cr}. 
For the HTLpt results a one-loop running coupling constant was used.
}
\label{cssq_1l}
\end{figure}

In Fig.~(\ref{cssq_1l}) we plot the NNLO HTLpt speed of sound for $\mu_B=0$ (left) and $\mu_B=400$ MeV (right)
together with lattice data from Refs.~\cite{Borsanyi:2010cj} and \cite{Borsanyi:2012cr}.  As we can see from this figure, 
there is quite good agreement between the NNLO HTLpt speed of sound and the lattice data when the central value of the scale is used.

\section{Quark number susceptibilities}
\label{sec:qns}

Having the full thermodynamic potential as a function of chemical potential(s) and temperature allows us 
to compute the quark number susceptibilities.  In general, one can introduce a separate chemical potential 
for each quark flavor giving a $N_f$-dimensional vector $\bm{\mu}\equiv(\mu_1,\mu_2,...,\mu_{N_f})$.  
By taking derivatives of the pressure with respect to chemical potentials in this set, we obtain the quark 
number susceptibilities\,\footnote{We have specified that the derivatives should be evaluated at $\bm{\mu}=0$. 
In general, one could define the susceptibilities at $\bm{\mu} = \bm{\mu}_0$.}
\be
\chi_{ijk\,\cdots}\left(T\right)&\equiv& \left. \frac{\partial^{i+j+k+ \, \cdots}\; {\cal P}\left(T,\bm{\mu}\right)}
{\partial\mu_u^i\, \partial\mu_d^j \, \partial\mu_s^k\, \cdots} \right|_{\bm{\mu}=0} \, .
\label{qnsdef}
\ee
Below we will use a shorthand notation for the susceptibilities by specifying derivatives by a string of quark flavors 
in superscript form, e.g. $\chi^{uu}_2 = \chi_{200}$, $\chi^{ds}_2 = \chi_{011}$, $\chi^{uudd}_4 = \chi_{220}$, etc.

When computing the derivatives with respect to the chemical potentials we treat $\Lambda$ as being a constant 
and only put the chemical potential dependence of the $\Lambda$ in after the derivatives are taken.  We have done
this in order to more closely match the procedure used to compute the susceptibilities using resummed dimensional
reduction \cite{Mogliacci:2013mca}.\footnote{One could instead put the chemical potential dependence of the $\Lambda$
in prior to taking the derivatives with respect to the chemical potentials.  If this is done, the central lines obtained
are very close to the ones obtained using the fixed-$\Lambda$ prescription, however, the scale variation typically
increases in this case.}

\subsection{Baryon number susceptibilities}
%
We begin by considering the baryon number susceptibilities.  The $n^{\rm th}$-order baryon number susceptibility is defined as
\be
\chi_B^n(T) \equiv \left.\frac{\partial^n {\cal P}}{\partial \mu_B^n}\right|_{\mu_B=0} \, .
\ee
For a three flavor system consisting of $(u,d,s)$, the baryon number susceptibilities can be related to the quark number 
susceptibilities~\cite{Petreczky:2012rq}
\be
\chi_2^B=\frac{1}{9}\[\chi_2^{uu}+\chi_2^{dd}+\chi_2^{ss}+2\chi_2^{ud}+2\chi_2^{ds}+2\chi_2^{us}\] \, ,
\label{gen_chi2}
\ee
and
\be
\chi_4^B &=& \frac{1}{81}\bigg[\chi_4^{uuuu}+\chi_4^{dddd}+\chi_4^{ssss}+4\chi_4^{uuud}+4\chi_4^{uuus}\nn
&&
\hspace{8mm}+\ 4\chi_4^{dddu}+4\chi_4^{ddds}+4\chi_4^{sssu}+4\chi_4^{sssd}+6\chi_4^{uudd}\nn
&&
\hspace{8mm}+\ 6\chi_4^{ddss}+6\chi_4^{uuss}+12\chi_4^{uuds}+12\chi_4^{ddus}+12\chi_4^{ssud} \bigg] \, .
\hspace{5mm}
\label{gen_chi4}
\ee
If we treat all quarks as having the same chemical potential $\mu_u=\mu_d=\mu_s=\mu=\frac{1}{3}\mu_B$, 
Eqs.~(\ref{gen_chi2}) and (\ref{gen_chi4}) reduce to $\chi_2^B=\chi_2^{uu}$ and $\chi_4^B=\chi_4^{uuuu}$. 
This allows us to straightforwardly compute the baryon number susceptibility by computing derivatives of
(\ref{finalomega1}) with respect to $\mu$.
\begin{figure}[tbh]
\begin{center}
\subfigure{
\hspace{-10mm}
\includegraphics[width=7.20cm]{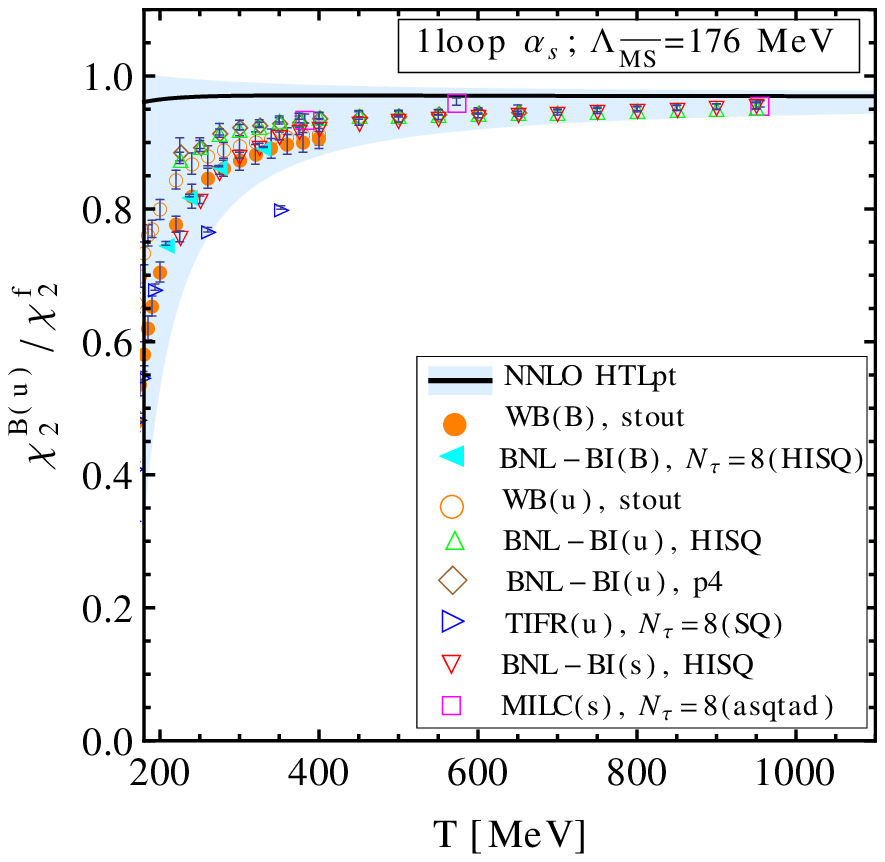}}
\subfigure{
\includegraphics[width=7.20cm]{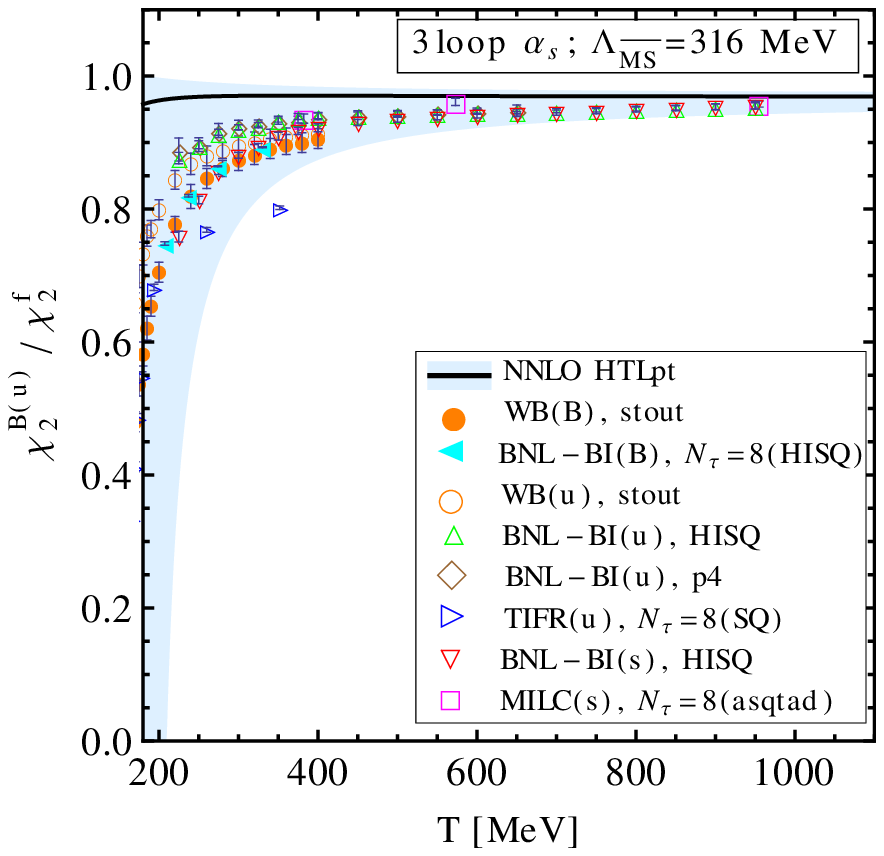}}
\caption[The scaled second order baryon number susceptibility compared with various lattice data using 
one loop running (left) and three-loop running (right).]{
The scaled second order baryon number susceptibility compared with various lattice data using 
one loop running (left) and three-loop running (right). The lattice data labeled WB, BNL-BI(B), 
BNL-BI(u,s), MILC, and TIFR come from Refs.~\cite{Borsanyi:2011sw}, \cite{Bazavov:2013dta}, \cite{Bazavov:2013uja},
\cite{Bernard:2004je}, 
and \cite{Datta:2014zqa}, respectively.
}
\label{qns2_1l}
\end{center}
\end{figure}
In Fig.~(\ref{qns2_1l}) we compare the NNLO HTLpt result for the second order baryon number susceptibility with lattice data
from various groups.  In the left panel of this figure we used the one-loop running and on the right we used the three-loop
running.  As one can see, for this quantity, the size of the light-blue band becomes larger if one uses the three-loop running,
however, the central value obtained is very close in both cases.  

Comparing to the lattice data we see that the NNLO HTLpt prediction is approximately 10\% higher than the lattice data at $T=250$
MeV and approximately 2\% higher at $T = 800$ MeV.  We note in this context that recently the four-loop second-order baryon number
susceptibility has been computed in Ref.~\cite{Mogliacci:2013mca} using the resummed dimensional reduction method.  The result from this 
approach lies within the NNLO HTLpt scale variation band and is even closer to the lattice data with the error at $T=250$ MeV
being approximately 2\% and $\lesssim 1\%$ at  $T = 800$ MeV.  Our result, taken together with the resummed dimensional reduction
results seem to indicate that the quark sector of the QGP can be quite accurately described using resummed perturbation theory 
for temperatures above approximately 300 MeV.
\begin{figure}[tbh]
\begin{center}
 \subfigure{
\hspace{-4mm}
\includegraphics[width=7.20cm]{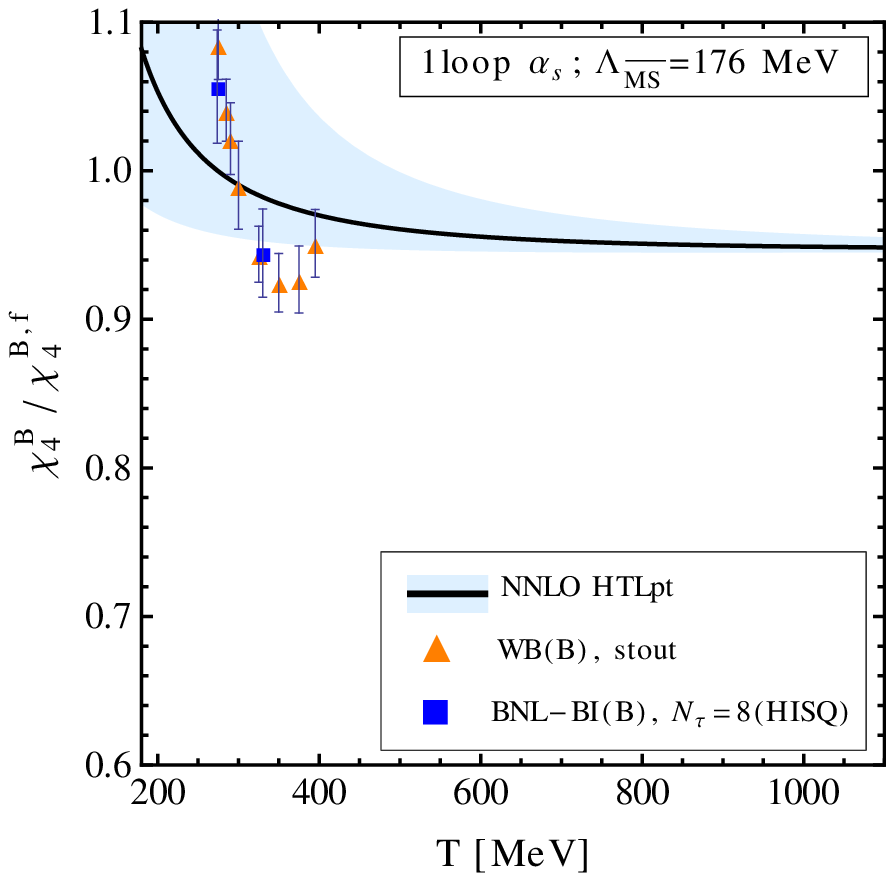}}
\subfigure{
\includegraphics[width=7.20cm]{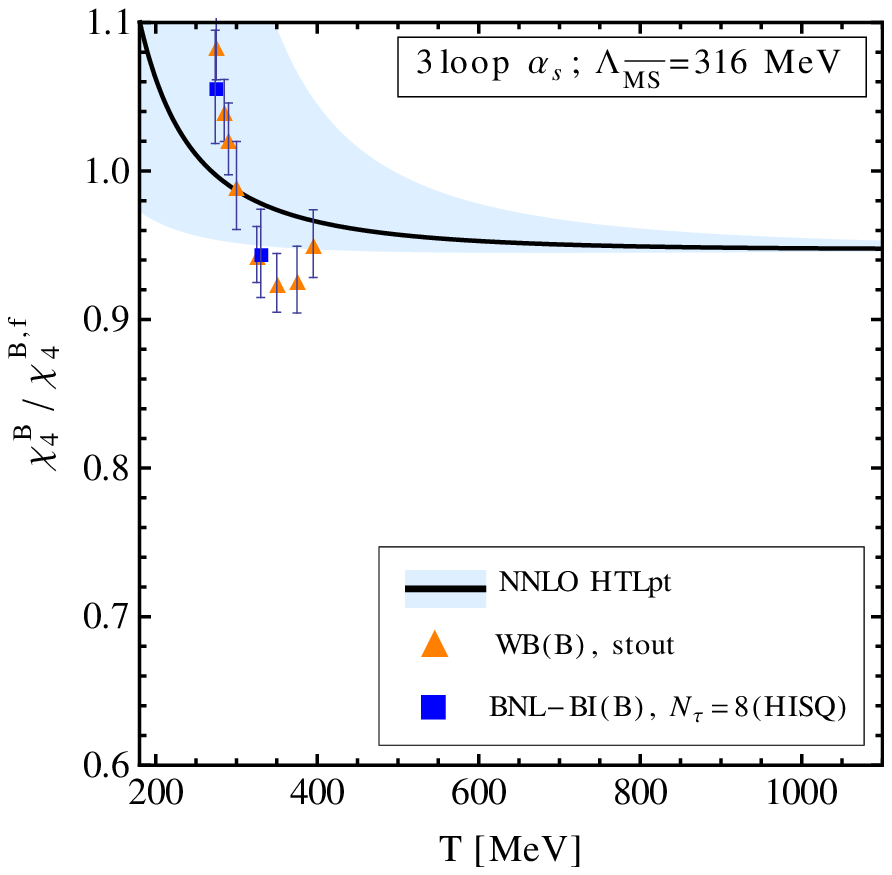}}
\caption[The scaled fourth order baryon number susceptibility compared with various lattice data using one loop running (left)
and three-loop running (right).]{
The scaled fourth order baryon number susceptibility compared with various lattice data using one loop running (left)
and three-loop running (right). The lattice data labeled WB, BNL-BI(B), BNL-BI(u,s), MILC, and TIFR come from 
Refs.~\cite{Borsanyi:2011sw}, \cite{Bazavov:2013dta}, \cite{Bazavov:2013uja}, \cite{Bernard:2004je}, and \cite{Datta:2014zqa},
respectively.}
\label{qns4_1l}
\end{center}
\end{figure}
In Fig.~(\ref{qns4_1l}) we compare the NNLO HTLpt result for the fourth order baryon number susceptibility with lattice data.
Once again we show in the left and right panels, the result obtain using the one-loop running coupling and three-loop running 
coupling, respectively.  Both the one- and three-loop running results are consistent with the lattice data shown; however, the 
lattice error bars on this quantity are somewhat large and the data are restricted to temperatures below 400 MeV, making it 
difficult to draw firm conclusions from this comparison.  That being said, HTLpt makes a clear prediction for the temperature 
dependence of the fourth order baryon number susceptibility.  It will be very interesting to see if future lattice data agree 
with this prediction.
\begin{figure}[tbh!]
\centerline{\includegraphics[width=8.5cm]{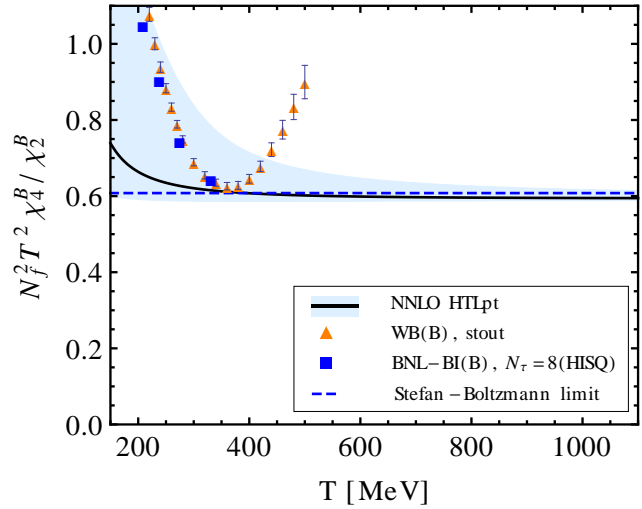}}
\caption[Comparison of the $N_f=2+1$ NNLO HTLpt ratio of the fourth to second order baryon susceptibility with lattice data.]{
Comparison of the $N_f=2+1$ NNLO HTLpt ratio of the fourth to second order baryon susceptibility with lattice data.  
For the HTLpt results a one-loop running coupling constant was used.  
The data labeled WB and BNL-BI(B) come from Refs.~\cite{Borsanyi:2012rr,Borsanyi:2013hza} and \cite{Bazavov:2013dta}, respectively.
}
\label{qnsrat_1l}
\end{figure}

In Fig.~(\ref{qnsrat_1l}) we plot the scaled ratio of the fourth and second order baryon number susceptibilities as a function 
of temperature along with lattice data for this ratio.  As we can see from
this figure, this ratio very rapidly approaches the Stefan-Boltzmann limit if one considers the central NNLO HTLpt line.  
Comparing with the lattice data we see that the NNLO HTLpt result is below the lattice data for temperatures less than approximately 
300 MeV.  Without lattice data at higher temperatures, it's hard to draw a firm conclusion regarding the temperature at which HTLpt 
provides a good description of this quantity.
\begin{figure}[tbh]
\centerline{\includegraphics[width=8.5cm]{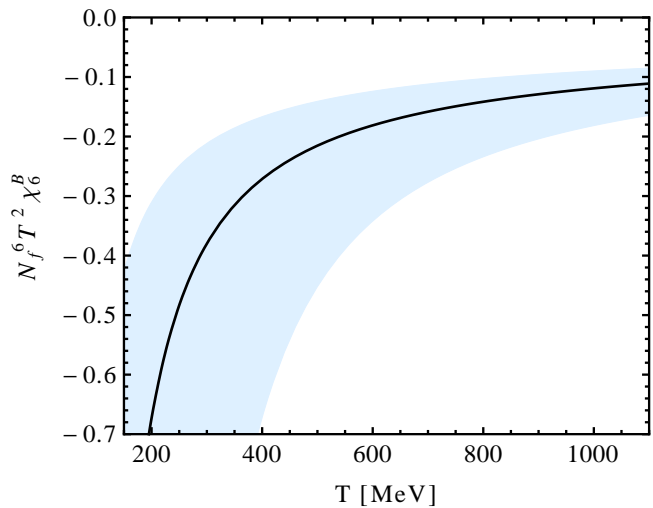}}
\caption{
The $N_f=2+1$ NNLO HTLpt scaled sixth-order baryon susceptibility as a function of temperature.
}
\label{qns6_1l}
\end{figure}
In Fig.~(\ref{qns6_1l}) we show the NNLO HTLpt prediction for the sixth order baryon number susceptibility.
To the best of our knowledge there is currently no publicly available lattice data for this quantity. 
It will be very interesting to see if these NNLO HTLpt predictions agree with lattice data as they becomes available.
\subsection{Single quark number susceptibilities}
\begin{figure}[tbh]
\subfigure{
 \includegraphics[width=7.2cm]{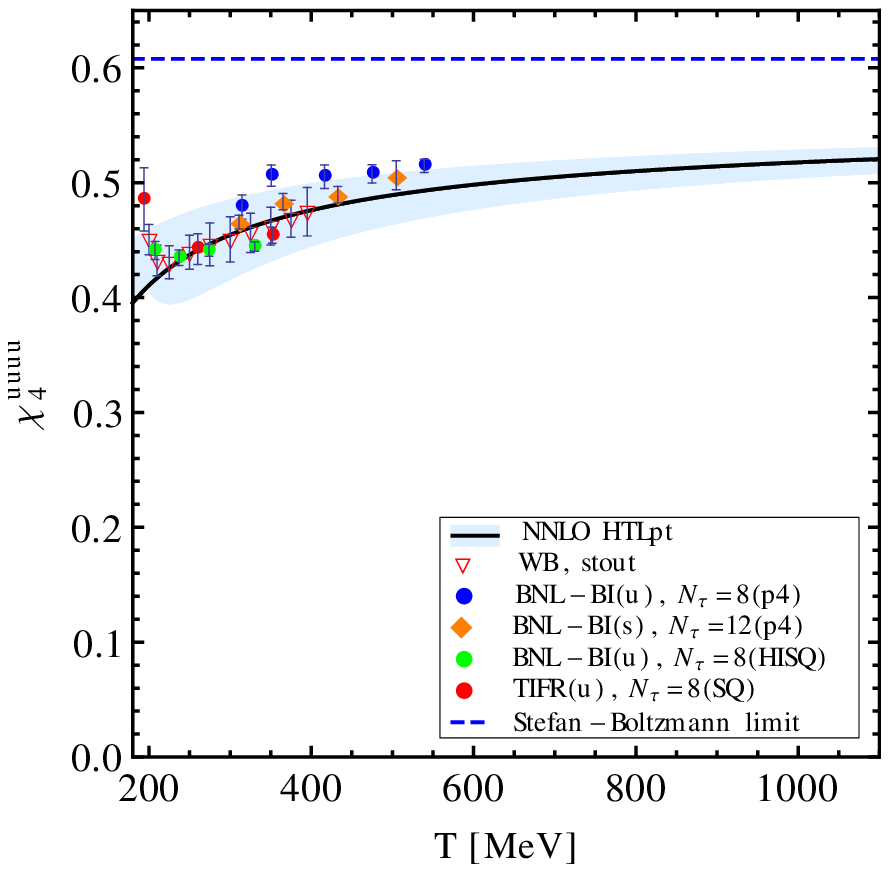}}
\subfigure{
\includegraphics[width=7.2cm]{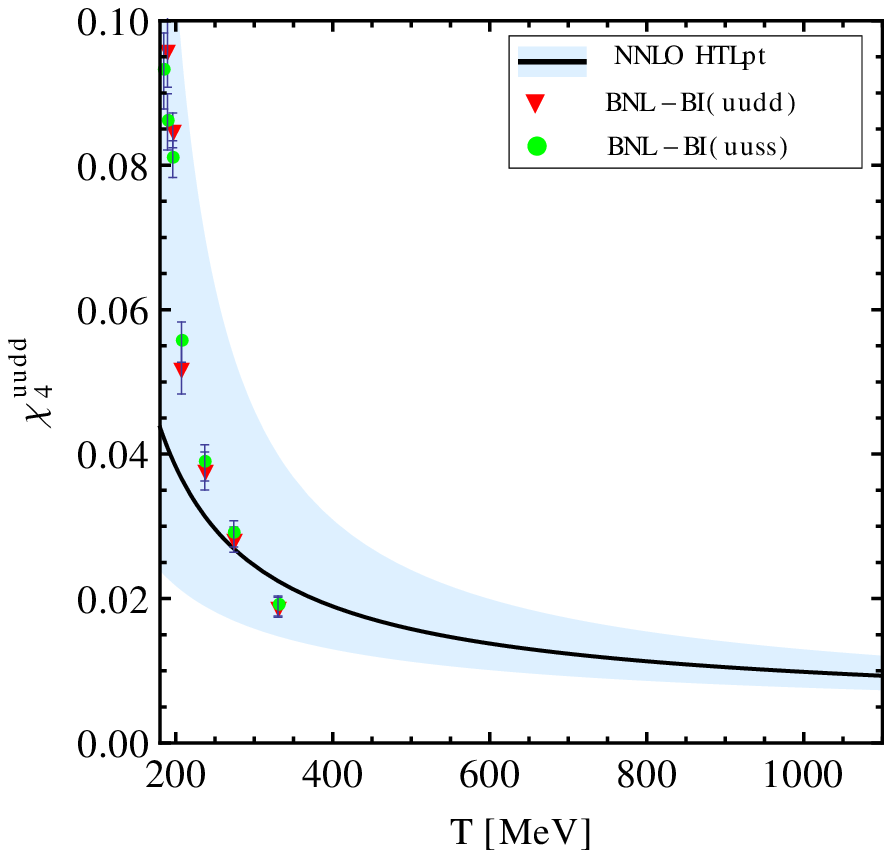}}
\caption[Comparison of the $N_f=2+1$ NNLO HTLpt ratio of the fourth order diagonal single quark number susceptibility 
(left) and the only non-vanishing fourth order off-diagonal quark number susceptibility (right) with lattice data.]{
Comparison of the $N_f=2+1$ NNLO HTLpt ratio of the fourth order diagonal single quark number susceptibility 
(left) and the only non-vanishing fourth order off-diagonal quark number susceptibility (right) with lattice data.
In the left figure the dashed blue line indicates the Stefan-Boltzmann limit for this quantity.  For the HTLpt 
results a one-loop running coupling constant was used.  The data labeled BNL-BI(uudd), BNL-BI(u,s), BNL-BI(uuss),
and TIFR come from Refs.~\cite{Bazavov:2013dta}, \cite{Bazavov:2013uja}, \cite{Bazavov:2012vg}, and \cite{Datta:2014zqa},
respectively.}
\label{figsingleq}
\end{figure}
\begin{figure}[tbh]
\vspace{3mm}
\centerline{\includegraphics[width=8.5cm]{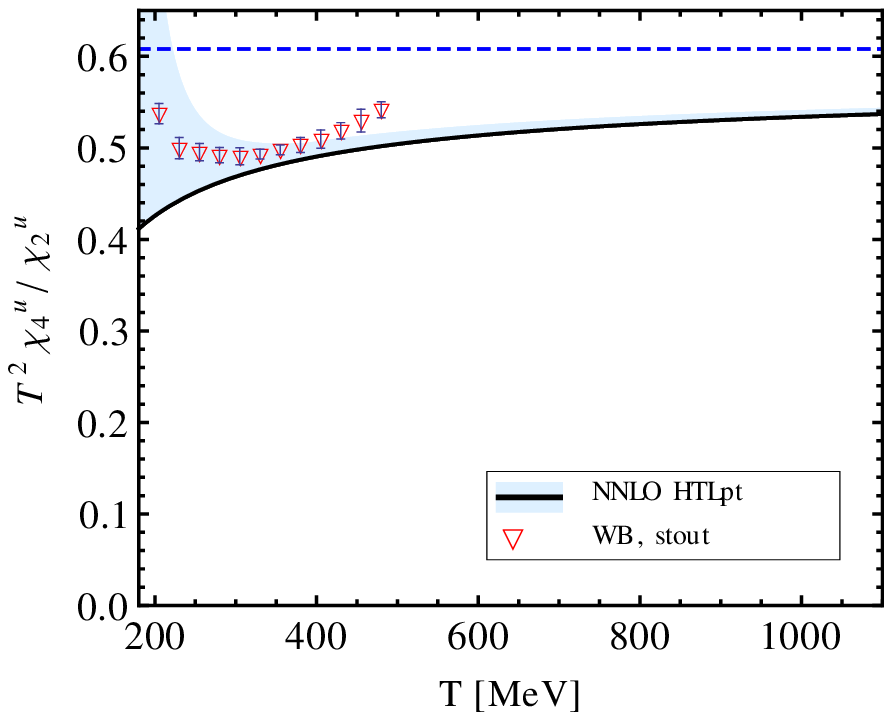}}
\caption[Comparison of the $N_f=2+1$ NNLO HTLpt ratio of the fourth to second order single quark susceptibility with lattice data.]
{Comparison of the $N_f=2+1$ NNLO HTLpt ratio of the fourth to second order single quark susceptibility with lattice data.  
For the HTLpt results a one-loop running coupling constant was used. The data labeled WB come from Refs.~\cite{Borsanyi:2012rr,
Borsanyi:2013hza}.}
\label{qnsrat2_1l}
\end{figure}
We now consider the single quark number susceptibilities (\ref{qnsdef}).  For these we use the general expression for the 
NNLO thermodynamic potential with different quark chemical potentials (\ref{finalomega}). The resulting susceptibilities
can either be diagonal (same flavor on all derivatives) or off-diagonal (different flavor on some or all indices).
In HTLpt there are off-diagonal susceptibilities emerging explicitly from graphs ${\cal F}_{3c}^f$ 
and ${\cal F}_{3j}^f$; however, the latter  vanishes when we use the variational mass prescription for the
quark mass ($m_q=0$), so we need only consider the ${\cal F}_{3c}^f$ graph.  Additionally, there are potential off-diagonal
contributions coming from all HTL terms since the Debye mass receives contributions from all quark flavors.  In practice,
however, because we evaluate derivatives with respect to the various chemical potentials and then take $\mu_i \rightarrow 0$,
one finds that all off-diagonal second order susceptibilities vanish in HTLpt.  Therefore, for the three-flavor case one has
\be
\chi_2^{ud}=\chi_2^{ds}=\chi_2^{su}=0 \, ,
\ee
and, as a result, the single quark second order susceptibility is proportional to the baryon number susceptibility
\be
\chi_2^{uu}=\frac{1}{3}\chi_2^B.
\ee
For the fourth order susceptibility, there is only one non-zero off-diagonal susceptibility,
namely $\chi_4^{uudd}=\chi_4^{uuss}=\chi_4^{ddss}$, which is related to the diagonal susceptibility,
e.g. $\chi_4^{uuuu}=\chi_4^{dddd}=\chi_4^{ssss}$, as 
\be
\chi_4^{uuuu}=27\chi_4^B-6\chi_4^{uudd}.
\label{u4rel}
\ee
As a consequence, one can compute $\chi_4^{uuuu}$ directly from (\ref{finalomega}) or by computing $\chi_4^B$ using
(\ref{finalomega1}) and $\chi_4^{uudd}$ using (\ref{finalomega}) and applying the above relation.  
In our final plots we compute $\chi_4^{uuuu}$ directly from (\ref{finalomega}), however, we have 
checked that we obtain the same result if we use (\ref{u4rel}) instead.

In Fig.~(\ref{figsingleq}) (left) we plot our result for the fourth order single quark susceptibility 
$\chi_4^{uuuu}$ compared to lattice data from Refs.~\cite{Bazavov:2013uja}, \cite{Bazavov:2013dta}, \cite{Bazavov:2012vg},
and \cite{Datta:2014zqa}.
  As we can see from this figure, for the fourth order susceptibility there is very good agreement with available lattice data.  
In addition, the scale variation of the HTLpt result is quite small for this particular quantity.  In Fig.~(\ref{figsingleq}) 
(right) we plot our result for the fourth order off-diagonal single quark susceptibility $\chi_4^{uudd}$ compared to lattice data. 
From this right panel we also see reasonably good agreement between the NNLO HTLpt result and the available lattice data.  

In Fig.~(\ref{qnsrat2_1l}) we plot the scaled ratio of the fourth- and second-order single quark susceptibilities.  
Once again we see good agreement between the NNLO HTLpt result and lattice data.  Once again, for both Figs.~(\ref{figsingleq}) 
and (\ref{qnsrat2_1l}), the lattice data are confined to relatively low temperatures. It will be interesting to compare higher 
temperature lattice data with the NNLO HTLpt prediction as they become available.

\begin{figure}[tbh]
\begin{center}
\includegraphics[width=10cm]{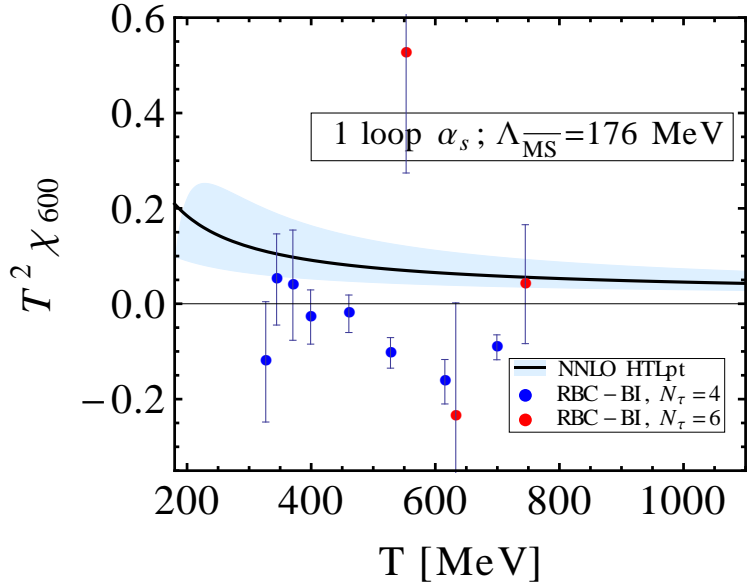}
\end{center}
\caption[
The $N_f=2+1$ NNLO HTLpt scaled sixth-order diagonal single quark susceptibility as a function of temperature.]{
The $N_f=2+1$ NNLO HTLpt scaled sixth-order diagonal single quark susceptibility $\chi_{600}$ as a function of temperature.
}
\label{qns6single_1l}
\end{figure}

Finally, in Fig.~(\ref{qns6single_1l})  we plot the diagonal  sixth-order quark number susceptibilities 
$\chi_{600}$ with available lattice data. In this figure we show lattice data available from the RBC-Bielefeld
collaboration~\cite{Petreczky:2009cr}. At this point in time the lattice sizes are small and the errors bars 
for $\chi_{600}$ are large, so it is hard to draw a firm conclusion from this comparison.
\begin{figure}[tbh]
\includegraphics[width=7.35cm]{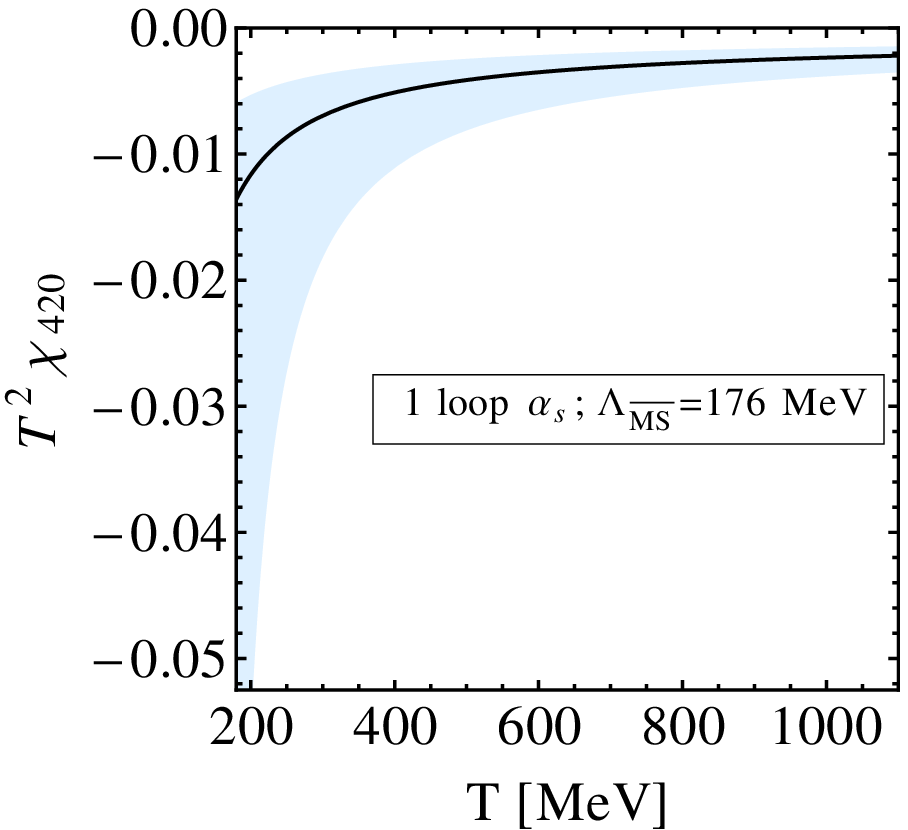}
\includegraphics[width=7.65cm]{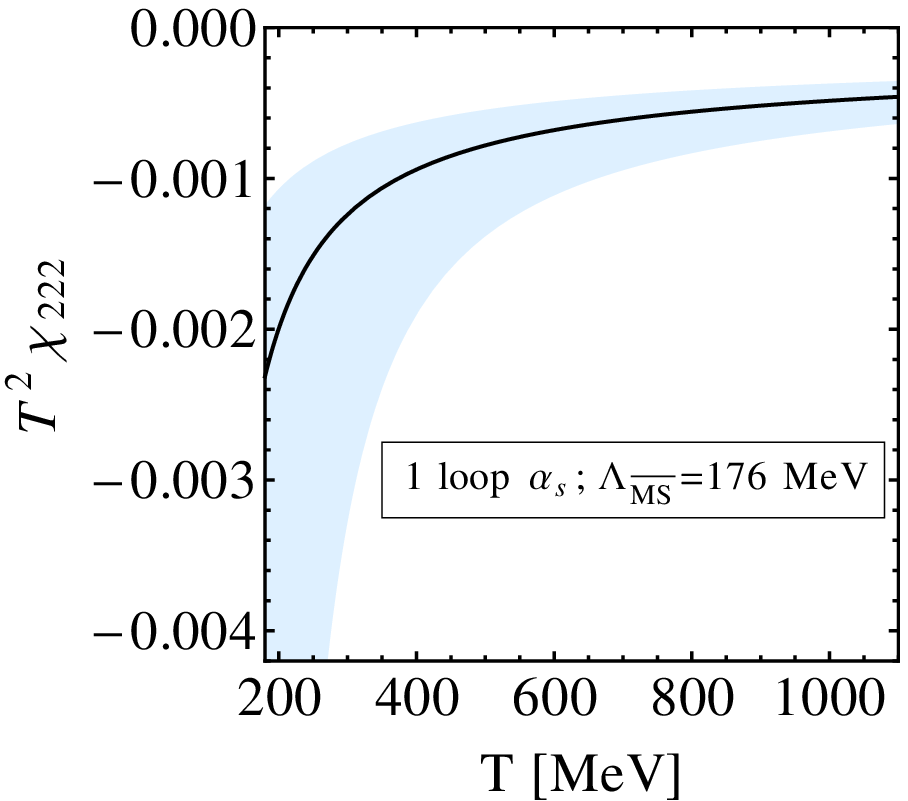}
\caption[The $N_f=2+1$ NNLO HTLpt scaled sixth-order off-diagonal single quark susceptibilities as a function 
of temperature.]{The $N_f=2+1$ NNLO HTLpt scaled sixth-order off-diagonal single quark susceptibilities $\chi_{420}$ (left),
and $\chi_{222}$ (right) as a function of temperature.
}
\label{qns6single_off}
\end{figure}

In Fig.~(\ref{qns6single_off}) we plot off-diagonal sixth-order quark number susceptibilities $\chi_{420}$ (left),
and $\chi_{222}$ (right). We are unaware of any lattice data for off-diagonal susceptibilities.
As before, it will be very interesting to see if these NNLO HTLpt predictions agree with lattice data as they becomes available.

\section{Conclusions and outlook}
\label{outlook}

In this chapter, we presented the results of a NNLO (three-loop) HTLpt calculation of the thermodynamic potential of QCD at
finite temperature and chemical potential(s).  Our final result (\ref{finalomega}) is completely analytic and 
should be valid in the region of the phase diagram for which $\mu_i \lesssim 2 \pi T$.  Based on the resulting thermodynamic potential 
we proceeded to calculate the pressure, energy density, entropy density, trace anomaly, and speed of sound of the QGP.  In all cases we 
found very good agreement between the results obtained using the central values of the renormalization scales and available lattice data.
Additionally, we have made predictions for the diagonal and off-diagonal sixth-order baryon number and single quark susceptibilities.

Looking to the future there are still many avenues for improvement in the HTLpt approach:  (1) inclusion of the effects of
finite quark masses (2) extension of results to $\mu_i \gtrsim 2 \pi T$ and eventually to $T=0$, and (3) to potentially 
resum logarithms in order to reduce the scale variation of the final results (light-blue bands in all figures).  Of these
three, the second task is the most straightforward; however, in order to make more definitive and constrained statements 
it now seems necessary to start moving in directions (1) and (3) as well. In closing, we emphasize that HTLpt provides a
gauge invariant reorganization of  perturbation theory for calculating static quantities in thermal field theory.
Since the NNLO HTLpt results are in good agreement with lattice data for various thermodynamic quantities down to temperatures
that are relevant for LHC, it would therefore be interesting and challenging to apply HTLpt to the calculation of dynamic
quantities, especially transport coefficients, at these temperatures.



%% file: text/dilepton.tex
\chapter{Dileption Production Rate}
\label{chapter:dilepton}
In this chapter we study the low mass dilepton rate from deconfined state of matter
using both perturbative and non-perturbative method. This chapter is based on: 
{\it Low Mass Dilepton Rate from the Deconfined Phase},
C.~Greiner, N.~Haque, M.~G.~Mustafa and M.~H.~Thoma, {\bf Phys.Rev. C83 (2011) 014908}.
\section{Introduction}
The prime intention for ultra relativistic heavy-ion collisions is to study
the behavior of nuclear or hadronic matter at extreme conditions like very
high temperatures and/or high densities. A particular goal lies in the
identification of a new state of matter formed in such collisions, the
quark-gluon plasma (QGP), where the quarks and gluons are deliberated from 
the nucleons and move freely over an extended space-time region. Various
measurements taken in CERN-SPS~\cite{Heinz:2000bk} and 
BNL-RHIC~\cite{Arsene:2004fa,Adcox:2004mh,Back:2004je,Adams:2005dq,
Adare:2009qk,Adler:2006yt,Adare:2006ti,Adcox:2002au,Chujo:2002bi} do lead to
{\textquoteleft  circumstantial evidence\textquoteright} for the formation of QGP. Evidence is (or can only be) 
{\textquoteleft circumstantial\textquoteright}  because only indirect diagnostic probes exist. 

Electromagnetic probes, such as real photon and dileptons, are a 
particular example, and accordingly thermal dileptons have been 
theoretically proposed long time ago~\cite{McLerran:1984ay}. At SPS
energies~\cite{Agakishiev:1995xb,Agakishiev:1997au,Masera:1995ck,Drees:1998rn}
there was an indication for an enhancement of the dilepton 
production at low invariant mass ($0.2\le M(\mbox{GeV}) \le 0.8$ ) compared 
to all known sources of electromagnetic decay of the hadronic particles 
and the contribution of a radiating simple hadronic fireball (for
comprehensive reviews see Refs.~\cite{Rapp:1999ej,Rapp:2009yu,Cassing:1999es}). 
One of the possible explanations of this is the modification of 
the in-medium properties of the vector meson ({\em viz.,} $\rho$-meson) by
rescattering in a hadronic phase along with only the lowest order perturbative
rate, {\em i.e.}, $q\bar q$ annihilation from a
QGP~\cite{Rapp:1999ej,Rapp:2009yu,Cassing:1999es,Brown:1991kk,Friman:1997tc,
Rapp:1997fs,Rapp:1995zy,Gale:1993zj,Rapp:1999qu,Klingl:1997kf,Peters:1997va,
Cassing:1997jz,Post:2000qi,Srivastava:1996vi,Pal:1996xn,Srivastava:1996wr,Pal:1998jr,
Srivastava:1996qd,Alam:1999sc,Alam:2001ar}. Also at RHIC energies~\cite{Adare:2009qk}
a substantial amount of excess of electron pairs was reported in the
low invariant mass region. Models taking into account
in-medium properties of hadrons with various ingredients 
(see for details~\cite{Dusling:2007su,Bratkovskaya:2008bf}) can not explain the data from RHIC in the
range $0.15 \le M (\mbox{GeV})\le 0.5$, whereas they fit the SPS data
more satisfactorily, indicating that a possible non-hadronic source
becomes important at RHIC.
 
On the other hand, the higher order perturbative calculations~\cite{Aurenche:1998nw} 
are also not very reliable at temperatures 
within the reach of the heavy-ion collisions. Moreover, perturbative 
calculations of the dilepton rate seem not to converge even in 
small coupling ($g$) limit.  Nevertheless, the lowest order perturbative 
$q\bar q$ annihilation is the only dilepton rate from the QGP phase that
is extensively used in the literatures. However, at large invariant
mass this contribution should be dominant but not at low invariant 
mass, where nonperturbative effects should play an important role. 
Unfortunately, the lattice data~\cite{Karsch:2001uw} due to its limitations 
also could not shed any light on the low mass dileptons.  However, the 
lattice calculations~\cite{Boyd:1996bx,Allton:2003vx,Allton:2005gk,
Bazavov:2009zn,Bernard:2004je,Petreczky:2009at,Petreczky:2010xg} provide evidence for 
the existence of nonperturbative effects associated with the bulk 
properties of the deconfined phase, in and around the deconfined 
temperature, $T_c$. Also, indications have been found that the 
QGP at RHIC energies behaves more as a strongly coupled liquid 
than a weakly coupled gas~\cite{Thoma:2004sp}. Thus,  a nonperturbative 
analysis of the dilepton rate from the deconfined phase is essential. 

The dilepton emission at low invariant mass from the deconfined phase is 
still an unsettled issue in heavy-ion collisions at SPS and RHIC energies 
and, in particular, would be an important question for LHC energies and 
for compact baryonic matter formation in future FAIR energies, 
and also for the quark-hadron duality~\cite{Rapp:1999ej,Rapp:2009yu,Rapp:1999us} that 
entails a reminiscence to a simple perturbative lowest order quark-antiquark 
annihilation rate~\cite{Cleymans:1986na}. In this chapter we reconsider the dilepton 
production rates within the perturbative QCD, and non-perturbative models 
based on lattice inputs and phenomenological
$\rho - q$ interaction in the deconfined phase. The analysis suggests that 
the nonperturbative dilepton rates are indeed important at the low 
invariant mass regime. 

This chapter is organised in following way.
In section~(\ref{dilep_rate}) we discuss the dilepton production rate from the deconfined phase
based on both perturbative and non-perturbative models. In section~(\ref{mom_int_rate})
we compare the momentum integrated rates from both QGP and Hadron gas (HG).
We discuss the quark-hadron duality in section~(\ref{qh_duality}), and conclude in section~(\ref{conclusion_dilepton}).

\section{Dilepton Rate From Deconfined Phase} 
\label{dilep_rate}
The dilepton production rate is related with the imaginary part of
the photon self-energy~\cite{McLerran:1984ay,Gale:1990pn} as

\begin{equation}
\frac{dR}{d^4xd^4P} = -\frac{\alpha_{\rm em}}{12\pi^4} \frac{1}{e^{E/T}-1} \ 
\frac {\mbox{Im} \Pi_\mu^\mu(P)}{M^2} , \label{dilep_static}
\end{equation}
where $\alpha_{\rm em }=e^2/4\pi$ is fine structure constant and $P$ is four momentum of the virtual photon, 
$E$ is its energy, and we use
the notation $P\equiv (p_0=E,{\vec {\mathbf p}})$ and $p=|{\vec {\mathbf p}}|$. 
The square of the invariant mass of dilepton pair is $M^2=p_0^2-p^2$.  

\subsection{Born Rate }
To the lowest order the dilepton rate follows from one-loop photon self
energy containing bare quark propagators. This rate corresponds to
a dilepton production by the annihilation of bare quarks and antiquarks
of the QGP. Alternatively, this so-called Born-rate can also be obtained 
from the matrix element of the basic annihilation process folded with the
thermal distribution functions of quarks. In the case of massless lepton 
pairs in a QGP with two massless quark flavors with chemical potential 
one finds~\cite{Cleymans:1986na} 
\begin{equation}
\frac{dR}{d^4xd^4P}= \frac{5\alpha_{\rm em}^2}{36\pi^4} \frac{T}{p} \frac{1}{e^{\beta E}-1}
\ln \frac{\left(x_2+\exp[-\beta(E+\mu)]\right)\left(x_1+
\exp[-\beta\mu]\right )}
{\left(x_1+\exp[-\beta(E+\mu)]\right)\left(x_2+
\exp[-\beta\mu]\right )} \ , \label{born_gen}
\end{equation}
where $\beta=1/T$, $x_1=\exp[-\beta(E+p)/2]$ and $x_2=\exp[-\beta(E-p)/2]$. A finite quark mass 
can easily be included.

For $\mu=0$ the dilepton rate becomes
\begin{equation}
\frac{dR}{d^4xd^4P}= \frac{5\alpha_{\rm em}^2}{18\pi^4} \frac{T}{p} \frac{1}{e^{E/T}-1}
\ln \left (\frac{\cosh\frac{E+p}{4T}}{\cosh\frac{E-p}{4T}}\right) \ , 
\label{rate_born}
\end{equation}
whereas that for total three momentum ${\vec {\mathbf p}}=0$ is given as
\begin{equation}
\frac{dR}{d^4xd^4P}= \frac{5\alpha_{\rm em}^2}{36\pi^4}\ n(E/2-\mu) \ n(E/2+\mu)
\ , \label{born_mup0}
\end{equation}
with $n(y)=(\exp(\beta y)+1)^{-1}$, the Fermi-Dirac distribution function.

\subsection{Rate using Hard Thermal Loop perturbation theory}\label{sec_htlrate}
In order to judge the reliability of the lowest order result, one should 
consider higher order corrections. These corrections involve quarks and gluons 
in the photon self energy beyond the one-loop approximation. Using bare 
propagators at finite temperature, however, one encounters infrared 
singularities and gauge dependent results. These problems can be resolved, at
least partially, by adopting the Hard-Thermal Loop (HTL) resummation 
scheme~\cite{Braaten:1989mz,Braaten:1989kk}.
The key point of this method is the distinction between the soft momentum
scale ($\sim gT$) and the hard one ($\sim T$), which is possible in the weak
coupling limit ($g\ll1$). Resumming one-loop self energies, in which the loop
momenta are hard (HTL approximation), effective propagators and vertices 
are constructed, which are as important as bare propagators if the momentum 
of the quark or gluon is soft. In HTLpt the bare $N$-point functions (propagator and 
vertices) are replaced by those effective $N$-point HTL functions which 
describe medium effects in the QGP such as the thermal masses for quarks 
and gluons and Landau damping.

The importance of the medium and other higher order effects on the dilepton
rate depends crucially on the invariant mass and the momenta of the virtual
photon. Therefore, we will discuss now the different kinematical regimes:

\begin{figure}[!tbh]
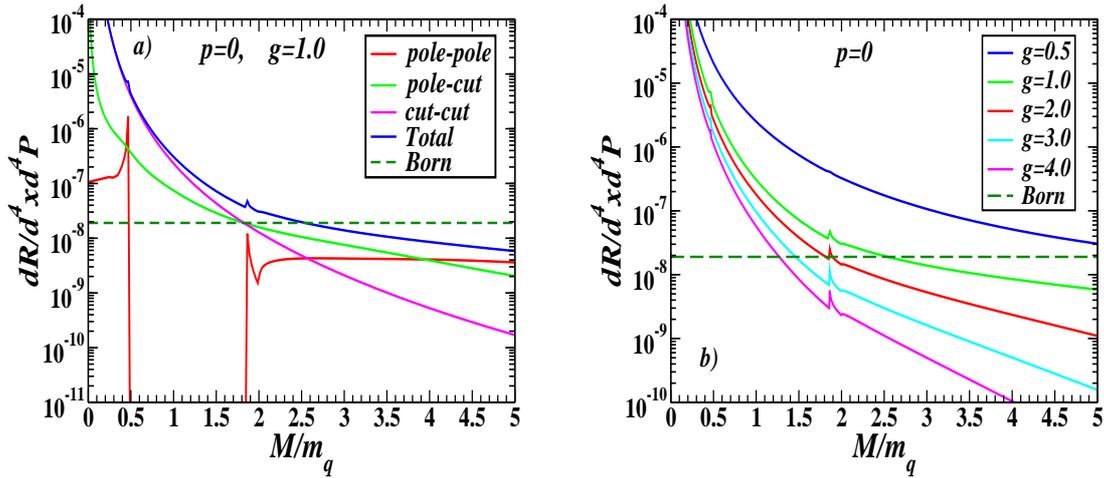

\vspace*{0.40in}
\subfigure{
\includegraphics[height=0.42\textwidth, width=0.45\textwidth]{figures/dilepton/rate_1loop_g1.eps}}
\hspace{.5cm}
\subfigure{
\includegraphics[height=0.42\textwidth, width=0.45\textwidth]{figures/dilepton/rate_1loop.eps}}
\vspace*{-0.08in}
\caption[One-loop dilepton rate from HTLpt for small
 invariant masses  at zero momentum alongwith born-rate 
versus the scaled invariant photon mass]{{\em Left panel (a):} 1-loop dilepton rate for small
 invariant masses $M\sim gT$ at zero momentum and Born-rate (dashed line) 
versus the scaled invariant photon mass $M/m_q$ for $g=1$. The van Hove peaks 
and energy gap are evident in the 1-loop rate. {\em Right panel (b):} Total
1-loop rate for various $g$ values.
}
\label{fig_1loop}
\end{figure}

\subsubsection*{Soft Rate ($M\sim gT$ and $p\sim gT$)}
For soft invariant masses\footnote{Note that for ultrasoft $M\sim g^2T$ and
arbitrary momentum the rate is non-perturbative and cannot be calculated even 
within the HTL improved perturbation theory. This observation holds in 
particular for real hard photon~\cite{Aurenche:1999tq}.} and momenta of order $gT$ 
one has to use HTL quark propagators and vertices in the one-loop photon 
self energy. These corrections are of same order as the Born-term~\cite{Braaten:1990wp}. 
Physically these corrections correspond to two different processes. First 
the poles of the HTL resummed quark
propagators describe quasiparticles in the QGP with an effective thermal 
quark mass of the order of $gT$. Hence dileptons are generated by the 
annihilation of collective quark modes instead of bare quarks. In particular
the HTL quark dispersion contains a so called plasmino branch which exhibits
a minimum at finite momentum. This nontrivial dispersion leads to sharp
structures (van Hove singularities and energy gap) in the dilepton production 
rate\footnote{For a discussion of van Hove singularities in the QGP at 
${\vec {\mathbf p}}=0$ see Refs.~\cite{Braaten:1990wp,Mustafa:2002pb,Peshier:1999dt,Karsch:2000gi} and also 
Ref.~\cite{Wong:1991be} for 
${\vec {\mathbf p}}\ne 0$.} in contrast to smooth Born-rate.
Secondly, the imaginary part of the HTL quark self energy containing 
effective HTL $N$-point (propagators and quark-photon vertex) functions 
corresponds to processes involving the absorption or emission of thermal 
gluons. 

The 1-loop dilepton rate in HTL approximation for zero momentum can be given from\cite{Braaten:1990wp} as
\be
\frac{dR}{d^4xd^4P}&=&\frac{10\alpha_{\rm em}^2}{9\pi^4E^2}\int\limits_0^\infty k^2 dk\int\limits_{-\infty}^\infty\frac{dk_0}{2\pi}
\int\limits_{-\infty}^\infty\frac{dk_0'}{2\pi} n(k_0)n(k_0')\nn
&\times&\delta(E-k_0-k_0')\Bigg[4\lb1-\frac{k_0^2-k_0'^2}{2kE}\rb^2\rho_+(k_0,k)\rho_-(k_0',k)\nn
&+&\lb1+\frac{k_0^2+k_0'^2-2k^2-2m_q^2}{2kE}\rb^2\rho_+(k_0,k)\rho_+(k_0',k)\nn
&+&\lb1-\frac{k_0^2+k_0'^2-2k^2-2m_q^2}{2kE}\rb^2\rho_-(k_0,k)\rho_-(k_0',k)\nonumber
\ee
\be
&+&\Theta(k^2-k_0^2)\frac{m_q^2}{4kE^2}\lb1-\frac{k_0^2}{k^2}\rb\bigg[\lb1+\frac{k_0}{k}\rb\rho_+(k_0',k)\nn
&+&\lb1-\frac{k_0}{k}\rb\rho_-(k_0',k)\bigg]\Bigg]
\label{rate_1loop}
\ee
where $\rho_\pm(k_0',k)$ is spectral functions of quark propagator $D_\pm((k_0',k))$ which is defined in Eqn.~(\ref{Dpm})
of chapter~(\ref{chapter:htl}).

In Fig.~(\ref{fig_1loop}) the 1-loop dilepton rate for zero momentum, 
containing such processes using Eq.~(\ref{rate_1loop}), is displayed as a function of the scaled invariant 
mass with the thermal quark mass and is also compared with the Born-rate. 
In the left panel of Fig.~(\ref{fig_1loop}$a$) the van Hove singularities 
due to the nontrivial dispersion 
of quarks in a medium are evident in pole-pole contributions whereas the
pole-cut and cut-cut contributions\footnote{These are due to the space-like
($k^2>k_0^2$) part of the $N$-point HTL functions that acquire a cut 
contribution from below the light cone.} are smooth representing absorption 
and emission of gluons in the medium. The right panel of
Fig.~(\ref{fig_1loop}$b$) displays the total one-loop contribution 
for a set of values of $g$, where the energy gaps are smoothened due to
the pole-cut and cut-cut contributions. Also the structures due to
the van Hove singularities become also less prominent in the total 
contributions. The HTL rate, in particular, due to the cut contributions 
is also singular at $M\rightarrow 0$ because the
HTL quark-photon vertex is inversely proportional to photon energy.

However, these corrections are not sufficient and two-loop diagrams within 
HTL perturbation scheme contribute to the same order and are even larger 
than the one-loop results~\cite{Aurenche:1998nw}. The total one-
and two-loop rate at ${\vec{\mathbf p}}=0$ and $M\ll T$ in the leading 
logarithm, {\em i.e.}, $\ln(1/g)$ approximation reads~\cite{Aurenche:1998nw,Aurenche:1999ec}
\begin{eqnarray}
\frac{dR}{d^4xd^4P}&=&\frac{5\alpha_{\rm em}^2}{9\pi^6}\frac{m_q^2}{M^2} \Bigg [
\frac{\pi^2m_q^2}{4M^2}\ln \frac{T^2}{m_q^2}+ \frac{3m_q^2}{M^2}
\ln\frac{T^2}{m_g^2}+\frac{\pi^2}{4}\ln \left (\frac{MT}{M^2+m_q^2} \right ) 
\nonumber\\
&&\hspace{5cm}+2\ln \left (\frac{MT}{M^2+m_g^2} \right ) 
\Bigg ] \ , \label{rate_2loop}
\end{eqnarray}
where the thermal gluon mass is given by $m_g^2=8m_q^2/3$ with 
$m_q=gT/\sqrt 6$. It is worth to mention here that the two-loop rate in~\ref{rate_2loop} is not complete,
they have consider only gluon in two-loop photon self energy diagram as soft but other fermions as hard.
To compute complete two-loop dilepton rate in HTL perturbation theory, one need to consider all the loop momenta as
effective. Note that this expression is of the same order in $g$ as 
the Born-term for soft $M\sim gT$. Now the Born-term for ${\vec{\mathbf p}}=0$ and $M\ll T$ 
is simply given by 
\begin{equation}
\frac{dR}{d^4xd^4P}=\frac{5\alpha_{\rm em}^2}{144\pi^4}=1.90\times 10^{-8} \ .
\label{born_p0}
\end{equation}

 \begin{figure}[!tbh]
\vspace*{-.0cm}\begin{center}
\includegraphics[height=0.5\textwidth, width=0.7\textwidth]{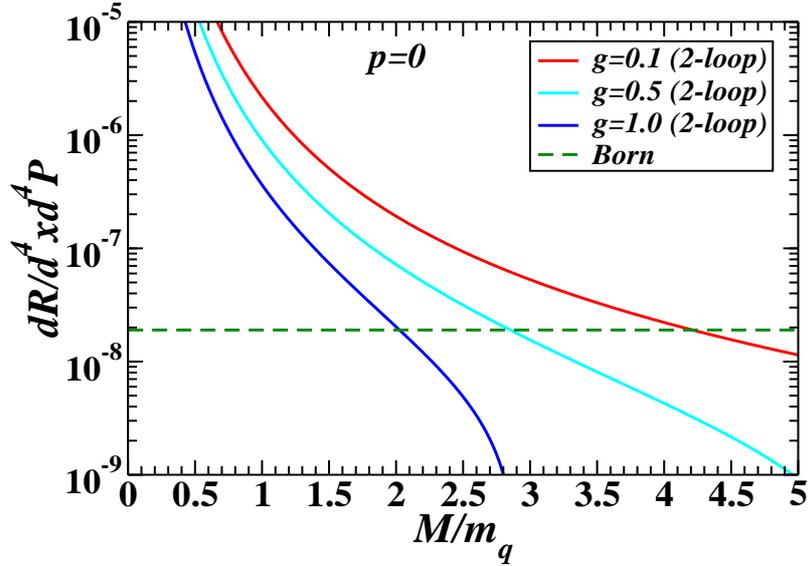}
\vspace*{-0.08in}
\caption[Complete 2-loop dilepton rate for small invariant masses $M\sim gT$ 
at zero momentum and Born-rate (dashed line) versus the scaled invariant 
photon mass]{Complete 2-loop dilepton rate for small invariant masses $M\sim gT$ 
at zero momentum and Born-rate (dashed line) versus the scaled invariant 
photon mass $M/m_q$ with the thermal quark mass $m_q$.
}
\label{fig_2loop}
\end{center}
\end{figure}
\vspace*{-.50cm}

In Fig.~(\ref{fig_2loop}) the Born-rate and the complete two-loop rate 
for a set of values of $g$ are compared. It is evident from 
Fig.~(\ref{fig_2loop}) that the $2$-loop rate dominates in the perturbative
regime ($g\leq 1$) over the Born-term for low mass domain, $M/m_q\leq 2$.  
However, the van Hove singularities contained in one-loop do not appear as they are 
washed out due to the leading logarithm approximation within the two-loop
HTL perturbation theory.

\subsubsection*{Semi-hard Rate ($M\sim T$ and $p\gg  T$)}

For $M$ of the order of $T$ and hard momenta ($p\gg T$), the 
$\alpha_s$-correction to the Born-rate has been 
calculated~\cite{Thoma:1997dk} 
within the HTLpt method as

\begin{equation}
\frac{dR}{d^4xd^4P}=\frac{5\alpha_{\rm em}^2\alpha_s}{27\pi^3}\frac{T^2}{M^2} e^{-E/T}
\left (\ln \frac{T(m_q+k^*)}{m_q^2}+C\right ) \ ,
\label{rate_alphas}
\end{equation}
where $k^*\approx |Em_q^2/M^2-m_q^2/(4E)| < (E+p)/2$ and $C\approx -0.5$
depends weakly on $M$. In Ref.\cite{Carrington:2007gt,Aurenche:2002wq,Aurenche:2002pc} it has been shown that further corrections
to the rate (\ref{rate_alphas}) are necessary. However, numerical results showed only 
a slight modification.

\begin{figure}[tbh]
\vspace*{0.45in}
\begin{center}
 \includegraphics[height=0.53\textwidth, width=0.7\textwidth]{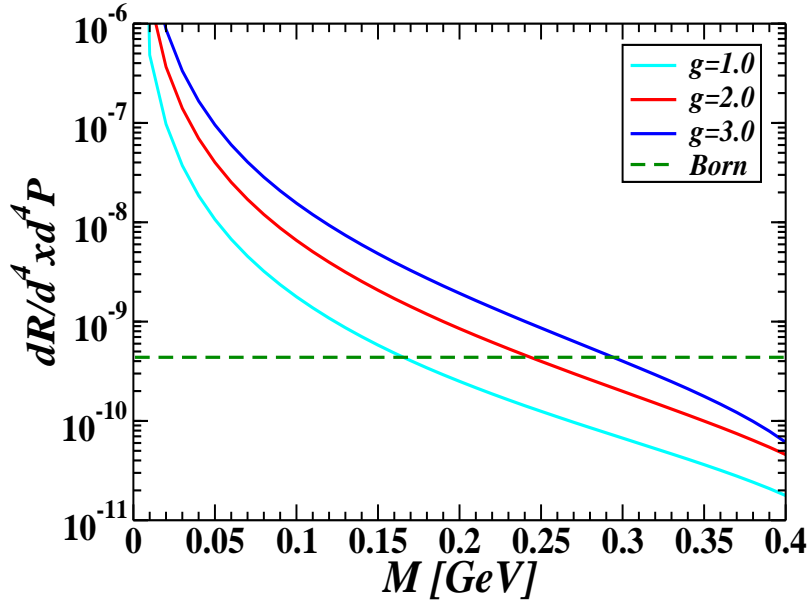}
\vspace*{-0.0in}
\caption[$\alpha_s$-correction to the dilepton rate and Born-rate 
(dashed line) versus the invariant photon mass scaled with the 
thermal quark mass]{$\alpha_s$-correction to the dilepton rate and Born-rate 
(dashed line) versus the invariant photon mass $M$ scaled with the 
thermal quark mass for $T=200$ MeV and $E=1$ GeV.}
\label{fig_alphas}
\end{center}
\end{figure}

Assuming  typical values of the strong coupling constant
and temperature, $T=200$ MeV, these corrections dominate over the Born term
for invariant masses below $300$ MeV as shown in Fig.~(\ref{fig_alphas}).
Similar results have been obtained using bare quark propagators~\cite{Altherr:1992th}.
However, the calculation within naive perturbation theory~\cite{Cleymans:1992gb,Cleymans:1993jm} 
resulted in $\alpha_s$-corrections which are of similar size as the 
Born-rate in the regime $M$ and $p$ of the order of $T$.

\subsubsection*{Hard Rate ($M \gg  T$)}
For $M\gg  T$ naive perturbation theory using bare propagators and vertices is 
sufficient. This is in contrast to the production of real photons, where one 
encounters an infrared singularity from bare quark propagator~\cite{Kapusta:1991qp,Baier:1991em}. 
For finite $M$, however, this singularity cancels~\cite{Altherr:1989jc}. Bare two-loop 
calculations~\cite{Altherr:1989jc,Cleymans:1992gb,Cleymans:1993jm} showed that the $\alpha_s$-corrections are 
negligible in this regime. However, a recent calculation of the 
$\alpha_s$-corrections~\cite{Kapusta:2000an} for large invariant mass $M\gg T$ and
small momenta $p\ll T$ yielded important corrections to the Born-rate for 
invariant masses below $(2-3)T$. However, this work has also been criticized 
\cite{Aurenche:2000xn}.

The main problem in applying perturbative results discussed above to realistic
situations is the fact that $g$ is not small but rather we have 
$g\sim 1.5-2.5$. Close to the critical temperature, $T_c$, even $g$ 
could be as high as $6$~\cite{Peshier:1994zf,Peshier:1995ty,Levai:1997yx}.
Hence the different momentum scales are not distinctly separated in the real
sense and, even if one still believes in perturbative results (see Fig.~(\ref{fig_1loop}),
\ref{fig_2loop} and \ref{fig_alphas}) at least qualitatively, it is not clear which of the above rates applies to 
heavy-ion collisions. However, in all cases there are substantial corrections 
to the Born-rate.  The perturbative rates within their uncertainties in 
various regime probably suggest that the Born-rate may not be sufficient 
for describing the low mass dilepton spectrum.  

%

\subsection[Rate using Gluon Condensate]{Rate using Gluon Condensate within the Green Function}
\label{rate_gc}

An important issue towards the understanding the phase structure of QCD
is to understand the various condensates, which serve as order parameters
of the broken symmetry phase. These condensates are non-perturbative in nature
and lattice provides a connection with bulk properties of QCD matter. However,
the quark condensate has a rather small impact on the bulk properties,
{\em e.g.,} on the equation of state of QCD matter, compared to the gluon 
condensate~\cite{Boyd:1996bx}. The relation of the gluon condensate to the 
bulk properties such as equation of states, in principle, can be tested 
through hydrodynamic or transport properties sensitive to the equation
of states, but is a non-trivial task.

\begin{figure}[!tbh]
\vspace*{0.47in}
\begin{center}
 \includegraphics[height=0.6\textwidth, width=0.7\textwidth]{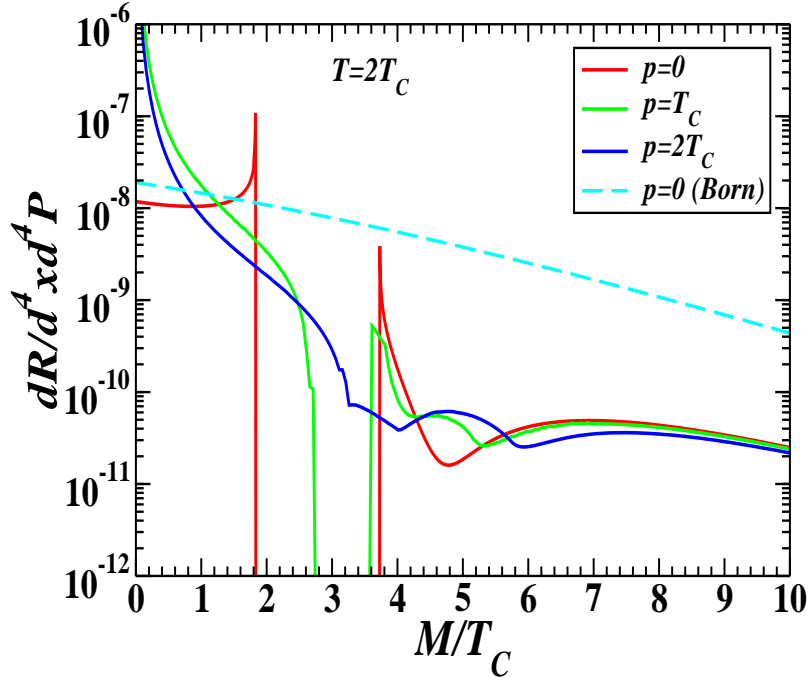}
\vspace*{-0.0in}
\caption[Van Hove singularities in the dilepton rate in the presence of gluon
condensate as a function of invariant mass scaled with $T_c$.]{Van Hove singularities in the dilepton rate in the presence of gluon
condensate as a function of invariant mass scaled with $T_c$ for a set of 
momenta at $T=2T_c$. The dashed curve is for Born-rate at zero momentum.}
\label{fig_gc}
\end{center}
\end{figure}

A semi-empirical way to consider nonperturbative aspects,{\em e.g}, gluon 
condensate has been suggested by combining lattice results with Green 
function in momentum space~\cite{Schaefer:1998wd,Mustafa:1999jz}. In this approach the effective 
$N$-point functions~\cite{Schaefer:1998wd,Mustafa:1999jz} have been constructed which
contain the gluon condensate in the deconfined phase, 
measured in lattice QCD~\cite{Boyd:1996bx}. The resulting quark dispersion relation 
with a mass $m_q\sim 1.15T_c$~\cite{Schaefer:1998wd} 
in the medium shows qualitatively the same behavior as the HTL dispersion, 
leading again to sharp structures (van Hove singularities, energy gap) in 
the dilepton production rates~\cite{Mustafa:1999cp,Mustafa:1999dt}, indicating that this features are
universal in relativistic plasmas independent of the approximation used \cite{Mustafa:2002pb,Peshier:1999dt}.
In Fig.~(\ref{fig_gc}) the dilepton production rate using gluon condensate
is displayed for various values of momentum at $T=2T_c$ and also compared
with the Born-rate. At very low invariant mass 
($M/T_c\le  2$; for $T_c\sim 165$ MeV, $M\le 330$ MeV) 
with realistic momentum the dilepton rate with gluon condensate dominates 
over the Born-rate. This rate will be important at very low invariant mass as
it has non-perturbative input from lattice QCD that describes the bulk 
properties of the deconfined phase, and is of course free from  
any uncertainty related to the strong coupling $g$ associated with the
perturbative rates discussed in Sec.~\ref{sec_htlrate}.

We, however, also note that the rate deviates from the Born-rate at high
$M/T_c$ ($\ge 4$). The difference at high $M/T_c$ has the origin  
in the asymptotic limit (large momentum $k$) of the quark dispersion
relation with gluon condensates. In this limit it is found that the normal 
quark mode behaves like $w_+=k+c$, where $c$ contains still the non-zero 
contribution from the condensates. The reason for which 
is the use of the momentum independent condensate values. This fact has 
crept in the dilepton rate at high $M/T_c$.  One way out could be to use an 
{\em ad hoc} separation scale ($M/T_c\sim 2-3$) up to which one may employ the 
non-perturbative quark dispersion associated with the gluon condensate
and beyond which a free dispersion is adopted. Alternatively, one could use a momentum 
dependent condensate, which is again beyond the scope of our calculation 
and has to be provided by the lattice analysis. To date we are not aware 
of such analysis. Nonetheless, we note that the nonperturbative contribution 
is important only at low invariant mass as we would see later in Sec.~\ref{mom_int_rate}.

\subsection[Rate from quark and rho-meson Interaction]{Rate from quark and  $\rho^0$-meson Interaction ($\rho$-meson in QGP)}
\label{rate_rhoq}

We assume that $\rho$-meson like states ($q\bar q$ correlator in the 
$\rho$-meson channel) can exist in a deconfined phase like QGP. Then 
there will also be a contribution from $\rho$-meson channel to the
dilepton pairs ($l^+l^-$) in addition to the perturbative production. 
In order to consider such a channel 
phenomenologically an interaction of $\rho-q$ coupling is 
introduced through the Lagrangian~\cite{Thoma:1999nm} 
\begin{equation}
{\cal L}= -\frac{1}{4} \rho_{\mu\nu}^a\rho^{\mu\nu}_a
+ \frac{1}{2}m_\rho^2\rho_\mu^a\rho^\mu_a 
+ \bar q \left (i\gamma_\mu\del^\mu -m_q+G_\rho \gamma^\mu \frac{\tau_a}{2}
\rho_\mu^a\right )q ,  \label{lag_rq}
\end{equation}
where $q$ is the quark field, $m_q$ is the quark mass,  $a$ is the isospin
or flavor index, and $\tau_a$ is the corresponding isospin matrix. 
The $\rho - q$ coupling, $G_\rho$, can be obtained in the same spirit as the
$4$-point 
interaction, $G_2(\bar q\gamma_\mu\tau_a q)^2$, in NJL-model. This suggests
$G_\rho=\sqrt{8m_\rho^2G_2} \sim 6 $, by taking $G_2$ from the literature. The
similar value for $G_\rho$ can be obtained by simply assuming that the $\rho$-meson
couples in a universal way to nucleons, pions and quarks~\cite{Thoma:1999nm}.

Now using the Vector Meson Dominance (VMD)~\cite{Gale:1990pn} the photon self-energy 
is related to the $\rho^0$ meson propagator, $D_{\mu\nu}(P)$, by
\begin{equation}
{\mbox{Im} \Pi_\mu^\mu(P)}= \frac{e^2}{G_\rho^2} m_\rho^4 \ {\mbox{Im} 
D_\mu^\mu(P)} \ \ . \label{vdm_rel}
\end{equation}

Then the thermal dilepton production rate from the $\rho$-meson 
can be written as
\begin{equation}
\frac{dR}{d^4x \, d^4P}  \ = -\ \frac{1}{3\pi^3} \ \frac{\alpha_{\rm em}^2}
{G_\rho^2} \ \frac{m_\rho^4}{M^2} \  \ \frac{1}{e^{E_p/T}-1} \ 
\left ({\cal A}_\rho^L + 2 {\cal A}_\rho^T \right) \ \ , \label{rate_rpp}
\end{equation}
and the spectral functions for $\rho$-meson can be obtained from the
self-energy of $\rho-$meson as
\begin{eqnarray}
{\cal A}_\rho^L(P) &=& \frac {{\rm {Im}} {\cal F}} { \left ( M^2 -m_\rho^2 - 
{\rm {Re} {\cal F}}\right )^2 + ({{\rm {Im}} {\cal F}})^2}  \ ,\\
{\cal A}_\rho^T (P) &=& \frac {{\rm {Im}} {\cal G}} { \left ( M^2 -m_\rho^2 - 
{\rm {Re} {\cal G}}\right )^2 + ({{\rm {Im}} {\cal G}})^2}\ ,
\end{eqnarray}
where ${\cal F}= - \frac{P^2}{p^2}\Pi^{00}(P)$ and ${\cal G}=\Pi_T(P)$ with
$L$ and $T$ stand for longitudinal and transverse modes, respectively.

Going beyond the HTL approximation, the integral
expression for the matter part of the one-loop photon self energy 
for assymetric charges in the deconfined phase ({\em viz.}, with  non-zero 
chemical potential, $\mu$, which would be appropriate for FAIR 
energies can be obtained easily by extending the results 
of Ref.~\cite{Thoma:1999nm} to finite chemical potential as,
\begin{eqnarray}
{\mbox{Re}\ {\cal F}}\!\!\! &=& \!\!\!
\frac{3G^2}{4\pi^2}\frac{M^2}{p^2} \int \limits_{0}^\infty dk \ k
\bigg[\frac{M^2+4\omega_k^2}{4p\omega_k}\ln |a| 
+ \frac{p_0}{p} \ln |b| -\frac{2k}{\omega_k}\bigg ]\nn
&&\hspace{4cm}\times \left[n(\omega_k-\mu)+n(\omega_k+\mu)\right]\ , \nonumber \\
{\mbox{Im}\ {\cal F}}\!\!\! &=&\!\!\!
\frac{3G^2}{4\pi}\frac{M^2}{p^3} \int \limits_{k_-}^{k_+} dk \ k
\left [ p_0-\omega_k-\frac{M^2}{4\omega_k}\right]\nn
&&\hspace{4cm}\times \left[n(\omega_k-\mu)+n(\omega_k+\mu) \right]\ ,\nonumber
\ee
\be
{\mbox{Re}\ {\cal G}} \!\!\! &=& \!\!\!
\frac{3G^2}{4\pi^2} \int \limits_{0}^\infty dk \ \frac{k^2}{\omega_k} 
\left [ 
-\left( \frac{\omega_k^2M^2}{2p^3k}+\frac{M^2}{4pk}
+\frac{M^4}{8p^3k}+\frac{m_q^2}{2pk} \right)
\ln |a|\right.\nn
&&\left.\hspace{1cm}+2+\frac{M^2}{p^2} 
-\frac{p_0M^2\omega_k}{2p^3k} \ln |b| \right]\left[n(\omega_k-\mu)+n(\omega_k+\mu)\right]\ , \nonumber \\
{\mbox{Im}\ {\cal G}} \!\!\! &=& \!\!\! 
\frac{3G^2}{8\pi p} \int \limits_{k_-}^{k_+}  dk k
\left[
-\omega_k+\frac{m_q^2}{\omega_k}+\frac{p_0^2}{p^2}\omega_k 
+\frac{M^2}{2\omega_k}+\frac{M^4}{4\omega_kp^2}-\frac{p_0M^2}{p^2}
\right]\nn
&&\hspace{5.6cm}
\times  \left[n(\omega_k-\mu)+n(\omega_k+\mu) \right]\ , \label{rate_rq} 
\end{eqnarray}
along with 
\begin{eqnarray}
a \ = \ \frac{(M^2+2pk)^2 - 4p_0^2\omega_k^2}
{(M^2-2pk)^2 - 4p_0^2\omega_k^2} \ \ , &&
b \ = \ \frac{M^4 - 4(pk+p_0\omega_k)^2}
{M^4 - 4(pk-p_0\omega_k)^2} \ \ , \nonumber \\
k_- \ = \frac{1}{2}\left | p_0 \sqrt{1-\frac{4m_q^2}{M^2}} 
- p \right| \ , &&
k_+ \ =  \frac{1}{2}\left ( p_0 \sqrt{1-\frac{4m_q^2}{M^2}} + p\right ) \ ,
\nonumber
\end{eqnarray}
where $\omega_k=\sqrt{k^2+m_q^2}\ .$

\begin{figure}[!tbh]
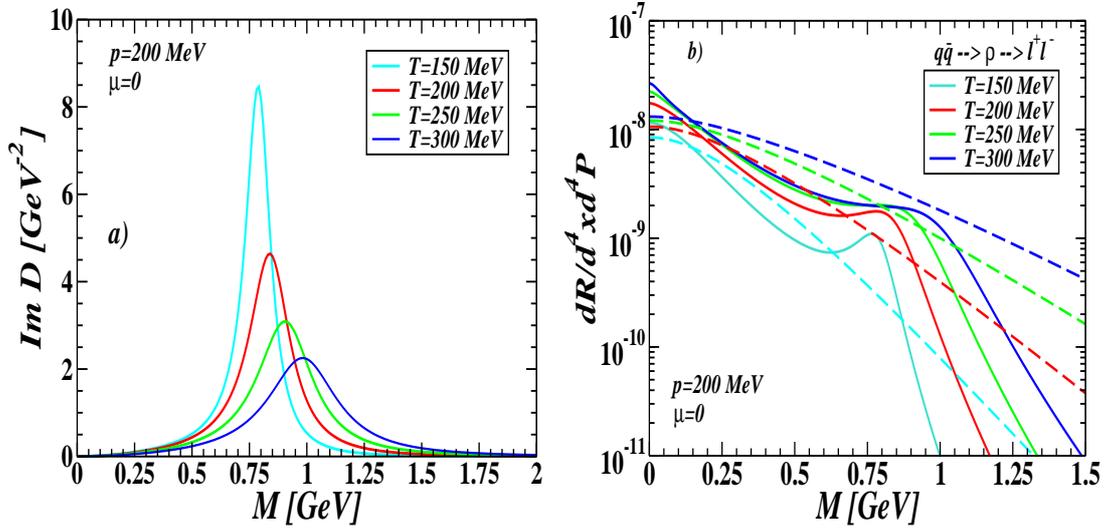

\subfigure{
\includegraphics[height=0.46\textwidth, width=0.47\textwidth]{figures/dilepton/spec_rho_p02_mu0.eps}}
\subfigure{
\includegraphics[height=0.465\textwidth, width=0.47\textwidth]{figures/dilepton/rate_rho_p02_mu0.eps}}

\caption[{\em Left panel:} $\rho$-meson spectral function as a function of the invariant mass for a set 
of values $T$. {\em Right panel:}
The dilepton rate from $\rho$-meson in a QGP as a function of 
$M$. The dashed lines are corresponding Born-rates.]{{\em Left panel:} Imaginary part of $\rho$-meson propagator
(spectral function) as a function of the invariant mass $M$ for a set 
of values $T$. {\em Right panel:}
The dilepton rate from $\rho$-meson in a QGP as a function of 
$M$. The dashed lines are corresponding Born-rates. We have used 
$G_\rho=6$.} 
\label{fig_rho_t}
\end{figure}

In Fig.~(\ref{fig_rho_t}) the $\rho$-meson spectral function related to
the imaginary part of the $\rho$-meson propagator (left panel) in 
(\ref{vdm_rel}) and the dilepton rate (right panel) are displayed for 
various temperature with $\mu=0$ and $p=200$ MeV. As the temperature 
increases the peak in the imaginary part of the $\rho$-meson propagator 
$D$ becomes broader and is also reflected in the dilepton rate. 
When the invariant mass is low ($\leq 1$ GeV), the rate is comparable with the 
Born-rate.
\vspace*{.80cm}
\begin{figure}[!tbh]
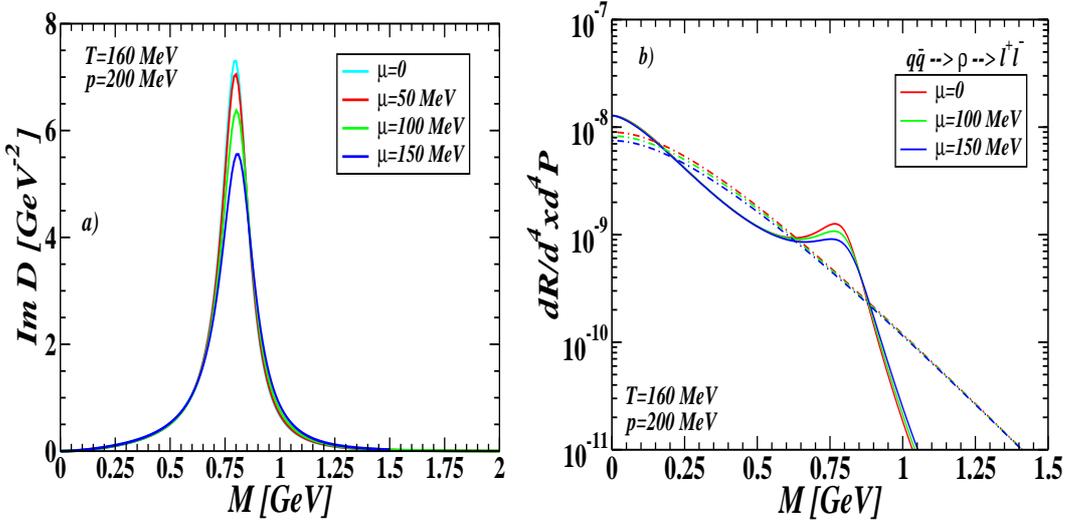

\vspace*{0.30in}
\subfigure{
\includegraphics[height=0.45\textwidth, width=0.44\textwidth]{figures/dilepton/spec_rho_t016_p02.eps}}
\subfigure{
\includegraphics[height=0.46\textwidth, width=0.47\textwidth]{figures/dilepton/rate_t016_p02.eps}}
\vspace*{-0.in}
\caption{{\em Left panel:} Imaginary part of $\rho$-meson propagator
(spectral function) as a function of $M$ for a set of values of $\mu$. 
{\em Right panel:}
The dilepton rate from $\rho$-meson in a QGP as a function of 
$M$. We chose $G_\rho=6$. The dash-dotted lines are corresponding Born-rates.} 
\label{fig_rho_mu}
\end{figure}

In Fig.~(\ref{fig_rho_mu}) the $\rho$-meson spectral function (left panel)
and the dilepton rate (right panel) are displayed for various $\mu$ at
$T=160$ MeV and $p=200$ MeV, which could be appropriate in the perspective 
of FAIR energies. The effect of broadening of the $\rho$-meson is far less
pronounced with increasing $\mu$ than increasing $T$, indicating that the
$\rho$-meson is not completely melted in the case of a system with finite
baryon density such as expected at FAIR energies even above the phase 
transition.  However, dilepton rates from 
$\rho$-meson as shown in Figs.~(\ref{fig_rho_t}) and (\ref{fig_rho_mu}) are 
comparable with the Born-rate in QGP in the low mass region ($M\le 1$ GeV), 
may be an indication for chiral restoration~\cite{Rapp:1999ej,Rapp:2009yu,Thoma:1999nm}. 
In addition this rate would be
important for invariant masses below $1$ GeV.

We also note that if one includes higher mass vector mesons such as 
$\phi$-meson within VMD, then there will be a peak corresponding to an
invariant mass of the order of $\phi$-meson mass but in low mass region
($M\leq 1$ GeV) there should be a very little change (less than $5\%$) in the
dilepton rate. Since we are interested in the low mass region, we have 
not discussed $\phi$-meson here.

\subsection{Rate from Lattice Gauge Theory}
The thermal dilepton rate describing the production of lepton pairs
with energy $\omega$ and momentum ${\vec {\mathbf p}}$ is related to the
Euclidian correlation function~\cite{Karsch:2000gi} of the vector current, 
$J_V^\mu={\bar {\psi}}(\tau,{\vec {\mathbf x}})\gamma^\mu \psi(\tau,{\vec 
{\mathbf x}})$, which can be calculated numerically in the framework of lattice
gauge theory. The thermal two-point vector correlation function in coordinate
space, ${\cal G}_V(\tau, {\vec {\mathbf x}})$, is defined as
\begin{equation}
{\cal G}_V(\tau, {\vec {\mathbf x}})=\langle J_V(\tau, {\vec {\mathbf x}})
J_V^\dagger(\tau, {\vec {\mathbf x}})\rangle =T\sum_{n=-\infty}^{\infty} \int
\frac{d^3p}{(2\pi)^3} e^{-i(w_n\tau-{\vec{\mathbf p}}
\cdot {\vec{\mathbf x}})} \chi_V(w_n,{\vec{\mathbf p}}) \ , \label{vec_cor}
\end{equation}
where the Euclidian time $\tau$ is restricted to the interval 
$[0,\beta=1/T]$, and 
the Fourier transformed correlation function $\chi_V$ is given at the discrete 
Matsubara modes, $w_n=2\pi n T$. The imaginary part of the momentum space
correlator gives the spectral function $\sigma_V(\omega, {\vec{\mathbf p}})$, as
\begin{equation}
\chi_V(w_n,{\vec{\mathbf p}}) = - \int\limits_{-\infty}^{\infty} 
d\omega\frac{\sigma_V(\omega,{\vec{\mathbf p}})}{iw_n-\omega} \ \Rightarrow \
\sigma_V(\omega,{\vec{\mathbf p}})= \frac{1}{\pi} {\mbox{Im}}
\ \chi_V(\omega,{\vec{\mathbf p}}) \ . \label{spec_mom_dilep}
\end{equation}

In coordinate space, the spectral representation of
the thermal correlation functions at fixed momentum can be obtained using (\ref{vec_cor})
and (\ref{spec_mom_dilep}) as
\begin{equation}
{\cal G}(\tau, {\vec {\mathbf p}})=\int\limits_0^\infty d\omega \ 
\sigma_V (\omega, {\vec {\mathbf p}}) \
\frac{\cosh[\omega(\tau-\beta/2)]}{\sinh[\omega\beta/2]} \ . \label{corr_mom}
\end{equation}

The vector spectral function, $\sigma_V$, is related to the differential
dilepton production rate~\cite{Karsch:2000gi}\footnote{A factor of $2$ differs 
from that of Ref.~\cite{Karsch:2001uw}} as
\begin{equation}
\sigma_V(\omega, {\vec {\mathbf p}})=\frac{18\pi^2N_c}{5\alpha_{\rm em}^2} \omega^2 
\ (e^{\omega/T}-1)\ \frac{dR}{d^4xd^4P} \ , \label{rel_dilep_spec}
\end{equation}
where $N_c$ is the number of color degree of freedom.

\begin{figure}[!tbh]
\vspace*{0.45in}
\begin{center}
 \includegraphics[height=0.7\textwidth, width=0.8\textwidth]{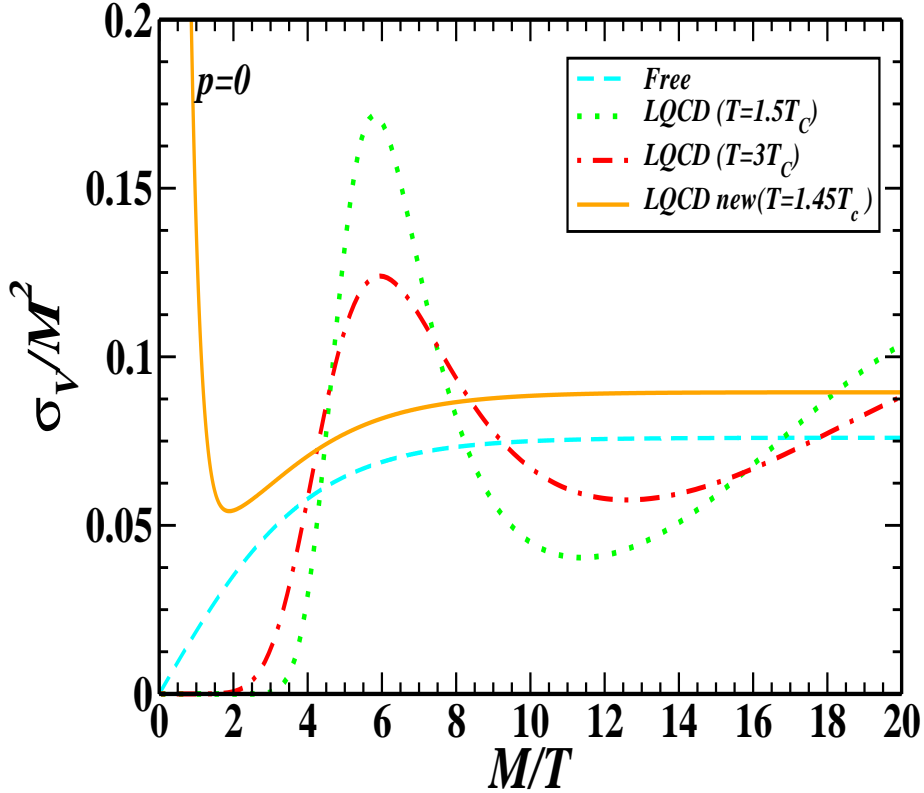}
\vspace*{-0.0in}
\caption[The zero momentum vector spectral function within lattice gauge theory]{The zero momentum (${\vec {\mathbf p}}=0$) vector spectral 
function, reconstructed from the correlation function~\cite{Karsch:2001uw,Ding:2010ga} within 
lattice gauge theory in quenched QCD using MEM, scaled with $M^2$ as 
a function of $M/T$ compared with that of the free one above the deconfinement
temperature $T_c$. Lattice data labeled as LQCD($T\!\!=\!\!1.5T_c$) and LQCD($T\!\!=\!\!3T_c$) come from~\cite{Karsch:2001uw} whereas 
LQCD new($T\!\!=\!\!1.45T_c$) come from Ref.~\cite{Ding:2010ga}.}
\label{fig_spec_lat}
\end{center}
\end{figure}

A finite temperature lattice gauge theory calculation 
can not calculate spectral function $\sigma_H(\omega,{\vec{\mathbf p}})$
directly from Eq.~(\ref{spec_mom_dilep}), 
instead it uses Eq.~(\ref{corr_mom}) to extract spectral function as discussed in Chapter~\ref{chapter:1loop}.
 In Fig.~(\ref{fig_spec_lat}) such a extracted spectral 
function~\cite{Karsch:2001uw,Ding:2010ga} scaled with $M^2$ (equivalently $\omega^2$ for $\vec{\mathbf p}=0$) 
is displayed 
as a function of $M/T$. The vector spectral functions above the deconfinement 
temperature ({\em viz.}, $T=1.5T_c \ {\mbox{and}} \ 3T_c$) show an 
oscillatory behavior compared to the free one. The spectral functions are
also found to be vanishingly small for $M/T\le 4$ due to the sharp cut-off 
used in the reconstruction. 

\begin{figure}[!tbh]
\vspace*{0.49in}
\begin{center}
\includegraphics[height=12cm, width=13cm]{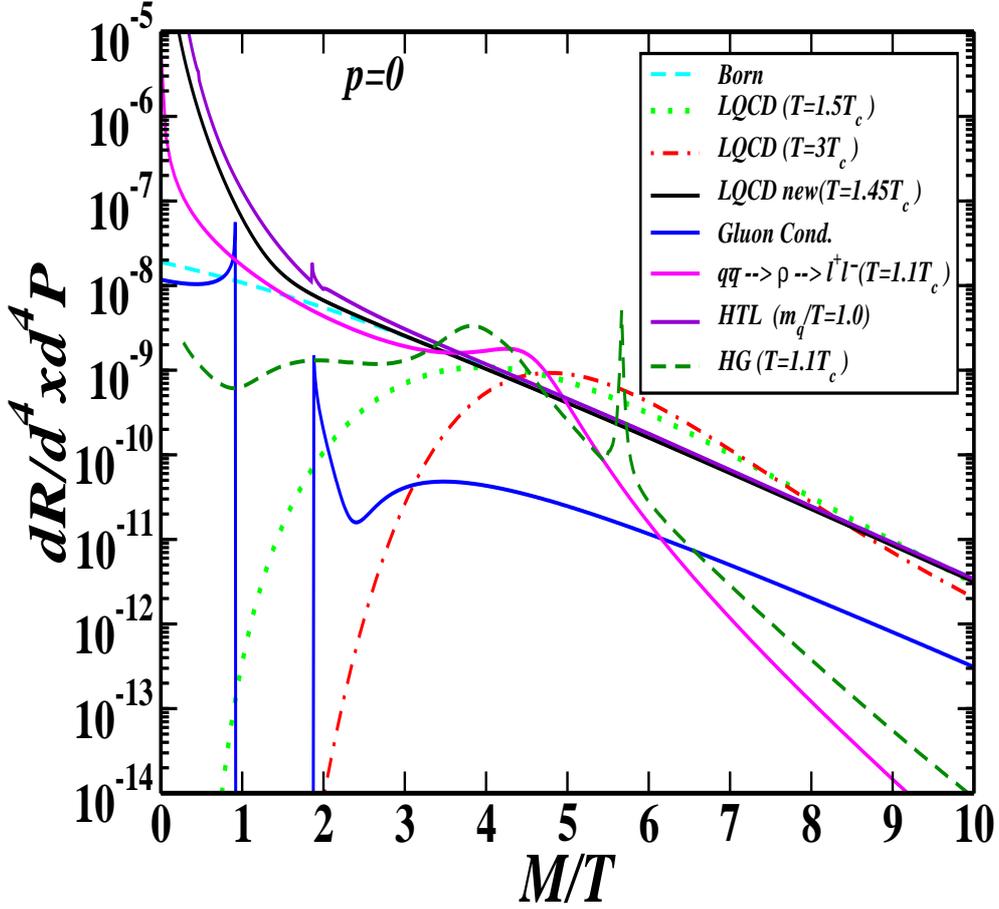}
\vspace*{-0.in}
\caption[Comparison of various dilepton rates in QGP and hadron gas.]{Comparison of various dilepton rates in a QGP 
and in a hadron gas (HG) as a function of 
$M/T$ for momentum ${\vec {\mathbf p}}=0$. The  critical temperature
is $165$ MeV~\cite{Petreczky:2009at,Petreczky:2010xg} and the value of $G_\rho$ is chosen as 6. 
The in-medium HG rate is from the recent calculations of Ref.~\cite{Ghosh:2010wt}.}
\label{fig_comp}
\end{center}
\end{figure}

A direct calculation of the differential dilepton rate 
using Eq.~(\ref{rel_dilep_spec}) above the deconfined temperature ($T_c$)
at ${\vec{\mathbf p}}=0$ was first time done in Ref.~\cite{Karsch:2001uw} within 
the lattice gauge theory in quenched QCD using the MEM. Later this calculation was
improved in Ref.~\cite{Ding:2010ga} by choosing different ansatz for the vector spectral
function. In Fig.~(\ref{fig_comp}) the lattice dilepton rates at ${\vec {\mathbf p}}=0$
from~\cite{Karsch:2001uw} for two temperatures ($T=1.5T_c$ and $3T_c$) and also from~\cite{Ding:2010ga}
for temperature $T=1.45T_c$ are displayed as a function of
the scaled invariant mass with temperature and $M/T=\omega/T$, the energy of 
the dileptons. We have also compared the perturbative, non-perturbative 
and in-medium hadrons rates within the same normalization as shown in 
the plot. We note that the rate with gluon condensate perfectly scales with
the temperature whereas that of HTL one depends on the choice of the
effective coupling, $m_q/T\sim g/\sqrt 6$. 
The lattice results 
are comparable within a factor of $2$ with the Born-rate as well as that of 
HTLpt at high invariant mass $M/T\ge 4$. The absence of peak structures 
around the $\rho$-mass and also at higher $M$ in the lattice dilepton rate 
probably constrain the broad resonance structures in the dilepton rates. 
However, for invariant mass 
below $M/T\le 4$ the lattice dilepton rate falls off very fast. This is due 
to the fact that the sharp cut-off is used to reconstruct the spectral function 
from the correlation function and the finite volume restriction in the lattice 
analysis. The lattice analysis is also based on rather small statistics.  
These lattice artefacts are related to the smaller invariant masses which 
in turn indicate that it is not yet very clear whether there will be any 
low mass thermal dileptons from the deconfined phase within the lattice 
gauge theory calculation. Future analysis could improve the situation in 
this low mass regime. One cannot rule out~\cite{Karsch:2001uw} the existence of 
van Hove singularities and energy gap, which are general features of 
massless fermions in a relativistic plasma~\cite{Mustafa:2002pb,Peshier:1999dt}, in the low 
mass dileptons. This calls for a further investigations on the lattice
gauge theory side by improving and refining the lattice ingredients and
constraints.

On the other hand, in HTLpt, apart from the 
uncertainty in the choice of $g$, the low mass ($M\rightarrow 0$, 
vanishing photon energy) one-loop dilepton rate obtained from vector
meson spectral function analysis~\cite{Karsch:2000gi} diverges because the
quark-photon 
vertex is inversely proportional to the photon energy. This also requires 
a further improvement of the HTLpt. However, we assume that 
the perturbative rate could also be reliable for $M\ge 200$ MeV 
with $T\ge 200$ MeV and $g\ge 2$. 
The other two phenomenological models, {\em viz.}, gluon condensate 
measured in 
lattice~\cite{Boyd:1996bx} and $\rho-q$ interaction in the deconfined phase 
as discussed respectively above in subsec.~\ref{rate_gc} and~\ref{rate_rhoq}, 
for non-perturbative dilepton production at low mass regime are at 
least cleaner than the perturbative rates which depend weakly on the
choice of the strong coupling constant. The rate with gluon condensate
is free from strong coupling whereas that from $\rho-q$ interaction does not
depend strongly on the choice of the coupling (see below in 
Fig.~(\ref{fig_mom_intg})).  
In addition to the perturbative rate 
these two together could also provide a realistic part of the dilepton 
rate at low mass regime ($\le 1$ GeV) from the deconfined phase, as
also can be seen in the next section. As a 
comparison, we have also shown the recent rate from in-medium hadrons 
of Ref.~\cite{Ghosh:2010wt}, where the analytic structure of $\rho$-meson 
propagator has been used due to its interaction with thermal mesons.

\section{Momentum Integrated Rate}
\label{mom_int_rate}
The momentum integrated dilepton rate can be obtained as
\begin{equation}
\frac{dR}{d^4xdM^2} = \int \frac{d^3p}{2p_0} \ \frac{dR}{d^4xd^4P} \ .
\label{mi}
\end{equation}

\begin{figure}[!tbh]
\vspace*{0.49in}
\begin{center}
 \includegraphics[height=0.6\textwidth, width=0.7\textwidth]{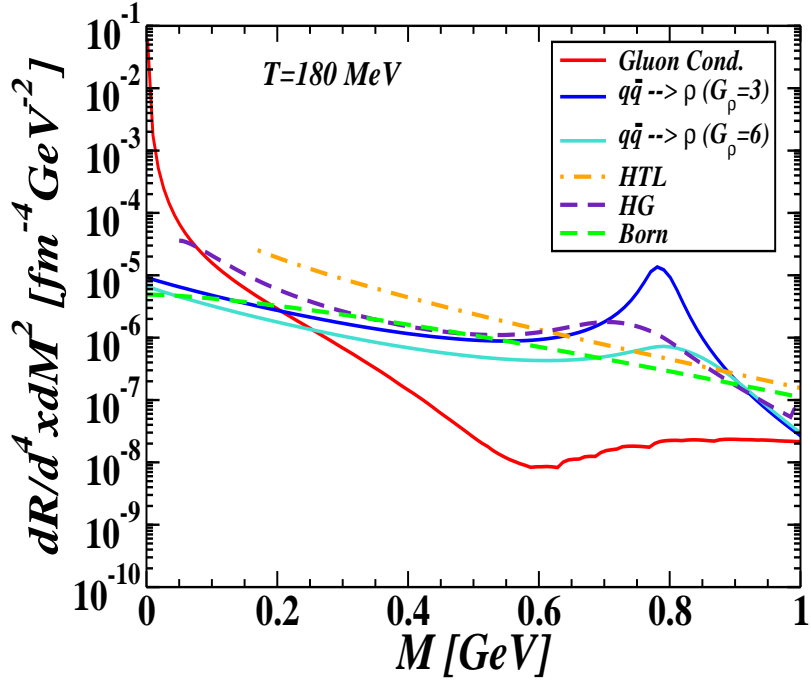}
\vspace*{0.1in}
\caption[Momentum integrated dilepton rate]{Momentum integrated dilepton rate as a function of the invariant mass 
$M$. We have used $T_c=165$ MeV for the nonperturbaive rate with gluon 
condensate. The in-medium hadronic rate (HG) is from Ref.~\cite{Ghosh:2010wt}.}
\label{fig_mom_intg}
\end{center}
\end{figure}
In Fig.~(\ref{fig_mom_intg}) the momentum integrated dilepton rates from QGP and in-medium 
hadrons are displayed as a function of invariant mass. As can be seen that at very low invariant mass
($M\le 200$ MeV) the non-perturbative contribution using gluon condensate becomes important as 
this domain is beyond reach of any reliable
perturbative calculations.
The non-perturbative rate is indeed important with input from
the first principle calculations~\cite{Boyd:1996bx} that describe the bulk
propertirs of the deconfined phase. 
 The rate from $\rho-q$ interaction is almost of the same order as 
that of the Born-rate as well as the in-medium hadrons for $M\le 600$ MeV
 whereas it is higher than the perturbative one in the domain 
$600\le M (\mbox{MeV})\le \ 800$  due to the broadening of the $\rho$ peak 
in the medium.  We also note that this rate has a weak dependence on the 
realistic range of values of the $\rho-q$ coupling $(2-6)$.  In addition the 
higher order perturbative rate from HTL, as discussed above, becomes 
reliable for $M\ge 200$ MeV and also becomes of the order of 
Born-rate for $M\ge 500$ MeV. We also note that the momentum integrated 
HTL rate used here has been obtained recently 
by Rapp et al.~\cite{Rapp:2009yu} through a parametrization of the prefactor of 
the zero momentum 1-loop HTL rate~\cite{Braaten:1990wp} with a temperature dependent $g$, 
which is claimed to reproduce the Born-rate
 in (\ref{born_gen}) within the appropriate limit.
Now for a comparison, we have also shown the recent 
rate from the in-medium hadrons of Ref.~\cite{Ghosh:2010wt}. 
It is now clear that for low invariant mass ($\le 1$ GeV) only 
the Born-rate from the QGP is not realistic as well as insufficient 
for describing the dilepton rate. 
Instead we suggest that the non-perturbative rate with gluon condensate 
should be important for $M\le 200$ MeV whereas the rates from 
$\rho-q$ interaction and HTLpt are important for $M\ge 200$ MeV.
Below we discuss some aspects
of the quark-hadron duality hypothesis~\cite{Rapp:1999us}.

\section{Thoughts on the Quark-Hadron Duality Hypothesis}
\label{qh_duality}
It is advocated~\cite{Rapp:1999ej,Rapp:1999us} that due to the potential 
broadening of the $\rho$-meson resonance suffering in a dense hadronic 
environment the overall (momentum integrated) dilepton rate out of 
the hadronic gas becomes equivalent to that from deconfined phase 
as 
\begin{equation}
\frac{dR_H}{d^4xdM^2} \approx  \frac{dR_Q}{d^4xdM^2} \ \ \ , \label{duality_rw}
\end{equation}
which entails a reminiscence to a simple perturbative $q\bar q$ 
annihilation in the vicinity of the expected QGP phase transition.
This hypothesis of 'extended' quark-hadron duality for the thermal 
source of low mass dileptons has been claimed as an indication for 
chiral symmetry restoration~\cite{Rapp:1999ej,Rapp:2009yu,Rapp:1999us} in the
deconfined phase.  
However, we would like to note that in this hypothesis the volume of 
QGP and hadronic gas was assumed to be same in a given instant of time 
and therefore, the dileptons shine equally bright from  both phases 
at a given instant of time per unit volume. This denotion of quark-hadron 
duality should be carefully re-addressed on its general validity, as the 
suggestive conclusion is indeed far-reaching. A more realistic way to 
look into it is envisaged below.

The momentum integrated rate in (\ref{mi}) shall be gauged to the adequate 
degrees of freedom in a particular phase. A certain measure is given by
the corresponding entropy density. Hence we suggest that for duality to hold
one approximately should have
\begin{equation}
\frac{1}{s_H}\frac{dR_H}{d^4xdM^2} \approx  
\frac{1}{s_Q} \frac{dR_Q}{d^4xdM^2} \ \ \ , \label{duality_us}
\end{equation}
where $s_i\ (i=H,Q)$ is the entropy density of the respective phase.
For an isoentropic crossing over the phase transition, one has 
$s_H dV_H \approx s_Q dV_Q$. Hence if one takes into account the 
respective volume of both phases at a given
instant of time, then instead of (\ref{duality_rw}) one should ask for 
\begin{equation}
{dV_H}\frac{dR_H}{d^4xdM^2} \approx {dV_Q} \frac{dR_H}{d^4xdM^2}
\ \ \ , \ \label{duality_carst}
\end{equation}
where $dV_i$ ($i=Q, \ H$) is the volume of the respective phase. 
Now, at a given instant of time this can lead to
\begin{equation}
\frac{dR_H}{dtdM} \approx  \frac{dR_Q}{dtdM}
\ \ , 
\label{duality_carst1}
\end{equation}
where ${dR_i/dtdM}$ is the total yield per time from total phase $i$ in
the system at any instant of time. Therefore, equation (\ref{duality_carst1}) 
means that the fireball emits the same number of dileptons per unit time
either if described by a hadronic or by a deconfined partonic description. 
This could likely be a more realistic way to look into
the quark-hadron duality.
Now, even if the momentum integrated rates in (\ref{mi}) from both
phases are same in some kinematic domain ({\em e.g.,} see 
Fig.~(\ref{fig_mom_intg})) may not necessarily imply a quark-hadron duality
as given by (\ref{duality_carst1}) because hadronic volume is expected to 
be larger than that of QGP by at least a factor of 4 to 5. Furthermore, 
we also note that the quark-hadron duality should also be true for any 
momentum at a given instant of time.

\section{Conclusion}
\label{conclusion_dilepton}
We have discussed the low mass dilepton production rate from the 
deconfined phase within various models, {\em viz.}, perturbative and 
non-perturbative, and compared with that of first principle 
calculations based on lattice gauge theory and in-medium hadrons.
We also have discussed in
details the limitations and uncertainties of all those models at
various domains of the invariant mass. It turns out that at very 
low invariant mass ($\le 200$ MeV) the non-perturbative rate using 
gluon condensate measured in lattice becomes important as this
domain is beyond reach of any reliable perturbative calculations. 
The other non-perturbative contribution from $\rho-q$ interaction
also becomes important below $1$ GeV as it is almost of same order as those 
of the Born and in-medium hadrons. We also note that these two rates
are at least cleaner than the perturbative
rates, in the sense that the gluon condensate rate has non-perturbative 
input from lattice equation of states and is thus free from any coupling 
uncertainties whereas the $\rho-q$ interaction rate does not depend strongly
on the choice of its coupling.  We also 
discussed the $\rho-q$ interaction in the perspective of FAIR scenario. 

On the other hand the perturbative contribution, within its various 
uncertainties, becomes steady and reliable beyond $M> 200$ MeV and 
also becomes comparable with the Born-rate and the LQCD rate for 
$M\ge 500$ MeV. The LQCD rate also constrains the broad resonance 
structure at large invariant mass. 
More specifically, the rate with gluon condensate is important 
for $M\le 200$ MeV whereas those from the $\rho-q$ interaction 
and HTLpt would be important for $M\ge 200$ MeV for the deconfined 
phase in heavy-ion collisions. Instead of considering only the Born-rate  
the various nonperturbative and perturbative  rates 
from appropriate domains of the invariant mass below $1$ GeV 
would comprise a more realistic rate for low mass dileptons from
the deconfined phase created in heavy-ion collisions. We hope that
more elaborate future lattice gauge theory studies on dileptons 
above the deconfined temperature can provide a more insight than  
present LQCD calculations on the low mass region, which could then verify 
the various model calculations on low mass dileptons above the
deconfined temperatures. Finally, we also have discussed a more 
realistic way to look into the quark-hadron duality hypothesis 
than it is advocated in the literature.


%% file: text/sum.tex
%
%
\chapter{Summary and Outlook}\label{chapter:sum}
This thesis is mainly devoted to study the thermodynamics of hot and dense nuclear
matter at finite temperature and finite chemical potential. To study the thermodynamical
quantities of hot and dense matter produced in heavy ion collision, one would prefer first 
principle numerical lattice QCD method. But due to sign problem, lattice QCD can be applied
to arbitrary temperature but at zero chemical potential. Finite temperature lattice QCD can
be extended to calculate thermodynamic potential at small chemical potential by expanding in Taylor
series and by calculating coefficients of Taylor series up to some finite order. This expansion up to
finite order term is valid
for small chemical potential. So the thermodynamic calculations at any chemical potential or at least 
at any chemical potential which is same order of temperature was essential. To perform these calculations at finite
chemical potential one was left with usual perturbation theory.
But the poor convergence of conventional perturbation theory at finite temperature/chemical potential
has been the main obstacle in the practical application of thermal QCD for decades. To improve the results of usual perturbation theory,
a considerable effort has been put into reorganizing the perturbative series at phenomenologically
relevant temperatures. The application of HTL perturbation theory to the problem carried
out in this thesis leads to laudable results for non-Abelian theories at finite temperature
and at finite chemical potential. Till date thermodynamical quantities using hard thermal loop perturbation
theory was available only at finite temperature and zero chemical potential. In this thesis I have systematically
calculated most of the relevant thermodynamical quantities in all loop order at finite temperature and finite 
chemical potential. This was essential to study the hot and dense nuclear medium
produced in RHIC BNL and also expected to produce in future experiment in FAIR GSI. Finite temperature and finite 
chemical thermodynamical calculations were also essential to study various order conserved density
fluctuations. 

In Chapter~\ref{chapter:1loop} we have discussed about leading order thermodynamical quantities {\em viz.}
number density, entropy density, pressure and hence quark number susceptibility and temporal correlation function in
hard thermal loop perturbation theory. In Chapter~\ref{chapter:2loop} we have discussed about next-to-leading order pressure 
in HTL perturbation theory. We have also discussed about NLO second and fourth order susceptibilities using NLO pressure in HTL
perturbation theory in Chapter~\ref{chapter:2loop}.

In Chapter~\ref{chapter:3loop} we have extended the NLO calculations as discussed in  
Chapter~\ref{chapter:2loop} to NNLO at finite temperature and finite chemical potential. In this chapter we have presented the NNLO
pressure, energy density, entropy density, trace anomaly, speed of sound at finite temperature and chemical potential and compared 
those results with available lattice QCD data. We have also presented the NNLO results of various order conserved charged
fluctuations {\em viz.} diagonal and off-diagonal second-, fourth- and sixth order quark number susceptibilities.
In all cases we found very good agreement between the results obtained using the central values of the renormalization
scales and available lattice data.

Along-with the computation of thermodynamic quantities of hot and dense matter,
we have also discussed in Chapter~\ref{chapter:dilepton} about low mass dilepton rate
from hot and dense medium within various models, {\em viz.},
perturbative and non-perturbative, and compared with that of first principle 
calculations based on lattice gauge theory and in-medium hadrons gas.  It turns out that at very 
low invariant mass ($\le 200$ MeV) the non-perturbative rate using 
gluon condensate measured in lattice becomes important as this
domain is beyond reach of any reliable perturbative calculations. 
The other non-perturbative contribution from $\rho-q$ interaction
also becomes important below $1$ GeV as it is almost of same order as those 
of the Born and in-medium hadrons. On the other hand the perturbative contribution
from HTL perturbation theory, within its various 
uncertainties, becomes steady and reliable beyond $M> 200$ MeV and 
also becomes comparable with the Born-rate and the LQCD rate for 
$M\ge 500$ MeV. The LQCD rate also constrains the broad resonance 
structure at large invariant mass. 

The HTL perturbation theory is a gauge invariant reorganization of usual
perturbation theory and it  shifts the expansion from usual
perturbation theory  around an ideal gas of massless particles 
to that of massive quasiparticles.
The HTL Feynman rules as discussed in this thesis for scalar as well as for gauge theory show
clearly that the propagators and vertices become dressed systematically by the thermal medium, 
as a result the interactions also get screened in the medium which can be seen that the coupling
strength gets screened by the thermal mass term explicitly. Therefore the expansion in terms
of the HTL Feynman rules are self-consistently around non-interacting quark gluon gas. The fact that the 
mass parameters are not arbitrary but a function of coupling constant $g$, temperature $T$, and chemical 
potential $\mu$ and determined variationally or perturbatively also indicates that HTLpt doesn't
modify the original gauge theory but just reorganizes the usual perturbation series. Although the 
renormalizability of HTL perturbation is not yet proven, 
the fact that it can be renormalized at every order using only known counterterms {\em viz.} vacuum energy,
debye and thermal quark mass and running coupling counterterm shows promising 
results till now.

So far, thermodynamics for hot and dense nuclear medium has been studied intensively, 
both in leading or in higher order perturbation theory or numerically on the lattice, however 
real-time dynamics is still in its very early stage of development. Transport coefficients 
such as coefficients of viscosity, conductivity, diffusion rate etc.  
are of great interest since they are theoretically clean and well defined non-equilibrium 
dynamical quantities. A considerable efforts have been devoted to calculate transport coefficients 
at leading order in perturbation theory~\cite{Hosoya:1983id,
Hosoya:1983xm,Svetitsky:1987gq,Baym:1990uj,Thoma:1990fm,Braaten:1991we,Jeon:1994if,Jeon:1995zm,Arnold:2000dr
,Arnold:2001ms,Arnold:2002ja}. 
However the only known transport coefficients in next-to-leading order are shear viscosity in scalar 
$\phi^4$ theory~\cite{Moore:2007ib}, heavy quark diffusion rate in QCD and ${\cal N}=4$ supersymmetric 
Yang-Mills theory~\cite{CaronHuot:2007gq,CaronHuot:2008uh}, 
and transverse diffusion rate $\hat{q}$ in QCD~\cite{CaronHuot:2008ni}, and all of them exhibit
poor convergence as bad as pressure in bare perturbation theory.
Since dynamical quantities are still not well described by lattice gauge theory, the computation
transport coefficients using HTL perturbation theory would interesting in order to achieve a better
understanding of that quantities.


Further this thesis was focused to calculate various quantities with vanishing quark mass and $\mu\lesssim 2\pi T$,
it would be very interesting to include the quark mass effect in higher order calculation using HTL perturbation theory.
This would be required to investigate various order conserved density fluctuations for massive quark along-with 
massless quarks and also to investigate chiral susceptibility. Also it would be very interesting to extend the
thermodynamic calculation at $\mu > 2\pi T$ and eventually at zero temperature to study the EoS of Neutron star
and other astrophysical objects.
 

%% file: text/aleph.tex
\chapter[Properties of the aleph functions]{Properties of the $\aleph$ functions}
\label{app:aleph}
\vspace{-.5cm}
For some frequently occurring combinations of special functions in Chapter~\ref{chapter:3loop} 
we will apply the following abbreviations
\be
\zeta'(x,y) &\equiv& \partial_x \zeta(x,y) \, , \label{}\\
\aleph(n,z) &\equiv& \zeta'(-n,z)+\lb-1\rb^{n+1}\zeta'(-n,z^{*}) \, , \\
\aleph(z) &\equiv& \Psi(z)+\Psi(z^*) \, ,
\ee
where $n$ is assumed to be a non-negative integer and $z$ is a general complex number given here by $z=1/2-i\hmu$.
Above $\zeta$ denotes the Riemann zeta function, and $\Psi$ is the digamma function
\be
\Psi(z)&\equiv&\frac{\Gamma '(z)}{\Gamma(z)} \, .
\ee
Below we list Taylor expansions of the function $\aleph$ and $\aleph(n,z)$ for values of $n$ necessary
for calculation of the susceptibilities presented in the main text.  For general application we evaluate
the $\aleph$ functions exactly using \emph{Mathematica}.
\be
 \aleph(z)\!\!\!&=&\!\!\!-2\gamma_E-4\ln 2+14\zeta(3)\hmu^2-62\zeta(5)\hmu^4+254\zeta(7)\hmu^6+{\cal O}(\hmu^8),\hspace{1cm}
\ee
\be
 \aleph(0,z)&=& 2\lb2\ln2+\gamma_E\rb i \hmu - \frac{14}{3}\zeta(3)i\hmu^3+\frac{62}{5}\zeta(5)i\hmu^5
+{\cal O}(\hmu^7) \, ,
\ee
\be
 \aleph(1,z)&=&-\frac{1}{12}\lb\ln2-\Za\rb - \lb1-2\ln2-\gamma_E\rb\hmu^2-\frac{7}{6}\zeta(3)\hmu^4\nn
&&+\ \frac{31}{15}\zeta(5)\hmu^6
+{\cal O}(\hmu^8) \, ,  
\ee
\be
\aleph(1,z+z') &=& -\frac{1}{6}\frac{\zeta'(-1)}{\zeta(-1)}-\lb1-\gamma_E\rb\lb\hmu+\hmu'\rb^2-
\frac{\zeta(3)}{6}\lb\hmu+\hmu'\rb^4\nn
&&+\ \frac{\zeta(5)}{15}\lb\hmu+\hmu'\rb^6+\mathcal{O}(\hmu^8,\hmu'^8)\, , 
\ee
\be
\aleph(2,z)&=&\frac{1}{12}\lb1+2\ln2-2\Za\rb i\hmu +\frac{1}{3}\lb3-2\gamma_E-4\ln2\rb i \hmu^3\nn
&&+\ \frac{7}{15}\zeta(3)i\hmu^5
+{\cal O}(\hmu^7) \, ,
\ee
\be
\aleph(2,z+z') &=& -\frac{1}{6}\Big(1-2\,\frac{\zeta'(-1)}{\zeta(-1)}\Big)i\lb\hmu+\hmu'\rb +
\frac{1}{3}\lb3-2\gamma_E\rb i \lb\hmu+\hmu'\rb^3\nn
&&+\ \frac{\zeta(3)}{15}i\lb\hmu+\hmu'\rb^5+\mathcal{O}(\hmu^7,\hmu'^7),
\ee
\be
\aleph(3,z)&=&\frac{1}{480}\lb\ln2-7\Zc\rb +\frac{1}{24} \lb5+6\ln2-6\Za\rb\hmu^2 
\nn
&+&
\frac{1}{12}\lb11-6\gamma_E-12\ln2\rb\hmu^4+\frac{7}{30}\zeta(3)\hmu^6
+{\cal O}(\hmu^8) \, , 
\ee
\be
\aleph(3,z+z') &=& \frac{1}{60}\frac{\zeta'(-3)}{\zeta(-3)}-\frac{1}{12}\Big(5-6\,\frac{\zeta'(-1)}{\zeta(-1)}\Big)
\lb\hmu+\hmu'\rb^2\nn
&+&
\frac{1}{12}\lb11-6\gamma_E\rb\lb\hmu+\hmu'\rb^4+\frac{\zeta(3)}{30}\lb\hmu+\hmu'\rb^6+\mathcal{O}(\hmu^8,\hmu'^8).\ 
\ee
where $\gamma_E=$Euler gamma.

%% file: thesis.bbl
\begin{thebibliography}{100}

\bibitem{Arsene:2004fa}
BRAHMS Collaboration, I.~Arsene {\em et~al.},
\newblock Nucl. Phys. {\bf A757}, 1 (2005).

\bibitem{Adcox:2004mh}
PHENIX Collaboration, K.~Adcox {\em et~al.},
\newblock Nucl. Phys. {\bf A757}, 184 (2005).

\bibitem{Back:2004je}
B.~Back {\em et~al.},
\newblock Nucl. Phys. {\bf A757}, 28 (2005), arXiv:nucl-ex/0410022.

\bibitem{Adams:2005dq}
STAR Collaboration, J.~Adams {\em et~al.},
\newblock Nucl. Phys. {\bf A757}, 102 (2005).

\bibitem{Fodor:2004nz}
Z.~Fodor and S.~Katz,
\newblock JHEP {\bf 0404}, 050 (2004), arXiv:hep-lat/0402006.

\bibitem{Gupta:2011wh}
S.~Gupta, X.~Luo, B.~Mohanty, H.~G. Ritter, and N.~Xu,
\newblock Science {\bf 332}, 1525 (2011), arXiv:1105.3934.

\bibitem{Boyd:1996bx}
G.~Boyd {\em et~al.},
\newblock Nucl. Phys. {\bf B469}, 419 (1996), arXiv:hep-lat/9602007.

\bibitem{Borsanyi:2010cj}
S.~Borsanyi {\em et~al.},
\newblock JHEP {\bf 1011}, 077 (2010), arXiv:1007.2580.

\bibitem{Borsanyi:2012uq}
S.~Borsanyi {\em et~al.},
\newblock JHEP {\bf 1208}, 126 (2012), arXiv:1205.0440.

\bibitem{Borsanyi:2012cr}
S.~Borsanyi {\em et~al.},
\newblock JHEP {\bf 1208}, 053 (2012), arXiv:1204.6710.

\bibitem{Borsanyi:2011sw}
S.~Borsanyi {\em et~al.},
\newblock JHEP {\bf 1201}, 138 (2012), arXiv:1112.4416.

\bibitem{Kajantie:2002wa}
K.~Kajantie, M.~Laine, K.~Rummukainen, and Y.~Schroder,
\newblock Phys. Rev. {\bf D67}, 105008 (2003), arXiv:hep-ph/0211321.

\bibitem{Vuorinen:2003fs}
A.~Vuorinen,
\newblock Phys. Rev. {\bf D68}, 054017 (2003), arXiv:hep-ph/0305183.

\bibitem{Haque:2010rb}
N.~Haque and M.~G. Mustafa,
\newblock (2010), arXiv:1007.2076.

\bibitem{Haque:2012my}
N.~Haque, M.~G. Mustafa, and M.~Strickland,
\newblock Phys. Rev. {\bf D87}, 105007 (2013), arXiv:1212.1797.

\bibitem{Haque:2013sja}
N.~Haque, J.~O. Andersen, M.~G. Mustafa, M.~Strickland, and N.~Su,
\newblock Phys. Rev. {\bf D89}, 061701 (2014), arXiv:1309.3968.

\bibitem{Haque:2014rua}
N.~Haque {\em et~al.},
\newblock JHEP {\bf 1405}, 027 (2014), arXiv:1402.6907.

\bibitem{Haque:2011iz}
N.~Haque, M.~G. Mustafa, and M.~H. Thoma,
\newblock Phys. Rev. {\bf D84}, 054009 (2011), arXiv:1103.3394.

\bibitem{Haque:2011vt}
N.~Haque and M.~G. Mustafa,
\newblock Nucl. Phys. {\bf A862-863}, 271 (2011), arXiv:1109.0799.

\bibitem{Haque:2013qta}
N.~Haque, M.~G. Mustafa, and M.~Strickland,
\newblock JHEP {\bf 1307}, 184 (2013), arXiv:1302.3228.

\bibitem{McLerran:1984ay}
L.~D. McLerran and T.~Toimela,
\newblock Phys. Rev. {\bf D31}, 545 (1985).

\bibitem{Greiner:2010zg}
C.~Greiner, N.~Haque, M.~G. Mustafa, and M.~H. Thoma,
\newblock Phys. Rev. {\bf C83}, 014908 (2011), arXiv:1010.2169.

\bibitem{Gross:1973id}
D.~J. Gross and F.~Wilczek,
\newblock Phys. Rev. Lett. {\bf 30}, 1343 (1973).

\bibitem{Carminati:2004fp}
ALICE Collaboration, F.~Carminati {\em et~al.},
\newblock J. Phys. {\bf G30}, 1517 (2004).

\bibitem{Alessandro:2006yt}
ALICE Collaboration, B.~Alessandro {\em et~al.},
\newblock J. Phys. {\bf G32}, 1295 (2006).

\bibitem{Friman:2011zz}
B.~Friman {\em et~al.},
\newblock Lect. Notes Phys. {\bf 814}, 1 (2011).

\bibitem{Heinz:2000bk}
U.~W. Heinz and M.~Jacob,
\newblock (2000), arXiv:nucl-th/0002042.

\bibitem{Adare:2009qk}
PHENIX Collaboration, A.~Adare {\em et~al.},
\newblock Phys. Rev. {\bf C81}, 034911 (2010), arXiv:0912.0244.

\bibitem{Adler:2006yt}
PHENIX Collaboration, S.~Adler {\em et~al.},
\newblock Phys. Rev. Lett. {\bf 98}, 012002 (2007), arXiv:hep-ex/0609031.

\bibitem{Adare:2006ti}
PHENIX Collaboration, A.~Adare {\em et~al.},
\newblock Phys. Rev. Lett. {\bf 98}, 162301 (2007), arXiv:nucl-ex/0608033.

\bibitem{Adcox:2002au}
PHENIX Collaboration, K.~Adcox {\em et~al.},
\newblock Phys. Rev. Lett. {\bf 89}, 092302 (2002), arXiv:nucl-ex/0204007.

\bibitem{Chujo:2002bi}
PHENIX Collaboration, T.~Chujo,
\newblock Nucl. Phys. {\bf A715}, 151 (2003), arXiv:nucl-ex/0209027.

\bibitem{Aamodt:2010pa}
ALICE Collaboration, K.~Aamodt {\em et~al.},
\newblock Phys. Rev. Lett. {\bf 105}, 252302 (2010), arXiv:1011.3914.

\bibitem{Aamodt:2010jd}
ALICE Collaboration, K.~Aamodt {\em et~al.},
\newblock Phys. Lett. {\bf B696}, 30 (2011), arXiv:1012.1004.

\bibitem{Chatrchyan:2011sx}
CMS Collaboration, S.~Chatrchyan {\em et~al.},
\newblock Phys. Rev. {\bf C84}, 024906 (2011), arXiv:1102.1957.

\bibitem{Aamodt:2010cz}
ALICE Collaboration, K.~Aamodt {\em et~al.},
\newblock Phys. Rev. Lett. {\bf 106}, 032301 (2011), arXiv:1012.1657.

\bibitem{Aamodt:2010pb}
ALICE Collaboration, K.~Aamodt {\em et~al.},
\newblock Phys. Rev. Lett. {\bf 105}, 252301 (2010), arXiv:1011.3916.

\bibitem{ALICE:2011ab}
ALICE Collaboration, K.~Aamodt {\em et~al.},
\newblock Phys. Rev. Lett. {\bf 107}, 032301 (2011), arXiv:1105.3865.

\bibitem{Hatsuda:1994pi}
T.~Hatsuda and T.~Kunihiro,
\newblock Phys. Rept. {\bf 247}, 221 (1994), arXiv:hep-ph/9401310.

\bibitem{Kunihiro:1991qu}
T.~Kunihiro,
\newblock Phys. Lett. {\bf B271}, 395 (1991).

\bibitem{Hatta:2002sj}
Y.~Hatta and T.~Ikeda,
\newblock Phys. Rev. {\bf D67}, 014028 (2003), arXiv:hep-ph/0210284.

\bibitem{Sasaki:2006ws}
C.~Sasaki, B.~Friman, and K.~Redlich,
\newblock Phys. Rev. {\bf D75}, 054026 (2007), arXiv:hep-ph/0611143.

\bibitem{Fukushima:2003fm}
K.~Fukushima,
\newblock Phys. Rev. {\bf D68}, 045004 (2003), arXiv:hep-ph/0303225.

\bibitem{Fukushima:2003fw}
K.~Fukushima,
\newblock Phys. Lett. {\bf B591}, 277 (2004), arXiv:hep-ph/0310121.

\bibitem{Ratti:2005jh}
C.~Ratti, M.~A. Thaler, and W.~Weise,
\newblock Phys. Rev. {\bf D73}, 014019 (2006), arXiv:hep-ph/0506234.

\bibitem{Ghosh:2006qh}
S.~K. Ghosh, T.~K. Mukherjee, M.~G. Mustafa, and R.~Ray,
\newblock Phys. Rev. {\bf D73}, 114007 (2006), arXiv:hep-ph/0603050.

\bibitem{Ghosh:2007wy}
S.~K. Ghosh, T.~K. Mukherjee, M.~G. Mustafa, and R.~Ray,
\newblock Phys. Rev. {\bf D77}, 094024 (2008), arXiv:0710.2790.

\bibitem{Mukherjee:2006hq}
S.~Mukherjee, M.~G. Mustafa, and R.~Ray,
\newblock Phys. Rev. {\bf D75}, 094015 (2007), arXiv:hep-ph/0609249.

\bibitem{Roessner:2006xn}
S.~Roessner, C.~Ratti, and W.~Weise,
\newblock Phys. Rev. {\bf D75}, 034007 (2007), arXiv:hep-ph/0609281.

\bibitem{Sasaki:2006ww}
C.~Sasaki, B.~Friman, and K.~Redlich,
\newblock Phys. Rev. {\bf D75}, 074013 (2007), arXiv:hep-ph/0611147.

\bibitem{Bhattacharyya:2010wp}
A.~Bhattacharyya, P.~Deb, S.~K. Ghosh, and R.~Ray,
\newblock Phys. Rev. {\bf D82}, 014021 (2010), arXiv:1003.3337.

\bibitem{Bhattacharyya:2010jd}
A.~Bhattacharyya, P.~Deb, A.~Lahiri, and R.~Ray,
\newblock Phys. Rev. {\bf D82}, 114028 (2010), arXiv:1008.0768.

\bibitem{Bhattacharyya:2010ef}
A.~Bhattacharyya, P.~Deb, A.~Lahiri, and R.~Ray,
\newblock Phys. Rev. {\bf D83}, 014011 (2011), arXiv:1010.2394.

\bibitem{Bhattacharyya:2011na}
A.~Bhattacharyya, S.~K. Ghosh, S.~Majumder, and R.~Ray,
\newblock Phys. Rev. {\bf D86}, 096006 (2012), arXiv:1107.5941.

\bibitem{Bhattacharyya:2012up}
A.~Bhattacharyya {\em et~al.},
\newblock Phys. Rev. {\bf C89}, 064905 (2014), arXiv:1212.6134.

\bibitem{Steinheimer:2014kka}
J.~Steinheimer and S.~Schramm,
\newblock (2014), arXiv:1401.4051.

\bibitem{Ratti:2007jf}
C.~Ratti, S.~Roessner, and W.~Weise,
\newblock Phys. Lett. {\bf B649}, 57 (2007), arXiv:hep-ph/0701091.

\bibitem{Schaefer:2006ds}
B.-J. Schaefer and J.~Wambach,
\newblock Phys. Rev. {\bf D75}, 085015 (2007), arXiv:hep-ph/0603256.

\bibitem{Schaefer:2009ui}
B.-J. Schaefer, M.~Wagner, and J.~Wambach,
\newblock Phys. Rev. {\bf D81}, 074013 (2010), arXiv:0910.5628.

\bibitem{Peshier:1994zf}
A.~Peshier, B.~Kampfer, O.~Pavlenko, and G.~Soff,
\newblock Phys. Lett. {\bf B337}, 235 (1994).

\bibitem{Peshier:1995ty}
A.~Peshier, B.~Kampfer, O.~Pavlenko, and G.~Soff,
\newblock Phys. Rev. {\bf D54}, 2399 (1996).

\bibitem{Peshier:1999ww}
A.~Peshier, B.~Kampfer, and G.~Soff,
\newblock Phys. Rev. {\bf C61}, 045203 (2000), arXiv:hep-ph/9911474.

\bibitem{Peshier:2002ww}
A.~Peshier, B.~Kampfer, and G.~Soff,
\newblock Phys. Rev. {\bf D66}, 094003 (2002), arXiv:hep-ph/0206229.

\bibitem{Bluhm:2007nu}
M.~Bluhm, B.~Kampfer, R.~Schulze, D.~Seipt, and U.~Heinz,
\newblock Phys. Rev. {\bf C76}, 034901 (2007), arXiv:0705.0397.

\bibitem{Bluhm:2007cp}
M.~Bluhm and B.~Kampfer,
\newblock Phys. Rev. {\bf D77}, 034004 (2008), arXiv:0711.0590.

\bibitem{Bannur:2006ww}
V.~M. Bannur,
\newblock JHEP {\bf 0709}, 046 (2007), arXiv:hep-ph/0604158.

\bibitem{Bannur:2007tk}
V.~M. Bannur,
\newblock Phys. Rev. {\bf C78}, 045206 (2008), arXiv:0712.2886.

\bibitem{Gardim:2009mt}
F.~Gardim and F.~Steffens,
\newblock Nucl.Phys. {\bf A825}, 222 (2009), arXiv:0905.0667.

\bibitem{Skokov:2010uh}
V.~Skokov, B.~Friman, and K.~Redlich,
\newblock Phys. Rev. {\bf C83}, 054904 (2011), arXiv:1008.4570.

\bibitem{Herbst:2013ail}
T.~K. Herbst, J.~M. Pawlowski, and B.-J. Schaefer,
\newblock Phys. Rev. {\bf D88}, 014007 (2013), arXiv:1302.1426.

\bibitem{Herbst:2013ufa}
T.~K. Herbst, M.~Mitter, J.~M. Pawlowski, B.-J. Schaefer, and R.~Stiele,
\newblock Phys. Lett. {\bf B731}, 248 (2014), arXiv:1308.3621.

\bibitem{Kim:2006ut}
K.~Jo, Y.~Kim, H.~K. Lee, and S.-J. Sin,
\newblock JHEP {\bf 0811}, 040 (2008), arXiv:0810.0063.

\bibitem{Kim:2010ag}
K.-i. Kim, Y.~Kim, S.~Takeuchi, and T.~Tsukioka,
\newblock Prog. Theor. Phys. {\bf 126}, 735 (2011), arXiv:1012.2667.

\bibitem{Kim:2010zg}
Y.~Kim, Y.~Matsuo, W.~Sim, S.~Takeuchi, and T.~Tsukioka,
\newblock JHEP {\bf 1005}, 038 (2010), arXiv:1001.5343.

\bibitem{Cui:2011wb}
L.-X. Cui, S.~Takeuchi, and Y.-L. Wu,
\newblock Phys. Rev. {\bf D84}, 076004 (2011), arXiv:1107.2738.

\bibitem{lqcd1}
T.~DeGrand and C.~DeTar,
\newblock {\em Lattice Methods for Quantum Chromodynamics} (World Scientific,
  2006).

\bibitem{lqcd2}
M.~Creutz,
\newblock {\em Quarks, gluons and lattices} (Cambridge University Press,
  Cambridge, 1985).

\bibitem{Berges:2004hn}
J.~Berges, S.~Borsanyi, U.~Reinosa, and J.~Serreau,
\newblock Phys. Rev. {\bf D71}, 105004 (2005), arXiv:hep-ph/0409123.

\bibitem{Borsanyi:2012ve}
S.~Borsanyi, G.~Endrodi, Z.~Fodor, S.~Katz, and K.~Szabo,
\newblock JHEP {\bf 1207}, 056 (2012), arXiv:1204.6184.

\bibitem{Borsanyi:2013bia}
S.~Borsanyi {\em et~al.},
\newblock Phys. Lett. {\bf B730}, 99 (2014), arXiv:1309.5258.

\bibitem{Borsanyi:2012xf}
S.~Borsanyi {\em et~al.},
\newblock Phys. Lett. {\bf B713}, 342 (2012), arXiv:1204.4089.

\bibitem{Borsanyi:2012rr}
S.~Borsanyi,
\newblock Nucl. Phys. {\bf A904-905}, 270c (2013), arXiv:1210.6901.

\bibitem{Ratti:2010kj}
Wuppertal-Budapest Collaboration, C.~Ratti {\em et~al.},
\newblock Nucl. Phys. {\bf A855}, 253 (2011), arXiv:1012.5215.

\bibitem{Borsanyi:2007bf}
S.~Borsanyi and U.~Reinosa,
\newblock Phys. Lett. {\bf B661}, 88 (2008), arXiv:0709.2316.

\bibitem{Borsanyi:2011bm}
Wuppertal-Budapest Collaboration, S.~Borsanyi {\em et~al.},
\newblock J. Phys. {\bf G38}, 124060 (2011), arXiv:1109.5030.

\bibitem{Borsanyi:2011kg}
S.~Borsanyi {\em et~al.},
\newblock PoS {\bf LATTICE2011}, 209 (2011), arXiv:1111.3500.

\bibitem{Borsanyi:2010zh}
S.~Borsanyi {\em et~al.},
\newblock (2010), arXiv:1011.4229.

\bibitem{Aoki:2006we} 
  Y.~Aoki, G.~Endrodi, Z.~Fodor, S.~D.~Katz and K.~K.~Szabo,
 \newblock Nature {\bf 443}, 675 (2006)
  [hep-lat/0611014].

\bibitem{Borsanyi:2011zzc}
S.~Borsanyi {\em et~al.},
\newblock J. Phys. {\bf G38}, 124101 (2011).

\bibitem{Borsanyi:2010bp} 
  S.~Borsanyi {\it et al.}  [Wuppertal-Budapest Collaboration],
\newblock JHEP {\bf 1009}, 073 (2010)
  [arXiv:1005.3508 [hep-lat]].

\bibitem{Aoki:2006br} 
  Y.~Aoki, Z.~Fodor, S.~D.~Katz and K.~K.~Szabo,
\newblock  Phys.\ Lett.\ B {\bf 643}, 46 (2006)
  [hep-lat/0609068].

\bibitem{Aoki:2009sc} 
  Y.~Aoki, S.~Borsanyi, S.~Durr, Z.~Fodor, S.~D.~Katz, S.~Krieg and K.~K.~Szabo,
 \newblock JHEP {\bf 0906}, 088 (2009)
  [arXiv:0903.4155 [hep-lat]].

\bibitem{Borsanyi:2013cga}
S.~Borsanyi {\em et~al.},
\newblock PoS {\bf LATTICE2013}, 155 (2013), arXiv:1312.2193.

\bibitem{Datta:2010sq}
S.~Datta and S.~Gupta,
\newblock Phys. Rev. {\bf D82}, 114505 (2010), arXiv:1006.0938.

\bibitem{Gupta:2014qka}
S.~Gupta, N.~Karthik, and P.~Majumdar,
\newblock (2014), arXiv:1405.2206.

\bibitem{Gavai:2008zr}
R.~Gavai and S.~Gupta,
\newblock Phys. Rev. {\bf D78}, 114503 (2008), arXiv:0806.2233.

\bibitem{Datta:2009tj}
S.~Datta and S.~Gupta,
\newblock Nucl. Phys. {\bf A830}, 749C (2009), arXiv:0906.3929.

\bibitem{Gavai:2005da}
R.~V. Gavai, S.~Gupta, and S.~Mukherjee,
\newblock Pramana {\bf 71}, 487 (2008), arXiv:hep-lat/0506015.

\bibitem{Gavai:2004se}
R.~V. Gavai, S.~Gupta, and S.~Mukherjee,
\newblock Phys. Rev. {\bf D71}, 074013 (2005), arXiv:hep-lat/0412036.

\bibitem{Gavai:2003nn}
R.~Gavai, S.~Gupta, and R.~Ray,
\newblock Prog. Theor. Phys. Suppl. {\bf 153}, 270 (2004), arXiv:nucl-th/0312010.

\bibitem{Gavai:2003mf}
R.~V. Gavai and S.~Gupta,
\newblock Phys. Rev. {\bf D68}, 034506 (2003), arXiv:hep-lat/0303013.

\bibitem{Gavai:2002kq}
R.~V. Gavai and S.~Gupta,
\newblock Phys. Rev. {\bf D65}, 094515 (2002), arXiv:hep-lat/0202006.

\bibitem{Gavai:2001ie}
R.~V. Gavai, S.~Gupta, and P.~Majumdar,
\newblock Phys. Rev. {\bf D65}, 054506 (2002), arXiv:hep-lat/0110032.

\bibitem{Gavai:2001fr}
R.~V. Gavai and S.~Gupta,
\newblock Phys. Rev. {\bf D64}, 074506 (2001), arXiv:hep-lat/0103013.

\bibitem{Gavai:2014lia}
R.~V. Gavai and S.~Sharma,
\newblock (2014), arXiv:1406.0474.

\bibitem{Gavai:2011uk}
R.~Gavai and S.~Sharma,
\newblock Phys. Rev. {\bf D85}, 054508 (2012), arXiv:1112.5428.

\bibitem{Banerjee:2008ii}
D.~Banerjee, R.~Gavai, and S.~Sharma,
\newblock Phys. Rev. {\bf D78}, 014506 (2008), arXiv:0803.3925.

\bibitem{Gavai:2008ea}
R.~Gavai and S.~Sharma,
\newblock Phys. Rev. {\bf D79}, 074502 (2009), arXiv:0811.3026.

\bibitem{Allton:2005gk}
C.~Allton {\em et~al.},
\newblock Phys. Rev. {\bf D71}, 054508 (2005), arXiv:hep-lat/0501030.

\bibitem{Bazavov:2009zn}
A.~Bazavov {\em et~al.},
\newblock Phys. Rev. {\bf D80}, 014504 (2009), arXiv:0903.4379.

\bibitem{Bernard:2004je}
MILC Collaboration, C.~Bernard {\em et~al.},
\newblock Phys. Rev. {\bf D71}, 034504 (2005), arXiv:hep-lat/0405029.

\bibitem{Petreczky:2009at}
P.~Petreczky,
\newblock Nucl. Phys. {\bf A830}, 11C (2009), arXiv:0908.1917.

\bibitem{Gottlieb:1987ac}
S.~A. Gottlieb, W.~Liu, D.~Toussaint, R.~Renken, and R.~Sugar,
\newblock Phys. Rev. Lett. {\bf 59}, 2247 (1987).

\bibitem{Gottlieb:1988cq}
S.~A. Gottlieb, W.~Liu, D.~Toussaint, R.~Renken, and R.~Sugar,
\newblock Phys. Rev. {\bf D38}, 2888 (1988).

\bibitem{Gavai:1989ce}
R.~Gavai, J.~Potvin, and S.~Sanielevici,
\newblock Phys. Rev. {\bf D40}, 2743 (1989).

\bibitem{Petreczky:2009cr}
RBC-Bielefeld Collaboration, P.~Petreczky, P.~Hegde, and A.~Velytsky,
\newblock PoS {\bf LAT2009}, 159 (2009), arXiv:0911.0196.

\bibitem{Cheng:2009zi}
M.~Cheng {\em et~al.},
\newblock Phys. Rev. {\bf D81}, 054504 (2010), arXiv:0911.2215.

\bibitem{Petreczky:2012rq}
P.~Petreczky,
\newblock J.Phys. {\bf G39}, 093002 (2012), arXiv:1203.5320.

\bibitem{Borsanyi:2013hza}
S.~Borsanyi {\em et~al.},
\newblock Phys. Rev. Lett. {\bf 111}, 062005 (2013), arXiv:1305.5161.

\bibitem{Sharma:2013hsa}
S.~Sharma,
\newblock Adv. High Energy Phys. {\bf 2013}, 452978 (2013), arXiv:1403.2102.

\bibitem{Karsch:2000ps}
F.~Karsch, E.~Laermann, and A.~Peikert,
\newblock Phys. Lett. {\bf B478}, 447 (2000), arXiv:hep-lat/0002003.

\bibitem{Bazavov:2013dta}
A.~Bazavov {\em et~al.},
\newblock Phys. Rev. Lett. 111, {\bf 082301}, 082301 (2013), arXiv:1304.7220.

\bibitem{Bazavov:2013uja}
A.~Bazavov {\em et~al.},
\newblock Phys. Rev. {\bf D88}, 094021 (2013), arXiv:1309.2317.

\bibitem{Bazavov:2012vg}
A.~Bazavov {\em et~al.},
\newblock Phys. Rev. Lett. {\bf 109}, 192302 (2012), arXiv:1208.1220.

\bibitem{Bazavov:2012jq}
HotQCD Collaboration, A.~Bazavov {\em et~al.},
\newblock Phys. Rev. {\bf D86}, 034509 (2012), arXiv:1203.0784.

\bibitem{Cheng:2008zh}
M.~Cheng {\em et~al.},
\newblock Phys. Rev. {\bf D79}, 074505 (2009), arXiv:0811.1006.

\bibitem{Brambilla:2014aaa}
N.~Brambilla {\em et~al.},
\newblock (2014), arXiv:1404.3723.

\bibitem{Freedman:1976dm}
B.~A. Freedman and L.~D. McLerran,
\newblock Phys. Rev. {\bf D16}, 1147 (1977).

\bibitem{Freedman:1976ub}
B.~A. Freedman and L.~D. McLerran,
\newblock Phys. Rev. {\bf D16}, 1169 (1977).

\bibitem{Shuryak:1977ut}
E.~V. Shuryak,
\newblock Sov. Phys. JETP {\bf 47}, 212 (1978).

\bibitem{Chin:1978gj}
S.~Chin,
\newblock Phys. Lett. {\bf B78}, 552 (1978).

\bibitem{Kapusta:1979fh}
J.~I. Kapusta,
\newblock Nucl. Phys. {\bf B148}, 461 (1979).

\bibitem{Toimela:1982hv}
T.~Toimela,
\newblock Phys. Lett. {\bf B124}, 407 (1983).

\bibitem{Frenkel:1992az}
J.~Frenkel, A.~Saa, and J.~Taylor,
\newblock Phys. Rev. {\bf D46}, 3670 (1992).

\bibitem{Arnold:1994ps}
P.~B. Arnold and C.-X. Zhai,
\newblock Phys. Rev. {\bf D50}, 7603 (1994), arXiv:hep-ph/9408276.

\bibitem{Arnold:1994eb}
P.~B. Arnold and C.-x. Zhai,
\newblock Phys. Rev. {\bf D51}, 1906 (1995), arXiv:hep-ph/9410360.

\bibitem{Parwani:1994xi}
R.~R. Parwani and C.~Coriano,
\newblock Nucl.Phys. {\bf B434}, 56 (1995), arXiv:hep-ph/9409269.

\bibitem{Parwani:1994zz}
R.~Parwani and H.~Singh,
\newblock Phys. Rev. {\bf D51}, 4518 (1995), arXiv:hep-th/9411065.

\bibitem{Braaten:1995cm}
E.~Braaten and A.~Nieto,
\newblock Phys. Rev. {\bf D51}, 6990 (1995), arXiv:hep-ph/9501375.

\bibitem{Parwani:1994je}
R.~R. Parwani,
\newblock Phys. Lett. {\bf B334}, 420 (1994), arXiv:hep-ph/9406318.

\bibitem{Andersen:1995ej}
J.~O. Andersen,
\newblock Phys. Rev. {\bf D53}, 7286 (1996), arXiv:hep-ph/9509409.

\bibitem{Zhai:1995ac}
C.-x. Zhai and B.~M. Kastening,
\newblock Phys. Rev. {\bf D52}, 7232 (1995), arXiv:hep-ph/9507380.

\bibitem{Braaten:1995ju}
E.~Braaten and A.~Nieto,
\newblock Phys. Rev. Lett. {\bf 76}, 1417 (1996), arXiv:hep-ph/9508406.

\bibitem{Braaten:1995jr}
E.~Braaten and A.~Nieto,
\newblock Phys. Rev. {\bf D53}, 3421 (1996), arXiv:hep-ph/9510408.

\bibitem{Ipp:2006ij}
A.~Ipp, K.~Kajantie, A.~Rebhan, and A.~Vuorinen,
\newblock Phys. Rev. {\bf D74}, 045016 (2006), arXiv:hep-ph/0604060.

\bibitem{Vuorinen:2002ue}
A.~Vuorinen,
\newblock Phys. Rev. {\bf D67}, 074032 (2003), arXiv:hep-ph/0212283.

\bibitem{Gynther:2007bw}
A.~Gynther, M.~Laine, Y.~Schroder, C.~Torrero, and A.~Vuorinen,
\newblock JHEP {\bf 0704}, 094 (2007), arXiv:hep-ph/0703307.

\bibitem{Andersen:2009ct}
J.~O. Andersen, L.~Kyllingstad, and L.~E. Leganger,
\newblock JHEP {\bf 0908}, 066 (2009), arXiv:0903.4596.

\bibitem{Linde:1980ts}
A.~D. Linde,
\newblock Phys. Lett. {\bf B96}, 289 (1980).

\bibitem{Braaten:1989mz}
E.~Braaten and R.~D. Pisarski,
\newblock Nucl. Phys. {\bf B337}, 569 (1990).

\bibitem{Blaizot:2003tw}
J.-P. Blaizot, E.~Iancu, and A.~Rebhan,
\newblock (2003), arXiv:hep-ph/0303185.

\bibitem{Kraemmer:2003gd}
U.~Kraemmer and A.~Rebhan,
\newblock Rept. Prog. Phys. {\bf 67}, 351 (2004), arXiv:hep-ph/0310337.

\bibitem{Andersen:2004fp}
J.~O. Andersen and M.~Strickland,
\newblock Annals Phys. {\bf 317}, 281 (2005), arXiv:hep-ph/0404164.

\bibitem{Chiku:1998kd}
S.~Chiku and T.~Hatsuda,
\newblock Phys. Rev. {\bf D58}, 076001 (1998), arXiv:hep-ph/9803226.

\bibitem{Braaten:1991gm}
E.~Braaten and R.~D. Pisarski,
\newblock Phys. Rev. {\bf D45}, 1827 (1992).

\bibitem{Braaten:1989kk}
E.~Braaten and R.~D. Pisarski,
\newblock Phys. Rev. Lett. {\bf 64}, 1338 (1990).

\bibitem{Blaizot:2001vr}
J.~Blaizot, E.~Iancu, and A.~Rebhan,
\newblock Phys. Lett. {\bf B523}, 143 (2001), arXiv:hep-ph/0110369.


\bibitem{Jiang:2010jz}
Y.~Jiang, H.-x. Zhu, W.-m. Sun, and H.-s. Zong,
\newblock J. Phys. {\bf G37}, 055001 (2010), arXiv:1003.5031.

\bibitem{Blaizot:1999ap}
J.~Blaizot, E.~Iancu, and A.~Rebhan,
\newblock Phys. Lett. {\bf B470}, 181 (1999), arXiv:hep-ph/9910309.

\bibitem{Blaizot:2000fc}
J.~Blaizot, E.~Iancu, and A.~Rebhan,
\newblock Phys. Rev. {\bf D63}, 065003 (2001), arXiv:hep-ph/0005003.

\bibitem{Andersen:2002jz}
J.~O. Andersen and M.~Strickland,
\newblock Phys. Rev. {\bf D66}, 105001 (2002), arXiv:hep-ph/0206196.


\bibitem{Blaizot:1999ip}
J.~Blaizot, E.~Iancu, and A.~Rebhan,
\newblock Phys. Rev. Lett. {\bf 83}, 2906 (1999), arXiv:hep-ph/9906340.


\bibitem{Andersen:1999fw}
J.~O. Andersen, E.~Braaten, and M.~Strickland,
\newblock Phys. Rev. Lett. {\bf 83}, 2139 (1999), arXiv:hep-ph/9902327.

\bibitem{Andersen:1999sf}
J.~O. Andersen, E.~Braaten, and M.~Strickland,
\newblock Phys. Rev. {\bf D61}, 014017 (2000), arXiv:hep-ph/9905337.

\bibitem{Andersen:1999va}
J.~O. Andersen, E.~Braaten, and M.~Strickland,
\newblock Phys. Rev. {\bf D61}, 074016 (2000), arXiv:hep-ph/9908323.

\bibitem{Chakraborty:2001kx}
P.~Chakraborty, M.~G. Mustafa, and M.~H. Thoma,
\newblock Eur. Phys. J. {\bf C23}, 591 (2002), arXiv:hep-ph/0111022.

\bibitem{Chakraborty:2002yt}
P.~Chakraborty, M.~G. Mustafa, and M.~H. Thoma,
\newblock Phys. Rev. {\bf D67}, 114004 (2003), arXiv:hep-ph/0210159.

\bibitem{Chakraborty:2003uw}
P.~Chakraborty, M.~G. Mustafa, and M.~H. Thoma,
\newblock Phys. Rev. {\bf D68}, 085012 (2003), arXiv:hep-ph/0303009.

\bibitem{Blaizot:2002xz}
J.~Blaizot, E.~Iancu, and A.~Rebhan,
\newblock Eur. Phys. J. {\bf C27}, 433 (2003), arXiv:hep-ph/0206280.

\bibitem{Andersen:2012wr}
J.~O. Andersen, S.~Mogliacci, N.~Su, and A.~Vuorinen,
\newblock Phys. Rev. {\bf D87}, 074003 (2013), arXiv:1210.0912.

\bibitem{Su:2012iy} 
  N.~Su,
\newblock Commun.\ Theor.\ Phys.\  {\bf 57}, 409 (2012), arXiv:1204.0260.

\bibitem{Andersen:2002ey}
J.~O. Andersen, E.~Braaten, E.~Petitgirard, and M.~Strickland,
\newblock Phys. Rev. {\bf D66}, 085016 (2002), arXiv:hep-ph/0205085.

\bibitem{Andersen:2003zk}
J.~O. Andersen, E.~Petitgirard, and M.~Strickland,
\newblock Phys. Rev. {\bf D70}, 045001 (2004), arXiv:hep-ph/0302069.

\bibitem{Andersen:2009tc}
J.~O. Andersen, M.~Strickland, and N.~Su,
\newblock Phys. Rev. Lett. {\bf 104}, 122003 (2010), arXiv:0911.0676.

\bibitem{Andersen:2010ct}
J.~O. Andersen, M.~Strickland, and N.~Su,
\newblock JHEP {\bf 1008}, 113 (2010), arXiv:1005.1603.

\bibitem{Andersen:2009tw}
J.~O. Andersen, M.~Strickland, and N.~Su,
\newblock Phys. Rev. {\bf D80}, 085015 (2009), arXiv:0906.2936.

\bibitem{Andersen:2010wu}
J.~O. Andersen, L.~E. Leganger, M.~Strickland, and N.~Su,
\newblock Phys. Lett. {\bf B696}, 468 (2011), arXiv:1009.4644.

\bibitem{Andersen:2011sf}
J.~O. Andersen, L.~E. Leganger, M.~Strickland, and N.~Su,
\newblock JHEP {\bf 1108}, 053 (2011), arXiv:1103.2528.


\bibitem{Andersen:2011ug}
J.~O. Andersen, L.~E. Leganger, M.~Strickland, and N.~Su,
\newblock Phys. Rev. {\bf D84}, 087703 (2011), arXiv:1106.0514.

\bibitem{Yukalov:1976pm}
V.~Yukalov,
\newblock Teor. Mat. Fiz. {\bf 26}, 403 (1976).

\bibitem{Stevenson:1981vj}
P.~M. Stevenson,
\newblock Phys. Rev. {\bf D23}, 2916 (1981).

\bibitem{Duncan:1988hw}
A.~Duncan and M.~Moshe,
\newblock Phys. Lett. {\bf B215}, 352 (1988).

\bibitem{Duncan:1992ba}
A.~Duncan and H.~Jones,
\newblock Phys. Rev. {\bf D47}, 2560 (1993).

\bibitem{Sisakian:1994nn}
A.~Sisakian, I.~Solovtsov, and O.~Shevchenko,
\newblock Int.J.Mod.Phys. {\bf A9}, 1929 (1994).

\bibitem{Janke:1995zz}
W.~Janke and H.~Kleinert,
\newblock Phys. Rev. Lett. {\bf 75}, 2787 (1995).

\bibitem{Karsch:1997gj}
F.~Karsch, A.~Patkos, and P.~Petreczky,
\newblock Phys. Lett. {\bf B401}, 69 (1997), arXiv:hep-ph/9702376.

\bibitem{Andersen:2000yj}
J.~O. Andersen, E.~Braaten, and M.~Strickland,
\newblock Phys. Rev. {\bf D63}, 105008 (2001), arXiv:hep-ph/0007159.

\bibitem{Andersen:2001ez}
J.~Andersen and M.~Strickland,
\newblock Phys.Rev. {\bf D64}, 105012 (2001), arXiv:hep-ph/0105214.

\bibitem{Andersen:2008bz}
J.~O. Andersen and L.~Kyllingstad,
\newblock Phys. Rev. {\bf D78}, 076008 (2008), arXiv:0805.4478.



\bibitem{Agakishiev:1995xb}
CERES Collaboration, G.~Agakichiev {\em et~al.},
\newblock Phys. Rev. Lett. {\bf 75}, 1272 (1995).

\bibitem{Agakishiev:1997au}
CERES/NA45 Collaboration, G.~Agakishiev {\em et~al.},
\newblock Phys. Lett. {\bf B422}, 405 (1998), arXiv:nucl-ex/9712008.

\bibitem{Masera:1995ck}
HELIOS Collaboration, M.~Masera,
\newblock Nucl. Phys. {\bf A590}, 93C (1995).

\bibitem{Drees:1998rn}
A.~Drees,
\newblock Nucl. Phys. {\bf A630}, 449C (1998).

\bibitem{Rapp:1999ej}
R.~Rapp and J.~Wambach,
\newblock Adv. Nucl. Phys. {\bf 25}, 1 (2000), arXiv:hep-ph/9909229.

\bibitem{Rapp:2009yu}
R.~Rapp, J.~Wambach, and H.~van Hees,
\newblock (2009), arXiv:0901.3289.

\bibitem{Cassing:1999es}
W.~Cassing and E.~Bratkovskaya,
\newblock Phys. Rept. {\bf 308}, 65 (1999).

\bibitem{Brown:1991kk}
G.~Brown and M.~Rho,
\newblock Phys. Rev. Lett. {\bf 66}, 2720 (1991).

\bibitem{Friman:1997tc}
B.~Friman and H.~Pirner,
\newblock Nucl. Phys. {\bf A617}, 496 (1997), arXiv:nucl-th/9701016.

\bibitem{Rapp:1997fs}
R.~Rapp, G.~Chanfray, and J.~Wambach,
\newblock Nucl. Phys. {\bf A617}, 472 (1997), arXiv:hep-ph/9702210.

\bibitem{Rapp:1995zy}
R.~Rapp, G.~Chanfray, and J.~Wambach,
\newblock Phys. Rev. Lett. {\bf 76}, 368 (1996), arXiv:hep-ph/9508353.

\bibitem{Gale:1993zj}
C.~Gale and P.~Lichard,
\newblock Phys. Rev. {\bf D49}, 3338 (1994), arXiv:hep-ph/9307363.

\bibitem{Rapp:1999qu}
R.~Rapp and C.~Gale,
\newblock Phys. Rev. {\bf C60}, 024903 (1999), arXiv:hep-ph/9902268.

\bibitem{Klingl:1997kf}
F.~Klingl, N.~Kaiser, and W.~Weise,
\newblock Nucl. Phys. {\bf A624}, 527 (1997), arXiv:hep-ph/9704398.

\bibitem{Peters:1997va}
W.~Peters, M.~Post, H.~Lenske, S.~Leupold, and U.~Mosel,
\newblock Nucl. Phys. {\bf A632}, 109 (1998), arXiv:nucl-th/9708004.

\bibitem{Cassing:1997jz}
W.~Cassing, E.~Bratkovskaya, R.~Rapp, and J.~Wambach,
\newblock Phys. Rev. {\bf C57}, 916 (1998), arXiv:nucl-th/9708020.

\bibitem{Post:2000qi}
M.~Post, S.~Leupold, and U.~Mosel,
\newblock Nucl. Phys. {\bf A689}, 753 (2001), arXiv:nucl-th/0008027.

\bibitem{Srivastava:1996vi}
D.~Srivastava, B.~Sinha, and C.~Gale,
\newblock Phys. Rev. {\bf C53}, 567 (1996).

\bibitem{Pal:1996xn}
D.~Pal, D.~Srivastava, and K.~Haglin,
\newblock Phys. Rev. {\bf C54}, 1366 (1996).

\bibitem{Srivastava:1996wr}
D.~Srivastava, B.~Sinha, D.~Pal, C.~Gale, and K.~Haglin,
\newblock Nucl. Phys. {\bf A610}, 350C (1996).

\bibitem{Pal:1998jr}
D.~Pal and M.~G. Mustafa,
\newblock Phys. Rev. {\bf C60}, 034905 (1999), arXiv:nucl-th/9808049.

\bibitem{Srivastava:1996qd}
D.~K. Srivastava, M.~G. Mustafa, and B.~Muller,
\newblock Phys.Rev. {\bf C56}, 1064 (1997), arXiv:nucl-th/9611041.

\bibitem{Alam:1999sc}
J.~Alam, S.~Sarkar, P.~Roy, T.~Hatsuda, and B.~Sinha,
\newblock Annals Phys. {\bf 286}, 159 (2001), arXiv:hep-ph/9909267.

\bibitem{Alam:2001ar}
J.-e. Alam, P.~Roy, S.~Sarkar, and B.~Sinha,
\newblock Phys. Rev. {\bf C67}, 054901 (2003), arXiv:nucl-th/0106038.

\bibitem{Dusling:2007su}
K.~Dusling and I.~Zahed,
\newblock Nucl. Phys. {\bf A825}, 212 (2009), arXiv:0712.1982.

\bibitem{Bratkovskaya:2008bf}
E.~Bratkovskaya, W.~Cassing, and O.~Linnyk,
\newblock Phys. Lett. {\bf B670}, 428 (2009), arXiv:0805.3177.

\bibitem{Aurenche:1998nw}
P.~Aurenche, F.~Gelis, R.~Kobes, and H.~Zaraket,
\newblock Phys. Rev. {\bf D58}, 085003 (1998), arXiv:hep-ph/9804224.

\bibitem{Karsch:2001uw}
F.~Karsch, E.~Laermann, P.~Petreczky, S.~Stickan, and I.~Wetzorke,
\newblock Phys. Lett. {\bf B530}, 147 (2002), arXiv:hep-lat/0110208.

\bibitem{Allton:2003vx}
C.~Allton {\em et~al.},
\newblock Phys. Rev. {\bf D68}, 014507 (2003), arXiv:hep-lat/0305007.

\bibitem{Petreczky:2010xg}
P.~Petreczky,
\newblock Mod. Phys. Lett. {\bf A25}, 3081 (2010), arXiv:1009.5935.

\bibitem{Thoma:2004sp}
M.~H. Thoma,
\newblock J. Phys. {\bf G31}, L7 (2005), arXiv:hep-ph/0503154.

\bibitem{Rapp:1999us}
R.~Rapp and J.~Wambach,
\newblock Eur. Phys. J. {\bf A6}, 415 (1999), arXiv:hep-ph/9907502.

\bibitem{Cleymans:1986na}
J.~Cleymans, J.~Fingberg, and K.~Redlich,
\newblock Phys. Rev. {\bf D35}, 2153 (1987).

\bibitem{Blaizot:2003iq}
J.~Blaizot, E.~Iancu, and A.~Rebhan,
\newblock Phys. Rev. {\bf D68}, 025011 (2003), arXiv:hep-ph/0303045.

\bibitem{Vuorinen:2004rd}
A.~Vuorinen,
\newblock (2004), arXiv:hep-ph/0402242.

\bibitem{Laine:2006cp}
M.~Laine and Y.~Schroder,
\newblock Phys. Rev. {\bf D73}, 085009 (2006), arXiv:hep-ph/0603048.

\bibitem{Rebhan:2003wn}
A.~Rebhan and P.~Romatschke,
\newblock Phys. Rev. {\bf D68}, 025022 (2003), arXiv:hep-ph/0304294.

\bibitem{Cassing:2007nb}
W.~Cassing,
\newblock Nucl. Phys. {\bf A795}, 70 (2007), arXiv:0707.3033.

\bibitem{Kurkela:2009gj}
A.~Kurkela, P.~Romatschke, and A.~Vuorinen,
\newblock Phys. Rev. {\bf D81}, 105021 (2010), arXiv:0912.1856.

\bibitem{Bazavov:2012ka}
A.~Bazavov {\em et~al.},
\newblock Phys. Rev. {\bf D86}, 114031 (2012), arXiv:1205.6155.

\bibitem{Beringer:1900zz}
Particle Data Group Collaboration, J.~Beringer {\em et~al.},
\newblock Phys. Rev. {\bf D86}, 010001 (2012).

\bibitem{vanRitbergen:1997va}
T.~van Ritbergen, J.~Vermaseren, and S.~Larin,
\newblock Phys. Lett. {\bf B400}, 379 (1997), arXiv:hep-ph/9701390.

\bibitem{Politzer:1973fx}
H.~D. Politzer,
\newblock Phys. Rev. Lett. {\bf 30}, 1346 (1973).

\bibitem{Bethke:2006ac}
S.~Bethke,
\newblock Prog.Part.Nucl.Phys. {\bf 58}, 351 (2007), arXiv:hep-ex/0606035.

\bibitem{kapusta}
J.~I. Kapusta and C.~Gale,
\newblock {\em Finite Temperature Field Theory Principle and Applications}, 2nd
  ed. (Cambridge University Press, Cambridge, 1996).

\bibitem{lebellac}
M.~LeBellac,
\newblock {\em Thermal Field Theory}, 1st ed. (Cambridge University Press,
  Cambridge, England, 1996).

\bibitem{Pisarski:1990ds}
R.~D. Pisarski,
\newblock Nucl. Phys. {\bf A525}, 175 (1991).

\bibitem{Frenkel:1989br}
J.~Frenkel and J.~Taylor,
\newblock Nucl. Phys. {\bf B334}, 199 (1990).



\bibitem{Braaten:1990az}
E.~Braaten and R.~D. Pisarski,
\newblock Nucl. Phys. {\bf B339}, 310 (1990).

\bibitem{Taylor:1990ia}
J.~Taylor and S.~Wong,
\newblock Nucl. Phys. {\bf B346}, 115 (1990).

\bibitem{Kobes:1990dc}
R.~Kobes, G.~Kunstatter, and A.~Rebhan,
\newblock Nucl. Phys. {\bf B355}, 1 (1991).

\bibitem{silin}
V.~Silin,
\newblock Sov.Phys. JETP {\bf 11}, 1136 (1960).

\bibitem{Su:2011zv}
N.~Su,
\newblock (2011), arXiv:1104.3450.


\bibitem{Klimov:1981ka}
V.~Klimov,
\newblock Sov. J. Nucl. Phys. {\bf 33}, 934 (1981).

\bibitem{Klimov:1982bv}
V.~Klimov,
\newblock Sov. Phys. JETP {\bf 55}, 199 (1982).

\bibitem{Weldon:1982aq}
H.~A. Weldon,
\newblock Phys. Rev. {\bf D26}, 1394 (1982).

\bibitem{Mogliacci:2013mca}
S.~Mogliacci, J.~O. Andersen, M.~Strickland, N.~Su, and A.~Vuorinen,
\newblock JHEP {\bf 1312}, 055 (2013), arXiv:1307.8098.

\bibitem{Mogliacci:2014upa}
S.~Mogliacci and N.~Su,
\newblock EPJ Web Conf. {\bf 70}, 00031 (2014).

\bibitem{Su:2012bba}
N.~Su,
\newblock PoS {\bf ConfinementX}, 189 (2012).

\bibitem{Mogliacci:2013iwa}
S.~Mogliacci,
\newblock J. Phys. Conf. Ser. {\bf 503}, 012005 (2014), arXiv:1311.2101.

\bibitem{Braaten:1990wp}
E.~Braaten, R.~D. Pisarski, and T.-C. Yuan,
\newblock Phys. Rev. Lett. {\bf 64}, 2242 (1990).

\bibitem{Baier:1993zb}
R.~Baier, S.~Peigne, and D.~Schiff,
\newblock Z. Phys. {\bf C62}, 337 (1994), arXiv:hep-ph/9311329.

\bibitem{Mustafa:2004hf}
M.~G. Mustafa, M.~H. Thoma, and P.~Chakraborty,
\newblock Phys. Rev. {\bf C71}, 017901 (2005), arXiv:hep-ph/0403279.

\bibitem{Mustafa:2005je}
M.~G. Mustafa, P.~Chakraborty, and M.~H. Thoma,
\newblock J. Phys. Conf. Ser. {\bf 50}, 438 (2006), arXiv:hep-ph/0504174.

\bibitem{Chakraborty:2006md}
P.~Chakraborty, M.~G. Mustafa, and M.~H. Thoma,
\newblock Phys. Rev. {\bf D74}, 094002 (2006), arXiv:hep-ph/0606316.

\bibitem{Chakraborty:2007ug}
P.~Chakraborty, M.~G. Mustafa, R.~Ray, and M.~H. Thoma,
\newblock J. Phys. {\bf G34}, 2141 (2007), arXiv:0705.1447.

\bibitem{Laine:2006ns}
M.~Laine, O.~Philipsen, P.~Romatschke, and M.~Tassler,
\newblock JHEP {\bf 0703}, 054 (2007), arXiv:hep-ph/0611300.

\bibitem{Dumitru:2007hy}
A.~Dumitru, Y.~Guo, and M.~Strickland,
\newblock Phys. Lett. {\bf B662}, 37 (2008), arXiv:0711.4722.

\bibitem{Dumitru:2009ni}
A.~Dumitru, Y.~Guo, A.~Mocsy, and M.~Strickland,
\newblock Phys. Rev. {\bf D79}, 054019 (2009), arXiv:0901.1998.

\bibitem{Dumitru:2009fy}
A.~Dumitru, Y.~Guo, and M.~Strickland,
\newblock Phys. Rev. {\bf D79}, 114003 (2009), arXiv:0903.4703.

\bibitem{Thakur:2012eb}
L.~Thakur, N.~Haque, U.~Kakade, and B.~K. Patra,
\newblock Phys. Rev. {\bf D88}, 054022 (2013), arXiv:1212.2803.

\bibitem{Thakur:2013nia}
L.~Thakur, U.~Kakade, and B.~K. Patra,
\newblock (2013), arXiv:1401.0172.

\bibitem{Pisarski:1993rf}
R.~Pisarski,
\newblock Phys. Rev. {\bf D47}, 5589 (1993).

\bibitem{Peigne:1993ky}
S.~Peigne, E.~Pilon, and D.~Schiff,
\newblock Z. Phys. {\bf C60}, 455 (1993), arXiv:hep-ph/9306219.

\bibitem{Sarkar:2012fk}
S.~Sarkar and A.~K. Dutt-Mazumder,
\newblock Phys. Rev. {\bf D88}, 054006 (2013), arXiv:1205.4895.

\bibitem{Abada:2011cc}
A.~Abada and N.~Daira-Aifa,
\newblock JHEP {\bf 1204}, 071 (2012), arXiv:1112.6065.

\bibitem{Braaten:1990it}
E.~Braaten and R.~D. Pisarski,
\newblock Phys. Rev. {\bf D42}, 2156 (1990).

\bibitem{Braaten:1991jj}
E.~Braaten and M.~H. Thoma,
\newblock Phys. Rev. {\bf D44}, 1298 (1991).

\bibitem{Braaten:1991we}
E.~Braaten and M.~H. Thoma,
\newblock Phys. Rev. {\bf D44}, 2625 (1991).

\bibitem{Thoma:1990fm}
M.~H. Thoma and M.~Gyulassy,
\newblock Nucl. Phys. {\bf B351}, 491 (1991).

\bibitem{Romatschke:2003vc}
P.~Romatschke and M.~Strickland,
\newblock Phys. Rev. {\bf D69}, 065005 (2004), arXiv:hep-ph/0309093.

\bibitem{Romatschke:2004au}
P.~Romatschke and M.~Strickland,
\newblock Phys. Rev. {\bf D71}, 125008 (2005), arXiv:hep-ph/0408275.

\bibitem{Mustafa:2004dr}
M.~G. Mustafa,
\newblock Phys. Rev. {\bf C72}, 014905 (2005), arXiv:hep-ph/0412402.

\bibitem{Mustafa:2003vh}
M.~G. Mustafa and M.~H. Thoma,
\newblock Acta Phys. Hung. {\bf A22}, 93 (2005), arXiv:hep-ph/0311168.

\bibitem{Djordjevic:2006tw}
M.~Djordjevic,
\newblock Phys. Rev. {\bf C74}, 064907 (2006), arXiv:nucl-th/0603066.

\bibitem{Chakraborty:2006db}
P.~Chakraborty, M.~G. Mustafa, and M.~H. Thoma,
\newblock Phys. Rev. {\bf C75}, 064908 (2007), arXiv:hep-ph/0611355.

\bibitem{Qin:2007rn}
G.-Y. Qin {\em et~al.},
\newblock Phys. Rev. Lett. {\bf 100}, 072301 (2008), arXiv:0710.0605.

\bibitem{Qin:2009uh}
G.-Y. Qin, A.~Majumder, H.~Song, and U.~Heinz,
\newblock Phys. Rev. Lett. {\bf 103}, 152303 (2009), arXiv:0903.2255.

\bibitem{Qin:2009gw}
G.-Y. Qin and A.~Majumder,
\newblock Phys. Rev. Lett. {\bf 105}, 262301 (2010), arXiv:0910.3016.

\bibitem{Mrowczynski:2000ed}
S.~Mrowczynski and M.~H. Thoma,
\newblock Phys. Rev. {\bf D62}, 036011 (2000), arXiv:hep-ph/0001164.

\bibitem{Romatschke:2003ms}
P.~Romatschke and M.~Strickland,
\newblock Phys. Rev. {\bf D68}, 036004 (2003), arXiv:hep-ph/0304092.

\bibitem{Romatschke:2004jh}
P.~Romatschke and M.~Strickland,
\newblock Phys. Rev. {\bf D70}, 116006 (2004), arXiv:hep-ph/0406188.

\bibitem{Rebhan:2004ur}
A.~Rebhan, P.~Romatschke, and M.~Strickland,
\newblock Phys. Rev. Lett. {\bf 94}, 102303 (2005), arXiv:hep-ph/0412016.

\bibitem{Schenke:2006fz}
B.~Schenke and M.~Strickland,
\newblock Phys. Rev. {\bf D74}, 065004 (2006), arXiv:hep-ph/0606160.

\bibitem{Rebhan:2008uj}
A.~Rebhan, M.~Strickland, and M.~Attems,
\newblock Phys. Rev. {\bf D78}, 045023 (2008), arXiv:0802.1714.

\bibitem{Attems:2012js}
M.~Attems, A.~Rebhan, and M.~Strickland,
\newblock Phys. Rev. {\bf D87}, 025010 (2013), arXiv:1207.5795.

\bibitem{Kiessig:2011ga}
C.~Kiessig and M.~Plumacher,
\newblock JCAP {\bf 1209}, 012 (2012), arXiv:1111.1235.

\bibitem{Kiessig:2011fw}
C.~Kiessig and M.~Plumacher,
\newblock JCAP {\bf 1207}, 014 (2012), arXiv:1111.1231.

\bibitem{Hashimoto:1992np}
T.~Hashimoto, A.~Nakamura, and I.~Stamatescu,
\newblock Nucl. Phys. {\bf B400}, 267 (1993).

\bibitem{Boyd:1994np}
G.~Boyd, S.~Gupta, F.~Karsch, and E.~Laermann,
\newblock Z. Phys. {\bf C64}, 331 (1994), arXiv:hep-lat/9405006.

\bibitem{Karsch:2000gi}
F.~Karsch, M.~Mustafa, and M.~Thoma,
\newblock Phys. Lett. {\bf B497}, 249 (2001), arXiv:hep-ph/0007093.

\bibitem{Alberico:2006wc}
W.~Alberico, A.~Beraudo, P.~Czerski, and A.~Molinari,
\newblock Nucl. Phys. {\bf A775}, 188 (2006), arXiv:hep-ph/0605060.

\bibitem{Czerski:2008zz}
P.~Czerski,
\newblock Nucl. Phys. {\bf A807}, 11 (2008).

\bibitem{Arnold:2000dr}
P.~B. Arnold, G.~D. Moore, and L.~G. Yaffe,
\newblock JHEP {\bf 0011}, 001 (2000), arXiv:hep-ph/0010177.

\bibitem{Arnold:2001ms}
P.~B. Arnold, G.~D. Moore, and L.~G. Yaffe,
\newblock JHEP {\bf 0112}, 009 (2001), arXiv:hep-ph/0111107.

\bibitem{Arnold:2003zc}
P.~B. Arnold, G.~D. Moore, and L.~G. Yaffe,
\newblock JHEP {\bf 0305}, 051 (2003), arXiv:hep-ph/0302165.

\bibitem{Aurenche:2002wq}
P.~Aurenche, F.~Gelis, G.~Moore, and H.~Zaraket,
\newblock JHEP {\bf 0212}, 006 (2002), arXiv:hep-ph/0211036.

\bibitem{Appelshauser:1998vn}
NA49 Collaboration, H.~Appelshauser {\em et~al.},
\newblock Nucl. Phys. {\bf A638}, 91 (1998).

\bibitem{Appelshauser:1999ft}
NA49 Collaboration, H.~Appelshauser {\em et~al.},
\newblock Phys. Lett. {\bf B459}, 679 (1999), arXiv:hep-ex/9904014.

\bibitem{Reid:1999it}
NA49 Collaboration, J.~Reid,
\newblock Nucl. Phys. {\bf A661}, 407 (1999).

\bibitem{Stephanov:1998dy}
M.~A. Stephanov, K.~Rajagopal, and E.~V. Shuryak,
\newblock Phys. Rev. Lett. {\bf 81}, 4816 (1998), arXiv:hep-ph/9806219.

\bibitem{Stephanov:1999zu}
M.~A. Stephanov, K.~Rajagopal, and E.~V. Shuryak,
\newblock Phys. Rev. {\bf D60}, 114028 (1999), arXiv:hep-ph/9903292.

\bibitem{forster}
D.~Forster,
\newblock {\em Hydrodynamics Fluctuation, Broken Symmetry and Correlation
  Function} (Benjamin/Cummings, Menlo Park, CA, 1975).

\bibitem{Callen:1951vq}
H.~B. Callen and T.~A. Welton,
\newblock Phys. Rev. {\bf 83}, 34 (1951).

\bibitem{Kubo:1957mj}
R.~Kubo,
\newblock J. Phys. Soc. Jap. {\bf 12}, 570 (1957).

\bibitem{Ding:2010ga}
H.-T. Ding {\em et~al.},
\newblock Phys. Rev. {\bf D83}, 034504 (2011), arXiv:1012.4963.

\bibitem{Sheikholeslami:1985ij}
B.~Sheikholeslami and R.~Wohlert,
\newblock Nucl. Phys. {\bf B259}, 572 (1985).

\bibitem{Luscher:1996jn}
M.~Luscher, S.~Sint, R.~Sommer, and H.~Wittig,
\newblock Nucl. Phys. {\bf B491}, 344 (1997), arXiv:hep-lat/9611015.

\bibitem{Nakahara:1999vy}
Y.~Nakahara, M.~Asakawa, and T.~Hatsuda,
\newblock Phys. Rev. {\bf D60}, 091503 (1999), arXiv:hep-lat/9905034.

\bibitem{Asakawa:2000tr}
M.~Asakawa, T.~Hatsuda, and Y.~Nakahara,
\newblock Prog. Part. Nucl. Phys. {\bf 46}, 459 (2001), arXiv:hep-lat/0011040.

\bibitem{Wetzorke:2000ez}
I.~Wetzorke and F.~Karsch,
\newblock p. 193 (2000), hep-lat/0008008.

\bibitem{karsch_private}
F.~Karsch,
\newblock private communication.



\bibitem{Huovinen:2009yb}
P.~Huovinen and P.~Petreczky,
\newblock Nucl. Phys. {\bf A837}, 26 (2010), arXiv:0912.2541.

\bibitem{KorthalsAltes:1999xb}
C.~Korthals-Altes, A.~Kovner, and M.~A. Stephanov,
\newblock Phys. Lett. {\bf B469}, 205 (1999), arXiv:hep-ph/9909516.

\bibitem{Pisarski:2000eq}
R.~D. Pisarski,
\newblock Phys. Rev. {\bf D62}, 111501 (2000), arXiv:hep-ph/0006205.

\bibitem{KorthalsAltes:2000gs}
C.~Korthals-Altes and A.~Kovner,
\newblock Phys. Rev. {\bf D62}, 096008 (2000), arXiv:hep-ph/0004052.

\bibitem{Zwanziger:2004np}
D.~Zwanziger,
\newblock Phys. Rev. Lett. {\bf 94}, 182301 (2005), arXiv:hep-ph/0407103.

\bibitem{Vuorinen:2006nz}
A.~Vuorinen and L.~G. Yaffe,
\newblock Phys. Rev. {\bf D74}, 025011 (2006), arXiv:hep-ph/0604100.

\bibitem{deForcrand:2008aw}
P.~de~Forcrand, A.~Kurkela, and A.~Vuorinen,
\newblock Phys. Rev. {\bf D77}, 125014 (2008), arXiv:0801.1566.

\bibitem{Fukushima:2013xsa}
K.~Fukushima and N.~Su,
\newblock Phys. Rev. {\bf D88}, 076008 (2013), arXiv:1304.8004.

\bibitem{Datta:2014zqa}
S.~Datta, R.~Gavai, and S.~Gupta,
\newblock PoS {\bf LATTICE2013}, 202 (2014).

\bibitem{Gale:1990pn}
C.~Gale and J.~I. Kapusta,
\newblock Nucl. Phys. {\bf B357}, 65 (1991).

\bibitem{Aurenche:1999tq}
P.~Aurenche, F.~Gelis, and H.~Zaraket,
\newblock Phys. Rev. {\bf D61}, 116001 (2000), arXiv:hep-ph/9911367.

\bibitem{Mustafa:2002pb}
M.~G. Mustafa and M.~H. Thoma,
\newblock Pramana {\bf 60}, 711 (2003), arXiv:hep-ph/0201060.

\bibitem{Peshier:1999dt}
A.~Peshier and M.~H. Thoma,
\newblock Phys. Rev. Lett. {\bf 84}, 841 (2000), arXiv:hep-ph/9907268.

\bibitem{Wong:1991be}
S.~Wong,
\newblock Z. Phys. {\bf C53}, 465 (1992).

\bibitem{Aurenche:1999ec}
P.~Aurenche, F.~Gelis, R.~Kobes, and H.~Zaraket,
\newblock Phys. Rev. {\bf D60}, 076002 (1999).

\bibitem{Thoma:1997dk}
M.~H. Thoma and C.~T. Traxler,
\newblock Phys. Rev. {\bf D56}, 198 (1997), arXiv:hep-ph/9701354.

\bibitem{Carrington:2007gt}
M.~Carrington, A.~Gynther, and P.~Aurenche,
\newblock Phys. Rev. {\bf D77}, 045035 (2008).

\bibitem{Aurenche:2002pc}
P.~Aurenche, F.~Gelis, and H.~Zaraket,
\newblock JHEP {\bf 0207}, 063 (2002).

\bibitem{Altherr:1992th}
T.~Altherr and P.~Ruuskanen,
\newblock Nucl. Phys. {\bf B380}, 377 (1992).

\bibitem{Cleymans:1992gb}
J.~Cleymans and I.~Dadic,
\newblock Phys. Rev. {\bf D47}, 160 (1993).

\bibitem{Cleymans:1993jm}
J.~Cleymans, I.~Dadic, and J.~Joubert,
\newblock Phys. Rev. {\bf D49}, 230 (1994).

\bibitem{Kapusta:1991qp}
J.~I. Kapusta, P.~Lichard, and D.~Seibert,
\newblock Phys. Rev. {\bf D44}, 2774 (1991).

\bibitem{Baier:1991em}
R.~Baier, H.~Nakkagawa, A.~Niegawa, and K.~Redlich,
\newblock Z. Phys. {\bf C53}, 433 (1992).

\bibitem{Altherr:1989jc}
T.~Altherr and P.~Aurenche,
\newblock Z. Phys. {\bf C45}, 99 (1989).

\bibitem{Kapusta:2000an}
J.~I. Kapusta and S.~Wong,
\newblock Phys. Rev. {\bf C62}, 027901 (2000), arXiv:hep-ph/0003196.

\bibitem{Aurenche:2000xn}
P.~Aurenche {\em et~al.},
\newblock Phys. Rev. {\bf D65}, 038501 (2002), arXiv:hep-ph/0009074.

\bibitem{Levai:1997yx}
P.~Levai and U.~W. Heinz,
\newblock Phys. Rev. {\bf C57}, 1879 (1998).

\bibitem{Schaefer:1998wd}
A.~Schaefer and M.~H. Thoma,
\newblock Phys. Lett. {\bf B451}, 195 (1999).

\bibitem{Mustafa:1999jz}
M.~G. Mustafa, A.~Schaefer, and M.~H. Thoma,
\newblock Phys. Lett. {\bf B472}, 402 (2000), arXiv:hep-ph/9911398.

\bibitem{Mustafa:1999cp}
M.~Mustafa, A.~Schafer, and M.~Thoma,
\newblock Nucl. Phys. {\bf A661}, 653 (1999), arXiv:hep-ph/9906391.

\bibitem{Mustafa:1999dt}
M.~G. Mustafa, A.~Schafer, and M.~H. Thoma,
\newblock Phys. Rev. {\bf C61}, 024902 (2000), arXiv:hep-ph/9908461.

\bibitem{Thoma:1999nm}
M.~H. Thoma, S.~Leupold, and U.~Mosel,
\newblock Eur. Phys. J. {\bf A7}, 219 (2000), arXiv:nucl-th/9905016.


\bibitem{Ghosh:2010wt}
S.~Ghosh, S.~Sarkar, and J.-e. Alam,
\newblock Eur. Phys. J. {\bf C71}, 1760 (2011).

\bibitem{Hosoya:1983id}
A.~Hosoya, M.-a. Sakagami, and M.~Takao,
\newblock Annals Phys. {\bf 154}, 229 (1984).

\bibitem{Hosoya:1983xm}
A.~Hosoya and K.~Kajantie,
\newblock Nucl. Phys. {\bf B250}, 666 (1985).

\bibitem{Svetitsky:1987gq}
B.~Svetitsky,
\newblock Phys. Rev. {\bf D37}, 2484 (1988).

\bibitem{Baym:1990uj}
G.~Baym, H.~Monien, C.~Pethick, and D.~Ravenhall,
\newblock Phys. Rev. Lett. {\bf 64}, 1867 (1990).

\bibitem{Jeon:1994if}
S.~Jeon,
\newblock Phys. Rev. {\bf D52}, 3591 (1995), arXiv:hep-ph/9409250.

\bibitem{Jeon:1995zm}
S.~Jeon and L.~G. Yaffe,
\newblock Phys. Rev. {\bf D53}, 5799 (1996).

\bibitem{Arnold:2002ja}
P.~B. Arnold, G.~D. Moore, and L.~G. Yaffe,
\newblock JHEP {\bf 0206}, 030 (2002).

\bibitem{Moore:2007ib}
G.~D. Moore,
\newblock Phys. Rev. {\bf D76}, 107702 (2007).

\bibitem{CaronHuot:2007gq}
S.~Caron-Huot and G.~D. Moore,
\newblock Phys. Rev. Lett. {\bf 100}, 052301 (2008).

\bibitem{CaronHuot:2008uh}
S.~Caron-Huot and G.~D. Moore,
\newblock JHEP {\bf 0802}, 081 (2008).

\bibitem{CaronHuot:2008ni}
S.~Caron-Huot,
\newblock Phys. Rev. {\bf D79}, 065039 (2009).

\end{thebibliography}
